\documentclass[preprint,10pt]{elsarticle}
\usepackage{amssymb}
\usepackage{physics}
\usepackage{amsmath}
\usepackage{cancel}
\usepackage{physics}
\usepackage{overpic}
\usepackage{hyperref} 
\usepackage{minitoc} 
\usepackage{lipsum}% http://ctan.org/pkg/lipsum
\usepackage{graphicx}
\usepackage{lineno}
\usepackage{natbib}

\usepackage{xcolor}

\usepackage{newpxtext}
\usepackage{textcomp}
\def\EE{{\mathbb{E}}}

\def\bsigma{{\boldsymbol{\sigma}}}

\def\bPi{{\boldsymbol{\Pi}}}

\def\eps{\varepsilon}
\def\bm{{\bf m}}
\def\bk{{\bf k}}
\def\bx{{\bf x}}

\def\br{{\bf r}}

\journal{Physica D}

\begin{document}

\begin{frontmatter}



%\title{Statistical mechanics of nonlinear multimode optical fiber systems}

\title{Wave turbulence, thermalization and multimode locking in optical fibers}

%\title{Wave turbulence, thermalization and multimode locking in optical fibers}


%\author{Tutti} %% Author name

\author[1,2]{M. Ferraro}

\author[3]{K. Baudin}

\author[4,5]{M. Gervaziev}

\author[6]{A. Fusaro}

\author[3]{A. Picozzi}

\author[7]{J. Garnier}

\author[3]{G. Millot}

\author[4,5]{D. Kharenko}

\author[4,5]{E. Podivilov}

\author[4,5]{S. Babin}

\author[8]{F. Mangini}

\author[2]{S. Wabnitz}


\affiliation[1]{organization={Department of Physics, University of Calabria},
                addressline={Via P. Bucci}, 
                city={Rende},
%               citysep={}, % Uncomment if no comma needed between city and postcode
                postcode={87036}, 
                state={(CS)},
                country={Italy}}
\affiliation[2] {organization={Department of Information Engineering, Electronics, and Telecommunications, Sapienza University of Rome},%Department and Organization
            addressline={Via Eudossiana 18}, 
            city={Rome},
            postcode={00184}, 
            country={Italy}}
            
\affiliation[3] {organization={Universit\'e Bourgogne Europe, CNRS, Laboratoire Interdisciplinaire Carnot de Bourgogne ICB, UMR 6303},%Department and Organization
%            addressline={Via Eudossiana 18}, 
            city={21000 Dijon},
   %         postcode={91128}, 
            country={France}}

\affiliation[4] {organization={Department of Physics, Novosibirsk State University},%Department and Organization
            addressline={Pirogova 1}, 
            city={Novosibirsk},
            postcode={630090}, 
            country={Russia}}
\affiliation[5] {organization={Institute of Automation and Electrometry, SB RAS},%Department and Organization
  %          addressline={Via Eudossiana 18}, 
            city={Novosibirsk},
            postcode={630090}, 
            country={Russia}}

\affiliation[6] {organization={Universit\'e Paris-Saclay, CEA, LMCE},
            city={Bruy\`eres-le-Ch\^atel},
            postcode={91680}, 
            country={France}}

\affiliation[7] {organization={CMAP, CNRS, Ecole Polytechnique, Institut Polytechnique de Paris},%Department and Organization
%            addressline={Via Eudossiana 18}, 
            city={Palaiseau},
            postcode={91120}, 
            country={France}}
\affiliation[8] {organization={Department of Engineering, Niccolò Cusano University},%Department and Organization
%            addressline={Via Eudossiana 18}, 
            city={Rome},
            postcode={00166}, 
            country={Itali}}
%% Abstract
\begin{abstract}
We present a comprehensive overview of recent advances in theory and experiments on complex light propagation phenomena in nonlinear multimode fibers.
On the basis of the wave turbulence theory, we derive kinetic equations describing the out-of-equilibrium process of optical thermalization toward the Rayleigh-Jeans (RJ) equilibrium distribution. Our theory is applied to explain the effect of beam self-cleaning (BSC) in graded-index (GRIN) fibers, whereby a speckled beam transforms into a bell-shaped beam at the fiber output as the input peak power grows larger. Although the output beam is typically dominated by the fundamental mode of the fiber, higher-order modes (HOMs) cannot be fully depleted, as described by the turbulence cascades associated to the conserved quantities.
We theoretically explore the role of random refractive index fluctuations along the fiber core, and show how these imperfections may turn out to assist the observation of BSC in a practical experimental setting. This conclusion is supported by the derivation of wave turbulence kinetic equations that account for the presence of a time-dependent disorder (random mode coupling). The kinetic theory reveals that a weak disorder accelerates the rate of RJ thermalization and beam cleaning condensation. On the other hand, although strong disorder is expected to suppress wave condensation, the kinetic equation reveals that an out-of-equilibrium process of condensation and RJ thermalization can occur in a regime where disorder predominates over nonlinearity. In general, the kinetic equations are validated by numerical simulations of the generalized nonlinear Schrodinger equation. We outline a series of recent experiments, which permit to confirm the statistical mechanics approach for describing beam propagation and thermalization. For example, we highlight the demonstration of entropy growth, and point out that there are inherent limits to peak-power scaling in multimode fiber lasers. We conclude by pointing out the experimental observation that BSC is accompanied by an effect of modal phase-locking. From the one hand this explains the observed preservation of the spatial coherence of the beam, but also it points to the need of extending current descriptions in future research.
\end{abstract}



%% Keywords
\begin{keyword}


Nonlinear optics \sep Statistical mechanics \sep Wave turbulence \sep Multimode fibers \sep Beam self-cleaning \sep Wave thermalization \sep Wave condensation
\end{keyword}

\end{frontmatter}


%\bigskip
\section{Introduction and theoretical framework} \label{Sec_1}
Up to now, the vast majority of optical fiber technologies are based on the use of singlemode fibers (SMFs) \cite{agrawal2012fiber}. Historically, multimode fibers (MMFs) have been manufactured and studied well before SMFs: an interesting example is the first supercontinuum generation observation in an MMF, performed back in 1978 by by Lin et al. \cite{lin1978wideband}. The reason for SMFs to outplace MMFs can be easily explained. Owing to their reduced core diameter, SMFs only support one mode, so that the spatial of the guided beam are fixed by the fiber geometry. %and do not vary upon its propagation. 
Whereas an optical beam that is carried by an MMF is typically transported by hundreds or even thousands of spatial modes, whose relative occupancy may vary upon propagation owing to random linear mode coupling. Since each mode propagates with its own phase and group velocity, multimode interference at the output of a MMF will lead to highly unstable speckled intensity patterns with extremely low beam quality, which prevents their use for imaging applications 
%\textcolor{red}{
that involve focused beams.
%}
%In other words, at variance with MMFs, SMFs are a much simpler optical platform, since one does not need to bother about the spatial degrees of freedom. 
Most importantly, for lightwave communications the use of MMFs is limited to very short distances only. This is because modal dispersion largely overcomes chromatic dispersion on the one hand, and also because its post-compensation by digital means (e.g., by multiple-input, multiple-output (MIMO) techniques) is very complex unless just a few modes are present.

Nevertheless, a renewed interest in light propagation in nonlinear MMFs has emerged over the past two decades \cite{cristiani2022roadmap}. This is mostly driven by the exponential growth in optical network traffic, and the potential of MMFs for increasing transmission capacity via spatial division (or mode division) multiplexing. In addition, the possibility offered by MMFs (thanks to their large cross-section, hence increased power threshold for optical damage) of permitting high-energy beam delivery is of utmost interest for scaling up the power generated by fiber lasers, and for multiphoton imaging applications. 

In this respect, studying nonlinear beam propagation in MMF is very relevant \cite{horak_multimode_2012}. Indeed, nonlinearity may be detrimental for both telecommunication applications and for high-power fiber laser sources, and many theoretical and experimental studies have been carried out in order to mitigate, and possibly suppress, the impact of nonlinear effects in MMFs. In contrast, nonlinear effects may be beneficial for the development of supercontinuum light sources, as well as for achieving high-resolution biomedical imaging, e.g., in the field of endoscopy. 

Still, the fundamental issue of how to deal with being able of theoretically predicting the outcome of the complex nonlinear interaction among such a large number of modes, i.e., of degrees of freedom, holds. As a matter of fact, in the presence of third-order nonlinear effects, such as the Kerr effect, Raman and Brillouin scattering, the control of mode interactions can be achieved in ways that have no counterpart in the singlemode fiber case \cite{Wright2015R31,Wright2015R29}. Some examples of robust wave structures which emerge from complex nonlinear mode mixing in MMFs are multimode optical solitons \cite{Hasegawa} and ultra-wideband parametric sidebands \cite{Longhi2003R27}  \cite{Renninger2012R31,KrupaPRLGPI}.

Looking backwards in time, a similar complex problem was faced by the physicists of the 19th century, when addressing the description of the physical properties of a gas in the framework of purely classical mechanics. Indeed, an ideal gas is nothing more than a system of N particles, whose motion solves Newton's equation. As it is well known, a dynamical system with three interacting bodies may be chaotic. Of course, whenever N$\gg$1, solving the problem of motion by brute force numerical integration of a system of N differential equations may become virtually impossible. In a gas of classical particles, N is of the order of the Avogadro's number ($\sim 10^{23}$). With no supercomputer at hand, Boltzmann and his fellow 1800s physicists had to find a clever way to conciliate classical mechanics with thermodynamics, without spending their lifetime to solve a complex system of coupled equations. Such a clever way was the development of statistical mechanics: the macroscopic properties of a mechanical system with many degrees of freedom %(aka macrostate) 
are obtained by averaging over a statistical ensemble of the microscopic properties.% (aka microstate). 

In this way, one does not need to know the exact position and velocity of all particles in a gas, for describing its macroscopic state. Only average (i.e., thermodynamic) quantities matter. These are the temperature, the pressure, the chemical potential, the entropy, and so on. A similar approach can be used for nonlinear multimode photonics.

\subsection{The statistical mechanics theory}

One of main successes of the statistical mechanics theory was to provide a kinetic theory of gases. According to this theory, at thermal equilibrium (which is defined as the state of maximum entropy) %established whenever the entropy reaches its maximum value, 
the velocities of all particles, which are supposed to be distinguishable, follow a Maxwell-Boltzmann probability distribution.
Applying a statistical mechanics approach to nonlinear multimode \textit{classical} waves leads to a different statistics for the mode power distribution. According to the theory, as long as the nonlinearity can be considered \textit{weak} and the number of modes is finite \cite{aschieri2011condensation}, the particles that compose a guided optical beam (i.e., the photons) follow a Rayleigh-Jeans (RJ) law. This can be easily understood by the following argument. Photons are bosons, which are indistinguishable and obey the Bose-Einstein distribution. This boils down to the RJ distribution in the classical limit, i.e., whenever the number of photons per mode is sufficiently high. In this review, we disregard the quantum nature of light, and only consider the classical limit in which the modes of the waveguide are highly populated. As such, we will always refer to the RJ law as the thermal equilibrium distribution.
%\textcolor{red}{Note: at the moment I avoided writing equations in this section as I do not know how to harmonize the notations of all our papers...}

Interestingly, the statistical mechanics approach describing nonlinear beam propagation in MMF has been developed by following two different approaches.
First, in 2011, a weak wave-turbulence theory was introduced by Ascheri et al.. Their work involved the processes of classical wave thermalization and condensation in GRIN MMFs \cite{aschieri2011condensation}, whose experimental verification was only carried out nine years later by Baudin et al. \cite{Baudin2020}.
On the other hand, Wu et al. and Parto et al., proposed in 2019 a thermodynamic theory of highly multimode nonlinear optical systems, developing an equation of state at thermal equilibrium that links together all of the thermodynamic parameters \cite{Wu2019,parto19thermodynamic}. The results of that work, mostly based on probabilistic (i.e., entropic) concepts, were then refined by Makris et al., who developed a theoretical framework for describing the thermalization process as a consequence of photon-photon interactions \cite{makris2020statistical}, in analogy with the textbook statistical mechanics of ideal gases. The statistical mechanics approach has  been subsequently applied and extended to various different forms of multimode optical systems \cite{wright2022physics,efremidis21fundamental,Selim:23}, e.g., to study the thermalization in non-Hermitian optical lattices \cite{Pyrialakos22Thermalization}, or to refine the notion of optical pressure \cite{wu20Thermodynamic,efremidis21fundamental,efremidis22simple,ren23nature,Efremidis:24Statistical,Ren:24dalton}.

As long as equilibrium states are involved, the two approaches provide exactly the same results, i.e., they lead to the same RJ distribution. This is because the two theories rely on the very same pillars. Both theories consider a weakly nonlinear fiber supporting a finite number of guided modes; upon propagation, an optical beam conserves both its power (number of photons) and "momentum", i.e., the system's Hamiltonian (referred to either as energy \cite{aschieri2011condensation,garnier19} or as internal energy \cite{Wu2019,makris2020statistical}). As a matter of fact, in standard cylindrical MMF there is a third conserved quantity, i.e., the orbital angular momentum of light \cite{podivilov2022thermalization,wu2022thermalization}.
However, for the sake of simplicity, in this review we will limit ourselves to consider optical beams that carry no orbital angular momentum. 
The beam energy has two contributions, which can be labeled either linear and nonlinear, or kinetic and potential.
In any case, in the weakly nonlinear regime, i.e., whenever the power of the optical beam is sufficiently low, the potential energy (nonlinearity) can be neglected in comparison with the kinetic energy. Each conserved quantity is associated with a thermodynamic parameter. Specifically, the conservation of the number of photons and of the energy lead to the definition of an optical chemical potential and an optical temperature, respectively. Notice that statistical mechanics has been also extended beyond the weakly nonlinear regime, by computing  elaborate partition functions \cite{Rasmussen00,Jordan00self,Jordan00mean,Rumpf01,Rumpf03,Rumpf04}. Along this way, the statistical properties of incoherent optical fields have been also analyzed by applying methods inherited from complex systems and spin-glass theory \cite{leuzzi09,conti11complexity,Ghofraniha15,Pierangeli17}.

The true potency of the statistical mechanics approach for predicting the evolution of weakly nonlinear multimode waves is its simplicity and effectiveness: by a trivial algebraic calculation, one obtains the (average) spatial profile of a guided beam into an MMF, based on the mere knowledge of its injection conditions. This is the main reason behind the outburst of theoretical and experimental studies on this topic which appeared in recent literature. Without the statistical mechanics approach, determining the beam at the output of a highly multimode fiber would require solving the nonlinear wave equation numerically. With standard numerical tools, this may last even days of computation. The trick behind the statistical mechanics approach is somehow equivalent to ``measuring" the temperature and chemical potential of the beam for determining its properties. Analogously to Boltzmann's approach to classical particles gases, one does not need to know the amplitude and phase of all individual modes in order to know, for instance, its temperature. It has to be mentioned, however, that, at variance with classical thermodynamics, there is no thermometer available here to measure the beam temperature when dealing with optical beams. The optical temperature has a pure statistical meaning, which can even take negative values, owing to the finite energy spectrum of multimode optical systems, as originally predicted in Refs.\cite{Wu2019,parto19thermodynamic} and subsequently experimentally reported \cite{baudin23,marquesmuniz23}. In a loose sense, the optical temperature reflects the amount of randomness of a speckled beam at equilibrium, and it can be computed by the sole knowledge of the beam's conserved quantities (kinetic energy and number of particles), which univocally determine the equilibrium mode power distribution. 

At this point, one may object that the thermodynamic equivalence picture that we have discussed only strictly holds at thermal equilibrium. Therefore, one may naturally wonder whether there is a way, aside from equilibrium statistical mechanics approaches, to provide a detailed nonequilibrium description of the process of thermalization in the weakly nonlinear regime. This is indeed possible by recurring to the help of wave turbulence theory.

%within the framework of statistical mechanics, to describe the evolution of a multimode beam upon propagation toward thermalization. .

\subsubsection{The wave turbulence theory}

The wave turbulence theory occupies a rather special place on the roadmap of modern science, at the interface between applied mathematics, fluid dynamics, statistical physics, and engineering \cite{Zakharov92,Nazarenko11,Newell11,galtier22}. It has potential applications in a diverse range of subjects, including oceanography, plasma physics, and condensed matter physics. Optical wave turbulence also constitutes a growing field of research covering various topics in modern optics, e.g., superfluid turbulence \cite{eloy21,glorieux23turbulent}, rogue wave generation \cite{hammaniPLA10,kiblerPLA11,onorato2013rogue,dudley19rogue}, supercontinuum generation \cite{barviauOE09,barviauPRA09,kiblerPRE11,barviauPRA13,meng21intracavity}, integrable turbulence \cite{congy24statistics,suret24soliton}, light condensation \cite{connaughton05,aschieri2011condensation,sun2012observation,Bortolozzo09,Laurie12,chiocchetta16} and thermalization \cite{pitoisPRL06b,suretPRL10,santic18nonequilibrium}, nonequilibrium transport in multimode optical systems \cite{kottos24,lian24coupled}, or dissipative cavities and lasers \cite{michel201111thermalization,turitsyna2013laminar,Turitsyna09,Turitsyna12,Churkin15} -- see  Refs.\cite{picozzi14,TuritsynOWTbookNazarenko} for reviews. 

Originally, the wave turbulence theory has been the subject of intense investigations in the context of plasma physics \cite{Tsytovich}, in which it is often referred to the so-called ``random phase-approximation" approach \cite{Zakharov92,Tsytovich,Hasselmann1,Hasselmann2,Dyachenko92,Dias04}. This approach may be considered as a convenient way of interpreting the results of the more rigorous technique based on a multi-scale expansion of the cumulants of the nonlinear field, as originally formulated in Refs.~\cite{Benney,NewellWT,BenNew}. This theory has been reviewed in Ref.\cite{Newell01}, and studied in more detail through the analysis of the probability distribution function of the random field in Refs.\cite{Nazarenko11}. In a loose sense, the so-called 'random phase approximation' may be considered as being justified when phase information becomes irrelevant to the wave interaction, due to the strong tendency of the waves to decohere. The random phases can thus be averaged out, to obtain a weak turbulence description of the incoherent wave interaction, which is formally based on irreversible kinetic equations \cite{Zakharov92}. It turns out that, in spite of the formal reversibility of the equation governing wave propagation, the kinetic equation describes an irreversible evolution of the field towards its thermodynamic equilibrium. Such an equilibrium state refers to the Rayleigh-Jeans spectrum, whose tails are characterized by an equipartition of energy among the modes. The mathematical statement of such an irreversible process is the $H$-theorem of entropy growth, whose origin is analogous to the Boltzmann's $H$-theorem relevant for gas kinetics \cite{Huang}.

Referring back to light propagation in GRIN MMFs, the wave turbulence theory is capable of describing the mode distribution associated with the so-called beam self-cleaning (BSC) effect, a striking phenomenon that has been attracting a great deal of interest over the last few years.

\subsection{The beam self-cleaning effect}

\begin{figure}[h]
    \center
    \includegraphics[width=0.99\textwidth]{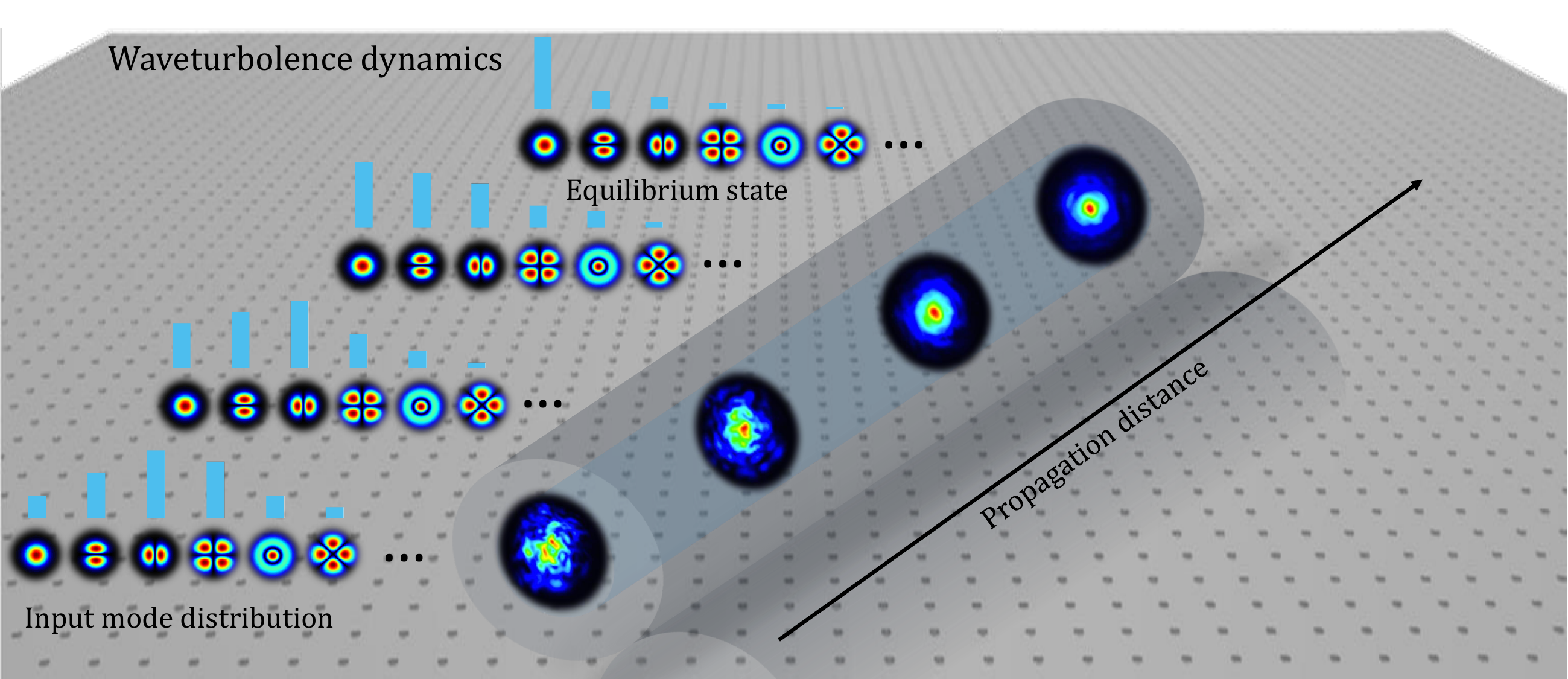}
    \caption{Toy figure of BSC as a result of classical wave thermalization/condensation.
    }
    \label{fig:toy}
\end{figure}

First reported by Krupa et al. in 2016~\cite{Krupa:16,Krupanatphotonics} for nanosecond pulses and later confirmed for femtosecond pulses by Lui et al.~\cite{LiuKerr}, the BSC effect in GRIN MMFs refers to the phenomenon whereby an input beam evolves, during its propagation through complex and spatially disordered states, to form a stable, well-defined output beam with a typical near-Gaussian profile. Such an effect can be numerically reproduced by using different approaches: e.g., the full-field model~\cite{Krupanatphotonics}, as well as the generalized multimode nonlinear Schr\"odinger equations (GMM-NLSE)~\cite{WrightNP2016}. Within the thermodynamic framework, the occurrence of BSC is seen as the establishment of a state of thermal equilibrium \cite{Ferraro2023apx}, as depicted in Fig. \ref{fig:toy}. A thermalized output beam, usually referred to as ``clean", turns out to be remarkably stable versus external perturbations. This has paved the way for the possible development of a novel class of MMF-based applications. 
As previouly discussed, most of current fiber technologies are based on SMFs. Since SMFs only support one mode, they inherently preserve the beam's spatial quality upon its propagation, even under the action of external perturbations, such as fiber banding and squeezing. 
When it comes to MMFs, the multitude of fiber modes that can interfere with each other do not ensure the preservation of the beam spatial profile, even in the absence of any onlinear effects. Moreover, the output beam may significantly vary if external perturbations are applied, e.g. temperature fluctuations or mechanical stress.

Indeed, preserving the spatial quality of the beam is a key feature when transporting it via optical fibers. For instance, who would buy a laser whose output beam is speckled and variable in time because of weak environmental variations? As a result of this rhetorical question, until a few years back most of the research on optical fibers was focused on SMFs. 
It is in this framework that the discovery of BSC was groundbreaking. Guaranteeing a high beam quality together with high-energy pulse delivery allows for scaling up the power of SMF lasers, as demonstrated by Te\u{g}in et al. \cite{teugin2020single} and for enhancing the resolution of multiphoton imaging devices as shown by Moussa et al. \cite{moussa2021spatiotemporal,wehbi2022continuous}.

%\bigskip
\subsection{The goal and the content of this review}

In essence, we aim to offer a comprehensive overview of nonlinear multimode fiber optics within a statistical mechanics framework. 
In particular, we focus on the wave turbulence approach, since this allows for capturing the whole process of thermalization, i.e., the evolution of a multimode laser beam through intermediate non-equilibrium states, eventually reaching RJ equilibrium. 
The wave turbulence theory specifically allows for the characterization of beam self-cleaning by providing a characteristic propagation length scale needed to achieve thermal equilibrium. This is important, in particular, when one considers the key role played in the thermalization process by random fluctuations of the MMFs' refractive index profile due to fabrication imperfections. 

We give particular emphasis to the case of BSC in GRIN MMFs. From the fundamental point of view, since its first observations it was clear that observing BSC requires the combined action of the Kerr and the spatial self-imaging (SI) effects \cite{Hansson2020}. SI arises from the unique parabolic refractive index profile of GRIN MMFs (BSC has not yet been clearly observed to occur in step-index fibers). These cause the modes to mix and redistribute their energy, eventually favoring an energy transfer towards the fundamental mode. The outcome of BSC is that the fundamental mode soaks up as much power as possible from the high-order modes (HOMs). However, according to the statistical mechanics theory, HOMs cannot be completely emptied, because this would result in a breach of the conservation laws of light power and
%momentum
kinetic energy. The fact that a BSC beam is not composed by the sole fundamental mode was experimentally verified by Leventoux in 2020 \cite{leventoux20203d}. This excluded possible explanations of BSC as a result of mode-dependent losses, thus strengthening the validity of the statistical mechanics picture, which, in contrast, is conservative. Indeed, the conservation laws drive the nonlinear beam dynamics in a way that thermal equilibrium is established, as a result of optical entropy maximization.

%In this work, we review at first the numerical and experimental observation that BSC can be described in analogy to a 2D hydrodynamic turbulence process. Two different mechanisms drive the mode power redistribution associated with BSC. In particular, the energy flows towards the fundamental mode (known in hydrodynamics as inverse cascade). At the same time, a second flow of energy (direct cascade) populates the HOMs \cite{PhysRevLett.122.103902}. This is the topic of Sec.~\ref{Sec_2}. \textcolor{red}{Note: actually, Sec. 2 is much broader than I see coming. We should see how to harmonize it with the rest of the manuscript.}

%Then, 
In this work, we largely focus on the role played by random disorder, which has been shown to facilitate the BSC process \cite{sidelnikov19}. We frame the role played by random disorder within wave turbulence theory. 
As a matter of fact, the wave turbulence theory has been essentially developed by ignoring the impact of disordered fluctuations of the propagating nonlinear medium \cite{Zakharov92,Newell11,Nazarenko11}.
However, the evolution of a turbulent wave system can be heavily affected by environmental disorder. This plays a fundamental role in ocean waves, where the scattering by disorder can originate from random variations of the bottom topography in oceans \cite{degueldreNP16}. In fiber optics, disorder originates from random fluctuations of the refractive index of the core material. In particular, disorder is known to impact light propagation in MMFs, due to imperfections and environmental perturbations, a feature that is relevant for instance to endoscopic imaging \cite{carmazzaNC19}.
So far, the impact of disorder on turbulent flows has been essentially studied by means of numerical simulations, while only few theoretical developments have considered purely time-independent disordered fluctuations \cite{cherroret15,cherroret21,wang20,nazarenko19,kottos23,Pyrialakos24}. 
On the other hand, light propagation in MMFs is characterized by random refractive index fluctuations that vary along the propagation, and are thus `time'-dependent. 
In Sec.~\ref{Sec_3} we review recent works \cite{Fusaro:PRL:2019,garnier19,berti22}, in which the wave turbulence theory has been extended to account for the presence of a time-dependent disorder, which is inherent to light propagation in MMFs. 
%}

In Sec.~\ref{Sec_4}, we present a selection of experimental investigations aimed at confirming the predictions of the statistical mechanics approach for nonlinear beam propagation in GRIN MMFs. We first consider the notable case of BSC. 
The validity of the statistical mechanics approach to BSC was first proved by the experiments of Baudin et al. in 2020 \cite{Baudin2020}. Then, thanks to the development of a field-projection mode decomposition (MD) based of self-referenced interferometric technique~\cite{Pariente2016np} and holographic MD~\cite{Gervaziev2021lpl} tools, a direct measurement of the RJ mode distribution was reported by Pourbeyram et al. \cite{pourbeyram2022direct} and, later, by Mangini et al. \cite{Mangini2022Statistical} in 2022. For a review of these early works, we refer the reader to Ref.\cite{Ferraro2023apx}. 
Here, instead, we focus on more recent advances. In particular, we highlight experimental evidences of the entropy growth during the thermalization of either one \cite{mangini2024maximization} or two waves \cite{ferraro2024calorimetry}. Next we discuss recent intriguing experiments: they demonstrate that, by increasing the temperature of a self-cleaned (i.e., at equilibrium) beam, one observe a progressive loss its beam quality. This result establishes a practical limit to the power up-scaling of MMF lasers.
Then, we focus on an aspect which has been largely debated within the optical statistical mechanics community, i.e., the framing of the statistical ensemble in experiments. 
Indeed, one of the pillars of the statistical mechanics theory is the averaging of physical observables over a meaningful statistical ensemble. As aforementioned, in the case of optical thermodynamics or wave turbulence, one generally averages over the mode phases. However, virtually all the BSC experimental observations did not involve a properly said phase average. In this sense, the only exception reported in the literature is Ref. \cite{Baudin2020}, where an average over an ensemble of injected speckle beams was considered. At last, we present the recent observation of a modal phase-locking mechanism that accompanies BSC \cite{mangini2023modal}.

Finally, in Sec.~\ref{Sec_5}, we provide our view on the BSC effect in light of recent results that have questioned its interpretation as a pure thermalization process. Then we conclude by tracing what will be, in our opinion, the foreseeable progress of research in the domain of the statistical mechanics of nonlinear MMF optics.

%\section{Basic principles and conservation laws}
%\label{Sec_2}
%\input{Sec2_Conservation}

\section{Wave turbulence in multimode fibers with disorder}
\label{Sec_3}
\subsection{Basic principles and conservation laws}
\label{subsec:basic}

Before introducing the technical formalism inherent to the wave turbulence theory, we aim to provide some simple physical insight into light thermalization in MMFs, grounded on basic principles and conservation laws. When considering the purely 2D spatial evolution of the optical field in a MMF, we may expand the monochromatic field amplitude $\psi(z,\br)=\sum_{\ell,m}a_{\ell,m}(z) u_{\ell,m}(\br) \exp(i \beta_{\ell,m} z)$ on the orthonormal basis of the eigenmodes $u_{\ell,m}(\br)=R_{\ell,m}(r) \exp(i m \phi)$ of the waveguide, where $R_{\ell,m}(r)$ are the radial components of the eigenmodes, with $z$ the spatial coordinate along the fiber axis and $\br=(r,\phi)$ the transverse coordinate. Here, $\beta_{\ell,m}$ denote the propagation constants [with $\beta_{\ell,m}= \beta_{0} (2 \ell + |m|)$ for the case of the parabolic graded-index fiber], while $|a_{\ell,m}|^2$ denote the power occupancy of the optical field, $(\ell,m)$ being the radial and azimuthal numbers (with $\ell=0, 1, 2,...$ and $m = 0, \pm 1, \pm 2,...$).

The optical field is characterized by the conservation of different quantities during its propagation through the fiber. The optical power, $N=\sum_{\ell,m} |a_{\ell,m}|^2$  (or ‘wave-action’ in the wave turbulence terminology, or particle number in a corpuscular picture) is a conserved quantity, as well as the orbital angular momentum, $\mathcal{M}=\sum_{\ell,m} m |a_{\ell,m}|^2$. Furthermore,  experiments demonstrating RJ thermalization in MMFs have thus far been conducted in a weakly nonlinear regime. Accordingly, the (kinetic) energy $E=\sum_{\ell,m} \beta_{\ell,m}|a_{\ell,m}|^2$, that is the linear contribution to the Hamiltonian, is also conserved through the propagation.

All conserved quantities ($N,E,\mathcal{M}$) are completely determined by the injection conditions of the laser beam into the MMF \cite{Podivilov2022prl}. During the propagation in the MMF, Kerr nonlinearity leads to energy exchanges among the fiber modes, similarly to particle collisions in a classical gas, thus shuffling the occupation of the fiber modes, $a_{\ell,m}(z)$. One could expect that the optical speckled beam reaches thermodynamic equilibrium over a finite "time" (i.e., a finite propagation length $z$), owing to the well-known Onsager's principle of detailed balance~\cite{doi:10.1098/rspl.1866.0039}. When applied to mode interactions, this principle implies that at thermal equilibrium (the thermalized state of an optical beam), every elementary four-wave mixing process that transfers energy among the modes is equally probable as its reverse process. At equilibrium, the average power occupation in a specific mode, $n_{\ell,m}^{eq}=\left< |a_{\ell,m}|^2\right>$, follows a RJ distribution. The detailed expression of the equilibrium distribution is obtained by maximizing the optical entropy, whose appropriate expression, $S=\sum_{\ell,m}  \log(n_{\ell,m})$, can be derived from a statistical mechanics approach \cite{Wu2019,makris2020statistical}, or alternatively from a $H-$theorem of entropy growth \cite{aschieri2011condensation}, as discussed below in the framework of the wave turbulence theory, see section~\ref{subsec:weak_disorder}. Maximizing the entropy $S[n_{\ell,m}]$ subject to the constraints of the conservation of the power $N=\sum_{\ell,m} n_{\ell,m}$, the energy $E=\sum_{\ell,m} \beta_{\ell,m} n_{\ell,m}$, and the angular momentum $\mathcal{M}=\sum_{\ell,m} m n_{\ell,m}$, readily gives~\cite{Podivilov2022prl,wu2022thermalization}:
\begin{equation}
n_{\ell,m}^{eq} = \frac{1}{a + b \, \beta_{\ell,m} + c \, m},
\label{eq:genRJ_mom}
\end{equation}
where $a,b$ and $c$ are three Lagrange’s multipliers associated to the three conserved quantities, $N,E$ and $\mathcal{M}$.

The conservation of angular momentum provided unforeseen insights into the physical mechanisms underlying spatial beam cleaning. For instance, an inspection of Eq.(\ref{eq:genRJ_mom}) reveals that the fundamental mode of the fiber is not necessarily the most populated in the presence of a non-vanishing angular momentum \cite{wu2022thermalization}, a remarkable feature that has been demonstrated experimentally in Ref.\cite{Podivilov2022prl}.

It should be noted that launching a beam carrying a non-zero angular momentum in a MMF requires specific injection conditions. In this section, we consider the general framework of the wave turbulence approach, and typically assume  initial conditions characterized by a random phase among the modes. More precisely, considering an average over a statistical ensemble of speckled beams with random phases, the average angular momentum vanishes at the fiber input, $\mathcal{M}=\sum_{\ell,m} m \left< |a_{\ell,m}|^2\right>=0$, which implies $c=0$ in the RJ equilibrium distribution (\ref{eq:genRJ_mom}). In this way, one usually poses $a=-\mu/T$ and $b=1/T$ following the standard procedure, where $\mu$ and $T$ are the chemical potential and the temperature in analogy to thermodynamics. The equilibrium RJ  distribution then recovers the standard form:
\begin{equation}
n_{\ell,m}^{eq} = \frac{T}{\beta_{\ell,m} -\mu}.
\label{eq:genRJ}
\end{equation}
The RJ distribution determines the equilibrium properties of the optical field. However, the equilibrium  approach discussed here provides no information about the characteristic "time" scales (i.e., propagation lengths)  needed to reach such an equilibrium state. In particular, random mode coupling can lead to power and energy exchanges among the modes, which strongly affect the rate of thermalization to equilibrium, as described by the wave turbulence kinetic theory.

\subsection{Impact of random mode coupling}

As discussed in the introduction section, the wave turbulence theory has been developed by ignoring the presence of a structural disorder in the system \cite{Zakharov92,Newell01,Newell11,Dias04,Nazarenko11,picozzi14,TuritsynOWTbookNazarenko}. So far, the impact of disorder on turbulent flows has been essentially studied by means of numerical simulations, while only few theoretical developments have considered purely time-independent (i.e., $z-$independent) disordered fluctuations \cite{cherroret15,cherroret21,wang20,nazarenko19,kottos23,Pyrialakos24}.
On the other hand, light propagation in MMFs is known to be affected by random fluctuations of the longitudinal and transverse profiles of the index of refraction, as a consequence of external factors such as bending, twisting, tensions, kinks, or core-size variations in the fabrication process of the fiber.
Such multiple physical perturbations introduce random polarization fluctuations as well as random coupling among the modes of the fiber. The specific mechanisms and models that describe how fiber imperfections impact light propagation in MMFs still remain an active topic of research \cite{Mecozzi:12,MecozziOpEx1,MumtazJLT12,Xiao:14,ho14,kaminow13}.

One usually assumes that for relative short propagation lengths (few dozen meters), the contribution of noise essentially arises from 'weak' disorder, which originates from polarization random coupling and random coupling among degenerate fiber modes \cite{ho14,agrawal_physics_2023}. The presence of polarization random coupling does not prevent the conservation of the three conserved quantities $(N,E,{\mathcal M})$, so that the expected RJ equilibrium has the form given in Eq.(\ref{eq:genRJ_mom}), while random coupling among degenerate modes modifies the angular momentum, so that the expected RJ distribution has the form (\ref{eq:genRJ}). On the other hand, the contribution of 'strong' disorder arises for larger propagation lengths (few hundred meters) from random coupling among non-degenerate modes \cite{ho14,agrawal_physics_2023}. The presence of strong disorder prevents the conservation of energy and angular momentum, so that the expected RJ equilibrium should reduce to an equipartition of the power among the modes, i.e., $b=c=0$ in Eq.(\ref{eq:genRJ_mom}), resulting in a parabolic intensity profile in graded-index fibers, and a rectangular intensity profile in step-index fibers~\cite{Kuznetsov2021sr,Kharenko2022ol}. In the following sections, we derive the kinetic equation for the case of weak disorder by considering polarization random fluctuations in Sec.~\ref{subsec:weak_disorder}, while the case of strong disorder will be addressed in Sec.~\ref{subsec:strong_disorder}. Unexpectedly, the kinetic theory reveals that thermalization to the RJ equilibrium (\ref{eq:genRJ}) may even occur in the presence of strong disorder, see Sec.~\ref{subsec:strong_disorder}.

\begin{figure}
\begin{center}
%\bigskip
\includegraphics[width=0.75\columnwidth]{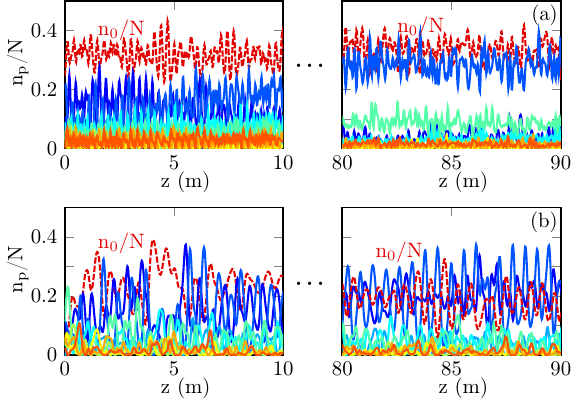}
\caption{
%{\bf preliminary figure}
Coherent regime of mode interaction without disorder:
Numerical simulations of the NLS Eq.(\ref{eq:nls_Ap}) showing the evolution of the modal components $n_p(z)=|{\bf a}_p(z)|^2$ (in the absence of disorder ${\bf D}_p=0$), starting from a coherent initial condition (a), starting from a 'speckle' beam with random phase among the modes (b). 
Evolutions of the modal populations $n_p/N$: fundamental mode $p=0$ (red dashed), $p=1$ (dark blue solid), $p=2$ (blue solid), $p=3$ (light blue solid), $p=4$ (cyan solid), $p=5$ (light green solid), $p=6$ (green solid), $p=7$ (yellow solid), $p=8$ (orange solid).
%light green p=5, green p=6, yellow  p=7, orange p=8, dark orange p=9
The system exhibits a persistent oscillatory dynamics among the low-order modes.
%, which reflects the discrete nature of the resonance manifold underlying wave turbulence in MMFs. 
%starting from a speckle  with random phases among the modes (b).
At complete thermal equilibrium the system would reach the condensate fraction $n_0^{eq}/N \simeq 0.68$ in (a), and $n_0^{eq}/N \simeq 0.58$ in (b).
The power is $N=47.5$ kW, $M=120$ modes, $R=26$ $\mu$m.
{\it Source:} From Ref.\cite{garnier19}.
}
\label{fig:sec3_no_dis}
\end{center}
\end{figure}

The impact of weak disorder in graded-index MMFs is also known to play an important role in the process of thermalization \cite{sidelnikov19,Fusaro:PRL:2019,garnier19}. In particular, it was shown in Ref.\cite{garnier19} that in the absence of disorder, a strong correlation among the modes is preserved for large propagation lengths, which leads to a phase-sensitive "coherent regime" of modal interaction. In this coherent regime, the modes experience a quasi-reversible exchange of power with each other, which tends to freeze the process of RJ thermalization, as illustrated in Fig.~\ref{fig:sec3_no_dis}. Note that such an oscillatory behavior of the modal components, and the slowing down of RJ thermalization, are related to the existence of Fermi-Pasta-Ulam recurrences, as recently discussed in Ref.\cite{biasi21,biasi23} in the framework of the weakly nonlinear regime of the 2D NLS equation with a parabolic trapping potential. We will see that the source of noise introduced by weak disorder (polarization random fluctuations) is sufficient to break the coherent regime of modal interaction. The resulting random phase dynamics then leads to a turbulent incoherent regime of the modal components, which can be described in the framework of the wave turbulence theory. 
\\

\subsection{Kinetic equation with weak disorder}
\label{subsec:weak_disorder}

Let us consider the case of weak disorder in the framework of polarization random fluctuations, and discuss its impact on the wave turbulence kinetic equation in MMFs.
We first treat the conventional "continuous" wave turbulence regime that may be relevant for very highly multimode fibers, and then the "discrete" wave turbulence regime, relevant to typical experiments of beam cleaning. On the basis of recent developments on finite size effects in wave turbulence \cite{Zakharov05,kartashova08,kartashova09,Lvov10,mordant18}, a detailed derivation of the discrete wave turbulence kinetic equation accounting for weak disorder is provided in Section~\ref{sec3:disc_wt}. 
We first consider a model of weak disorder, in which the modes experience an independent (decorrelated) noise. 
We note that, in the opposite limit where the modes experience the same disorder, the analysis reveals that the optical beam recovers a quasi-coherent regime of mode interactions, similar to that obtained in the absence of disorder.
In other words, the fully mode-correlated model of disorder appears ineffective, and in this limit the optical beam does not exhibit a fast process of condensation.
In this way, an intermediate model of partially correlated disorder is considered, where only degenerate modes experience the same noise.
The theory reveals that this partially mode-correlated disorder is sufficient to re-establish an efficient acceleration of condensation.

%\textcolor{blue}{Before delving into the details, we would like to highlight  that there are important differences that distinguish the regimes of propagation  presented in this section and  those covered  in the previous section~\ref{Sec_2}.  Accordingly, to avoid confusion, we have structured this section to be self-contained.}

\subsubsection{Model equation}
\label{sec3:model}

Let us consider the weak disorder regime, taking into account the main contribution of polarization random fluctuations \cite{MumtazJLT12,Xiao:14,agrawal_physics_2023}. 
Accordingly, we resort to a vector model of light propagation in a GRIN MMF that includes  polarization mode dynamics.
In the rectilinear polarization basis, the evolutions of the modal components ${\bf a}_p(z)=(a_{p,x},a_{p,y})^T$  are governed by 
\begin{eqnarray}
i \partial_z {\bf a}_p = \beta_p {\bf a}_p + {\bf D}_p(z) {\bf a}_p 
- \gamma  {\bf P}_p({\bf a}),
\label{eq:nls_Ap}
\end{eqnarray}
with the nonlinear term 
\begin{eqnarray}
{\bf P}_p ({\bf a})  =
% \sum_{l,m,n=0}^{M-1} 
\sum_{l,m,n} 
S_{plmn} \Big(\frac{1}{3}
{\bf a}_l^T {\bf a}_m {\bf a}_n^* +\frac{2}{3} {\bf a}_n^\dag  {\bf a}_m {\bf a}_l\Big) ,
\label{def:nlterm0}
\end{eqnarray}
where $\beta_p$ are the eigenvalues verifying 
$\beta_p u_p(\br)=-\alpha \nabla^2 u_p(\br) + V(\br)u_p(\br)$, with $\alpha=1/(2n_{\rm co}k_0)$, 
and $k_0=2\pi/\lambda$, $\lambda$ being the laser wavelength.
We assume that the refractive index potential is parabolic-shaped, $V({\bf r})=q |{\bf r}|^2$ for $|{\bf r}| \le R$, $R$ being the radius of fiber core, and $q=k_0 (n_{\rm co}^2-n_{\rm cl}^2)/(2n_{\rm co} R^2)$ with $n_{\rm co}$ and $n_{\rm cl}$ the refractive indexes of the core and the cladding, respectively.
Accordingly, the eigenvalues of the parabolic potential read $\beta_p=\beta_0(p_x+p_y+1)$, where $p$ labels the two integers $(p_x, p_y)$ that specify a mode in the Hermite-Gauss basis, and $\beta_0=2\sqrt{\alpha q}$.
In Eq.(\ref{def:nlterm0}), $S_{lmnp} = \int u_l(\br) u_m(\br) u_n^*(\br) u_p^*(\br) d\br/\int |u_0|^4(\br) d\br$, so that $S_{0000}=1$, and $\gamma = k_0 n_2/A_{eff}^0$, where $A_{eff}^0=1/\int |u_0|^4(\br) d\br$ is the effective area of the fundamental mode,  and $n_2$ the nonlinear Kerr coefficient. 
%(m$^2$/W).
% ($|{\bf a}_p|^2$ in Watts).
The sum in (\ref{def:nlterm0}) is carried over the mode indices $l=(l_x,l_y)$, $m=(m_x,m_y)$, $n=(n_x,n_y)$ of the $M$ propagative (or guided) modes.
The superscripts $*,T,\dag$ stand for the conjugate, the transpose, and the conjugate transpose operations.
%\textcolor{blue}{Before pursuing, we note that Eq.(\ref{eq:nls_Ap}) for the evolution of the modal components has a form similar to Eq.(\ref{Enumeric}) considered in the previous section~\ref{Sec_2}. As commented above, there are important differences that distinguish the regimes considered here and those  discussed in section~\ref{Sec_2}. We note in particular the vector nature of Eq.(\ref{eq:nls_Ap})  that accounts for polarization random fluctuations. Also note that Eq.(\ref{eq:nls_Ap}) is written in the Hermite-Gauss basis, whereas Eq.(\ref{Enumeric})  is written in the Laguerre-Gauss basis. Furthermore, the eigenvalues and eigenmodes considered here in Eq.(\ref{eq:nls_Ap}) refer to the NLS approximation of the more general Helmholtz-like equation considered in section~\ref{Sec_2}. At the end of this section, we will discuss the impact of the correction due to angular dispersion effects inherent to the Helmholtz form of the eigenvalues in the wave turbulence kinetic equation.}

The characteristic nonlinear length is $L_{nl} \sim 1/(\gamma N)$,
% corresponds to the nonlinear propagation length when all the power is confined in the fundamental mode. 
%Usually the optical beam populates many higher-order modes and the nonlinear length is $L_{nl}=1/(\gamma_0 P/S_{eff})$ where $S_{eff}$ is the effective surface section of the beam.
while the linear characteristic length from Eq.(\ref{eq:nls_Ap}) is $L_{lin}=1/\beta_0$.
%$L_{lin}=1/\delta \beta_{eff}$, where $\delta \beta_{eff}$ denotes the effective ('spectral') bandwidth of the optical beam in the mode space.
In usual experiments of beam self-cleaning in GRIN fibers we can have $\beta_0 \sim 5\times 10^{3}$ m$^{-1}$, so that the optical beam evolves in the weakly nonlinear regime
\begin{eqnarray}
L_{lin} \ll L_{nl}, 
\label{eq:weakly_nl_reg}
\end{eqnarray}
where $L_{lin} \lesssim 1$ mm, while $L_{nl}$ is typically  larger than 10 cm. 
As will be discussed below in more details, the weakly nonlinear regime of the experiments allows us to derive a kinetic equation in the framework of the weak turbulence theory.

Let us recall that longitudinal and transverse fluctuations in the refractive index profile of the MMF lead to a random coupling among the modes, as expressed by the coupled-mode theory \cite{kaminow13,MumtazJLT12}. 
However, by following this procedure one obtains a scalar perturbation that does not introduce coupling among the polarization components, a feature commented in particular in Ref.\cite{Xiao:14}. 
For this reason, random polarization fluctuations are usually introduced in a phenomenological way, see e.g.\cite{MumtazJLT12,Xiao:14,agrawal_physics_2023}. 
Here, we consider the most general form of polarization disorder that conserves the power $N=\sum_p |{\bf a}_p|^2$, and the (kinetic) energy $E=\sum_p \beta_p |{\bf a}_p|^2$, of the optical beam.
For this purpose, the Hermitian matrices ${\bf D}_p(z)$ are expanded into the Pauli matrices, which are known to form a basis for the vector space of 2$\times$2 Hermitian matrices.
The matrices then have the form
\begin{eqnarray}
\label{model:Dp0a}
{\bf D}_p(z) = \sum_{j=0}^3 \nu_{p,j} (z) \bsigma_j  ,
\end{eqnarray}
where $\bsigma_j$ ($j=1,2,3$) are the Pauli matrices and $\bsigma_0$ is the identity matrix.
%$$
%\bsigma_0=\begin{pmatrix} 1 & 0 \\ 0 & 1 \end{pmatrix} ,\quad
%\bsigma_1=\begin{pmatrix} 0 & 1 \\ 1 & 0 \end{pmatrix} ,\quad
%\bsigma_2=\begin{pmatrix} 0 & -i \\ i & 0 \end{pmatrix} ,\quad
%\bsigma_3=\begin{pmatrix} 1 & 0 \\ 0 & -1 \end{pmatrix} ,
%$$
The functions $\nu_{p,j}(z)$ are independent and identically distributed Gaussian real-valued  random processes, with 
\begin{eqnarray}
\left< \nu_{p,j}(z) \nu_{p',j'}(z') \right>
=  \sigma^2_\beta \delta^K_{pp'} \delta^K_{jj'} {\cal R}\Big( \frac{z-z'}{l_\beta}\Big) .
\label{eq:corr_nu}
\end{eqnarray}
Here $l_\beta$ is the correlation length of the random process and $\sigma^2_\beta$ is its variance.
The normalized correlation function ${\cal R}$ is such that ${\cal R}(0)=O(1)$,  $\int_{-\infty}^\infty {\cal R}(\zeta)d\zeta=O(1)$, and ${\cal R}( z/l_\beta) \to l_\beta \delta(z)$ in the limit $l_\beta \to 0$.
We will consider Ornstein-Uhlenbeck processes:
$$
d \nu_{p,j} = - \frac{1}{l_\beta} \nu_{p,j} dz + \frac{\sigma_\beta}{\sqrt{l_\beta}} dW_{p,j}(z)
$$
where $W_{p,j}$ are independent Brownian motions. This means that $\nu_{p,j}$ are  Gaussian processes with mean zero and covariance 
function of the form (\ref{eq:corr_nu}) with ${\cal R}(\zeta) = \exp(-|\zeta|) /2$. 
We introduce an effective parameter of disorder 
$\Delta \beta=\sigma^2_\beta l_\beta$ and the associated length scale 
\begin{eqnarray}
L_d=1/\Delta \beta.
\label{eq:L_d}
\end{eqnarray}
We assume that disorder effects dominate nonlinear effects 
\begin{eqnarray}
L_d \ll L_{nl},
\label{eq:Ld_small_Lnl}
\end{eqnarray}
and that $l_\beta \ll L_d$ (or equivalently $\sigma_\beta l_\beta \ll 1$).
%which also imposes that $l_\beta \ll L_{nl},z$.
Let us recall that usual experiments of beam self-cleaning are carried out in the weakly nonlinear regime [see Eq.(\ref{eq:weakly_nl_reg})], so that $\beta_0^{-1} \ll L_d$ (or $\Delta \beta \ll \beta_0$).
Aside from beam-cleaning experiments, in the following we will also consider a regime where the mode-spacing can be very small $\beta_0 \ll \Delta \beta$ (MMFs with huge number of modes $M \gg 1$), in order to address a regime described by a continuous wave turbulence approach. %description with $\Delta \beta \lesssim \beta_0$.}
%A kinetic equation without this latter assumption can also be derived, however its form would be more complicated.
Finally, note that since the disorder is ('time') $z-$dependent, the system is of a different nature from the case where the interplay of thermalization and Anderson localization is studied \cite{cherroret15,cherroret21,wang20,nazarenko19,kottos24,Pyrialakos24}.

\subsubsection{Vanishing correlations among modes}
\label{sec3:wd_vanish_corr}

First of all, we study the correlations among the modes through the analysis of the evolution of the second-order moments in the 2$\times$2 matrix $\left< {\bf a}_p^* {\bf a}_q^T\right>(z)$. 
The impact of disorder is treated by making use of the Furutsu-Novikov theorem.
This reveals that conservative disorder introduces an effective dissipation in the system.
In this respect, we recall that, in principle, the laser beam excites strongly correlated modes at the fiber input.
% as discussed above through the coherent modal regime of interaction.
It is the effective dissipation due to disorder that breaks such a strong modal phase-correlation.

%The analysis developed in Appendix~A (section~\ref{appA1}) completes that reported in the Supplemental of Ref.\cite{PRL19}, which was focused on correlations between different modes, $\left< {\bf a}_p^* {\bf a}_q^T\right>$ for $p \neq q$.
%The correlations within a single mode (i.e., among the orthogonal polarization components for $p=q$) is more delicate and it is reported in detail in Appendix~A (section~\ref{appA1}).
The theory developed in the \ref{app:weak_disorder_1} reveals that in the regime discussed in Sec.~\ref{sec3:model}, the correlations $\left< {\bf a}_p^* {\bf a}_q^T\right>(z)$ among different modes ($p \neq q$), or within a single mode ($p=q$), both have the form 
%\begin{eqnarray}
$$
\left< {\bf a}_p^* {\bf a}_q^T\right>(z) \sim \frac{ \gamma \Big({\bf P}_p({\bf a}(z))^* {\bf a}_q^T(z)
- {\bf a}_p^*(z) {\bf P}_q({\bf a}(z))^T \Big) }{4 \Delta \beta -i(\beta_p-\beta_q)},
$$
%\label{eq:correl}
%\end{eqnarray}
%where 
%$$
%{\bf G}_{pq}({\bf a}(z))=. 
%$$
Recalling that $L_d \ll L_{nl}$ we see that, at leading order, modal correlations vanish for propagation lengths larger than the nonlinear length, $\left< {\bf a}_p^* {\bf a}_q^T\right>(z) \simeq 0$ for $z \gg L_{nl}$.

\subsubsection{Closure of the moments equations}

In the following, we derive the kinetic equation by following the wave turbulence perturbation expansion procedure, where linear dispersive effects dominate nonlinear effects $L_{lin} \ll L_{nl}$. 
Accordingly, an effective large separation of the linear and the nonlinear lengths scales takes place \cite{Zakharov92,Newell01,Newell11,Dias04,Nazarenko11}.
When combined with disorder effects, the modes exhibit random phases with quasi-Gaussian statistics, which allows one to achieve the closure of the infinite hierarchy of the moment equations.
%Note that the statistics does not need to be Gaussian initially. 
%Because linear effects of dispersion and disorder dominate nonlinear effects, it is the linear behavior which brings the system close to Gaussianity.
More precisely, because of the nonlinear character of the NLS equation, the evolution of the second-order moment of the field depends on the fourth-order moment, while the equation for the fourth-order moment depends on the sixth-order moment, and so on. In this way, one obtains an infinite hierarchy of moment equations, in which the $n-$th order moment depends on the $(n+2)-$th order moment of the field. This makes the equations impossible to solve, unless some method can be found to truncate the hierarchy. The closure of the infinite hierarchy of moment equations can be obtained in the weakly nonlinear regime by virtue of the Gaussian moment theorem. We remark in this respect that the key assumption underlying the wave turbulence approach is the existence of a random phase among the modes, rather than a genuine Gaussian statistics, as recently discussed in detail in Refs.\cite{Nazarenko11,Chibbaro17,Chibbaro18}.
We emphasize that in the present work the random phase of the modes is induced by the structural disorder of the medium, which dominates nonlinear effects ($L_d \ll L_{nl}$).

As discussed here above, the nondiagonal components of the 2$\times$2 matrix $\left< {\bf a}_p^* {\bf a}_p^T\right>(z)$ vanish. 
Accordingly, in the following we derive an equation governing the evolutions of the diagonal components $w_p(z)=\frac{1}{2}\left< |{\bf a}_p(z)|^2\right>$.
Starting from the modal NLS Eq.(\ref{eq:nls_Ap}), we have 
\begin{align}
\partial_z w_p =&  \frac{1}{3}  \gamma \left< X_p^{(1)}\right> 
+ \frac{2}{3}  \gamma \left<  {X}_p^{(2)}\right> ,
\label{eq:pzwp1}\\
X_p^{(1)} =&{\rm Im}\Big\{ \sum_{l,m,n} S_{lmnp}^* ({\bf a}_l^\dag {\bf a}_m^*) ({\bf a}_n^T {\bf a}_p) \Big\}  ,
\label{def:Xp1}\\
{X}_p^{(2)} =&{\rm Im}\Big\{ \sum_{l,m,n} S_{lmnp}^* ({\bf a}_n^T {\bf a}_m^*) ({\bf a}_l^\dag {\bf a}_p) \Big\}  .
\label{def:Xp2}
\end{align}
The detailed derivation of the equations for the fourth order correlators $J_{lmnp}^{(1)}(z)=\left<({\bf a}_l^\dag {\bf a}_m^*) ({\bf a}_n^T {\bf a}_p)\right>$ and ${J}_{lmnp}^{(2)}(z)=\left<({\bf a}_n^T {\bf a}_m^*) ({\bf a}_l^\dag {\bf a}_p)\right>$ is given in the \ref{app:weak_disorder_2}.
As already noticed, as a result of the Furutsu-Novikov theorem, conservative disorder introduces an effective dissipation in the system, so that the evolutions of the fourth-order moments have the form of a forced-damped oscillator equation:
\begin{eqnarray}
\partial_z J_{lmnp}^{(j)} &=&
(-8\Delta \beta +i  \Delta \omega_{lmnp} ) J_{lmnp}^{(j)}
+ i \gamma \big<{Y}_{lmnp}^{(j)}\big> 
\label{eq:4th_order_moment}
%\partial_z {\tilde J}_{nmlp} &=&
%(-8\Delta \beta +i  \Delta \omega_{nmlp} ) {\tilde J}_{nmlp}
%+ i \gamma  \big<{\tilde Y}_{nmlp}\big> 
%\label{eq:4th_order_moment_2}
\end{eqnarray}
where $\Delta \omega_{lmnp}  = \beta_l+\beta_m-\beta_n-\beta_p$ is the resonance frequency, while $\big<{Y}_{lmnp}^{(j)}\big>$ denote the six-th order moments that have been split into products of second-order moments, by virtue of the factorizability property of statistical Gaussian fields (see the detailed expressions of $\big<{Y}_{lmnp}^{(j)}\big>$ for $j=1,2$ in Eq.(\ref{eq:Y_lmnp_1}) and Eq.(\ref{eq:Y_lmnp_2}) in the \ref{app:weak_disorder_2}).
The solution to Eq.(\ref{eq:4th_order_moment}) reads
\begin{eqnarray}
J_{lmnp}^{(j)}(z) &=& 
J_{lmnp}^{(j)}(0) G_{lmnp}(z) + {\cal I}_{lmnp}^{(j)}(z)
\label{eq:mom4pq}
\end{eqnarray}
where the convolution integral reads
\begin{eqnarray}
{\cal I}_{lmnp}^{(j)}(z) = i \gamma \int_0^z\left< {Y}_{lmnp}^{(j)}\right> (z-z') G_{lmnp}(z') dz'  
\label{eq:conv_int}
\end{eqnarray}
with the Green function 
\begin{eqnarray}
G_{lmnp}(z)=H(z) \exp(i \Delta \omega_{lmnp} z -8\Delta \beta z),
\label{eq:green}
\end{eqnarray}
and $H(z)$ the Heaviside function.  
We now arrive at the key point of the analysis, in which one makes a distinction between the continuous and the discrete wave turbulence approaches.

\subsubsection{Continuous wave turbulence}
\label{sec3:cont_wt}

In general, an exchange of power among four different modes does not need to satisfy the exact resonance condition $\Delta \omega_{lmnp}=0$. 
Indeed, it is sufficient for a mode quartet to verify the quasi-resonant condition 
\begin{eqnarray}
|\Delta \omega_{lmnp}| \lesssim 1/L_{kin}^{disor}
\label{eq:cont_turb}
\end{eqnarray}
to provide a non-vanishing contribution to the convolution integral (\ref{eq:conv_int}), where $L_{kin}^{disor}$ denotes the characteristic evolution length scale of the moments of the field $\left< {Y}_{lmnp}^{(j)}\right>(z)$ (see Eq.(\ref{eq:L_kin_disor}) below).
In the usual {\it continuous wave turbulence approach}, there is a large number of non-resonant mode quartets $\Delta \omega_{lmnp} \neq 0$ that verifies the quasi-resonant condition (\ref{eq:cont_turb}), and that contributes significantly to the evolution of the moments equations.
%Considering the example of the parabolic potential (GRIN-MMF), this means that the non-vanishing minimal value $\min(|\Delta \omega_{lmnp}|)=\beta_0$ can be of the order than $L_{kin}^{cont}$. 

We anticipate that the continuous wave turbulence regime does not correspond to usual experiments of optical beam self-cleaning \cite{Krupanatphotonics,WrightNP2016}, since we have typically $\beta_0 \sim 10^3$m$^{-1}$, so that $\beta_0 \gg 1/L_{kin}^{disor}$.
In other words, since any non-resonant mode quartet verifies $|\Delta \omega_{lmnp}| \geq \beta_0 \gg 1 /L_{kin}^{disor}$, the system does not exhibit quasi-resonances, but solely exact resonances $\Delta \omega_{lmnp}=0$.
As a matter of fact, experiments of beam self-cleaning are described by the discrete wave turbulence approach that will be discussed in detail in the next Sec.~\ref{sec3:disc_wt}.
Here, for the sake of clarity, we first discuss the conventional continuous wave turbulence regime \cite{Nazarenko11}, which can be relevant for MMFs characterized by a huge number of modes ($M  \gg 1$ such that $\beta_0 \ll \Delta \beta$).
The discussion of the usual continuous regime is also important, in that it enlightens the effect of acceleration of thermalization, owing to the presence of structural disorder in the system.

%Secondly, we consider the continuous regime in the case where disorder plays a dominant role with respect to nonlinear effects, \blue{$\beta_0^{-1} \lesssim L_d=(\Delta \beta)^{-1} \ll L_{kin}^{cont}$.}
%Following the usual wave turbulence approach, for large interaction ('times') lengths  $Y_{lmnp}^{(j)}(z)$ is slowly varying, and the convolution integral can be approximated for $z \gg L_d$
Recalling that $L_d=1/\Delta \beta \ll L_{nl}$, then the Green function decays on a length scale which is much smaller than the evolution length of $\left<Y_{lmnp}^{(j)}\right>(z)$, so that the convolution integral can be approximated for $z \gg L_d$
\begin{eqnarray}
{J}_{lmnp}^{(j)} \simeq \gamma \left< {Y}_{lmnp}^{(j)}\right> 
\frac{i8\Delta \beta-\Delta \omega_{lmnp}}{\Delta \omega_{lmnp}^2+(8\Delta \beta)^2}.
\label{eq:I_lmnp_damp}
\end{eqnarray}
By substitution of Eq.(\ref{eq:I_lmnp_damp}) in the fourth-order moments (\ref{eq:mom4pq}), one obtains the expressions of the averaged moments $\left< X_p^{(j)}\right>$ given in (\ref{def:Xp1}-\ref{def:Xp2}).
Collecting all terms in (\ref{eq:pzwp1}) gives the equation for the modal amplitudes $n_p(z)=2w_p(z)=\left< |{\bf a}_p(z)|^2\right>$.
Since the MMF supports a  large number of modes $M  \gg 1$, we consider the continuous limit where the discrete sums in (\ref{def:Xp1}-\ref{def:Xp2}) are replaced by continuous integrals. We obtain the continuous kinetic equation 
\begin{align}
\nonumber 
\partial_z \tilde{n}_{\bk_4}(z) =&
 \frac{4 \gamma^2}{3\beta_0^6} \iiint d\bk_{1,2,3}  \frac{\overline{\Delta \beta}}{\Delta {\tilde \omega}_{\bk_1 \bk_2 \bk_3 \bk_4}^2+\overline{\Delta \beta}^2} 
\times |\tilde{S}_{\bk_1 \bk_2 \bk_3 \bk_4}|^2 
\tilde{M}_{\bk_1 \bk_2 \bk_3 \bk_4}
\\
 &+\frac{32 \gamma^2}{9\beta_0^2} \int d\bk_1  \frac{\overline{\Delta \beta}}{\Delta \tilde{\omega}_{\bk_1 \bk_4}^2+\overline{\Delta \beta}^2} |\tilde{s}_{\bk_1 \bk_4}(\tilde{\bf n})|^2 ( \tilde{n}_{\bk_1}-\tilde{n}_{\bk_4})  
\label{eq:kin_contin_dis}
\end{align}
where $d\bk_{1,2,3}=d\bk_1 d\bk_2 d\bk_3$, $\overline{\Delta \beta}=8\Delta \beta$ and
\begin{equation}
{\tilde s}_{\bk_1 \bk_4} (\tilde{\bf n})= \frac{1}{\beta_0^2}\int d\bk' \, {\tilde S}_{\bk_1 \bk' \bk' \bk_4} \, {\tilde n}_{\bk'}.
\label{eq:kin_np_cont}
\end{equation}
The functions with a tilde refer to the natural continuum extension of the corresponding discrete functions, i.e., $\tilde{n}_{\bf k}(z)=n_{[\bf k/\beta_0]}(z)$, 
$\tilde{\beta}_\bk = \beta_{[\bk/\beta_0]}$, $\tilde{S}_{\bk_1 \bk_2\bk_3 \bk_4}= S_{[\bk_1/\beta_0][\bk_2/\beta_0][\bk_3/\beta_0][\bk_4/\beta_0]}$ and so on, where $[x]$ denotes the integer part of $x$.
With these notations we have  
$\tilde{M}_{\bk_1 \bk_2 \bk_3 \bk_4}=\tilde{n}_{\bk_1} \tilde{n}_{\bk_2}\tilde{n}_{\bk_3}\tilde{n}_{\bk_4}
\big( \tilde{n}_{\bk_4}^{-1}+ \tilde{n}_{\bk_3}^{-1}- \tilde{n}_{\bk_1}^{-1}-\tilde{n}_{\bk_2}^{-1} \big)$,
$\Delta {\tilde \omega}_{\bk_1 \bk_2 \bk_3 \bk_4}=\tilde{\beta}_{\bk_1}+\tilde{\beta}_{\bk_2}-\tilde{\beta}_{\bk_3}-\tilde{\beta}_{\bk_4}$, $\Delta \tilde{\omega}_{\bk_1 \bk_2}=\tilde{\beta}_{\bk_1}-\tilde{\beta}_{\bk_2}$, and 
$\tilde{\beta}_\bk= \kappa_x + \kappa_y + \beta_0$, with ${\bf \kappa}=\beta_0 (p_x, p_y)$ \cite{aschieri2011condensation}.

The main novelty of the kinetic Eq.(\ref{eq:kin_np_cont}) with respect to the case without any structural disorder \cite{aschieri2011condensation}, is that the mechanism of disorder-induced dissipation introduces a finite bandwidth into the four-wave resonances among the modes.
Accordingly, instead of the Dirac $\delta-$distribution that guarantees energy conservation at each four-wave interaction, here the kinetic equation involves a Lorentzian distribution.
We remark that this aspect was already discussed in different circumstances \cite{Dyachenko92}, in particular in recent work \cite{Churkin15} dealing with random fiber lasers in the presence of gain and losses.
Here, the originality is that the finite bandwidth of the Lorentzian distribution and the associated effective dissipation $\Delta \beta$ originate from the {\it conservative} structural disorder of the material.

Taking the formal limit $\Delta \beta \to 0$:
$\overline{\Delta \beta}/\big(\Delta {\tilde \omega}_{\bk_1 \bk_2 \bk_3 \bk_4}^2+\overline{\Delta \beta}^2 \big)
\rightarrow
\pi \delta(\Delta {\tilde \omega}_{\bk_1 \bk_2 \bk_3 \bk_4}),
$
%\label{eq:limit_lorentz}
we recover a continuous kinetic equation with a form similar to that derived in \cite{aschieri2011condensation} in the absence of disorder ($\Delta \beta = 0$) and in the absence of polarization effects (scalar limit ${\bf a}_p \to a_{p,x}$).
%Note in particular that the length scale $L_{kin}^{cont}$ given in Eq.(\ref{eq:L_kin_cont}) also characterizes the thermalization rate without any disorder.
However, the above limit $\Delta \beta \to 0$ is not physically relevant here, since we have assumed $L_d=1/\Delta \beta \ll L_{nl}$ in order to derive Eq.(\ref{eq:4th_order_moment}).
Here, we are considering the regime $\beta_0 \ll \Delta \beta$, so that it is the opposite limit that is physically relevant
$\overline{\Delta \beta}/\big(\Delta {\tilde \omega}_{\bk_1 \bk_2 \bk_3 \bk_4}^2+\overline{\Delta \beta}^2 \big)
\rightarrow 1/\overline{\Delta \beta}
$, and the kinetic equation takes the reduced form:
\begin{align}
\nonumber 
\partial_z \tilde{n}_{\bk_4}(z) =&
 \frac{\gamma^2}{6 \Delta \beta \beta_0^6} \iiint d\bk_{1,2,3}   
 |\tilde{S}_{\bk_1 \bk_2 \bk_3 \bk_4}|^2 
\tilde{M}_{\bk_1 \bk_2 \bk_3 \bk_4}
\\
 &+\frac{4 \gamma^2}{9 \Delta \beta \beta_0^2 } \int d\bk_1   |\tilde{s}_{\bk_1 \bk_4}(\tilde{\bf n})|^2 ( \tilde{n}_{\bk_1}-\tilde{n}_{\bk_4})  
\label{eq:kin_contin_dis2}
\end{align}
The second term in the right-hand side of (\ref{eq:kin_contin_dis2}) enforces an isotropic distribution of mode occupancies $\tilde{n}_{\bk}(z)$ for degenerate modes, while the first term enforces the mode occupancies to reach the most disordered equilibrium distribution. 
The kinetic equation (\ref{eq:kin_contin_dis2}) conserves the power, $N=\beta_0^{-2}\int \tilde{n}_{\bk} d\bk$, and exhibits a $H-$theorem of entropy growth, $\partial_z {\cal S}^{neq}(z) \ge 0$, where the nonequilibrium entropy is defined by ${\cal S}^{neq}(z) = \beta_0^{-2} \int \log(\tilde{n}_{\bk}) d\bk$. 
However, at variance with the conventional wave turbulence kinetic equation, the kinetic Eq.(\ref{eq:kin_contin_dis2}) does not conserve the energy, $E=\beta_0^{-2} \int {\tilde \beta}_{\bk} \tilde{n}_{\bk} d\bk$.
Then the equilibrium distribution maximizing the entropy, given the constraint of the conservation of power $N$, is given by the uniform distribution 
\begin{eqnarray}
\tilde{n}^{eq}_{\bk}= {\rm const}.
\label{eq:rj_cont_disor}
\end{eqnarray}
This equilibrium state denotes an equipartition of ('particles') power among all modes.
Recalling that $\beta_0 \ll \Delta \beta$, the Lorentzian distribution in the kinetic Eq.(\ref{eq:kin_contin_dis}) is dominated by disorder, so that the length scale characterizing the rate of thermalization toward the equilibrium distribution (\ref{eq:rj_cont_disor}) is given by
\begin{eqnarray}
L_{kin}^{disor} \sim \Delta \beta L_{nl}^2/{\bar S_{lmnp}^2},
%L_{kin}^{ord,pert} \sim b_0 L_{nl}^2/{\bar S_{lmnp}^2},
%L_{kin}^{ord,pert}/L_{kin}^{ord} \sim b_0/\beta_0
\label{eq:L_kin_disor}
\end{eqnarray}
where ${\bar S_{lmnp}^2}$ denotes the average square of the tensor $S_{lmnp}$ involving non-trivial resonances among nondegenerate modes.

The main conclusion is that the kinetic Eq.(\ref{eq:kin_np_cont}) does not describe a process of condensation characterized by a macroscopic occupation of the fundamental mode.
This indicates that weak disorder should prevent an effect of beam self-cleaning in MMFs possessing a huge number of modes. 

%%%%%%%%%%%%%%%%%%%%%%%%%%%%

\subsubsection{Discrete wave turbulence}
\label{sec3:disc_wt}

We have seen that, in the continuous wave turbulence regime, quasi-resonances verifying $\Delta \omega_{lmnp} \lesssim 1/L_{kin}^{disor}$ contribute to the convolution integral (\ref{eq:conv_int}).
At variance with the continuous regime, in the discrete case the non-vanishing minimum value of $\Delta \omega_{lmnp}$ is such that
\begin{eqnarray}
{\rm min}(|\Delta \omega_{lmnp}|) = \beta_0 \gg 1/L_{kin}^{disor}.
\label{eq:disc_turb}
\end{eqnarray}
Accordingly, only exact resonances $\Delta \omega_{lmnp} =0$ contribute to the integral (\ref{eq:conv_int}), while non-resonant mode quartets lead to a vanishing integral \cite{Nazarenko11}.
Note that this procedure can also be justified by a homogenization approach, as reported in \cite{Kuksin} in the presence of a non-conservative disorder accounting for gain and losses in the system.
%, an effective kinetic equation of the form (\ref{eq:kin_np_disc}) was obtained by a homogenization procedure retaining exact resonances and dicarding quasi-resonances \cite{kuksin}.
%Considering a non-conservative disorder accounting for gain and losses in the system, an effective kinetic equation of the form (\ref{eq:kin_np_disc}) was obtained by a homogenization procedure retaining exact resonances and dicarding quasi-resonances \cite{kuksin}.
%because of the presence of the dissipation due to the disorder.
As discussed above, usual experiments of optical beam self-cleaning are described by the discrete wave turbulence approach, since $\beta_0^{-1} \sim 10^{-3}$m and the above condition is well verified in the experiments \cite{Krupanatphotonics,WrightNP2016}.
%As discussed in \cite{PRL19}, in the discrete regime the averaged modal components exhibit an accelerated dynamics, which we denote by the characteristic length scale $L_{kin}^{disor}$, see below Eq.(\ref{eq:L_kin_disc}).

In the discrete wave turbulence regime we need to consider separately the cases of resonant and non-resonant mode interactions.
For mode quartets verifying $\Delta \omega_{lmnp}=0$, the Green function (\ref{eq:green}) decays on a length scale which is much smaller than the evolution of $n_p(z)$, because $L_{d}=(\Delta \beta)^{-1} \ll L_{kin}$, so that  (\ref{eq:conv_int}) can be approximated by $J_{lmnp}^{(j)}(z)  \simeq \frac{i\gamma}{8\Delta \beta } \left< {Y}_{lmnp}^{(j)}\right>(z)$.
For $\Delta \omega_{lmnp} \neq 0$, the Green function oscillates rapidly over a length scale smaller than $L_{lin}=\beta_0^{-1} \ll L_{kin}$.
Such a rapid phase rotation, combined with the fast decay of the Green function over a length $L_d \ll L_{nl}$, leads to a vanishing convolution integral in (\ref{eq:conv_int}).
%(\ref{eq:mom4pq}).
As a result, the fourth-order moment can be written in the form
\begin{eqnarray}
J_{lmnp}^{(j)}(z)  \simeq
\frac{i\gamma}{8\Delta \beta }\left< {Y}_{lmnp}^{(j)}\right>(z) \delta^K(\Delta \omega_{lmnp}),
\label{eq:mom4_disc_dis}
\end{eqnarray}
where $\delta^K(\Delta \omega_{lmnp})=1$ if $\Delta \omega_{lmnp}=0$, and zero otherwise.
Note that the discrete regime discussed here does not exactly correspond to the discrete regimes associated with finite size effects in homogeneous wave turbulence \cite{Nazarenko11} -- here the system is non-homogeneous ($V(\br) \neq $const) and the resonance for the momentum reflected by the tensor
$S_{lmnp}$ is not as rigid as the usual one involving the Dirac $\delta-$distribution in homogeneous turbulence.

We provide a detailed computation of the fourth-order moments $J_{lmnp}^{(j)}(z)$ and corresponding moments $\left< X_p^{(j)}\right>$ defined in (\ref{def:Xp1}-\ref{def:Xp2}) in the \ref{app:weak_disorder_2}, see Eq.(\ref{eq:momX_p1}) and (\ref{eq:momX_p2}).
Then collecting all terms in (\ref{eq:pzwp1}) gives the discrete kinetic equation for the modal amplitudes $n_p(z)=2w_p(z)$
\begin{align}
\nonumber
\partial_z n_p(z) =&  \frac{ \gamma^2}{6\Delta \beta} \sum_{l,m,n} |S_{lmnp}|^2 \delta^K(\Delta\omega_{lmnp}) M_{lmnp}({\bf n}) \\
&
+  \,  \frac{4\gamma^2}{9 \Delta \beta}  \sum_l  
 |  s_{lp}({\bf n}) |^2 \delta^K(\Delta\omega_{lp})   (n_l-n_p), \quad \quad
\label{eq:kin_np_disc}
\end{align}
with $s_{lp}({\bf n})=\sum_{m'} S_{lm'm'p} n_{m'}$, and $M_{lmnp}({\bf n})=  n_l n_m n_p+n_l n_m n_n -  n_n n_p n_m -n_n n_p n_l$, with '$n_m$' for '$n_m(z)$', $\Delta \omega_{lp}=\beta_l-\beta_p$.
According to the kinetic equation (\ref{eq:kin_np_disc}), the length scale characterizing the rate of thermalization is the same as that obtained in the continuous wave turbulence regime (\ref{eq:L_kin_disor}), namely 
%\begin{eqnarray}
$L_{kin}^{disor} \sim \Delta \beta L_{nl}^2/{\bar S_{lmnp}^2}$.
%\label{eq:L_kin_disc}
%\end{eqnarray}
%Considering a non-conservative disorder accounting for gain and losses in the system, an effective kinetic equation of the form (\ref{eq:kin_np_disc}) was obtained by a homogenization procedure retaining exact resonances and dicarding quasi-resonances \cite{kuksin}.

Aside from its discrete form, the kinetic Eq.(\ref{eq:kin_np_disc}) has a structure which is analogous to the conventional wave turbulence kinetic equation, so that it describes a process of wave condensation that occurs irrespective of the sign of the nonlinearity $\gamma$ (see the factor $\gamma^2$ in Eq.(\ref{eq:kin_np_disc}))\cite{connaughton05,Nazarenko11}.
We refer the reader to Refs.\cite{connaughton05,during09,Nazarenko11} for details on condensation in the homogeneous case ($V(\br)=0$), and to \cite{aschieri2011condensation,lvov03wave} for the non-homogeneous case in a waveguide potential ($V(\br) \neq 0$).
It is important to note that, in contrast to the kinetic Eq.(\ref{eq:kin_contin_dis2}) derived in the continuous wave turbulence regime, here Eq.(\ref{eq:kin_np_disc}) also conserves the energy $E=\sum_p \beta_p n_p(z)$ --  despite the presence of the dissipation effect $\Delta \beta$.
The reason for this is that only exact resonances contribute to the discrete turbulence regime, so that the discrete kinetic equation is not affected by the dissipation-induced resonance broadening which inhibits energy conservation in the continuous turbulence regime.
%At variance with the kinetic Eq.(\ref{eq:kin_contin_dis2}) considered above, here Eq.(\ref{eq:kin_np_disc}) also conserves the energy $E=\sum_p \beta_p n_p(z)$ in addition to the 'number of particles' $N=\sum_p n_p(z)$.
The kinetic Eq.(\ref{eq:kin_np_disc}) also conserves the 'number of particles' $N=\sum_p n_p(z)$ and exhibits a $H-$theorem of entropy growth 
\begin{eqnarray}
\partial_z {\cal S}^{neq} \ge 0,
\label{eq:h-theo}
\end{eqnarray}
for the nonequilibrium entropy
\begin{eqnarray}
{\cal S}^{neq}(z)=\sum_p \log\big(n_p(z)\big). 
\label{eq:ent_neq}
\end{eqnarray}
The growth of entropy saturates once equilibrium is reached, as discussed in a phenomenological way in section~\ref{subsec:basic}. Accordingly, the kinetic Eq.(\ref{eq:kin_np_disc}) describes an irreversible evolution to the (maximum entropy) thermodynamic Rayleigh-Jeans equilibrium 
\begin{eqnarray}
n^{eq}_p=T/(\beta_p - \mu).
\label{eq:wd_rj}
\end{eqnarray}
Note that this equilibrium distribution coincides with the equilibrium (\ref{eq:genRJ}) discussed in sec.\ref{subsec:basic}. Indeed, the equilibrium distribution (\ref{eq:wd_rj}) does not depend on disorder. However, disorder impacts the rate of thermalization to this equilibrium, a property that is discussed in the next section.
%The system exhibits a phase transition to condensation when $\mu \to \beta_0$ \cite{aschieri2011condensation}: for $E \ge E_{\rm crit}=\frac{NV_0}{2}(1+2\beta_0/V_0)$  there is no condensation $n_0^{eq}/N=0$, while for $E < E_{\rm crit}$ the fundamental mode of the MMF gets macroscopically populated
%\begin{eqnarray}
%\frac{n_0^{eq}}{N}=1-\frac{E-E_0}{E_{\rm crit}-E_0},
%\label{eq:n_eq_vs_E}
%\end{eqnarray} 
%where $E_0=N\beta_0$ is the minimum energy (all particles are in the fundamental mode).
%Here, $V_0=k_0 (n_{\rm co}^2-n_{\rm cl}^2)/(2n_{\rm co})$ denotes the depth of the parabolic potential, with $n_{\rm co}$ and $n_{\rm cl}$ the refractive indexes of the core and the cladding, respectively.
%Note that Eq.(\ref{eq:n_eq_vs_E}) only provides an approximation of the condensation curve $n_0^{eq}$ vs $E$, see Ref.\cite{aschieri2011condensation} for a more detailed discussion.
%As discussed in \cite{PRL19}, a stable self-cleaned shape of the intensity pattern $|\psi|^2(\br)$ can be interpreted as a consequence of the macroscopic population of the fundamental mode of the MMF (see the movie published in the Supplemental of \cite{PRL19}).

\begin{figure}
\begin{center}
%\bigskip
\includegraphics[width=0.55\columnwidth]{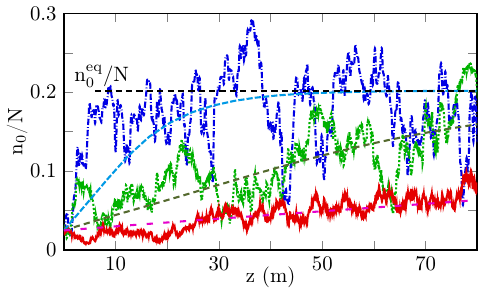}
\caption{
Scaling of acceleration of thermalization with mode-decorrelated disorder:
Numerical simulations of the NLS Eq.(\ref{eq:nls_Ap}) showing the evolutions of the fundamental mode $n_0(z)$, for different amounts of disorder $\Delta \beta$.
The dashed lines show the corresponding simulations of the discrete kinetic Eq.(\ref{eq:kin_np_disc}), starting from the same initial condition as the NLS simulations.
Parameters are: 
$\Delta \beta \simeq 2.6$ m$^{-1}$ ($2\pi/\sigma_\beta=2.1$ m, $l_\beta= 30$ cm) blue (dash-dotted);
$\Delta \beta \simeq 10.5$m$^{-1}$, ($2\pi/\sigma_\beta=26$ cm, $l_\beta= 1.88$ cm) green (dotted);
$\Delta \beta \simeq 42$m$^{-1}$, ($2\pi/\sigma_\beta=6.6$ cm, $l_\beta= 0.47$ cm) red (solid);
while the power is $N=47.5$ kW ($M=120$ modes, fiber core radius $R=26$ $\mu$m).
%$2 \pi /\sigma_\beta= 2.12$m (blue);
%$\Delta \beta \simeq 10.5$m$^{-1}$ (green), 
%$\delta n = 8\times 10^{-6}$ (green);
%$2 \pi /\sigma_\beta= 0.13$m;
%$\Delta \beta \simeq 42$m$^{-1}$ (red).
%$\delta n = 16\times 10^{-6}$ (red).
%$2 \pi /\sigma_\beta= 0.067$m  (red).
The curves eventually relax to the common theoretical equilibrium value $n_0^{eq}/N \simeq 0.2$ (dashed black line) with different rates, confirming the scaling of acceleration of condensation predicted by the theory in Eq.(\ref{eq:accel_thermal}), without using adjustable parameters.
{\it Source:} From Ref.\cite{garnier19}.
}
\label{fig:sec3_acc_dis}
\end{center}
\end{figure}

\subsubsection{Acceleration of thermalization mediated by disorder}
\label{sec3:acc_thermal}

The length scale characterizing the rate of thermalization in the absence of structural disorder is obtained from the continuous kinetic equation that was derived in Ref.\cite{aschieri2011condensation}, 
%\begin{eqnarray}
$L_{kin}^{ord} \sim \beta_0 L_{nl}^2/{\bar S_{lmnp}^2}$.
%\label{eq:L_kin_order}
%\end{eqnarray}
A similar scaling behavior is expected in the discrete wave turbulence regime, see \cite{garnier19}.
Hence, in regimes of beam cleaning experiments where $\beta_0 \gg \Delta \beta$, the weak disorder leads to a significant acceleration of the rate of thermalization and condensation 
\begin{eqnarray}
L_{kin}^{disor}/L_{kin}^{ord} \sim \Delta \beta / \beta_0.
\label{eq:accel_thermal}
\end{eqnarray}
Considering typical values of $\beta_0 \sim 10^3$m$^{-1}$ and $L_d =1/\Delta \beta$ larger than tens of centimeters, we see that we always have $L_{kin}^{disor}/L_{kin}^{ord} \ll 1$.
Accordingly, weak disorder is responsible for a significant acceleration of the rate of thermalization and condensation \cite{Fusaro:PRL:2019}.
Note that the presence of a perturbation on the dispersion relation can modify the above scaling, a feature that will be discussed later on.

\begin{figure}
\begin{center}
%\bigskip
\includegraphics[width=0.65\columnwidth]{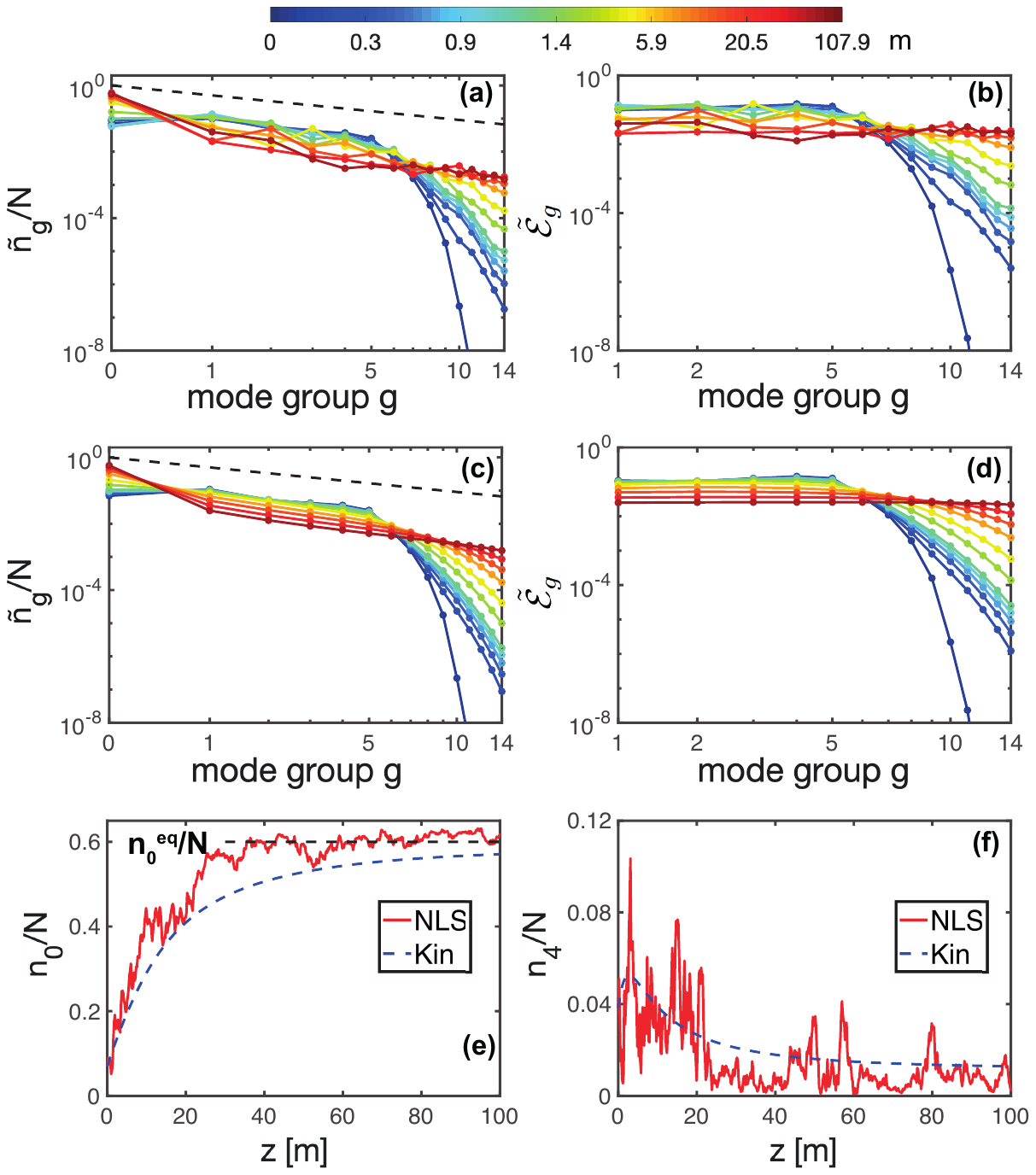}
\caption{
Energy and power flows underlying light thermalization: Numerical simulation of the modal NLS Eq.(\ref{eq:nls_Ap}) (a)-(b), and wave turbulence kinetic Eq.(\ref{eq:kin_np_disc}) (c)-(d): Evolutions of the modal distribution of the power ${\tilde n}_g$ (a)-(c), and energy ${\tilde {\cal E}}_g$ (b)-(d), for  $g_{max}=$15 groups of non-degenerate modes of the parabolic MMF (an average has been taken among the degenerate modes). The dashed black lines in (a) and (c) denote the RJ  power-law ${\tilde n}_g^{eq} \sim 1/g$. The thermalization to the RJ equilibrium is characterized by an energy flow toward the higher-order modes, and a power flow toward the fundamental mode and the higher-order modes. Evolutions of $n_0(z)$ (e) and $n_4(z)$ (f) obtained from simulation of the NLS Eq.(\ref{eq:nls_Ap}) (red line) and the kinetic Eq.(\ref{eq:kin_np_disc}) (dashed blue): The modal components thermalize to the theoretical equilibrium value predicted by the RJ theory (the dashed black line denotes $n_0^{eq}/N=0.6$).
Parameters: $N=$47.5kW, 
%$\Delta \beta=10.4$m$^{-1}$, 
$l_\beta=0.019$m, $2 \pi /\sigma_\beta =0.26$m, there is no average over the realizations for the NLS simulation. {\it Source:} From Ref.\cite{baudin21}.
}
\label{fig:sec3_cascades}
\end{center}
\end{figure}

We have confirmed the scaling (\ref{eq:accel_thermal}) by performing numerical simulations of the NLS Eq.(\ref{eq:nls_Ap}) for different amounts of disorder $\Delta \beta$.
%We refer the reader to the Supplemental of Ref.\cite{Fusaro:PRL:2019} for the numerical scheme used to solve the  NLS Eq.(\ref{eq:nls_Ap}) in the presence of disorder.
The results are reported in Fig.~\ref{fig:sec3_acc_dis}, and confirm the scaling of the rate of acceleration of thermalization and condensation as they are predicted by the theory.
More precisely, a remarkable quantitative agreement has been obtained between the simulations of the NLS Eq.(\ref{eq:nls_Ap}) and those of the discrete kinetic Eq.(\ref{eq:kin_np_disc}), without using any adjustable parameter.
Note that the equilibrium value of the condensate fraction $n_0^{eq}/N$ vs. the energy $E$ is obtained from the analysis of the RJ equilibrium distribution \cite{connaughton05,aschieri2011condensation,Nazarenko11}.
The condensation curve $n_0^{eq}/N$ vs. the energy $E$ is reported in explicit form in the \ref{app:condensate_fraction}, and its experimental study is discussed in section~\ref{sec4:subsecRJcond}. 
Also note that we have deliberately chosen a small value of the condensate fraction $n_0/N \simeq 0.2$, so as to avoid large deviations from Gaussianity for the fundamental mode -- though the theory has been validated even for large condensate fractions in \cite{Fusaro:PRL:2019} (also see Fig.~\ref{fig:sec3_cascades} here below).
We recall here the recent works showing that a key assumption of the wave turbulence approach is the existence of a random phase among the modes, rather than a genuine Gaussian statistics \cite{Chibbaro17,Chibbaro18}.
%The parameters in Fig.~\ref{fig:1} are $\Delta \beta \simeq 2.6$m$^{-1}$, $2 \pi /\sigma_\beta= 2.12$m (blue);
%$\Delta \beta \simeq 10.5$m$^{-1}$, $2 \pi /\sigma_\beta= 0.26$m (green);
%$\Delta \beta \simeq 42$m$^{-1}$, $2 \pi /\sigma_\beta= 0.067$m  (red).
%These parameters may be artificial, but the purpose here is to verify the scaling of thermalization acceleration.

\subsubsection{Light thermalization resulting from turbulence cascades}
\label{sec3:cascades}

Spatial beam self-cleaning in MMFs can also be understood in the context of turbulence cascades associated to conserved quantities, as initially explored in the broader framework of hydrodynamic turbulence in Ref.\cite{Podivilov2018prl}. To properly contextualize this problem within the wave turbulence theory \cite{Zakharov92,Nazarenko11,Newell11}, let us consider a generic dissipative system that is driven far from equilibrium by an external source. A typical illustrative example of forced physical system can be the excitation of hydrodynamic surface waves by the wind in oceans. In general, the frequency-scales of forcing and damping differ significantly. The nonlinear interaction leads to a spectral redistribution, which  eventually relaxes to a nonequilibrium stationary state characterized by a non-vanishing flux of the conserved quantities, typically the power (wave action), and the kinetic energy. The nonequilibrium stationary solutions of the wave turbulence kinetic equation were originally obtained by Zakharov in the mid-1960s \cite{zakharov65weak,zakharov67on}, and they can be considered as the analogue of the spectra of hydrodynamic turbulence proposed in 1941 by Kolmogorov \cite{kolmogorov41the}. 

Although nonequilibrium stationary spectra of turbulence  are sustained by a continuous forcing and damping, it is important to remark that they can also emerge  in the transient evolution of a purely conservative system, prior to reaching the RJ equilibrium distribution, see e.g.  Refs.\cite{Bortolozzo09,Laurie12,zhu2023self}. It is from this perspective that the irreversible process of light thermalization in MMFs has been addressed in Ref.\cite{baudin21}. In this article, numerical simulations of the NLS Eq.(\ref{eq:nls_Ap}), and the corresponding kinetic Eq.(\ref{eq:kin_np_disc}), have been realized by starting from the same initial mode distribution (except that independent random phases among the modes were considered for NLS simulations). The results are reported in Fig.~\ref{fig:sec3_cascades}. During light propagation in the MMF, the spectrum for the power distribution $n_p$ essentially flows toward the fundamental mode (inverse cascade), while a small fraction of $n_p$ flows toward the higher-order modes. For convenience, we have reported in Fig.~\ref{fig:sec3_cascades} the average power ${\tilde n}_g$ within each group of degenerate modes, where $g=0,\ldots,g_{max}-1$ indexes the mode group (in Fig.~\ref{fig:sec3_cascades} $g_{max}=15$ for a total $M=g_{max}(g_{max}+1)/2=120$ modes). The RJ power-law ${\tilde n}_g \sim 1/g$ is verified by the simulation of the kinetic and NLS equations. Note that, due to the extensive computation times required, averaging over the realizations of NLS simulations was not feasible, resulting in the spectrum's noisy structure of ${\tilde n}_g$ in Fig.~\ref{fig:sec3_cascades}(a)-(b).

The results for the power mode distribution are corroborated by those of the modal distribution of the energy, which exhibits a flow toward the higher-order modes (direct cascade). In this example, we considered a relatively small value of the conserved energy $E$, which is below the critical value of the transition to condensation $E_{\rm crit}$. In the condensed state, the chemical potential tends to the lowest energy level, $\mu \to \beta_0^-$ (see \ref{app:condensate_fraction}), so that the waves that started from an initial state with an excess energy in the low-energy modes, eventually tend to an equilibrium state displaying an energy equipartition among the modes ${\cal E}_p = (\beta_p-\beta_0) n_p \sim  T$ [or equivalently ${\tilde {\cal E}}_g = \beta_0 g {\tilde n}_g \sim  T$], as illustrated in Fig.~\ref{fig:sec3_cascades}(b)-(d). Then RJ thermalization is characterized by a macroscopic population of the fundamental mode, as illustrated in Fig.~\ref{fig:sec3_cascades}(e), where the condensate fraction relaxes toward the theoretical equilibrium value $n_0^{eq}/N \simeq 0.6$. Note that the good agreement between the simulations of the NLS and kinetic equations in Fig.~\ref{fig:sec3_cascades} is obtained without using adjustable parameters.

One may question whether the above power and energy flows can be described theoretically by means of nonequilibrium stationary solutions of the kinetic equation -- the so-called Zakharov-Kolmogorov spectra of turbulence \cite{Zakharov92}. In this regard, it is crucial to note that the kinetic Eq.(\ref{eq:kin_np_disc}) considered here, differs from the conventional wave turbulence kinetic equation in two key aspects: (i) It involves the tensor $|S_{plmn}|^2$ instead of the Dirac $\delta-$function over the wave-vectors, because the transverse trapping potential inherent to the MMF waveguide breaks the conservation of the transverse momentum. (ii) The kinetic Eq.(\ref{eq:kin_np_disc}) is discrete in frequencies. This latter property does not allow the application of the standard procedure based on the Zakharov conformal transformation to derive nonequilibrium stationary solutions featured by a non-vanishing flux of the conserved power and energy. These remarks appear consistent with the results of the  numerical simulations of the NLS and kinetic equations, which  do not evidence the formation of a nonequilibrium power-law spectrum in the transient evolution that precedes the formation of the equilibrium RJ spectrum, ${\tilde n}_g^{eq} \sim 1/g$.

We finally note that an experimental signature of the direct and inverse turbulence cascades underlying RJ thermalization in MMFs has been recently reported in Ref.\cite{baudin2023rayleigh}. More precisely, it was shown that light thermalization is characterized by a flow of the energy toward the higher-order modes, and a bi-directional redistribution of power from intermediate modes toward the fundamental mode and the higher-order modes.

%%%%%%%%%%%%%%%%%%%%%%%%%%%%%%%%%%%%
\begin{figure}
\begin{center}
%\bigskip
\includegraphics[width=.55\columnwidth]{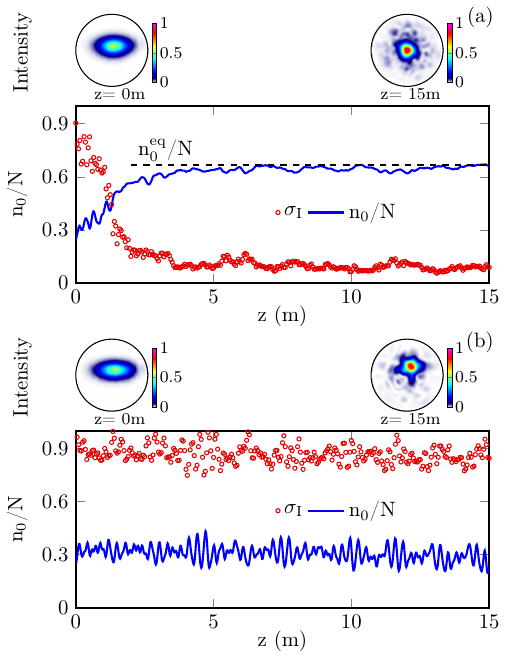}
\caption{
Disorder-induced beam cleaning:
Numerical simulations of the NLS Eq.(\ref{eq:nls_Ap}) showing the evolution of the relative variance of intensity fluctuations $\sigma_{I}(z)$ from Eq.(\ref{eq:variance}) (red circles), and of the condensate fraction $n_0(z)/N$ (blue line), in the presence of disorder (a), and in the absence of disorder (b).
The intensity patterns $I(\br)$ inside the fiber core (circles) are shown at $z=0$ and $z=15$m.
The horizontal dashed black line in (a) shows the condensate fraction at thermal equilibrium, $n_0^{eq}/N \simeq 0.68$.
The disorder induces a stable beam cleaning condensation, characterized by a significant reduction of intensity fluctuations down to $\sigma_{I} \simeq 0.1$ (a), which is in contrast with the case without disorder, where the variance of intensity fluctuations remain almost constant throughout propagation with $\sigma_{I} \simeq 0.9$.
Parameters are: $\Delta \beta=2.6$ m ($2\pi/\sigma_\beta=2.13$ m, $l_\beta= 30$ cm), $M=120$ modes, fiber core radius $R=26$ $\mu$m, $N=47.5$ kW.
{\it Source:} From Ref.\cite{garnier19}.
}
\label{fig:sec3_var}
\end{center}
\end{figure}

\subsubsection{Disorder-induced beam cleaning}
\label{sec3:dis_ind_bclean}

The amount of 'beam cleaning' can be quantified through the analysis of the fluctuations of the intensity during the propagation in the MMF.
To this aim, we consider the relative variance of intensity fluctuations relevant for a spatially non-homogeneous incoherent beam
\begin{eqnarray}
\sigma_{I}^2(z)= \frac{\int \left< I^2(\br,z) \right> - \left< I(\br,z) \right>^2 d\br }{\int \left< I(\br,z) \right>^2 d\br},
\label{eq:variance}
\end{eqnarray}
where the intensity is $I(\br,z)=|{\bf \psi}|^2(\br,z)$.
Note that for a beam with Gaussian statistics we have $\sigma_{I}^2=1$.
We report in Fig.~\ref{fig:sec3_var} the evolutions of the variance of intensity fluctuations in the presence and the absence of disorder, respectively.
The brackets $\left<.\right>$ in Eq.(\ref{eq:variance}) refers to an averaging over the propagation length $\Delta z=10$ mm, which is larger than the mode-beating length scale $\sim \beta_0^{-1}$($\simeq 0.2$mm in the example of Fig.~\ref{fig:sec3_var}). 
The simulations in Fig.~\ref{fig:sec3_var} clearly show that the presence of disorder induces a rapid condensation process, which in turn leads to a significant reduction of the relative standard deviation of intensity fluctuations $\sigma_{I} \simeq 0.1$.
Conversely, in the absence of disorder the multimode beam can exhibit an enhanced brightness at some propagation lengths (see the intensity pattern at $z=15$ m in Fig.~\ref{fig:sec3_var}(b)), however its oscillatory multimode nature prevents a stable beam-cleaning propagation, as evidenced by the relative standard deviation of intensity fluctuations that only slowly decreases below $\sigma_I \simeq 0.9$.

\subsubsection{Correlated and partially correlated disorder}
\label{sec3:corr_part_corr_dis}

In the previous section~\ref{sec3:model} we have considered a model of disorder that can be termed 'mode-decorrelated', in the sense that each individual mode of the MMF experiences a different noise, i.e., the functions $\nu_{p,j}(z)$ in (\ref{model:Dp0a}) are independent of each other.
Although this approach can be considered to be justified in different circumstances \cite{MumtazJLT12,Xiao:14,cao18}, 
%in particular for few mode fibers \cite{mumtaz13,xiao14}, 
one may question its validity for the case of an MMF exhibiting a large number of modes. 
%($\sim$ 100 modes) such as those used in the experiments of beam self-cleaning.
In the following we address this question by considering two different models of disorder, namely the fully mode-correlated disorder model, and the partially mode-correlated disorder model.

%\subsubsection{Mode-correlated noise}
\medskip
\noindent
{\it Mode-correlated noise:} We first consider the fully mode-correlated model that can be considered as the opposite limit of the 'mode-decorrelated' one, in the sense that all modes experience the same noise (more precisely, the same realization of the noise).
In this limit, the 2$\times$2 matrices describing the modal noise in (\ref{model:Dp0a}) reduce to ${\bf D}_p={\bf D}$ with
%\begin{eqnarray}
${\bf D} = \sum_{j=0}^3 \nu_j \bsigma_j$.
%\label{model:D_full_corr}
%\end{eqnarray}

\begin{figure}
\begin{center}
%\bigskip
\includegraphics[width=.55\columnwidth]{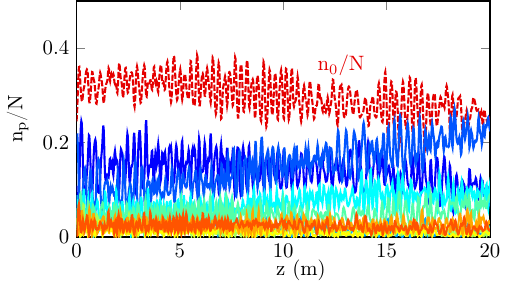}
\caption{
%{\bf preliminary figure}
Mode-correlated disorder:
Numerical simulations of the NLS Eq.(\ref{eq:nls_Ap}) showing the evolutions of the modal components $n_p(z)$, starting from a coherent initial condition: 
fundamental mode $p=0$ (red dashed), $p=1$ (dark blue solid), $p=2$ (blue solid), $p=3$ (light blue solid), $p=4$ (cyan solid), $p=5$ (light green solid), $p=6$ (green solid), $p=7$ (yellow solid), $p=8$ (orange solid).
%light green p=5, green p=6, yellow  p=7, orange p=8, dark orange p=9
%starting from a speckle  with random phases among the modes (b).
The mode-correlated disorder does not introduce an effective dissipation in the system ($\Delta \beta=0$): phase-correlations among the low-order modes are preserved, and lead to an oscillatory dynamics similar to that in the absence of disorder (see Fig.~\ref{fig:sec3_no_dis}).
%which significantly slow down the thermalization process 
At complete thermal equilibrium the system would reach the condensate fraction $n_0^{eq}/N \simeq 0.68$.
%; there is no acceleration of condensation 
Parameters are: 
$\Delta \beta \simeq 2.6$ m$^{-1}$ ($2\pi/\sigma_\beta=2.1$ m, $l_\beta= 30$ cm), 
the power is $N=47.5$ kW, $M=120$ modes, fiber core radius $R=26$ $\mu$m.
{\it Source:} From Ref.\cite{garnier19}.
}
\label{fig:sec3_mode_correl}
\end{center}
\end{figure}

The theory developed above for the decorrelated model of disorder can be extended to this fully correlated model.
The theory reveals a remarkable result in this case, namely, that the equations for the fourth-order moments $J_{lmnp}^{(j)}(z)$ do not exhibit an effective damping, i.e., $\Delta \beta=0$, see \ref{app:weak_disorder_3}. 
This is in marked contrast with the previously discussed decorrelated model of disorder, see Eqs.(\ref{eq:4th_order_moment}).
This result has a major consequence: the fully mode-correlated noise does not modify the regularization of wave resonances, and the system recovers a dynamics which is analogous to that obtained in the absence of any disorder. 
We illustrate this in Fig.~\ref{fig:sec3_mode_correl}, that reports the evolution of the modal components $n_p(z)$ obtained by simulation of the modal NLS Eq.(\ref{eq:nls_Ap}) in the presence of a mode-correlated disorder.
The initial condition is coherent (all modes are correlated with each other), and we see in Fig.~\ref{fig:sec3_mode_correl} that the low-order modes recover an oscillatory dynamics reflecting the presence of strong phase-correlations, as for the coherent regime discussed in the absence of disorder, see Fig.~\ref{fig:sec3_no_dis}.
This coherent dynamics is consistent with the intuitive idea that, in the absence of an effective dissipation ($\Delta \beta=0$), the phase-correlations are not forgotten during the evolution.

\medskip
\noindent
{\it Partially mode-correlated noise:} We have seen that the fully mode-correlated disorder does not introduce an effective dissipation ($\Delta \beta=0$), and thus leads to a coherent dynamics which is analogous to that obtained in the absence of disorder.
We have thus considered a 'partially correlated' model of disorder, in which modes that belong to different groups of degenerate modes experience a decorrelated noise, while degenerate modes of the same group experience the same noise.
This artificial model for MMFs can be considered as an intermediate model between the two limits of correlated and decorrelated models. It is mathematically tractable, and it illustrates the conjecture that the discrete kinetic Eq.(\ref{eq:kin_np_disc}) is robust as soon as disorder is not fully mode-correlated.
%Although it is difficult to properly evaluate the physical relevance of this model for MMFs used in experiments of beam self-cleaning, it may be considered as a natural intermediate model between the two limits of correlated and decorrelated models.

The theory developed above for the model of decorrelated disorder has been extended to address the partially correlated model, see \ref{app:weak_disorder_4}.
The theory reveals that (second-order) correlations among non-degenerate modes are vanishing small and can be neglected, as it was shown for the model of decorrelated disorder.
However, the computation of the fourth-order moments $J_{lmnp}^{(j)}$ is more delicate, because the mode-correlated noise introduces more terms in the calculation of the equations for the moments.
Different results for the fourth-order moments are obtained, which depend on the specific modes involved in the moments. 
Almost all of the fourth-order moments satisfy an evolution equation with a dissipation that is proportional to $\Delta \beta$.
This result is analogous to that obtained for the model of decorrelated disorder considered above, though the coefficients in front of $\Delta \beta$ are different, and their values depend on the specific modes involved in the fourth-order moment. 
In addition, there are also particular cases where the fourth-order moments do not exhibit any dissipation.
Such special cases do not contribute to the fast thermalization process described by the effective dissipation $\Delta \beta$, but instead they induce a reversible exchange of power within a group of degenerate modes, see \ref{app:weak_disorder_4}.

\begin{figure}
\begin{center}
%\bigskip
\includegraphics[width=.55\columnwidth]{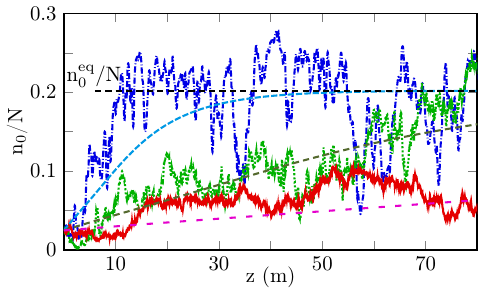}
\caption{
Scaling of acceleration of thermalization with a partially mode-correlated disorder:
Numerical simulations of the NLS Eq.(\ref{eq:nls_Ap}) showing the evolutions of the fundamental mode $n_0(z)$, for different amounts of disorder $\Delta \beta$.
The dashed lines show the corresponding simulations of the discrete kinetic Eq.(\ref{eq:kin_np_disc}), starting from the same initial condition as the NLS simulations.
Parameters are: 
$\Delta \beta \simeq 2.6$ m$^{-1}$ ($2\pi/\sigma_\beta=2.1$ m, $l_\beta= 30$ cm) blue (dashdotted);
$\Delta \beta \simeq 10.5$ m$^{-1}$, ($2\pi/\sigma_\beta=26$ cm, $l_\beta= 1.88$ cm) green (dotted);
$\Delta \beta \simeq 42$ m$^{-1}$, ($2\pi/\sigma_\beta=6.6$ cm, $l_\beta= 0.47$ cm) red (solid);
while the power is $N=47.5$ kW ($M=120$ modes, fiber radius $R=26$ $\mu$m).
%$2 \pi /\sigma_\beta= 2.12$m (blue);
%$\Delta \beta \simeq 10.5$m$^{-1}$ (green), 
%$\delta n = 8\times 10^{-6}$ (green);
%$2 \pi /\sigma_\beta= 0.13$m;
%$\Delta \beta \simeq 42$m$^{-1}$ (red).
%$\delta n = 16\times 10^{-6}$ (red).
%$2 \pi /\sigma_\beta= 0.067$m  (red).
The curves eventually relax to the common theoretical equilibrium value $n_0^{eq}/N$  with different rates,  and confirm the theoretical scaling of acceleration of condensation in Eq.(\ref{eq:accel_thermal}) for a partially mode-correlated disorder. {\it Source:} From Ref.\cite{garnier19}.
}
\label{fig:sec3_part_mode_correl}
\end{center}
\end{figure}

To summarize, 
%although we cannot derive a compact form of the discrete kinetic equation for the model of partially correlated disorder, 
the theoretical developments reported in \ref{app:weak_disorder_4} allow to infer that the kinetic equation is still of the form given by the discrete kinetic Eq.(\ref{eq:kin_np_disc}), and that the scaling of the rate of thermalization is still given by $L_{kin}^{disor} \sim  \Delta \beta L_{nl}^2/{\bar S_{lmnp}^2}$ in Eq.(\ref{eq:L_kin_disor}).
This scaling of the rate of thermalization has been confirmed by the numerical simulations of the NLS Eq.(\ref{eq:nls_Ap}).
The results are reported in Fig.~\ref{fig:sec3_part_mode_correl} for the same parameters of disorder as those considered in Fig.~\ref{fig:sec3_acc_dis}, except that a model of partial mode-correlated disorder has been considered. 
The  agreement with the discrete kinetic Eq.(\ref{eq:kin_np_disc}) confirms the scaling (\ref{eq:accel_thermal}) of the rate of acceleration of thermalization for the 'partially correlated' model of disorder.

\subsubsection{Corrections of the dispersion relation}

The wave turbulence approach developed in this section is based on the NLS equation with an ideal parabolic potential and the (linear) dispersion relation $\beta_p=\beta_0(p_x+p_y+1)$.
Several factors introduce perturbations to this expression of the dispersion relation, which can be  written in the form ${\tilde \beta}_p=\beta_p + b_p$, where the perturbation $b_p$ is a function of $(p_x,p_y)$ with $b_0/\beta_0 \ll 1$. An example of perturbation is the leading order correction due to angular dispersion effects in the Helmholtz equation, $b_p=(\beta_0^2/(2 k_0 n_0))(1+p_x+p_y)^{2}$, which typically can give $b_0/\beta_0 \sim 5\times 10^{-4}$ with usual beam-cleaning experimental parameters. As a matter of fact, such a  correction to the dispersion relation was considered in \cite{Podivilov2018prl} to study cascaded energy transfers among the fiber modes. Another example is provided by the well-known fact that a GRIN MMF usually exhibits deviations from the ideal parabolic shape, i.e., $V(\br) \sim |\br|^{\nu}$ with an exponent that deviates from $\nu=2$. The general expression of the eigenvalue is rather complicated: it is of the form ${\tilde \beta}_p \propto (1+p_x+p_y)^{2\nu/(\nu+2)}$ \cite{shemirani09principal}. Considering a deviation of a few percents from $\nu=2$ \cite{shen05compensation}, one has $b_0/\beta_0 \sim 5\times 10^{-3}$ with typical parameters of beam cleaning experiments.
In addition, we can notice that the truncation of the parabolic refractive index profile due to the fiber cladding can introduce perturbations of the higher-order propagation constants.
It is also important to note that the standard deviation of the fluctuations of the structural disorder of the MMF due to imperfections and external perturbations (term ${\bf D}_p {\bf a}_p$ in the modal NLS Eq.(\ref{eq:nls_Ap})) may be of the same order as the correction of the dispersion relation.

Let us discuss the possible impact of perturbations of the dispersion relations. In this respect, resonances that are exact at leading order ($\Delta \omega_{lmnp} =\beta_l+\beta_m-\beta_n-\beta_p=0$) exhibit a residual non-resonant contribution, i.e., $\Delta {\tilde \omega_{lmnp}} = {\tilde \beta}_l+{\tilde \beta}_m-{\tilde \beta}_n-{\tilde \beta}_p=\Delta b_{lmnp}$ with  $\Delta b_{lmnp}=b_l+b_m-b_n-b_p$.
The kinetic equation accounting for the correction on the dispersion relation can be derived with the above assumption  $L_d = 1/\Delta \beta \ll L_{nl} < L_{kin}^{disor}$.
The result of the convolution integral Eq.(\ref{eq:conv_int}) is approximated by 
\begin{eqnarray}
{J}_{lmnp}^{(j)} \simeq \gamma \left< {Y}_{lmnp}^{(j)}\right> 
\frac{i8\Delta \beta-\Delta b_{lmnp}}{\Delta b_{lmnp}^2+(8\Delta \beta)^2} \delta^K(\Delta \omega_{lmnp}).
\label{eq:I_lmnp_damp_Bp}
\end{eqnarray}
%where $\Delta b_{lmnp}=b_l+b_m-b_n-b_p$ and $b_p=b_0 f(p_x,p_y)$.
Proceeding as in section~\ref{sec3:disc_wt}, we obtain the discrete kinetic equation 
\begin{eqnarray}
\nonumber
\partial_z n_p(z) &=&  \frac{4 \gamma^2 \overline{\Delta \beta}}{3} \sum_{l,m,n} \frac{ \delta^K(\Delta \omega_{lmnp})}{\Delta b_{lmnp}^2+\overline{\Delta \beta}^2}   |S_{lmnp}|^2 M_{lmnp}({\bf n}) \quad  \\
&&
+  \,  \frac{32\gamma^2 \overline{\Delta \beta}}{9}  \sum_l  \frac{ \delta^K(\Delta \omega_{lp})}{\Delta b_{lp}^2+\overline{\Delta \beta}^2}
 |  s_{lp}({\bf n}) |^2 (n_l-n_p)  \quad 
\label{eq:kin_np_dsp_corr}
\end{eqnarray}
where we recall that $s_{lp}({\bf n})=\sum_{m'} S_{lm'm'p} n_{m'}$, $M_{lmnp}({\bf n})=  n_l n_m n_p+n_l n_m n_n -  n_n n_p n_m -n_n n_p n_l$, with $\Delta b_{lp}=b_l-b_p$ and $\overline{\Delta \beta}=8\Delta \beta$.
As already commented through Eq.(\ref{eq:kin_contin_dis}), the Lorentzian distribution reflects the finite bandwidth of the four-wave resonances owing to the effective dissipation $\Delta \beta$. 
Accordingly, the kinetic Eq.(\ref{eq:kin_np_dsp_corr}) conserves $E=\sum_p \beta_p n_p(z)$,  but not ${\tilde E}=\sum_p {\tilde \beta}_p n_p(z)$.
In addition, the conservation of power $N=\sum_p n_p(z)$ and the $H-$theorem of entropy growth for ${\cal S}(z)=\sum_p \log\big(n_p(z)\big)$ describe a relaxation to $n_p^{eq}=T/(\beta_p-\mu)$, i.e., the same equilibrium as in the absence of the correction on the dispersion relation ($b_p=0$). 
In summary, in the regime $\Delta \beta \gg b_0$, the Lorentzian distribution can be simplified 
$\overline{\Delta \beta}/\big(\Delta b_{lmnp}^2+\overline{\Delta \beta}^2 \big)
\rightarrow 1/\overline{\Delta \beta}$,  
and the kinetic Eq.(\ref{eq:kin_np_dsp_corr}) exactly recovers the previous  kinetic Eq.(\ref{eq:kin_np_disc}), i.e., the correction of the dispersion relation $b_p$ does not affect the rate of thermalization. On the other hand, in the regime $b_0 \sim \Delta \beta$, the correction $b_p$ leads to a deceleration of the rate of thermalization and condensation.

%%%%%%%%%%%%%%%%%%%%%%%%%%%%%%%%%%%%%%%%%%%%%%%%%%%
%%%%%%%%%%%%%%%%%%%%%%%%%%%%%%%%%%%%%%%%%%%%%%%%%%%
%%   STRONG DISORDER
%%%%%%%%%%%%%%%%%%%%%%%%%%%%%%%%%%%%%%%%%%%%%%%%%%%
%%%%%%%%%%%%%%%%%%%%%%%%%%%%%%%%%%%%%%%%%%%%%%%%%%%
\subsection{Kinetic equation with strong disorder}
\label{subsec:strong_disorder}

In this section we discuss wave turbulence in MMFs in the presence of a 'strong' disorder, that is a random coupling among non-degenerate modes of the fiber.
The regime of 'strong' disorder then differs from the previously analyzed regimes with polarization random coupling (section~\ref{subsec:weak_disorder}), which is referred to as the  'weak' disorder regime. A key distinction between the regimes of strong and weak disorder is that strong disorder prevents the conservation of beam energy during its propagation in the MMF.
On the basis of the wave turbulence theory, here we derive a kinetic equation that accounts for the presence of a time-dependent random mode coupling \cite{berti22}.
We start the analysis by considering the NLS equation with a disordered potential that conserves the total power of the light beam, i.e., we do not consider losses.
The theory describes the antagonist impacts of nonlinearity and disorder:
While strong disorder enforces a relaxation to the homogeneous equilibrium distribution of the modal components ('particle' equipartition, $n_p^{\rm eq} =\text{const}$), the nonlinear process of thermalization favours the emergence of a condensed state exhibiting
a macroscopic population of the fundamental mode ($n_0 \gg n_p$ for $p \neq 0$), and an energy equipartition among the other modes.
We will see that, despite a dominant strength of disorder, the system can exhibit a process of {\it non-equilibrium condensation} in the initial evolution stage, while the system eventually relaxes to the homogeneous equilibrium distribution dictated by strong disorder.
%The theory reveals that, aside from the expected inter-modal random coupling that inhibits condensation, there exists an intra-modal random coupling that counter-intuitively favours condensation and RJ thermalization. 
The theory is confirmed by numerical simulations of the NLS equation, which are found in quantitative agreement with the simulations of the derived kinetic equation \cite{berti22}.
%We report experiments in MMFs with an applied external stress to control the strength of disorder, which evidences the process of RJ thermalization and condensation in the presence of strong disorder.

%Aside from a recent study that considered a z-independent (i.e. stationary) disordered potential [28], fluctuations of the medium are usually introduced phenomenologically in the final kinetic equation [35]. We propose to develop a systematic method to tackle the impact of a coloured spatio-temporal disorder on the fourth-order moment equations, as well as its role on the regularization of wave resonances and the problem of achieving a closure of the hierarchy of moment equations in wave turbulence. V. M. Malkin and N. J. Fisch, Transition between inverse and direct energy cascades in multiscale optical turbulence, Phys. Rev. E 97 (2018), 032202. 14

\subsubsection{Model equation}

In order to render this section self-contained, we start the presentation of the model from the original NLS equation with a random potential: 
%governing the transverse spatial evolution of an optical beam propagating along the $z-$axis:
% of a waveguide modeled by
%a confining potential $V_0(\br)$ [with $\br = (x, y)$].
%light propagation in a nonlinear waveguide (trapping) potential
\begin{equation}
i \partial_z \psi= - \alpha \nabla^2 \psi +V(\br) \psi -\gamma |\psi|^2 \psi + \delta V(\br,z) \psi  .
\label{eq:psi}
\end{equation}
%It governs
%, in particular, 
The potential $V(\br)$ [$\br = (x, y)$] is the ideal transverse index profile of the MMF, while $\delta V(z,\br)$ is the 'time'-dependent random perturbation of the potential 
(refractive index fluctuations), which is assumed to be real-valued with zero mean, 
$\left< \delta V\right>=0$.
We recall that $\alpha=1/(2 k_0 n_{\rm co})$ and $\gamma$ denotes the nonlinear coefficient.
%The disorder being ('time') $z-$dependent, our system is of different nature than those studying the interplay of thermalization and Anderson localization \cite{cherroret15,nazarenko19,wang20,cherroret21,cherroret20b}.
%We consider the eigenvalues/eigenmodes $(\beta_j,u_j)$ solution of $\beta_j u_j = - \alpha \nabla^2 u_j +V_0(\bx) u_j$.
%We can take the eigenmodes to be real-valued.
%We assume that the $N$ modes are non degenerate, i.e. all $\beta_j$ are distinct.
We expand the field $\psi(z,\br) = \sum_{p} a_p(z) u_p(\br)$ on the basis of the $M$ real-valued eigenmodes $u_p(\br)$ (solution of $\beta_p u_p = - \alpha \nabla^2 u_p +V(\br) u_p$) of the unperturbed waveguide. 
The mode amplitudes $a_p(z)$ then satisfy 
\begin{equation}
i\partial_z a_p = \beta_p a_p -\gamma \sum_{l,m,n} S_{plmn} a_l a_m a_n^* +\sum_{l}C_{pl}(z) a_p ,
\label{eq:a_j}
\end{equation}
where 
%$S_{plmn}=\int u_p(\br) u_l(\br) u_m(\br)  u_n(\br) d\br$ denotes the mode overlap, and 
the random mode coupling matrix reads
\begin{equation}
C_{pl}(z) = \int u_p(\br) \delta V(z,\br)  u_l(\br) d\br .
\end{equation}
The stochastic NLS Eq.(\ref{eq:psi}) and the modal NLS Eq.(\ref{eq:a_j}) are fully equivalent.
They both conserve the total power $N=\int |\psi|^2 d\br = \sum_p |a_p|^2$, while the random potential $\delta V(\br,z)$ in Eq.(\ref{eq:psi}) (or the random matrix 
${\bf C}(z)$ in Eq.(\ref{eq:a_j})), prevents the conservation of the energy (Hamiltonian).

\subsubsection{Kinetic equation}

Let us consider the situation where the random potential is a weak perturbation with respect to linear propagation effects ($\delta V \ll V$), i.e. $L_{lin} = 1/\beta_0 \ll L_{dis} =1/\sigma_\beta$ and $L_{lin} \ll l_\beta$, 
%where $\beta_0$ is the fundamental mode eigenvalue and 
where we recall that $\sigma_\beta^2$ denotes the variance of the fluctuations of the random potential (i.e., 'strength' of disorder) and $l_\beta$ denotes the corresponding correlation length.
Note that this is the usual case in an optical waveguide configuration, e.g., in MMFs.
Furthermore,  we assume that the disorder dominates over nonlinear effects $L_{dis} \ll L_{nl})$.
Note that this assumption differs from that required in the theory developed for weak disorder (see Eq.(\ref{eq:Ld_small_Lnl})), since here $L_{dis}$ does not depend on the correlation length.

Combining the (discrete) wave turbulence theory \cite{Zakharov05,kartashova08,kartashova09,Lvov10,Nazarenko11,mordant18} 
with tools developed for the asymptotic analysis of randomly driven  linear differential equations (time-dependent disorder) \cite{fouque07}, we derive in the \ref{app:strong_disorder} the kinetic equation governing the evolution of the averaged modal components $n_p(z)=\left< |a_p(z)|^2\right>$:
%We extend the wave turbulence theory by using tools developed for the asymptotic analysis of randomly driven linear differential equations.  
%We develop a wave turbulence theory 
%\cite{zakharov92,Newell01,nazarenko11,Newell_Rumpf,shrira_nazarenko13,
%laurie12,Lvov10,onorato15,PR14} accounting for a time-dependent disorder by exploiting tools inherited from the asymptotic analysis of randomly driven ordinary differential equations \cite{fouque07}.  
%We derive in the \ref{app:strong_disorder} the kinetic equation governing the evolution of the averaged modal components $n_p(z)=\left< |a_p(z)|^2\right>$:
% On the basis of the wave turbulence theory and using tools inhererited from asymptotic analysis of randomly driven  differential equations (time-dependent disorder) \cite{fouque07}
\begin{equation}
\partial_z n_p  = 
\sum_{l\neq p} \Gamma_{pl}^{\rm OD} \big(n_l-n_p \big) + 8 \gamma^2   \sum_{l,m,n}  \frac{\delta_{plmn}^K  S_{plmn}^2}{G_{plmn}^{\rm D}} {R}_{plmn}[{\bf n}],
\label{eq:kin}
\end{equation}
where 
\begin{align*}
{R}_{plmn}[{\bf n}](z)=&n_l(z) n_m(z) n_p(z) + n_l(z) n_m(z) n_n(z) \\
& -  n_p(z) n_n(z) n_m(z) - n_p(z) n_n(z) n_l(z),
%\label{eq:kin}
\end{align*}
and the Kronecker symbol denotes a frequency resonance ($\delta_{plmn}^K=1$ if $\Delta \beta_{plmn}=\beta_p-\beta_l-\beta_m+\beta_n=0$, and zero otherwise).
For clarity, we assume that the modes are not degenerate. We refer the reader to the \ref{app:strong_disorder} for the kinetic equation accounting for mode degeneracies.

The kinetic Eq.(\ref{eq:kin}) unveils the interplay of nonlinearity and disorder:
it reveals that diagonal and off-diagonal elements of the random matrix ${\bf C}$ play fundamental different roles. The first term in the kinetic equation (\ref{eq:kin}) originates in off-diagonal elements of $C_{pl}$ ($p \neq l$):
\begin{equation}
\label{eq:defGammaOD}
\Gamma^{\rm OD}_{pl}= 2 \int_0^\infty\left<C_{pl}(0)C_{pl}(z)\right> \cos \big((\beta_p-\beta_l)z\big) dz.
\end{equation}
This equation describes an irreversible relaxation toward the homogeneous distribution, characterized by an equipartition of power ('particles') among the modes, $n_p^{\rm eq}=N/M=$const.
This process occurs over the typical propagation length 
\begin{equation}
{\cal L}^{\rm eq}_{\rm kin}\simeq  1/\overline{\Gamma_{jl}^{\rm OD}}.
\end{equation}
This is the well-known evolution of a system ruled by strong random mode coupling.
To be more precise, we note that ${\cal L}^{\rm eq}_{\rm kin}= 1/\lambda_2^{\rm OD}$ where $-\lambda_2^{\rm OD}$ is the second eigenvalue of the matrix $\tilde{\boldsymbol{\Gamma}}^{\rm OD}$, with $\tilde{\Gamma}_{jl}^{\rm OD}=\Gamma^{\rm OD}_{jl}$ for $j \neq l$ and $\tilde{\Gamma}^{\rm OD}_{jj}=-\sum_{l \neq j}\Gamma^{\rm OD}_{jl}$, the first eigenvalue being $\lambda_1^{\rm OD}=0$.

%The main result of our work is to 
We now show that this relaxation process mediated by strong disorder does not necessarily inhibit the nonlinear processes of thermalization and condensation.
This becomes apparent through the analysis of the collision term in the kinetic equation (\ref{eq:kin}), which exclusively involves the diagonal components $C_{pp}$:
\begin{equation}
\Gamma^{\rm D}_{pl}= 
 \int_0^\infty \left<C_{pp}(0)C_{ll}(z)\right> + \left<C_{ll}(0)C_{pp}(z)\right> dz.
\end{equation}
The matrix ${\bf \Gamma}^{\rm D}$ contributes to the tensor involved in the collision term (see the \ref{app:strong_disorder}): 
\begin{align*}
G_{plmn}^{\rm D}=\Gamma^{\rm D}_{ll}+\Gamma^{\rm D}_{mm}+\Gamma^{\rm D}_{nn}+\Gamma^{\rm D}_{pp}+2\Gamma^{\rm D}_{lm}
 -2\Gamma^{\rm D}_{ln}-2\Gamma^{\rm D}_{lp}-2\Gamma^{\rm D}_{mn}-2\Gamma^{\rm D}_{mp}+2\Gamma^{\rm D}_{np}.
\end{align*}
%$G_{plmn}^{\rm D}=\Gamma^{\rm D}_{ll}+\Gamma^{\rm D}_{mm}+\Gamma^{\rm D}_{nn}+\Gamma^{\rm D}_{pp}+2\Gamma^{\rm D}_{lm}
%-2\Gamma^{\rm D}_{ln}-2\Gamma^{\rm D}_{lp}-2\Gamma^{\rm D}_{mn}-2\Gamma^{\rm D}_{mp}+2\Gamma^{\rm D}_{np}$
To discuss the role of the collision term, let us forget for a while the first term in the kinetic equation (\ref{eq:kin}).
%\blue{Following the usual proof \cite{zakharov92}, it can be shown that} 
The collision term conserves $N=\sum_p n_p(z)$, $E=\sum_p \beta_p n_p(z)$, and exhibits a $H-$theorem of entropy growth $\partial_z S \ge 0$, with $S(z)=\sum_p \log[n_p(z)]$, see \ref{app:strong_disorder}.
Hence, the collision term describes a process of thermalization to the RJ distribution $n_p^{\rm RJ}=T/(\beta_p-\mu)$, exhibiting energy equipartition among the modes.
This occurs over a typical propagation length 
\begin{equation}
{\cal L}^{\rm RJ}_{\rm kin} \simeq L_{nl}^2 \overline{G_{plmn}^{\rm D}/S_{plmn}^2 }.
\end{equation}

\begin{figure}
\begin{center}
\includegraphics[width=.65\columnwidth]{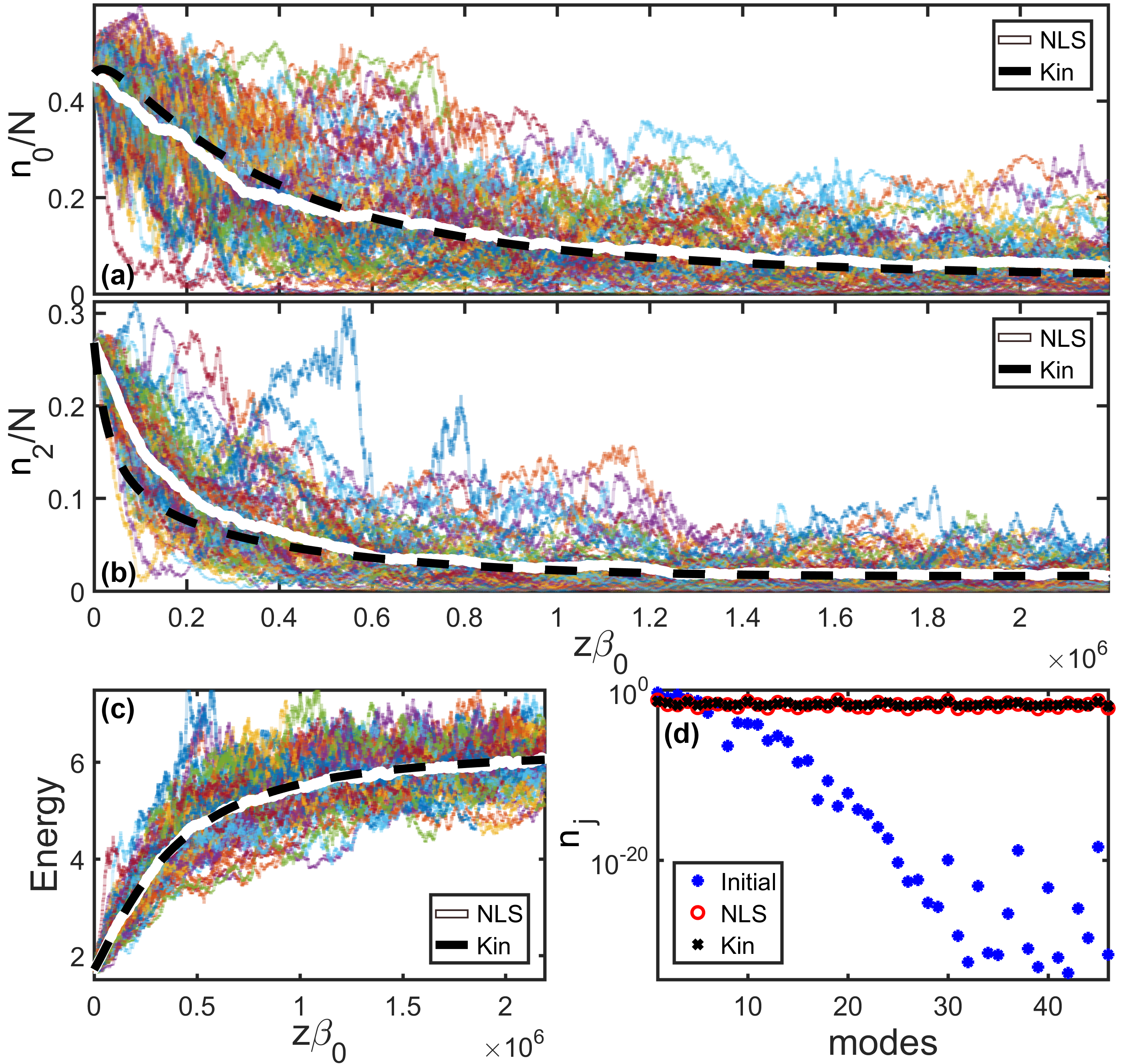}
\caption{
\baselineskip 10pt
Dynamics dominated by strong disorder ${\cal L}^{\rm RJ}_{\rm kin} \gg {\cal L}^{\rm eq}_{\rm kin}$: 
The system irreversibly relaxes toward the equilibrium $n_p^{\rm eq}$.
Evolutions of the fundamental mode $n_0(z)$ (a), and $n_2(z)$ (b), the energy $E(z)/(N \beta_0)$ (c), obtained from the numerical simulation of the NLS Eq.(\ref{eq:a_j}): 64 realizations are reported with colored lines; 
the bold white line is the corresponding empirical average;
the dashed black line is the prediction of the KE (\ref{eq:kin}).
(d) Modal distribution $n_p$ in the initial condition ($z=0$, blue) and at $z \beta_0=2 \times 10^6$ for the NLS simulation (red), and the kinetic equation (black).
Parameters: $L_{dis}/L_{lin}=7$, $L_{dis}/L_{nl}= 4.1 \times 10^{-4}$, $l_\beta \beta_0=42$.
{\it Source:} From Ref.\cite{berti22}.
% ${\cal L}^{\rm RJ}_{\rm kin} / {\cal L}^{\rm eq}_{\rm kin} \simeq 400$. 
%Parameters: 46 modes, $\beta_0 = 8371$m, $\sigma_\beta =1200$m$^{-1}$, $L_{NL0} = 2$m, $l_\beta = 5$mm, $b_x = 0.4$, $b_y = 0.5$, 64 realizations, $\beta_{0x} = \sqrt{2} \beta_{0y}$.
%[The parameters in the simulations should be given in term of $\beta_{p_x=0,p_y=0}=\beta_0=(\beta_{0,x}+\beta_{0,y})/2=\sqrt{\alpha}(\sqrt{q_x}+\sqrt{q_y})$: $\sigma_\beta/\beta_0=?$, $\beta_0 l_\beta=?$, $\beta_0 L_{nl} = ?$.]
}
\label{fig:sec3_strong1}
\end{center}
\end{figure}

\begin{figure}
\begin{center}
\includegraphics[width=.65\columnwidth]{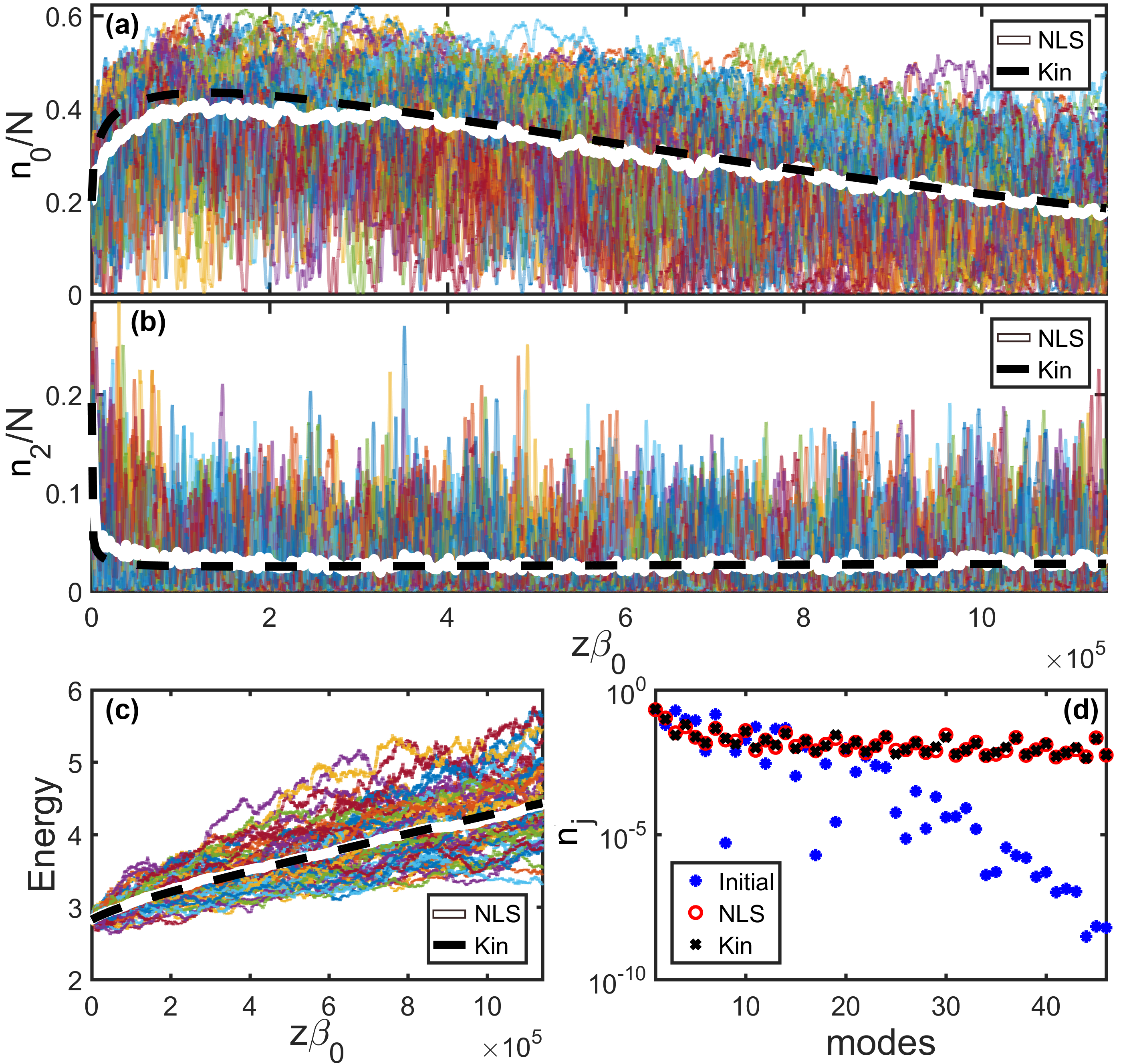}
\caption{
\baselineskip 10pt
Thermalization precedes equilibrium relaxation:
Same panels as in Fig.~\ref{fig:sec3_strong1}, but in the regime ${\cal L}^{\rm RJ}_{\rm kin} \lesssim {\cal L}^{\rm eq}_{\rm kin}$.
The system exhibits an incipient process of RJ thermalization and nonequilibrium condensation characterized by a growth of the condensate amplitude $n_0(z)$ for $z \beta_0 \lesssim 2\times 10^5$.
Disorder subsequently prevails, which induces a decay of $n_0(z)$ (and eventually brings the system to equilibrium $n_p^{\rm eq}$). 
%Evolutions of the fundamental mode $n_0(z)$ (a), and $n_2(z)$ (b), the energy $E(z)/(N \beta_0)$ (c), obtained from the numerical simulation of the NLS Eq.(\ref{eq:a_j}): 64 realizations are reported with colors and the corresponding average with the white line. Simulation of the kinetic Eq.(\ref{eq:kin}) (dashed lack line). (d) Modal distribution $w_j$ in the initial condition ($z=0$, blue) and at $z \beta_0=8 \times 10^5$ for the NLS simulation (red, averaged over the 64 realizations), and the kinetic equation (black).
Parameters: $L_{dis}/L_{lin}=7$, $L_{dis}/L_{nl}= 0.033$, $l_\beta \beta_0=167$.
{\it Source:} From Ref.\cite{berti22}.
% ${\cal L}^{\rm RJ}_{\rm kin} / {\cal L}^{\rm eq}_{\rm kin} \simeq 0.07$. 
%Parameters: 46 modes, $\beta_0 = 8371$m, $\sigma_\beta =1200$m$^{-1}$, $L_{NL0} = 25$mm, $l_\beta = 20$mm, $b_x = 0.4$, $b_y = 0.5$, 64 realizations, $\beta_{0x} = \sqrt{2} \beta_{0y}$.
}
\label{fig:sec3_strong2}
\end{center}
\end{figure}

\begin{figure}
\begin{center}
\includegraphics[width=.65\columnwidth]{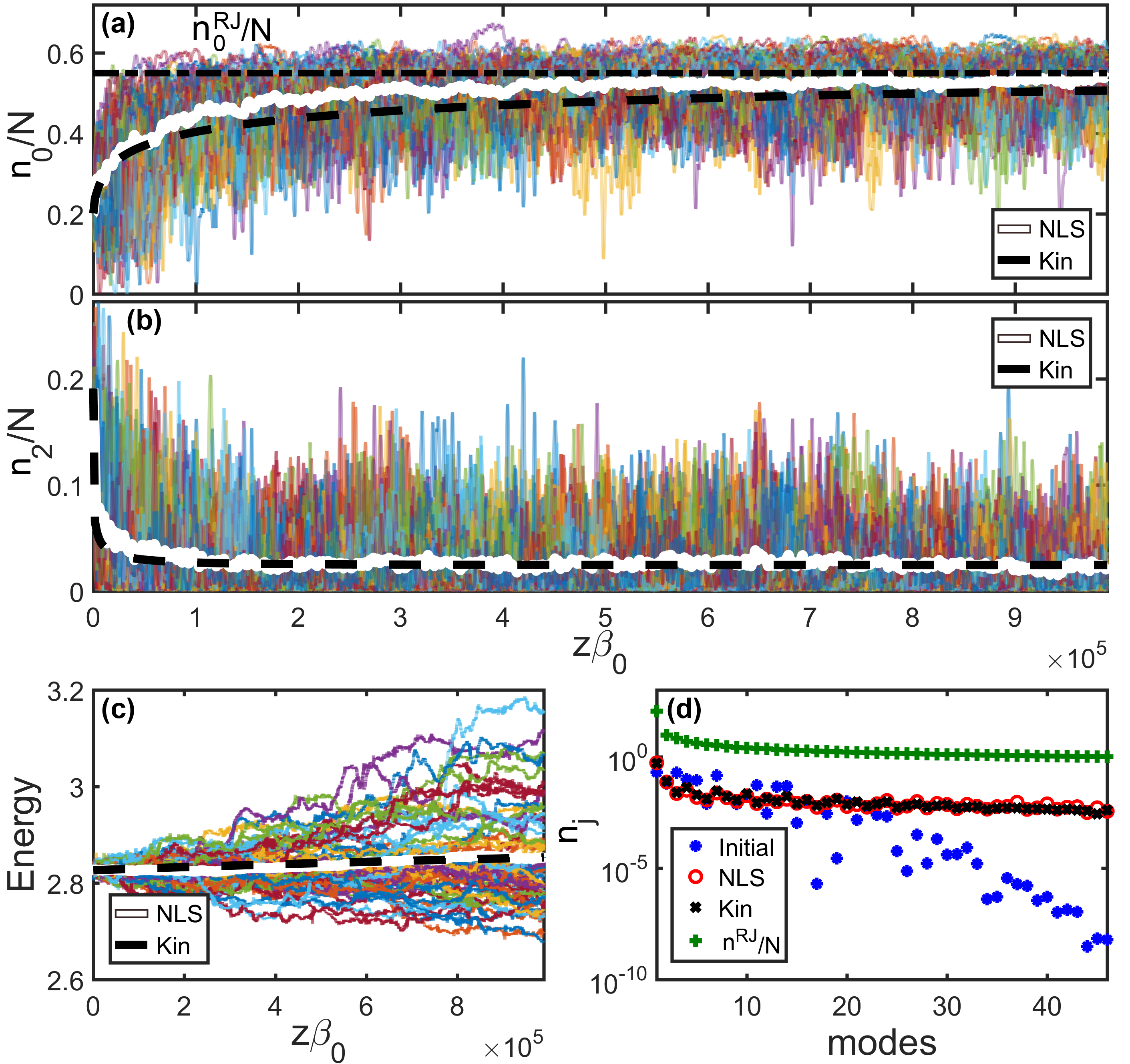}
\caption{
\baselineskip 10pt
Thermalization prevails over equilibrium relaxation:
Same panels as in Fig.~\ref{fig:sec3_strong1}, but in the regime ${\cal L}^{\rm RJ}_{\rm kin} \ll {\cal L}^{\rm eq}_{\rm kin}$.
The system exhibits a process of RJ thermalization and condensation characterized by a significant growth of the condensate amplitude $n_0(z)$ to the value predicted by the  RJ distribution, $n_0^{\rm RJ}/N \simeq 0.55$ (horizontal dashed-dotted black line) (a).
At variance with Figs.~\ref{fig:sec3_strong1}-\ref{fig:sec3_strong2}, the energy $E(z)$ is almost constant (c).
The modes approach the RJ distribution $n_p^{\rm RJ}$ (green) (d).
%Disorder subsequently prevails and eventually brings the system to equilibrium $w_j^{\rm eq}$. 
%Evolutions of the fundamental mode $n_0(z)$ (a), and $n_2(z)$ (b), the energy $E(z)/(N \beta_0)$ (c), obtained from the numerical simulation of the NLS Eq.(\ref{eq:a_j}): 64 realizations are reported with colors and the corresponding average with the white line. Simulation of the kinetic Eq.(\ref{eq:kin}) (dashed lack line). (d) Modal distribution $w_j$ in the initial condition ($z=0$, blue) and at $z \beta_0=8 \times 10^5$ for the NLS simulation (red, averaged over the 64 realizations), and the kinetic equation (black).
Parameters: $L_{dis}/L_{lin}=8.4$, $L_{dis}/L_{nl}= 0.04$, $l_\beta \beta_0=4 \times 10^3$.
{\it Source:} From Ref.\cite{berti22}.
% ${\cal L}^{\rm RJ}_{\rm kin} / {\cal L}^{\rm eq}_{\rm kin} \simeq 0.003$. 
%Parameters: 46 modes, $\beta_0 = 8371$m, $\sigma_\beta =1000$m$^{-1}$, $L_{NL0} = 25$mm, $l_\beta = 0.5$m, $b_x = 0.3$, $b_y = 0.4$, 64 realizations, $\beta_{0x} = \sqrt{2} \beta_{0y}$.
}
\label{fig:sec3_strong3}
\end{center}
\end{figure}

As discussed in the \ref{app:condensate_fraction}, for an energy smaller than the critical value $E \le E_{\rm crit}$, the RJ distribution $n_p^{\rm RJ}$ exhibits a phase transition to a condensed state \cite{Baudin2020}.
The condensate amplitude $n_{0}$ then constitutes the natural parameter that distinguishes the two antagonist regimes:\\
(i) For ${\cal L}^{\rm eq}_{\rm kin} \ll {\cal L}^{\rm RJ}_{\rm kin}$, the disorder dominates and 
$n_0(z) \to n_p^{\rm eq}=N/M=$const for $p=0,1,...,M-1$;\\
(ii) For ${\cal L}^{\rm eq}_{\rm kin} \gg {\cal L}^{\rm RJ}_{\rm kin}$, the dynamics is dominated by RJ thermalization, and  condensation leads to a macroscopic population of the fundamental mode $n_0(z) \to n_0^{\rm RJ} \gg n_{p}^{\rm RJ}$ for $p=1,...,M-1$.

\subsubsection{Numerical simulations}

We have performed extensive numerical simulations to test the validity of our theory.
We have considered the concrete example of a parabolic trapping potential of the form $V(\br)=q_x x^2 + q_y y^2$, with the fundamental mode eigenvalue $\beta_0=\sqrt{\alpha}(\sqrt{q_x}+\sqrt{q_y})$.
%Aside from Bose-Einstein condensates, the parabolic-shaped trapping models graded-index MMFs where 
%spatial beam cleaning \cite{krupa16,wright16,liu16,krupa17}, as well as 
%light thermalization has been recently observed experimentally \cite{PRL20,EPL21,wise_arxiv,mangini22}.
Let us consider a general model of disorder with a random potential of the form $\delta V(\br,z)=\mu(z) g(\br)$, where $\mu(z)$ is a real-valued stochastic function with zero mean and $\left< \mu(0) \mu(z)\right>= \sigma_\beta^2 \exp(-|z|/l_\beta)$.
In order to remove mode degeneracies, we consider in the simulations an elliptical parabolic potential ($q_x \neq q_y$). 
To compute the matrices ${\bf \Gamma}^{\rm OD}$ and ${\bf \Gamma}^{\rm D}$ in analytical form, we consider the random mode coupling 
$g(x,y)=\cos(b_x x/x_{0})\cos(b_y y/y_{0})$, where $(x_{0},y_{0})$ denote the radii of the fundamental elliptical mode, see \ref{app:strong_disorder}.

According to the theory, the two terms in the kinetic equation (\ref{eq:kin}) are antagonists and compete against each other.
If ${\cal L}^{\rm eq}_{\rm kin} \ll {\cal L}^{\rm RJ}_{\rm kin}$, disorder prevails and the system relaxes to the expected equilibrium $n_p^{\rm eq}=$const.
This is illustrated in Fig.~\ref{fig:sec3_strong1}, which reports the results of the numerical integration of the NLS Eq.(\ref{eq:a_j}) for 64 realizations (${\cal L}^{\rm RJ}_{\rm kin} / {\cal L}^{\rm eq}_{\rm kin} \simeq 400$).
The corresponding average over such realizations (bold white line) is in agreement with the  simulation of the kinetic equation (\ref{eq:kin}) (dashed black line) starting from the same initial condition.
Here and thereafter, the quantitative agreement between NLS and kinetic equation simulations is obtained without any adjustable parameter.

Unexpectedly, however, a {\it nonequilibrium} process of condensation and thermalization can be observed in the initial stage of propagation when ${\cal L}^{\rm RJ}_{\rm kin} \lesssim {\cal L}^{\rm eq}_{\rm kin}$ (see Fig.~\ref{fig:sec3_strong2} for ${\cal L}^{\rm RJ}_{\rm kin} / {\cal L}^{\rm eq}_{\rm kin} \simeq 0.07$), while asymptotically the system still relaxes to the homogeneous equilibrium state $n_p^{\rm eq}$.
The nonequilibrium property of condensation is reflected by the fact that the energy $E(z)=\sum_p \beta_p n_p(z)$ {\it is not conserved during the evolution}, see Fig.~\ref{fig:sec3_strong2}(c).
%Note that the parameters in Fig.~2 verify $L_{dis} \sim \sqrt{L_{nl} L_{lin}}$.
%\blue{Note that, the introduction of losses, distributed either homogeneously or non-homogeneously amongst the modes, does not significantly affect the condensate peak in Fig.~2 \cite{supplement}.}

We stress that the condensation processes may occur very efficiently by increasing the correlation length $l_\beta$, in such a way that ${\cal L}^{\rm RJ}_{\rm kin} \ll {\cal L}^{\rm eq}_{\rm kin}$, see Fig.~\ref{fig:sec3_strong3} for ${\cal L}^{\rm RJ}_{\rm kin} / {\cal L}^{\rm eq}_{\rm kin} \simeq 0.003$. 
In this regime, the energy is almost conserved $E \simeq$~const and RJ thermalization occurs almost completely, as confirmed by the modal populations that approach the RJ distribution $n_p^{\rm RJ}$ (Fig.~\ref{fig:sec3_strong3}(d)), and the condensate approaches the RJ prediction $n_0^{\rm RJ}/N \simeq 0.55$, see Fig.~\ref{fig:sec3_strong3}(a). 
Note that a similar effect of pre-thermalization and condensation has been identified recently in nonlinear disordered Floquet systems \cite{haldar24rayleigh}.
% in the absence of random mode coupling (disorder).

It is interesting to remark that, although random mode coupling does not affect the RJ equilibrium, it significantly accelerates the rate of thermalization to this equilibrium.
Indeed, the ratio of thermalization in the presence and the absence of disorder scales as $\sim {\bar G}^{\rm D}/ \beta_0 \ll 1$.
This effect of acceleration of thermalization is very similar to that pointed out in the previous section through the analysis of weak disorder, see Sec.~\ref{sec3:acc_thermal}.

\subsection{Conclusions of the wave turbulence approach}

On the basis of the wave turbulence theory, we have derived kinetic equations describing the nonequilibrium evolution of random waves in a regime where disorder dominates nonlinear effects ($L_d \ll L_{nl}$).
The theory revealed that the presence of a conservative weak disorder introduces an effective dissipation in the system, whose resonance broadening prevents the conservation of energy, which inhibits the effect of condensation within the usual continuous wave turbulence approach.
However, usual experiments of beam-cleaning are not described by the continuous wave turbulence theory, but instead by a discrete wave turbulence approach.
In this discrete turbulence regime only exact resonances contribute to the kinetic equation, which is no longer sensitive to the effect of dissipation-induced resonance broadening. 
Accordingly, the discrete kinetic equation conserves the energy, which re-establishes the process of wave condensation.
The main result is that the effective dissipation induced by disorder modifies the regularization of such discrete resonances, which leads to an acceleration of the rate of thermalization and condensation.

We have considered different models of weak disorder in MMFs.
The theory reveals that when all  modes experience the same (mode-correlated) noise, dissipation induced by disorder vanishes, and the system no longer exhibits a fast process of condensation. 
However, even a relative small decorrelation among the noise experienced by the modes is sufficient to re-establish a disorder-induced acceleration of condensation. 
The simulations are in quantitative agreement with the theory without using adjustable parameters, in a regime where disorder dominates nonlinear effects ($L_d \ll L_{nl}$). 
However, the impact of weak disorder due to polarization fluctuations in beam cleaning experiments is expected to be of the same order as that of nonlinear effects \cite{ho14}.
Accordingly, the impact of a moderate disorder ($L_d \gtrsim L_{nl}$) was considered in Ref.\cite{garnier19}. 
In this regime, the phase-correlations among the modes are not completely suppressed, and the system enters a mixed coherent-incoherent regime. 
Accordingly, the modal components can exhibit an oscillatory behavior that slows down the thermalization process, although the optical beam still exhibits a fast condensation process that is consistent with the experimental results of RJ thermalization. 
In addition, the analysis of polarization effects reported in \cite{garnier19} revealed that optical beam cleaning is responsible for an effective partial repolarization of the central part of the beam. 
This property was observed experimentally in Ref.\cite{Krupa:19} and it can be interpreted as a consequence of the condensation-induced macroscopic population of the fundamental mode of the MMF. 
Then considering the dominant contribution of weak disorder originating in polarization fluctuations, the nonequilibrium kinetic theory and simulations both provide a qualitative understanding of the effect of optical beam self-cleaning, in particular when a large number of modes are excited into the MMF, as in Ref.\cite{Baudin2020}.
On the other hand, the discrete nature of wave turbulence in MMFs is responsible for an effective freezing of thermalization and condensation when only a small number of modes is excited.
Such an effective freezing of condensation may also explain why beam cleaning has not been clearly observed in step-index MMFs \cite{garnier19}.

It is important to recall that, at variance with the experiments of beam cleaning where sub-nanosecond pulsed are injected into the MMF, in this section we considered a purely spatial model of light propagation in the fiber. 
This is justified by the fact that in the nanosecond regime temporal dispersion effects play a negligible role at leading order. 
Then although the theory can describe a mechanism of beam cleaning condensation, it cannot provide a complete  description of the experimental results.
%, which would require a detailed analysis of the temporal averaging effect inherent to the pulsed regime considered in the experimental measurements. 

In the second part of this section we have derived a wave turbulence kinetic equation in the presence of strong disorder. 
The theory  revealed that RJ thermalization and condensation can even take place when the dynamics is dominated by strong disorder ($L_{dis}/L_{nl} \ll 1$), provided that ${\cal L}_{\rm kin}^{ \rm RJ} \ll {\cal L}_{\rm kin}^{eq}$.
%Experiments in MMFs confirm, at a qualitative level, that RJ thermalization and condensation do occur in the presence of strong disorder (strong mode coupling).
%At variance with weak disorder, strong disorder introduces very large fluctuations of the modal components $w_j$ (see NLS simulations in Figs.~1-3), whose average over the realizations is found in quantitative agreement with the simulations of the kinetic Eq.(\ref{eq:kin}), {\it without using any adjustable parameter}.
Note that, at variance with weak disorder, strong disorder introduces very large fluctuations of the modal components $n_p$ (see NLS simulations in Figs.~\ref{fig:sec3_strong1}-\ref{fig:sec3_strong3}).
The comparison with the theory then require an average over the realizations, which is found in quantitative agreement with the simulations of the kinetic Eq.(\ref{eq:kin}), without using adjustable parameters.

Experiments carried out using relatively short pieces of MMFs ($\sim$12m) have been reported in Ref.\cite{berti22} to test the theory at a purely qualitative level.
Strong mode coupling was obtained by applying a stress to the MMF with clamps~\cite{cao16}. 
By adjusting the applied stress, the strength of mode coupling could be tuned.
Strong random mode coupling leads to an increase of the energy (Hamiltonian) -- the larger the strength of applied stress, the larger the increase of energy.
The experiments show that in the presence of significant applied stress, RJ thermalization and condensation are inhibited.
By decreasing the amount of applied stress, the experiments confirm that RJ condensation may still occur, even in the presence of strong disorder \cite{berti22}.

It is important to note that an effect of beam cleaning of different nature than RJ condensation has been recently observed experimentally in MMFs in the linear regime of propagation, over very large propagation distances of the order of the kilometer, see Ref.\cite{Zitelli:23}. With such large propagation lengths, strong random mode coupling in MMFs can no longer be neglected. Taking into account for the losses due to a linear coupling between the highest order guided modes and the radiative modes, a non-homogeneous solution predicting a larger population of the fundamental fiber mode was obtained \cite{olshansky75} (also see \cite{marcuse74}). This can provide a natural explanation for the remarkable effect of beam cleaning observed experimentally in the linear regime in Ref.\cite{Zitelli:23}.

The linear propagation of scalar waves in randomly perturbed open waveguides was also addressed in Ref.\cite{garnier20}. From first principles and by using a multiscale analysis, it was shown that the coupling between guided and radiative modes involves a perturbed form of 'particle' equipartition: the mean powers of the guided modes become proportional to the first eigenvector of an effective scattering matrix that is a refined version of (\ref{eq:defGammaOD}), and that is determined by the second-order statistics of the waveguide fluctuations. When the coupling from guided to radiative modes is negligible compared to the coupling between guided modes, this eigenvector is uniform and the standard homogeneous equipartition result is recovered. But in general the eigenvector is not uniform ,and it can exhibit a concentration close to the lowest eigenmode, for which the coupling towards the radiative modes is the smallest.

%\begin{align}
% f(x) &= (x+a)(x+b) \\
%      &= \sum_i=1^N{f(a)b^io}
%\end{align}

%\begin{figure*}[t]
%	\includegraphics[width=0.8\linewidth]{Figures_Sec3/Toy.pdf}
%	\centering
%	\caption{caption caption caption caption}
%	\label{fig:figure3}%
%\end{figure*}

\section{Experimental evidences}\label{Sec_4}
\begin{figure}[ht!]
\centering\includegraphics[width=12cm]{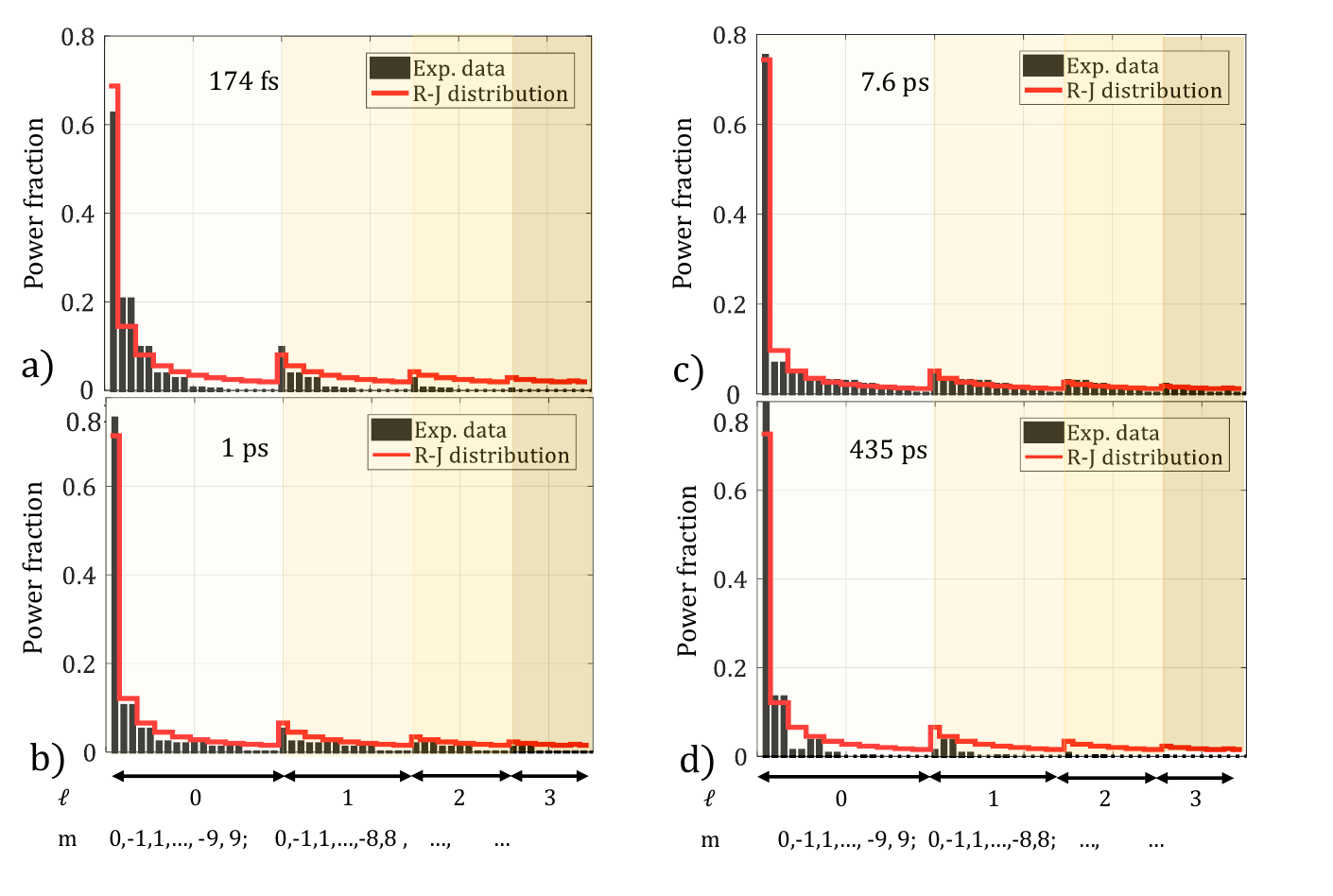}
\caption{Illustration of the mode power distribution when sorting the modes by their radial ($\ell$) and azimuthal ($m$) indices, for input pulse duration of 174 fs (a), 1 ps (b), 7.6 ps (c), and 435 ps (d). The black bars are the experimental values, while the red lines are the theoretical RJ distributions. Image from \cite{Mangini2022Statistical}.}
\label{radial}
\end{figure}

In this Section we present a series of recent experimental studies, aiming at validating the predictions of the thermodynamic or statistical mechanics approach. In these works, pulsed lasers with peak powers of a few tens of kW with pulse durations ranging between hundreds of femtoseconds to nanoseconds were used. In these conditions, it has been verified that thermodynamic equilibrium can be reached for beams propagating over the relatively short distances of a few meters of standard GRIN MMF. With such power levels and propagation distances, one fulfills the theoretical hypotheses of conservation of the optical beam power (or 'number of particles') and kinetic energy  upon propagation in the MMF, given that both linear and nonlinear optical losses can be neglected \cite{Agrawal2013,ferraro2021femtosecond}, as discussed in section~\ref{Sec_3}. However, working with short pulses may lead to additional nonlinear effects, that modify their temporal and spectral properties. The main effects here are self-phase modulation, Raman scattering, and geometric parametric instability, which produce a progressive beam spectral broadening upon propagation. In our statistical description of BSC, we have considered purely monochromatic waves, disregarding temporal aspects. To alleviate this, experimental measurements were typically carried out either by integrating pulses over time or, as an alternative, by inserting at the fiber output a narrow bandpass filter centered at the laser wavelength. In this regard, it is important to mention that Mangini et al. showed that the RJ distribution at thermal equilibrium may be reached, for input pulse durations ranging from hundredths of ps down to hundredths of fs (see Fig. \ref{radial}). 

Direct observations of the attainment of the RJ equilibrium were made possible by the development of holographic MD tools. A notable example is that of MD systems based on phase-only spatial light modulators, which allow for fully decomposing a beam in its modal components, i.e., both the mode amplitude and phase can be retrieved \cite{gervaziev2020mode}. Other approaches exploited, instead, interferometric setups, e.g., in Ref. \cite{pourbeyram2022direct}.
Generally speaking, holographic methods were quite successful in experimental studies of the statistical mechanics theory.  

Within this experimental framework, many investigations of the BSC effect were reported in the literature \cite{pourbeyram2022direct,Mangini2022Statistical,mangini2024maximization,ferraro2024calorimetry,mangini2023modal,mangini2023high,ferraro2023spatial,ferraro2023multimode}. Still, a few papers studied the thermalization process of spatially structured light (see Ref. \cite{baudin2023rayleigh} for a comparison between wave thermalization and BSC). In particular, some experiments have shown that, when increasing the beam power well above the self-cleaning threshold, a thermalized beam may still yield a speckled intensity pattern~\cite{mangini2023high}. In addition, the presence (and conservation) of the orbital angular momentum of light may lead to thermalized beams without a bell shape \cite{podivilov2022thermalization}. In this regard, in Fig. \ref{radial} one may note that each couple of bars corresponding to modes with the same value of $\ell$, and opposite signs of $m$ have the same height.
Furthermore, as an ultimate case, whenever a beam is associated with a negative temperature, the occupancy of the fundamental mode tends to vanish, thus making it virtually impossible to achieve a Gaussian-like intensity profile \cite{baudin2023rayleigh}.

When it comes to BSC, it may seem counterintuitive that as the beam cleans up, its associated entropy increases, i.e., the disorder grows larger. %a bell-shaped beam has a higher entropy (i.e., is more disordered) than a speckled beam. 
This has led to speculations that the whole statistical mechanics approach to BSC fails because of a breach of the second law of thermodynamics \cite{steinmeyer2023generalized}.
Nevertheless, the growth of entropy during BSC was observed in several experiments. Baudin et al. have determined the equilibrium entropy in the classical RJ condensation in a GRIN MMF, proving that the equilibrium entropy decreases with the kinetic energy \cite{Baudin2020}. 
%\textcolor{blue}{Actually, beam-cleaning condensation is  driven by an energy flow toward the higher-order modes (direct energy cascade), and a bi-directional redistribution of the power to the fundamental mode and to higher-order modes, see e.g., \cite{baudin21}.} 
%\textcolor{red}{I'm not sure that it is worth saying this here. This concept has already been said in Sec. 1 and (perhaps) in Sec. 2. Since here we are introducing experiments, I'd keep it as less theoretical as possible.}
%In their experiments, the input power $\mathcal{P}$ is kept constant, and the beam energy $\mathcal{U}$ is varied by controlling the spatial correlation of the speckle input beam with a diffuser. As a matter of fact, the thermodynamic theory considers $\mathcal{P}$ and $\mathcal{U}$ as constants of motion. However, 

Still, verifying that BSC is consistent with the second principle of thermodynamics means verifying that the entropy reaches a maximum value when thermal equilibrium is attained. This poses the challenges of experimentally measuring the entropy of out-of-equilibrium states. 

In the following, we will go through recent experimental results that have achieved such a goal \cite{mangini2024maximization, ferraro2024calorimetry}. Then, we will present the cases of wave thermalization in three significant cases, i.e., at high, near zero, and negative temperatures. Finally, we will shortly review the recent discovery of a modal phase-locking mechanism which accompanies BSC.

\subsection{The growth of optical entropy}
\label{sec_exp_entropy_growth}
A remarkably simple, and yet effective way of writing the Boltzmann entropy was proposed by Wu et al. in \cite{Wu2019}. This reads
\begin{equation}
S = \sum_p \ln n_p,
\label{boltzmann}
\end{equation}
where the sum runs over all guided modes. Strictly speaking, such a formula only holds at thermal equilibrium. However, it is interesting to note that the same expression of the entropy also holds out of equilibrium, where the modal populations depend on the propagation length, as discussed through the $H-$theorem of entropy growth in the previous section~\ref{Sec_3}, see Eqs.(\ref{eq:ent_neq}). Accordingly, the nonequilibrium entropy grows upon beam propagation, and its growth saturates as the field approaches thermal equilibrium, where the entropy reaches its equilibrium value. In this way, the entropy given in Eq.(\ref{boltzmann}) (with $n_p$ given by the RJ distribution) provides the upper bound of the out-of-equilibrium entropy. As discussed in section~\ref{Sec_3}, this occurs under the hypothesis that the fiber modes have Gaussian statistics, which may result from the presence of random linear mode coupling (i.e., fluctuations of the index of refraction).

According to (\ref{boltzmann}), to experimentally verify the entropy growth one needs to track the evolution of the occupation of the modes, i.e., the values of the $\{n_p\}$, over ``time". As discussed above, the role of time in nonlinear wave thermalization is played by the propagation distance $z$. However, the BSC effect is typically observed in standard GRIN MMFs by increasing the beam input power at a given fiber length, with fixed injection conditions. This is mostly because varying the power is a trivial experimental operation, whereas cut-back experiments where the fiber is progressively shortened may be difficult and tedious (especially when short fiber spans are used). In addition, one needs to realign the output fiber tip at each cut, which may be a tough process when operating with a holographic MD setup. Finally, it has to be noted that each cutback experiment requires a brand-new fiber. Interestingly, a specular problem is faced when running simulations: it is quite simple to add steps of integration of the nonlinear wave equation; whereas quasi-continuously varying the beam power requires running several simulations, thus being quite time-consuming.  

In standard experiments, increasing the beam power has the beneficial effect of facilitating its thermalization. This is because higher powers provide enhanced nonlinear effects, i.e., more efficient power and energy 
%and momentum 
transfers among the modes. Indeed, in the thermodynamic-analogy picture, one may see nonlinearity as the mechanism leading to interactions among the modes via photon exchanges.
Now, from (\ref{boltzmann}) it is evident that the entropy logarithmically scales with power. Indeed, if one doubles the power of all the modes, the entropy is enhanced by a $M\ln 2$ factor, being $M$ the number of modes. Therefore, it is convenient to separate the explicit and the implicit contributions of power to the entropy. This can be easily done by rewriting the Boltzmann entropy as 
\begin{equation} 
   S=M\ln N+\sum_{p=1}^M\ln |f_p|^2,
\label{eq:S-double}
\end{equation}
where $|f_p|^2=n_p/N$ is the power fraction of the $p$-th mode. In fact, in (\ref{eq:S-double}) the first term contains the logarithmic increases of $S$ with $N$, while the second term only implicitly depends on $N$. Within this expression, it appears evident that to prove the entropy growth during thermalization one can just compute the second term in (\ref{eq:S-double}), which is referred to as configuration entropy \cite{mangini2024maximization} (or entropy per particle in \cite{ferraro2024calorimetry}), 
\begin{equation}
   \Tilde{S}=\sum_{p=1}^M\ln |f_p|^2.
\label{eq:S-double2}
\end{equation}
Indeed, no matter whether wave thermalization is achieved by increasing the power at a fixed fiber length, or at a fixed input power and then increasing the length as it can be verified by cut-back experiments, the maximization of $\Tilde{S}$ directly proves the maximization of $S$. 

\begin{figure}[!ht]
  \centering
  \includegraphics[width=0.75\linewidth]{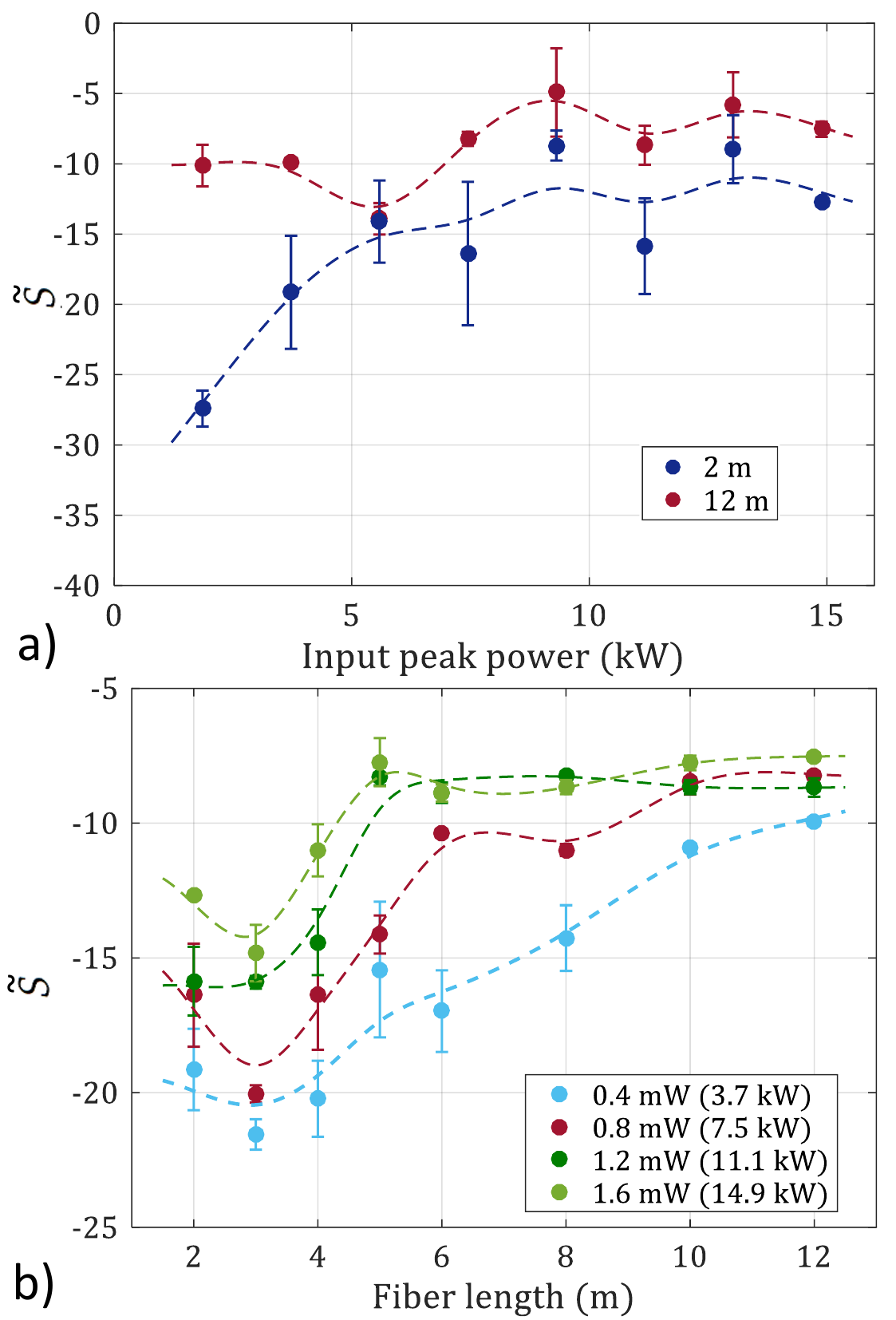}
\caption{a) Evolution of configuration entropy vs. input peak power during BSC for a fiber length of 2 (blue symbols) and 12 m (red symbols). b) Increase of the configuration entropy vs. fiber length which accompanies BSC, for different input beam powers (the legend shows the average and the peak power of the optical pulses, respectively). Symbols are experimental data, while the dashed curves are mere guides for the eye. The experiments were carried out with a standard 50/125 GRIN MMF, coiled around a $15\text{ cm}$ diameter plastic drum. Figure from \cite{mangini2024maximization}.}
\label{entropy}
\end{figure}

A recent work has reported both types of experiments \cite{mangini2024maximization}. The main results are illustratedd in Fig. \ref{entropy}. The value of $\Tilde{S}$ were calculated by eq.(\ref{eq:S-double2}) starting from the values of $n_p$ that were experimentally measured with the holographic MD method of Ref. \cite{gervaziev2020mode}.
In the first set of experiments, whose results are shown in Fig. \ref{entropy}a, a trend of growing $\Tilde{S}$ at a fixed length was observed, when increasing the input power above the self-cleaning power threshold. 
%A clear increase of entropy was always observed (see the red and blue dots in Fig. \ref{entropy}a): here the curves are mere guides for the eye. 
%Note that when out-of-equilibrium states are concerned, oscillations in the evolution of both $S$ and $\Tilde{S}$ are possible. 
For a short fiber length (2 m), a marked increase of $\Tilde{S}$ is observed when the power grows up to 5 kW (cfr. blue curve in Fig. \ref{entropy}a). At higher powers, $\Tilde{S}$ shows a saturation trend. This is expected when the beam is near thermal equilibrium, since the entropy has to reach its maximum. Interestingly, Fig. \ref{entropy}a also shows that at a longer fiber length (12 m), $\Tilde{S}$ experiences, instead, a relatively slow growth (cfr. red curve in Fig. \ref{entropy}a). This is because the output beam remains relatively close to its state of equilibrium at all input power values. 
%In Figs. \ref{entropy} and \ref{conservation}, error bars represent the standard deviation when repeating the same reconstruction algorithm by moving the center of the output beam of ±1 pixel in each direction with respect to the center position of the CCD camera of the MD setup (refer to \cite{ferraro2024calorimetry} for more details).

The results of the second type of experiments (cutback) are shown in Fig. \ref{entropy}b: as can be seen, here $\Tilde{S}$ exhibits a general trend of growing larger with distance. Consistently with the results of Fig. \ref{entropy}a, at the highest peak power of 14.9 kW, $\Tilde{S}$ remains essentially a constant for lengths longer than 4 meters. This can be explained by considering that at these power levels thermalization is achieved at relatively shorter distances. On the other hand, an overall rapid growth of $\Tilde{S}$ over the first 6 meters is observed at lower powers (see, for instance, the light blue curve in Fig. \ref{entropy}b). Whenever thermal equilibrium is reached, the entropy growth ends with a plateau. 
%Finally, it is worth noting that as far as out-of-equilibrium states are concerned, i.e., either at low powers or short fiber lengths, the trend of $\Tilde{S}$ may present oscillations since the growth of the entropy towards equilibrium is not necessarily monotonic. 
%\textcolor{blue}{Finally, note that a monotonic evolution of the entropy with the propagation (or the power) is expected when considering an average over an ensemble of realizations of beam-cleaning experiments.}

%% Bridge
So far, we have considered the growth of entropy of a single beam, i.e., its transition through non-equilibrium states to attain a final state thermal equilibrium. In a loose sense, this process can be interpreted as a consequence of the second law of thermodynamics, which has a probabilistic origin in the statistical mechanics framework. As such, thermodynamic parameters such as the temperature and the chemical potential merely acquire a statistical meaning. It is, indeed, when heat exchanges between two systems are allowed that one may provide $S$, $T$, and $\mu$ with a physical connotation, without recurring to probabilistic arguments. Indeed, the entropy growth can be seen as the driver of the unidirectional nature of the heat flow from hot to cold. Hence, in accordance with the zeroth principle of thermodynamics, one may demonstrate that $S$, $T$, and $\mu$ are true physical forces by observing an irreversible energy transfer between two systems with different temperatures via irreversible heat exchanges. This result was experimentally achieved in Ref. \cite{ferraro2024calorimetry}, which will be briefly reviewed in the following subsection.

\subsubsection{Two-beam thermalization and optical calorimetry}
\label{sec_calorimetry}

Thermodynamics involves the description of the properties of physical systems at thermal equilibrium via macroscopic parameters such as temperature  \cite{Huang}. On the other hand, calorimetry deals with thermodynamic processes under non-equilibrium conditions, e.g., via the exposition of a given object to an external heat source. The easiest calorimetry experiment that one can imagine involves two bodies with different temperatures, that are put in contact inside a calorimeter. In the latter, heat exchanges are allowed between the bodies, but there is no heat dissipation towards the outside world. In other words, the system composed of the two objects is adiabatic. Within this setup, the heat unidirectionally flows from the higher-temperature body towards the other body. An equilibrium is then established whenever the two bodies reach the same temperature, and heat exchanges no longer take place.

\begin{figure*}[!htb]
  \centering
  \includegraphics[width=12cm]{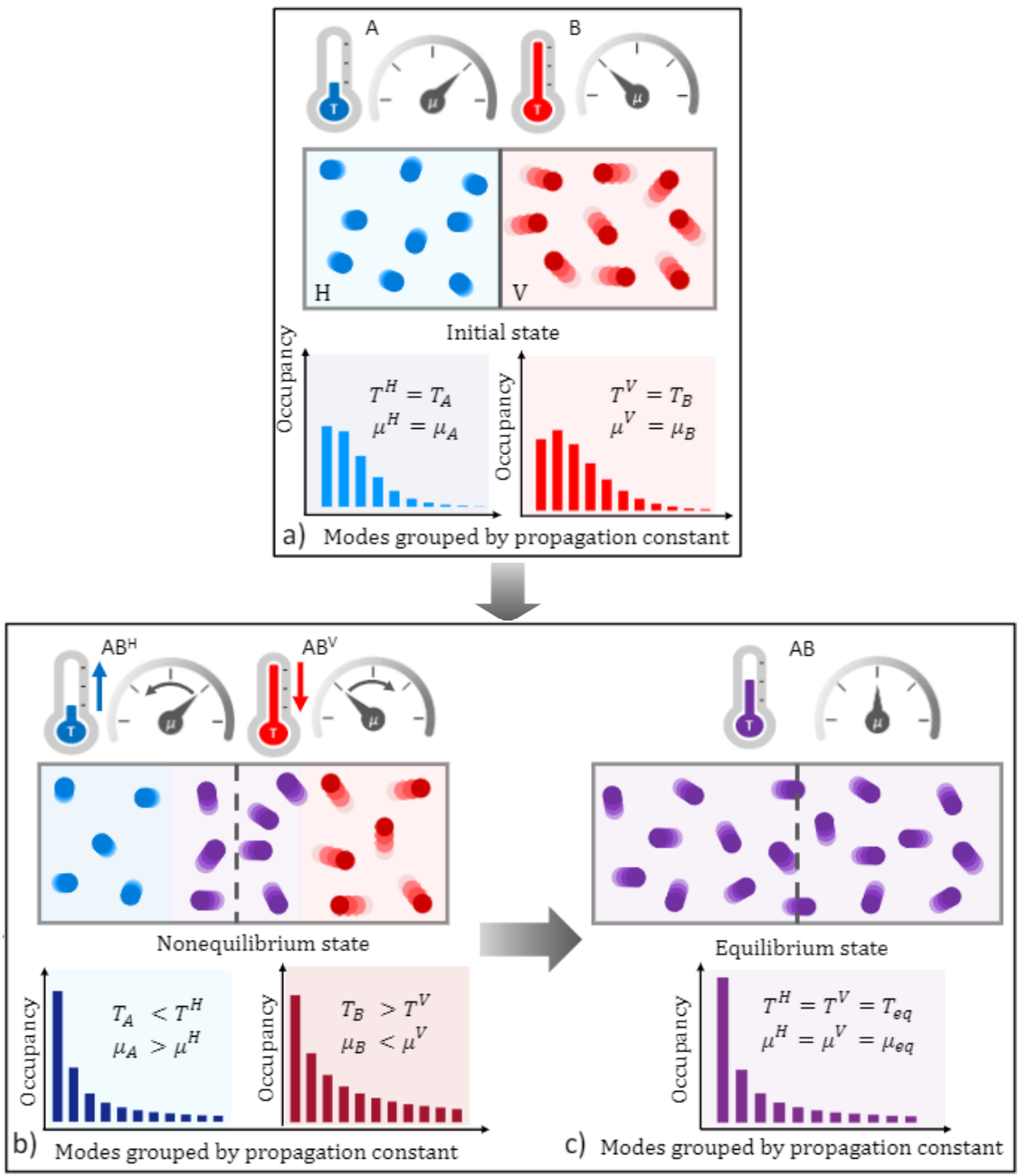}
\caption{Sketch of the thermalization process of two photon gases. (a) Initial state: two orthogonally polarized beams are prepared with different temperatures, chemical potentials, and mode distributions. (b) Nonequilibrium transient state: the difference in temperature and chemical potential among the left and right sides of the box progressively quenches, until their associated mode occupancy progressively approaches the RJ distribution, as a consequence of nonlinear beam propagation, i.e., of heat exchange between the two gases. (c) Equilibrium state: the two gases reach the same values of temperature and chemical potential, and their mode distribution obeys the RJ law. Figure from \cite{ferraro2024calorimetry}.}
  \label{fig-toy}
\end{figure*}

Such an experiment was recently reproduced within the multimode photonics platform. Two laser beams, representing the two bodies, were simultaneously injected into an MMF. The two beams are distinguished by their polarization state. The first beam (A), having a temperature $T_A$ and chemical potential $\mu_A$, is horizontally polarized; whereas the second beam (B), which is hotter ($T_B>T_A$, $\mu_B>\mu_A$), has a vertical state of polarization. The initial state of the system associated with this experimental setup can be depicted as two gases that occupy two halves of the total available mode volume (see Fig. \ref{fig-toy}a). On each half of the box, the color of the beams indicates their temperature. Specifically, the colder (hotter) gas is depicted in blue (red). This use of colors, however, is only meant to help the visualization of the phenomenon. Indeed, as well known, photons are indistinguishable particles.

It has to be noted that, in analogy with the previously discussed case of the entropy, it is possible to formally associate a beam with both an optical temperature and chemical potential even when it is not yet at equilibrium. Indeed, one may compute the values of $T$ and $\mu$ starting from the knowledge of the mode occupancies, even when these do not follow the RJ law, as depicted at the bottom of Fig. \ref{fig-toy}a. This is, strictly speaking, improper since the temperature and chemical potential do not possess a rigorous physical meaning whenever one is out of equilibrium. In this sense, one may refer to $T$ and $\mu$ as the values of temperature and chemical potential that a beam would eventually reach, once it is at thermal equilibrium. In the case of two beams considered here, $T$ and $\mu$ of each beam are those achieved in the absence of the other beam, i.e., with a single beam injection into the fiber. 

The initial state in Fig. \ref{fig-toy}a is defined by $T^H = T_A$, $\mu^H = \mu_A$, $T^V = T_B$, and $\mu^V = \mu_B$, where the superscript letters `H' and `V' are used to identify the parameters of the left and right sides of the box, respectively. Once the system evolves, the two photon gases interact, exchanging both energy and particles (cfr. Fig. \ref{fig-toy}b), until progressively reaching the same equilibrium values of temperature ($T_{eq}$) and chemical potential ($\mu_{eq}$), as depicted in Fig. \ref{fig-toy}c. The exchange of particles is due to the lack of polarization conservation during single beam propagation. Indeed, it was shown in Ref. \cite{ferraro2023spatial} that, within standard experimental conditions, an optical beam undergoes full depolarization.

The equilibrium values of temperature ($T_{eq}$) and chemical potential ($\mu_{eq}$) are first established at the center of the box, where the initial interactions between the two gases occur (refer to Fig. \ref{fig-toy}b). For the sake of simplicity, we indicate the equilibrium region by depicting violet particles in Figs. \ref{fig-toy}b-c. Thermal equilibrium is then gradually achieved at the periphery of the box (see Fig. \ref{fig-toy}c). Note, however, that the boxes in Fig. \ref{fig-toy} indicate a mode volume. Thus, defining a ``local" temperature is formally improper.
During the nonequilibrium transient state, each side of the box has a mode distribution that gradually approaches the RJ law (as depicted in the histograms at the bottom of Figs. \ref{fig-toy}b,c). The color map indicates that before equilibrium is reached, the temperatures on the two sides of the box differ. The left side of the box still contains the colder gas, resulting in a higher temperature for the horizontally polarized photon gas when compared with the vertically polarized photons, as the former still contains more energetic (red) particles (i.e., $T_A<T^H \leq T^V<T_B$, and $\mu_A>\mu^H \geq \mu^V>\mu_B$). The temperature difference between the left and right sides of the box diminishes as the input power increases, eventually disappearing when thermal equilibrium is reached (i.e., $T^H=T^V=T_{eq}$ and $\mu^H=\mu^V=\mu_{eq}$).

%% Experimental methods
This calorimetry picture was confirmed by experiments that rely on the same holographic MD technique used in the case of a single beam (see Ref. \cite{ferraro2024calorimetry} for details).
In particular, the growth of the optical entropy was experimentally verified, in pure analogy with the results presented in Sec. \ref{sec_exp_entropy_growth}. Here, however, the relative power of the two beams (i.e., the ratio of the mass of the two photon gases) provided an additional degree of freedom. Specifically, the mass ratio impacts the final equilibrium temperature. Moreover, the latter can be varied by tuning the temperature of only one of the two beams. This result, which was experimentally demonstrated, is of utmost importance for applications based on BSC. 
%% Beam quality improvements
Indeed, the success of optical calorimetry experiments showed that it is possible to modify the optical temperature, i.e., the quality or brightness of a laser beam, by exploiting the nonlinear interaction with another beam in an MMF. In this sense, these results pave the way for developing a new generation of photonic tools for all-optically controlling the spatial profile of intense laser beams.  
% BRIDGE
The fact that the temperature of a beam is associated with its spatial quality was further highlighted in Ref. \cite{mangini2023high}, whose content will be shortly reviewed below.

\subsection{High-temperature wave thermalization}
\label{sec_exp_temperature}
In Sec. \ref{sec_exp_entropy_growth} and, specifically, in Fig. \ref{entropy}, we have discussed the establishment of thermal equilibrium when increasing either the input power $P$ or the fiber length. However, as we have previously mentioned, these two methods are not equivalent. Indeed, on the one hand, the propagation distance plays the role of time. Therefore, the number of particles and the kinetic energy are conserved in cutback experiments: thus, one may properly associate the beam with a temperature $T$ and a chemical potential $\mu$. On the other hand, when achieving wave thermalization by increasing the input power, the number of particles and the kinetic energy are no longer conserved. Thus, the values of $T$ and $\mu$ continuously vary with $P$. Specifically, increasing $P$ leads to an increase of $T$, i.e., to a larger occupancy of the HOMs. As a result, too high input powers cannot provide BSC, even within the weak nonlinearity regime.

\begin{figure}[!htb]
\centering
\includegraphics[width=0.8\linewidth]{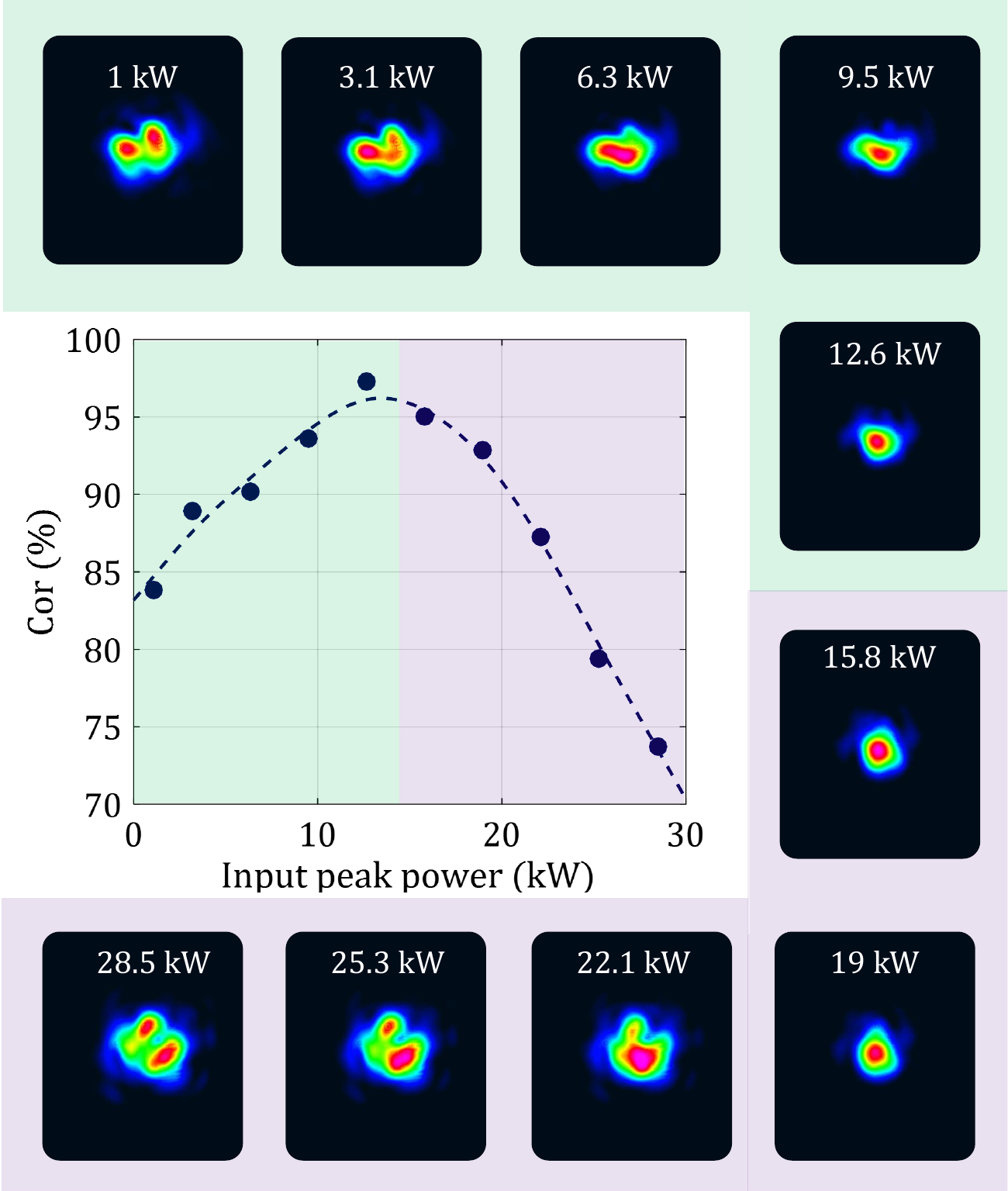}
\caption{Center panel: correlation between the output beam intensity profile and the fundamental GRIN fiber mode, vs. input peak power; insets show the experimental output beam intensity for increasing values of the peak power $\mathcal{P}$ of 7.8 ps input pulses, showing that the initial beam cleaning (green background region) is followed by beam spoiling at higher input power values (violet background region). {\it Source:} From Ref.\cite{mangini2023high}.}
\label{fig:patterns}
\end{figure}

This concept was experimentally demonstrated in Ref. \cite{mangini2023high}, by using the same holographic MD setup of the experiments discussed in Sec. \ref{sec_exp_entropy_growth}. The main result is illustrated in Fig. \ref{fig:patterns}. Specifically, the insets show the output transverse intensity profiles obtained for increasing values of the beam power. When the input peak power grows from 1 kW up to about 13 kW, the BSC effect takes place: the beam transforms from speckles to a bell shape. However, as the input power is further increased, a clear ``self-spoiling" of the output beam occurs.

A quantitative analysis of the intensity patterns ($I$) shown in Fig. \ref{fig:patterns} was carried out by computing their correlation with the intensity of the fundamental mode ($I_0 \propto \exp\{-2{(x^2+y^2)}/{w^2}\}$, $w$ = 6.33 $\mu$m) as
\begin{equation}
    Cor = \frac{\int I(x,y) I_0(x,y) dx dy}{\sqrt{\int I(x,y)^2 dx dy\int I_0(x,y)^2 dx dy}},
    \label{eq:cor}
\end{equation}
The experimental variation of $Cor$ with input beam power is reported in the central panel of Fig. \ref{fig:patterns}. A maximum value of $Cor\simeq 98\%$ is reached at the occurrence of BSC. On the contrary, at either higher or lower powers, $Cor$ rapidly drops. An analogous (and opposite) trend was observed for the $M^2$ parameter (not shown here, see Ref. \cite{mangini2023high}). In the linear regime, a value of $M^2 \simeq 3$ was measured. Whereas, near the BSC threshold, $M^2$ decreases below 2. Then, at higher powers, the beam quality worsens, and the associated $M^2$ grows even more than in the linear regime.

In this regard, it is important to mark the difference between the speckled patterns at low and high powers. The former corresponds to nonequilibrium states, whereas the latter are associated with thermalized waves at high temperatures. This was confirmed by MD experiments: the distribution of the mode occupancy at low input power only depends on the beam injection conditions; to the contrary, the degradation of the beam quality at powers above the BSC threshold, i.e., the beam self-spoiling, always leads to a mode power distribution that obeys the RJ law. 

%BRIDGE
From the application point of view, these findings establish a maximum brightness for multimode beam propagating in nonlinear GRIN MMF, at least in the weakly nonlinear regime. 

In summary, high optical temperatures lead to the loss of spatial beam quality, which is detrimental for most fiber-based applications. In contrast, as discussed in Sec. \ref{sec_calorimetry}, cooling down an optical beam increases the power fraction of the fundamental mode. The ultimate case would be $T=0$, i.e., the full wave condensation condition in which a single mode beam propagates undisturbed in an MMF. In contrast, whenever $T<0$ the fundamental mode is depleted in favor of HOMs. Both the $T$ near-zero and the negative temperature cases will be the topics of the next section.

%\bigskip
%{\bf OPTIONAL: The following section may be included to present experiments by varying the (kinetic) energy. I don't know if it is really appropriate. Feel free to remove it if it appers to complicate the presentation of the section 4.}

\subsection{Light thermalization and condensation by varying the beam energy}
\label{sec4:cond}

In this section, we discuss experiments in MMFs that have been performed by varying the kinetic energy of the launched optical beam (we remind here that the kinetic energy refers to the linear contribution of the Hamiltonian, see section~\ref{subsec:basic}). In this way, we present two different phenomenologies, the RJ condensation of classical waves, and the thermalization toward negative temperature equilibrium states.  

\subsubsection{Rayleigh-Jeans condensation}
\label{sec4:subsecRJcond}

In spite of the physical differences that can distinguish RJ condensation and BE condensation of quantum particles (atoms, molecules, polaritons, photons...), the underlying mathematical origin is similar, because of the common singular behavior (vanishing denominator) of the equilibrium Bose distribution for quantum particles, and the equilibrium Rayleigh-Jeans (RJ) distribution for classical waves, as recently discussed in Ref.\cite{zanaglia2024bridging}. The RJ condensation of classical waves has been studied since a long time \cite{davis2001simulations,blakie2005projected,connaughton05,aschieri2011condensation}, and we refer the reader to the \ref{app:condensate_fraction} for a succinct summary of the equilibrium properties of the phase transition of wave condensation. 

As commented in previous sections in relation with RJ thermalization, light propagation in MMFs is essentially a conservative system, in which the power (`particle number') and the kinetic energy are conserved through propagation. In other terms, there is no thermostat in the optical experiments: By keeping constant the  power (``number of particles") $N$, the energy $E$ plays the role of the control parameter in the transition to condensation for this microcanonical statistical ensemble. Here, we discuss the experimental observation of the transition to condensation by varying the energy $E$ of the beam launched into the MMF, while keeping constant the power $N$ \cite{Baudin2020}. In order to vary the kinetic energy, the laser beam is passed through a diffuser to generate a speckle beam, which is subsequently injected in the fiber. In this way, the amount of randomness of the speckle beam can be varied by keeping a constant power, i.e., the larger the randomness of the speckle beam, the higher the energy. Following this procedure, the experimental results are averaged over an ensemble of different realizations of speckle beams with the same power and the same energy. The importance of this aspect can be understood by recalling the fact that the RJ distribution is inherently a statistical distribution, whose verification requires, in principle, to consider an average over a statistical ensemble of speckle beams.

The experiments reported in Ref.\cite{Baudin2020} have been performed by using a conventional GRIN MMF of 12 m long that  guides $M \simeq 120$ modes (fiber radius $R=26$ $\mu$m). In the experiments, the averages over ensembles of speckle beams have been performed from a total number of 2$\times$1000 measurements of the near-field and far-field intensity distributions recorded for different energies $E$ and a fixed power of $N=7$ kW. The ensemble of realizations of speckle beams was partitioned within small energy intervals $\Delta E$, and then averaged in each individual energy interval, see Ref.\cite{Baudin2020} for details. 

At variance with homogeneous condensation in a bulk medium (i.e., without a trapping potential $V(\br) = 0$) 
\cite{connaughton05,Nazarenko11}, the presence of the parabolic potential reestablishes condensation in the `thermodynamic limit' in 2D.
There exists a (non-vanishing) critical energy $E_{\rm crit}^*=NV_0/2$ such that $\mu=\beta_0$, where $V_0$ denotes the depth of the parabolic trapping potential associated with the finite number of modes of the GRIN MMF, see \cite{aschieri2011condensation} or \ref{app:condensate_fraction}.
At this critical point the denominator of the RJ distribution {\it vanishes exactly} and the singularity of the RJ distribution is regularized by the macroscopic population of the fundamental mode.
This leads to the following expression of the condensate fraction as a function of the energy:
%At this critical point the denominator of the RJ distribution {\it vanishes exactly} for the fundamental mode \cite{PRL05}. 
%The singularity is regularized by the macroscopic population of this fundamental mode \cite{aschieri2011condensation,supplemental}:
%$n_0^{eq}/N =0$ for $E \ge E_{\rm crit}^*$, while for $E < E_{\rm crit}^*$ the fundamental mode gets macroscopically populated \cite{aschieri2011condensation,supplemental}:
\begin{eqnarray}
\frac{n_0^{eq}}{N}=1-  \frac{E-E_0}{E_{\rm crit}^*-E_0}.
\label{eq:n_0_TL}
\end{eqnarray}
%$E_0=N \beta_0$ being the minimum energy where $n_0^{eq}/N=1$.
Then $n_0^{eq}$ vanishes at $E_{\rm crit}^*$, and $n_0^{eq}/N \to 1$ as $E$ reaches the minimum value of the energy $E_0=N\beta_0$, where $\beta_0$ is the fundamental mode eigenvalue, see \ref{app:condensate_fraction}.

\begin{figure}[]
\begin{center}
%\bigskip
\includegraphics[width=0.8\columnwidth]{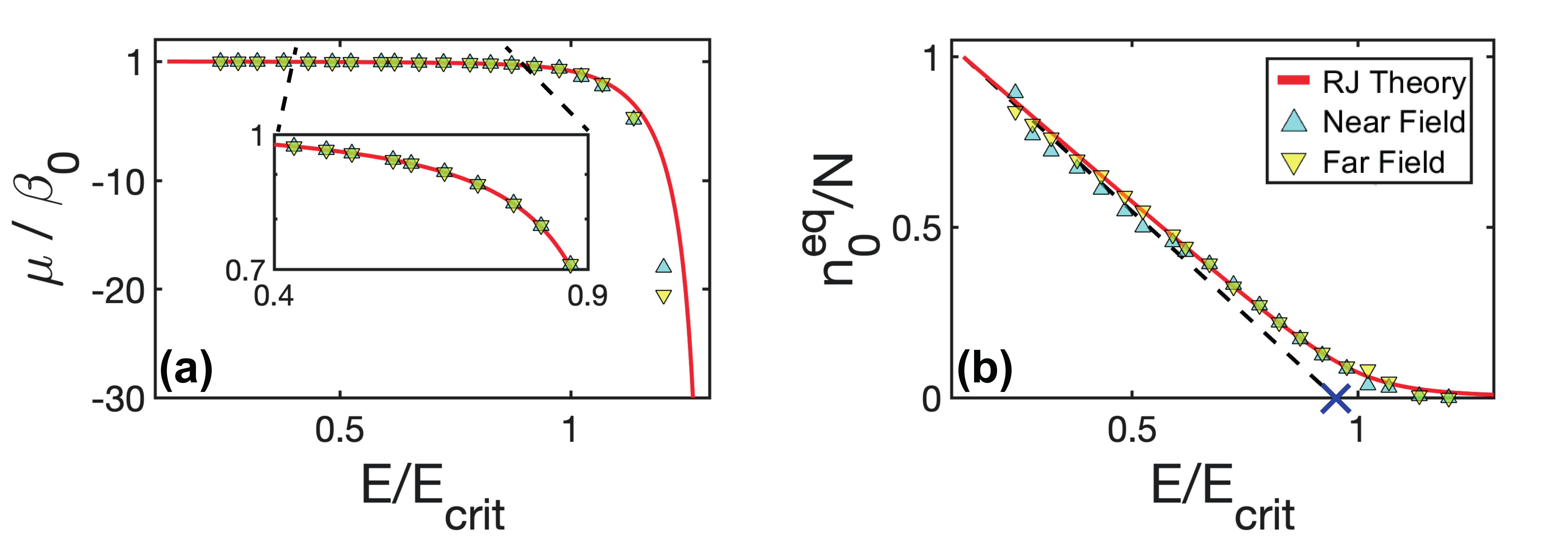}
\caption{
Rayleigh-Jeans condensation:
(a) Chemical potential vs energy: For $E \lesssim E_{\rm crit}$, $\mu \to  \beta_0^{-}$, 
which leads to the macroscopic population of the fundamental mode, see $n_0^{eq}/N$ vs $E/E_{\rm crit}$ (b).
The blue (yellow) triangles report the experimental results from the NF (FF) intensity distributions (averaged over the realizations).
The red lines report the RJ theory for the MMF used in the experiment.
By decreasing the energy $E$ below the critical value $E_{\rm crit}$, the chemical potential $\mu \to  \beta_0^{-}$ (inset shows a zoom), which leads to the macroscopic population of the fundamental mode.
%$R=26\mu$m is the fiber radius, $k_c=1.15 \mu$m$^{-1}$ the frequency cut-off.
%The blue cross in (e) denotes $E_{\rm crit}^*/E_{\rm crit}=0.95$: the experiment is close to the thermodynamic limit.
The dashed black line in (b) refers to the thermodynamic limit (Eq.(\ref{eq:n_0_TL})) and the blue cross denotes $E_{\rm crit}^*/E_{\rm crit}\simeq 0.95$: The experiment is relatively `close' to the thermodynamic limit. {\it Source:} From Ref.\cite{Baudin2020}.
%The dashed black line in (e) is the condensate fraction in the thermodynamic limit (Eq.(\ref{eq:n_0_TL})) and the blue cross denotes $E_{\rm crit}^*/E_{\rm crit}\simeq 0.95$: The experiment is `close' to the thermodynamic limit.
}
\label{fig:sec_rj_cond_2}
\end{center}
\end{figure}

Because of finite size effects, the experiment reported in Ref.\cite{Baudin2020} does not occur in the strict thermodynamic limit.
%The theory of RJ condensation beyond this limit (i.e. accounting for finite size effects) gives the critical energy $E_{\rm crit} = E_0 \big(1+(M-1)/\varrho \big)$, where $\varrho=\sum_{p \neq 0} (p_x+p_y)^{-1}$ \cite{aschieri2011condensation,PR14,supplemental}.
The theory of RJ condensation accounting for finite size effects gives the critical energy $E_{\rm crit} = E_0 \big(1+(M-1)/\varrho \big)$, where $\varrho$ is a constant number that tends to $V_0/\beta_0$ in the thermodynamic limit, see \ref{app:condensate_fraction}.
Considering the experimental parameters, we obtain $E_{\rm crit}^*/E_{\rm crit} \simeq 0.95$ (blue cross in Fig.~\ref{fig:sec_rj_cond_2}b), so that the experiment is relatively `close' to the thermodynamic limit.

In order to compare the theory of RJ condensation with the experiments, the condensate amplitude $n_0^{eq}$ and the chemical potential $\mu$ need to be extracted from the experimental data.
In Ref.\cite{Baudin2020}, this was done by implementing a fitting procedure.
%: Using the experimental intensities averaged over the realizations, $(n_0^{eq},\mu)$ is retrieved from the RJ intensity distributions [Eqs.(\ref{eq:I_NF_fit_cond}-\ref{eq:I_FF_fit_inc})] by a least square method \cite{Baudin2020}.
%For this purpose, the temperature $T$ in Eqs.(\ref{eq:I_NF_fit_inc}-\ref{eq:I_FF_fit_inc}) has been expressed in term of $(n_0^{eq},\mu)$ by using $N=n_0^{eq} + T\sum_{p\neq 0} (\beta_p - \mu)^{-1}$.
%In this way $(n_0^{eq},\mu)$ can be extracted independently from either the NF or FF intensity distributions.
%The validity of the procedure is confirmed by a small least square residual error \cite{supplemental}.
Figure~\ref{fig:sec_rj_cond_2}a reports the chemical potential $\mu$ vs $E$: by decreasing $E$, $\mu$ increases and condensation occurs when $\mu \to \beta_0^{-}$ for $E=E_{\rm crit}$.
Below the transition ($E \le E_{\rm crit}$) the fundamental mode gets macroscopically populated, $n_0^{eq} \gg n_{p \neq 0}^{eq}$, see Fig.~\ref{fig:sec_rj_cond_2}b.
%The corresponding RJ equilibrium theory (red line) is in quantitative agreement with the experimental results \cite{supplemental}.
%The corresponding projections $\mu(E)$ and $n_0^{eq}(E)$ are reported in Fig.~1d-e with the experimental results averaged over the realizations.
The triangles report the experimental data retrieved from the fitting procedure from  the near-field and far-field intensity distributions, see \cite{Baudin2020} for details. The red line reports the RJ theory accounting for finite size effects (i.e., beyond the thermodynamic limit) for the MMF used in the experiment, see \ref{app:condensate_fraction}.
The experimental results in Fig.~\ref{fig:sec_rj_cond_2} are in quantitative agreement with the RJ theory, where the parameters $\beta_0$ and $M$ are fixed by the fiber used in the experiment.
Furthermore, the experimental results are close to the thermodynamic limit given by Eq.(\ref{eq:n_0_TL}), see Fig.~\ref{fig:sec_rj_cond_2}b.
Notice that, as usual, finite size effects make the transition to condensation `smoother', as evidenced by the red line in Fig.~\ref{fig:sec_rj_cond_2}b.

\subsubsection{Light thermalization to negative temperature equilibrium states}

\begin{figure}[!ht]
\begin{center}
%\bigskip
\includegraphics[width=0.8\columnwidth]{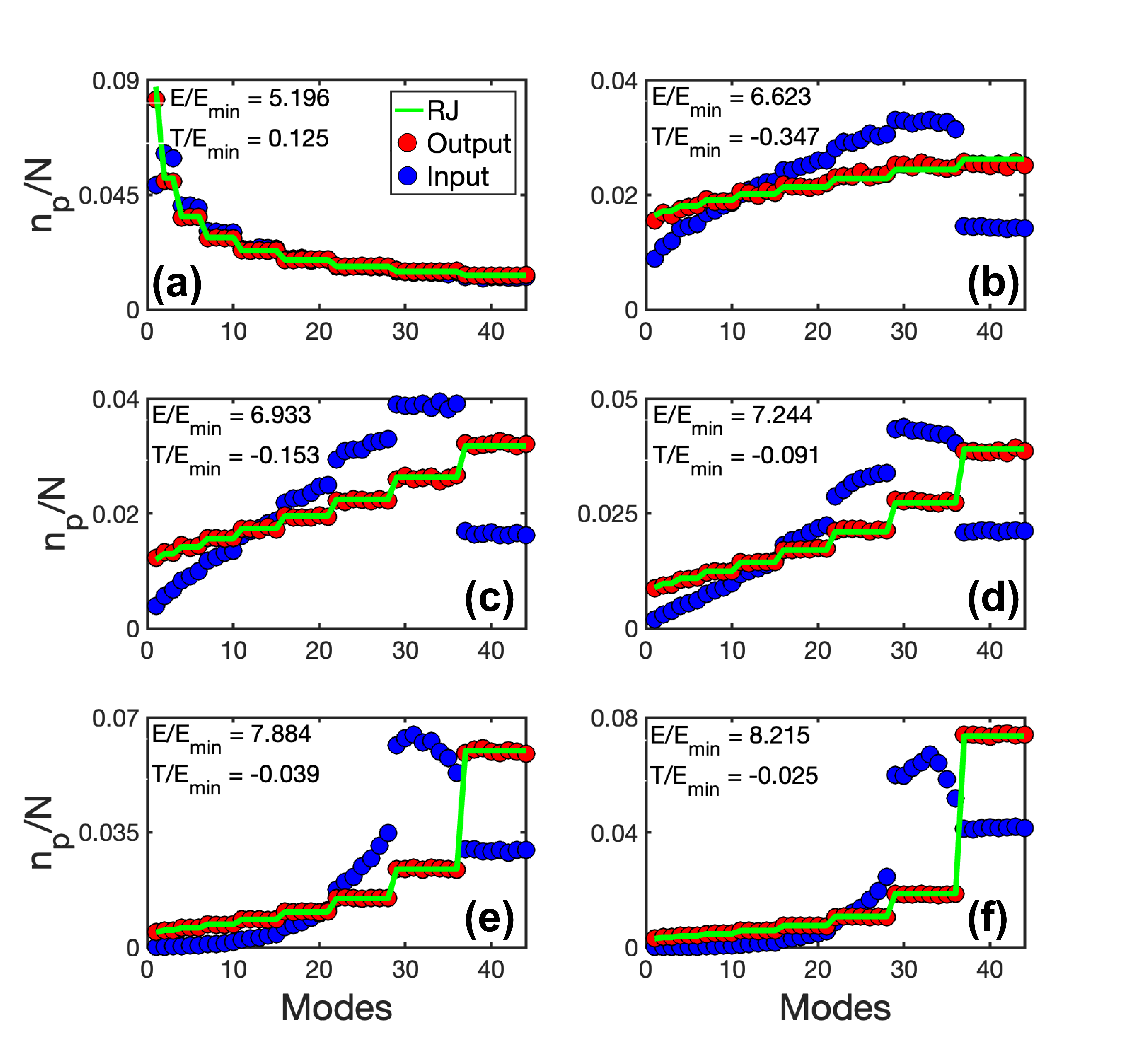}
\caption{
Thermalization to negative temperature equilibria. 
Experimental modal distributions averaged over realizations.
%$n_p^{\rm exp}=\left<|a_p^{\rm exp}|^2\right>$. 
Initial modal distributions at the fiber input (blue), and modal distributions 
at the fiber output (red).
Corresponding RJ equilibrium distributions $n_p^{\rm RJ}$ (green).
% featured by an inverted modal population.
The six panels correspond to different values of the energy (Hamiltonian) $E$, or equivalently different temperatures $T$ ($T>0$ in (a), $T<0$ in (b-f)). The modal distribution peaked on the lowest mode for $T>0$ (a), gets inverted for $T<0$ (b-f).
An average over $\simeq35$ realizations of speckle beams is considered for each panel.
The fiber modes are sorted in the Hermite-Gauss basis from the fundamental one ($\beta_0$) to the highest mode group (nine-fold degenerate with $\beta_{\rm max}=9\beta_0$).
Degenerate modes are equally populated at equilibrium, leading to a staircase distribution at equilibrium.
{\it Source:} From Ref.\cite{baudin23}.
}
\label{fig:sec_rj_cond_4}
\end{center}
\end{figure}

The experimental procedure implemented in Ref.\cite{Baudin2020} to study light thermalization with different energies has been extended to observe the thermalization to negative temperature equilibrium states. Indeed, equilibrium states with an energy larger than a threshold value $E_*$ ($E_*=N \left<\beta_p\right>$, where $\left<\beta_p\right>$ is the mean of the propagation constants, see \ref{app:condensate_fraction}), are characterized by a negative temperature, i.e.,  an equilibrium state exhibiting an inverted modal population, where  high-order modes are more populated than low-order modes.
Originally conceived by Onsager \cite{onsager49}, and Ramsey \cite{ramsey56}, negative temperatures created their own share of confusion for many decades, until being broadly accepted \cite{baldovin21,onorato23}, in line with recent experimental observations with cold atoms \cite{braun13} and 2D quantum superfluids \cite{gauthier19,johnstone19}.
More recently, negative temperatures have been predicted for classical multimode optical wave systems in the weakly nonlinear regime \cite{Wu2019,parto19thermodynamic}, and subsequently observed in these systems with light waves in Refs.\cite{baudin23,marquesmuniz23}. 
Here, we summarize the experiments relevant to MMFs which have been reported in Ref.\cite{baudin23}.
In these experiments, a GRIN MMF of 12 m was used, which guides nine groups of degenerate
modes (with a fiber radius of $\simeq 15$ $\mu$m). As described above in the experiments of RJ condensation, an average over ensembles of speckle beams was considered: $2 \times 300$ realizations of the near-field and far-field intensity distributions were recorded for the same power. The ensemble of realizations was then partitioned within small energy intervals, and then averaged in each interval to get the modal distributions reported in Fig.~\ref{fig:sec_rj_cond_4}. The resolved modal distribution was obtained from the well-known Gerchberg-Saxton algorithm \cite{fienup82phase,shechtman15phase}, which allows to retrieve the transverse phase profile of the speckle field from the near-field and far-field intensity distributions measured in the experiments, see Refs.\cite{Baudin2020,baudin2023rayleigh} for details. The procedure was applied at the fiber output (12 m), as well as the fiber input (after 20 cm of propagation).
The results are reported in Fig.~\ref{fig:sec_rj_cond_4}, which evidence the process of light thermalization to negative temperature equilibria that are characterized by an inverted modal population. 

\begin{figure}[!ht]
\begin{center}
%\bigskip
\includegraphics[width=0.8\columnwidth]{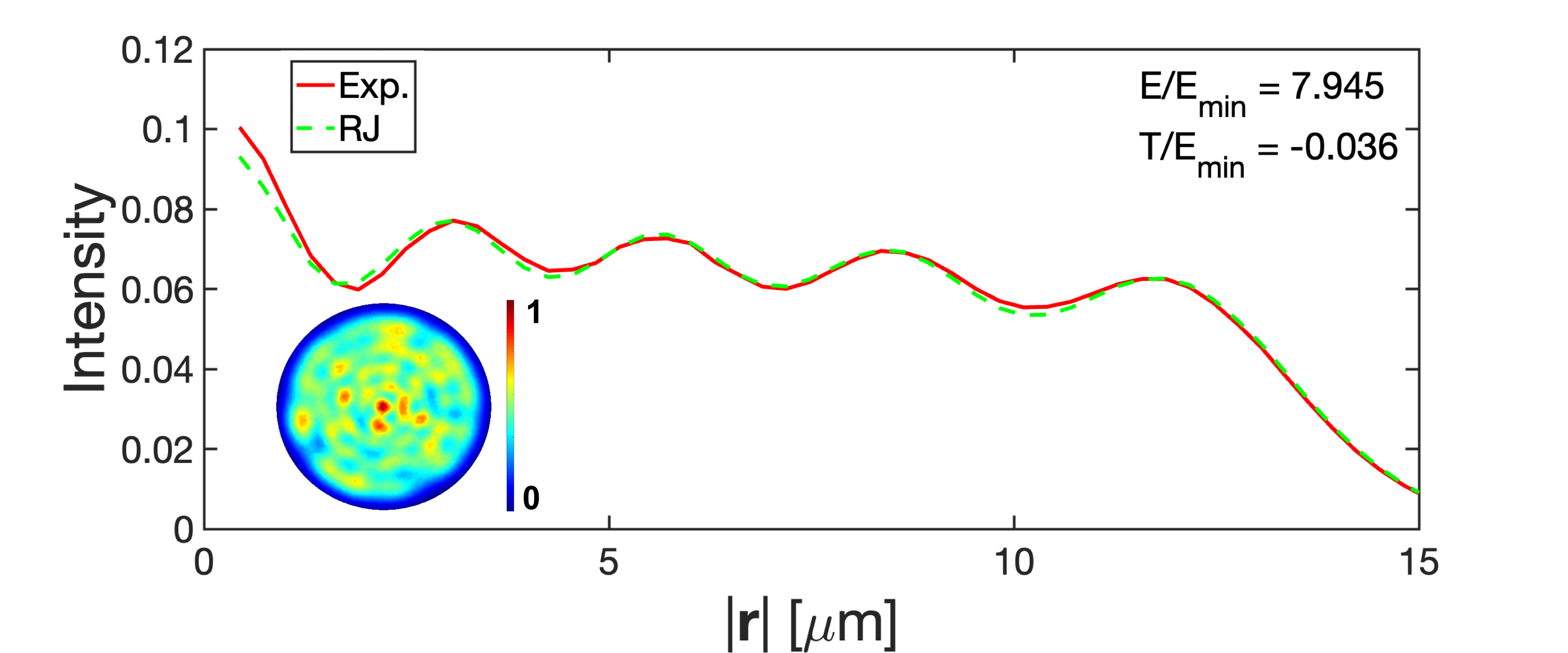}
\caption{
Oscillating radial intensity distribution at negative temperature.
Experimental averaged intensity distribution as a function of the radial (angle-averaged) distance $|{\bf r}|$ (red).
%The intensity is averaged over 89 realizations (over a small energy interval $\Delta E$ centred in $E/E_{\rm min}=7.9$).
Note the good agreement with the theoretical RJ intensity distribution (dashed green).
%$I^{\rm RJ}(|\br|)$ in Eq.(\ref{eq:int_rj}) (dashed green).
% is obtained without adjustable parameters.
The oscillating behavior of the intensity distribution is a signature of the negative temperature equilibrium.
Inset: corresponding 2D intensity averaged over the realizations (the radius of the circle is the fiber radius).
{\it Source:} From Ref.\cite{baudin23}.
}
\label{fig:sec_rj_cond_5}
\end{center}
\end{figure}

A distinguishing property of negative temperature equilibrium states is that the radial near-field intensity distribution exhibits an oscillating behavior. Indeed, the intensity distribution of usual positive-temperature equilibria is, in average, a monotonic decreasing function with the radial distance. This is consistent with the intuitive idea that low-order modes localized near the fiber center are the most populated ones. In  contrast, the inverted modal population of negative temperature equilibria is characterized by an oscillating behavior of the radial intensity, as illustrated in Fig.~\ref{fig:sec_rj_cond_5}. Note that the number of radial oscillations in Fig.~\ref{fig:sec_rj_cond_5} is given by the most oscillating mode of the fiber, that is the mode LP$_{04}$ that exhibits five oscillations.

\subsection{Beyond thermodynamics: the modal phase-locking}
\label{sec_exp_phase}
As discussed in the previous sections, the whole statistical mechanics approach to nonlinear multimode fiber systems relies on the pillar of averaging over a meaningful statistical ensemble. As such, experiments of classical wave thermalization and condensation involve results which are averaged over several realizations. These have shed light on the description of the phenomenon of BSC within the statistical mechanics framework, as discussed in Sec. \ref{sec4:cond}.

At the same time, most of the experimental demonstrations of BSC did not involve an explicit averaging process. Such an approach, which was used in the experiments reported in Sec. \ref{sec_exp_entropy_growth} and Sec. \ref{sec_exp_temperature}, is justifiable by the fact that averaging over several realizations is not practical for applications based on BSC. Remarkably, however, the results of experiments carried out without averaging over several realizations still lead to an excellent agreement with some theoretical predictions, e.g., the match of the experimentally retrieved mode occupancy with the RJ law. In this regard, a notable exception is the entropy growth discussed in Sec. \ref{sec_exp_entropy_growth}. Indeed, according to the statistical mechanics theory, and specifically to the H-theorem, a monotonic evolution of the entropy with the propagation (or the power) is expected when considering an average over an ensemble of realizations of beam-cleaning experiments. Whereas, the experimental results presented in Sec. \ref{sec_exp_entropy_growth} show that the trend of the configuration entropy $\Tilde{S}$ exhibits oscillations, i.e., it is not monotonic, at least as far as out-of-equilibrium states are concerned (see Fig. \ref{entropy}). This seems to rule out the hypothesis that the averaging operation is somehow intrinsically due to the pulse-to-pulse variance of the mode-locked lasers used, for instance, in Ref. \cite{pourbeyram2022direct} and Ref. \cite{Mangini2022Statistical}. Still, one may object that the oscillations of $\Tilde{S}$ are not very pronounced. In addition, the accuracy of holographic MD tools is questionable when dealing with HOMs, which, at thermal equilibrium, provide the largest contribution to the entropy. 

In this regard, an important experimental proof that BSC, in the absence of average over a statistical ensemble, cannot be fully identified as a wave thermalization phenomenon was presented in Ref. \cite{mangini2023modal}. In that work, Mangini et al. showed that BSC involves a mechanism of power-induced phase-locking among the \textit{lowest-order} modes, that carry most of the beam power. 
This observation collides with the statistical mechanics approach that only considers incoherent waves, i.e., it does not take into account the relative phase among modes. 

\begin{figure}[t!]
\centering
\includegraphics[width=0.80\linewidth]{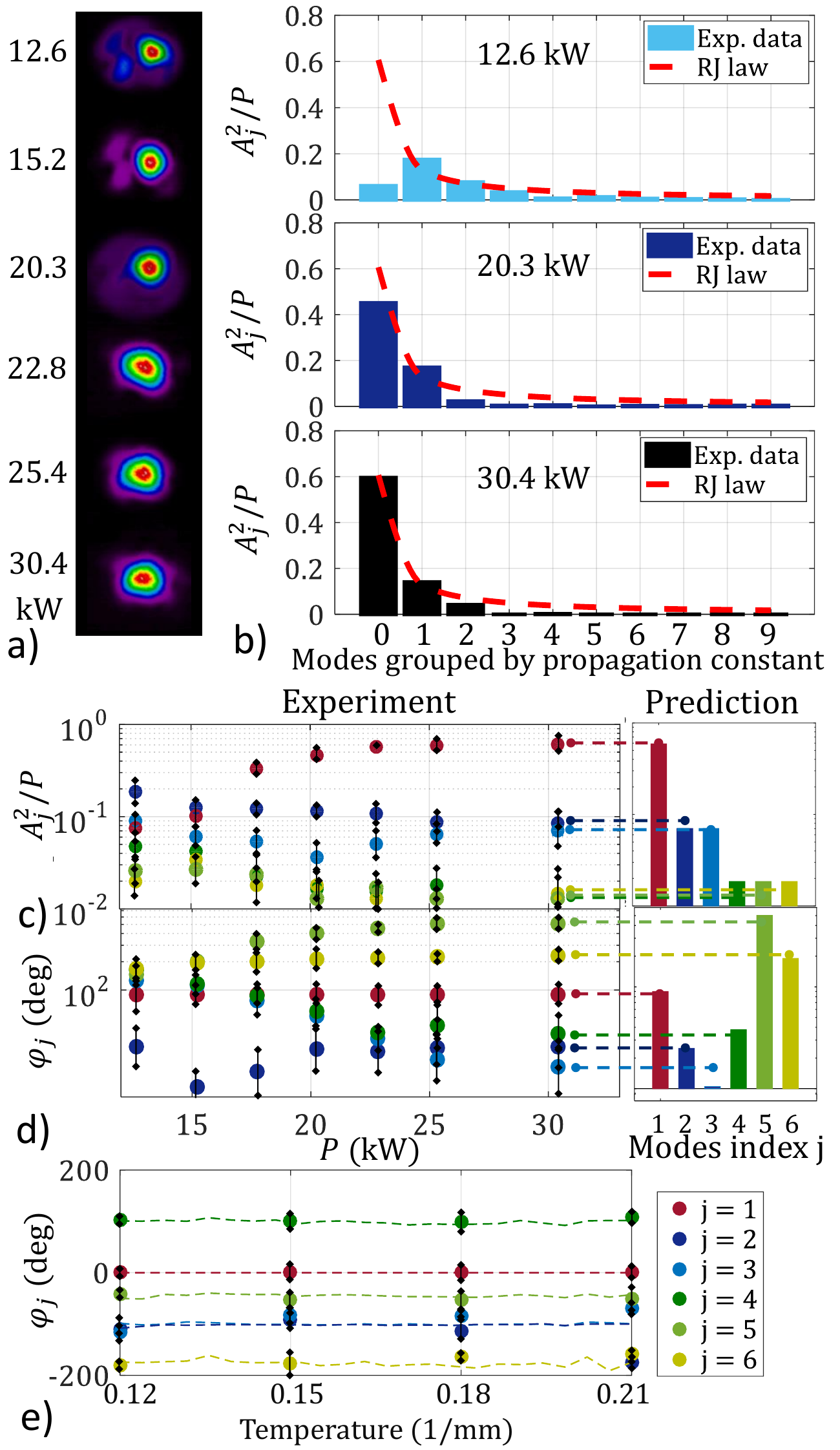}
\caption{(a) Experimental beam intensity profile when increasing $\mathcal{P}$. (b) Comparison of the corresponding mode power distribution with the RJ law. (c,d) Comparison between experimental data (left) and predictions of the mode amplitude and phase, respectively. (e) Experimental (dots) and numerical (solid lines) values of the phase vs. beam temperature.}
\label{fig:exp}
\end{figure}

Still, consistently with previous observations carried out with the same experimental setup \cite{Mangini2022Statistical}, the mode occupancy distribution is in agreement with the statistical theory. This is shown in Fig. \ref{fig:exp}a,b, where the beam intensity profile and the mode power fraction ($A_j^2/P$, $A_j$ being the mode amplitude) at different input power levels are shown. 
As it can be seen, at low powers the output beam is speckled, meaning that thermal equilibrium was not yet reached. Whereas at high power levels the BSC effect takes place and the experimental mode amplitude agrees with the RJ law. %for $T=0.11$ mm$^{-1}$ and $\mu+\beta_0 =-51.06$ mm$^{-1}$ (red dashed line).

%So one naturally wonders how to re-conciliate the experimental evidence of BSC with statistical mechanics approaches. In other words, how can we move beyond statistical mechanics to provide an exhaustive theoretical understanding of BSC? To answer this question, it is necessary to determine the evolution of the nonlinear spatial phase ($\varphi_j$), henceforth \lq\lq phase\rq\rq, which is a challenging task in the experiments.

The establishment of RJ thermal equilibrium is nothing more than a mechanism of mode amplitude locking. Indeed, the amplitudes of the modes reach steady values as the input power grows larger. The left panel of Fig. \ref{fig:exp}c shows the amplitude locking effect. The equilibrium values of the amplitude are in agreement with the statistical mechanics predictions, as illustrated in the right panel of Fig. \ref{fig:exp}d. A similar locking effect for the mode amplitude was observed (cfr. left panel of Fig. \ref{fig:exp}d).
Interestingly, by recurring to statistical mechanics considerations, it is possible to predict the values of the mode phases at thermal equilibrium. In particular, the experimental steady value of the mode phases turn out to be in excellent agreement with the predictions made with the following minimization problem
\begin{equation}
\min_{\{\varphi_j\}}\left\{ \varepsilon\right\}=\min_{\{\varphi_j\}}\left\{\left|\Psi(\theta,r)\right|^2-P\cdot|\psi_1(\theta,r)|^2\right\},
\label{eq:minim}
\end{equation}
where $\Psi$ is the output intensity profiles of the total field and $\psi_1$ is fundamental mode intensity.
As discussed in the previous sections, in fact, the beam temperature $T$ has to be sufficiently low, in order to ensure that a large amount of the optical power is stored in the fundamental mode at thermal equilibrium. 
Hence, one can reasonably assume that at $T \simeq 0^+$, the field $\Psi$ is close to the sole fundamental mode, which is the rationale behind the assumption of eq. (\ref{eq:minim}).

However, it has to be mentioned that the predictive capability of the numerical minimization (\ref{eq:minim}) turns out to be quite limited. In fact, the numerical procedure diverges whenever a relatively large number of modes is taken into account. Still, a unique solution can be found by limiting the problem to the six lowest-order modes (i.e., the first three mode groups), which, anyway, carry most of the power at thermal equilibrium in accordance to Fig. \ref{fig:exp}b. The remarkable effectiveness of eq. (\ref{eq:minim}) in predicting the experimental outcomes can be 
visibly checked by comparing the left and right panel of Fig. \ref{fig:exp}d. In particular, it is remarkable to ascertain that modes belonging to the same group, which have almost the same amplitude, are nevertheless associated with different phases. 
As a final note, it is important to mention that an excellent agreement between experiments and numerical predictions was found  for temperatures lower than  0.2 mm${}^{-1}$, as illustrated in Fig. \ref{fig:exp}e.

In conclusion, BSC in GRIN MMFs results from a complex mode-locking phenomenon, where both mode amplitudes and spatial nonlinear phases remain simultaneously locked. The occurrence of a phase locking mechanism, in particular, is associated with the well-known spatial coherence preservation during BSC \cite{Krupanatphotonics,fabert2020coherent}. This property is critically important for various applications, such as mode-locked fiber lasers and beam delivery in high-resolution nonlinear imaging, to name a few.

\section{Conclusions and perspectives}\label{Sec_5}
In this review we presented a comprehensive analysis of optical beam propagation in nonlinear multimode fibers, through the lens of a statistical mechanics framework, with particular focus on the wave turbulence approach. This perspective has allowed us to elucidate the process of thermalization in GRIN MMFs, by capturing the intricate evolution of multimode laser beams through intermediate non-equilibrium states towards their final state of RJ equilibrium. 

From the theoretical point of view, we reviewed recent works on the impact of random disorder, by extending wave turbulence theory to account for longitudinal refractive index fluctuations inherent to MMFs. This approach bridges the gap between theoretical predictions and the complex realities of light propagation in disordered environments, providing deeper insights into the role of environmental perturbations in beam dynamics.

As far as experimental investigations are concerned, we reviewed a number of key studies that have validated the statistical mechanics predictions for nonlinear beam propagation. Specifically, we  discussed recent direct observation of RJ mode distributions at positive, negative, and near-zero (wave condensation) temperatures. We also addressed the process of entropy growth during thermalization of either a single beam or two beams (optical calorimetry). In particular, the latter consists of the establishment of an equilibrium as a consequence of the heat exchange between two beams, which undoubtedly proves that the thermodynamic parameters associated with the RJ mode distribution have a true physical meaning.

Interestingly, while thermodynamics and wave turbulence models effectively predict that mode amplitudes conform to the RJ law during BSC, statistical mechanics arguments inherently fall short of capturing certain aspects of self-cleaning, namely, the preservation of spatial coherence and the enhancement of beam brightness. These, in fact, turn out to be associated with a modal phase-locking mechanism, which can be remarkably well described by a simple minimization problem (\ref{eq:minim}). In this regard, we emphasize that the entropy does not depend on the modal phases. Therefore, one may read Eq. (\ref{eq:minim}) as being analogous to a principle of minimum ``energy'' of a system with fixed entropy. Moreover, BSC can be seen as the superposition of two thermodynamic effects within two disjoint thermodynamic frameworks. The first involves the modal amplitudes, and it relies on the maximization of entropy at fixed energy (the Hamiltonian $H$). The second involves the modal phases, and it relies on the minimization of an "energy'', which is defined in (\ref{eq:minim}) at fixed entropy. 
On the other hand, we note that certain aspects of the WT theory have been recently extended to account for the emergence of strongly correlated fluctuations in turbulence flows \cite{guasoni17incoherent,onorato23}. This could be an appropriate approach to develop a detailed explanation of the effect of modal phase-locking observed experimentally.

In terms of perspectives, many challenges await us. For instance, we foresee that the next natural step will be the investigation of the statistical mechanics approach in the strongly nonlinear regime. Indeed, virtually all experiments that demonstrated the process of RJ thermalization have been carried out in the weak nonlinear regime, so far. Even a very recent experimental demonstration of the Joule-Thomson expansion, which intrinsically deals with strong nonlinearities, eventually relies on beam propagation in the weakly nonlinear regime \cite{pyrialakos2024optical}. 
The reason behind the researchers' fondness for the weakly nonlinear regime is easy to guess: the presence of strong nonlinearity complicates the description of the beam evolution toward thermal equilibrium, as well as the definition of equilibrium state itself. Let us consider the case of wave condensation, which is relatively simple to describe in cases where the nonlinearities are weak: in the strong nonlinear regime, the condensed component forms a distinct phase, characterized by different properties when compared to the thermal (uncondensed) component.
For instance, the presence of a strong condensate modifies the linear dispersion relation of the thermal component, so that the condensate amplitude $n_0^{eq}$ is not simply given by the RJ distribution $n_p^{eq}$ taken at $p=0$. One needs to resort to a Bogoliubov-like transformation~\cite{bogoliubov1947on} in order to properly describe the condensate fraction at equilibrium, see Ref.\cite{connaughton05}. Notice that this approach has been also extended to light propagation in GRIN MMFs, which confirmed that the Bogoliubov correction is negligible for the weakly nonlinear regime that it has been explored so far in optical thermalization experiments \cite{baudin21}. 
It is also interesting to note that the modified Bogoliubov dispersion relation has been recently measured experimentally in an optical system in a bulk configuration~\cite{fontaine2018observation}. Such a dispersion relation is at the origin of the peculiar physical properties which have been experimentally investigated in fluids of light. Among these, we may cite superfluidity \cite{frisch1992transition,carusotto2014superfluid,michel2018,braidotti22}, and the turbulence flow of superfluid light past an obstacle \cite{eloy21,glorieux23turbulent}, which are related to quantum vortex dynamics and the Berezinskii-Kosterlitz-Thouless transition \cite{situ2020dynamics}.

As a future perspective, it would also be interesting to carry out optical experiments aimed at studying the transition to condensation beyond the weakly nonlinear regime explored so far in MMFs, in relation with the spontaneous formation of strongly nonlinear coherent structures inherent to the focusing regime (spatial solitons, rogue waves...), see e.g., \cite{Rumpf09}. %We note in this respect the recent experiment in Ref.\cite{zhang2024bose} that has reported the observation of spatial beam-cleaning in a dual-core optical fiber   characterized by the formation of a spatial soliton in the fundamental fiber mode. 

Another aspect which has been largely disregarded so far and, we believe, will be investigated in the near future, is the role that time plays in the nonlinear beam dynamics which leads to thermal equilibrium, e.g., in the case of BSC. Indeed, since the results reported in Sec. \ref{sec_exp_phase} rule out the idea that modal phases define the statistical ensemble that leads to the experimental observation of the RJ distribution in standard BSC experiments, it is reasonable to suppose that time averaging is the mechanism which is responsible for the agreement between experimental observations and predictions of theories based on statistical approaches. However, such hypothesis has not yet been experimentally proved. As a matter of fact, only a few experimental works on BSC have considered the time domain so far. Besides the results by Mangini et al. discussed in Sec. \ref{Sec_4}, the experimental observations of Krupa et al. \cite{PhysRevA.97.043836} and Leventoux et al. \cite{leventoux20203d}, as well as Labaz and Sidorenko \cite{labaz2024spatial}, indicate that the temporal features of self-cleaned beam are non-trivial. Specifically, in 
\cite{PhysRevA.97.043836, leventoux20203d} it was
observed that spatial features of optical pulses that undergo BSC are quite different when considering the center of the pulse or its tails; whereas the Authors of \cite{labaz2024spatial} have shown that the spatial characteristics of self-cleaned beams depend on the temporal profile of the input beam \cite{labaz2024spatial}. 
%\textcolor{red}{
In addition, it has to be noted that all experiments reported so far in the literature do not take into account the intrinsic bandwidth of the measurement devices. Only recently, Goery et al. presented at a conference a single-shot characterization of the complex spatial amplitude of beams at the output of multimode fibers in the nonlinear regime \cite{genty2025real}. Unexpectedly, these results show that, when using ultrashort pulses, the measurement bandwidth has a significant influence on the output beam quality.
%}

As far as temporal aspects are involved, it is worth emphasizing that the wave turbulence theory discussed in Sec.~\ref{Sec_3} can be extended to study spatio-temporal effects in MMFs \cite{Wright2015R31,PhysRevLett.115.223902,KrupaPRLGPI,WrightNP2016,Shtyrina2018,PhysRevA.97.023803}. The development of a spatio-temporal kinetic theory would also be important to study complex incoherent behaviors, such as the generation of supercontinuum radiation in MMFs \cite{Wright2015R31,WrightNP2016,Eftekhar:17,KrupaLuot:16,Galmiche:16}, in relation with the wave turbulence approach that has been developed to study supercontinuum generation in singlemode fibers \cite{barviauOE09,barviauPRA09,barviauPRA13,xu2017origin}. The discrete wave turbulence approach discussed in this work can also be extended to study the impact of disorder on turbulence cascades, from both the theoretical and the experimental point of view, in relation with the recent study in MMFs \cite{baudin21}.

Finally, a last research line that will undoubtedly be pursued is the improvement of the accuracy of MD techniques and the development of novel MD devices. As a matter of fact, state-of-the-art MD tools have rather limited accuracy when measuring the population of HOMs. This creates several issues. A remarkable example in this regard is the divergence of the optical entropy \cite{Ferraro2023apx, ferraro2024calorimetry}. Besides accuracy, future directions of MD involve the development of fast devices and methods that, ideally, will allow to carry out the MD analysis of single pulses as well as of time-resolved MD tools. In this regard, invaluable help is likely to be provided by artificial intelligence algorithms, whose impact on all fields of the scientific research has been exponentially growing over the last few years.

%%%%%%%%%%%%%%%%%%%%%%%%%%%%%%%%%%%%%%%%%%%%%%%%%%%
\clearpage

\tableofcontents

\newpage

\appendix
%\section{Example Appendix Section}
%\label{app1}
%\subsection{Sub Appendix}
%\label{sec:mode_decompositions}

%In this appendix,...
%\begin{equation}
%\mathcal{E}_{\rm in}(x,y) = \frac{1}{\sigma\sqrt{\pi}}e^{-\frac{(x-x_0)^2+(y-y_0)^2}{2\sigma^2}},
%\label{eq:input_beam}
%\end{equation}
%where $\sigma$ is related to the beam full width at half maximum (FWHM) of its intensity $|\mathcal{E}_{\mathrm{in}}(x,y)|^2$, or beam size $w=\sigma/1.665$, under the normalization condition:

%%%%%%%%%%%%%%%%%%%%%%%%%%%%%%%%%%%%%%%%%%%%%%%%%%%%%%%%%%%%%%%%%%%%%%%
%%%%%%%%%%%%%%%%%%%%%%%%%%%%%%%%%%%%%%%%%%%%%%%%%%%%%%%%%%%%%%%%%%%%%%%
%%%%%%%%%%%%%%%%%%%%%%%   Appendix Weak Disorder
%%%%%%%%%%%%%%%%%%%%%%%%%%%%%%%%%%%%%%%%%%%%%%%%%%%%%%%%%%%%%%%%%%%%%%%
%%%%%%%%%%%%%%%%%%%%%%%%%%%%%%%%%%%%%%%%%%%%%%%%%%%%%%%%%%%%%%%%%%%%%%%

\section{Derivation of the discrete kinetic equation with weak disorder [Eq.(\ref{eq:kin_np_disc})].}
\label{app:weak_disorder}

\subsection{Modal correlations}
\label{app:weak_disorder_1}

To study correlations among the modes, we derive an equation for the moments of the $2\times 2$ matrix $\left<{\bf a}_p^* {\bf a}_q^T\right>(z)$.
%The computation for distinct modes ($p \neq q$) was reported in the Supplemental of Ref.\cite{PRL19}.
We first compute the correlations within a specific mode ($p=q$), and next for distinct modes.

The second moments satisfy:
$$
\partial_z \left< {\bf a}_p^* {\bf a}_p^T\right> = i \left< {\bf D}_p^* {\bf a}_p^* {\bf a}_p^T\right>
-i \left< {\bf a}_p^* {\bf a}_p^T {\bf D}_p^T \right> 
- i\gamma \big< {\bf G}_{pp}({\bf a})\big>  .
$$
where ${\bf G}_{pq}({\bf a}(z))={\bf P}_p({\bf a})^* {\bf a}_q^T(z)
- {\bf a}_p^* {\bf P}_q({\bf a})^T (z)$.
According to the Furutsu-Novikov theorem:
$$
 \left<  \nu_{p,j} \bsigma_j^* {\bf a}_p^* {\bf a}_p^T\right> = 
 \int_0^z \left< \frac{\delta \big( \bsigma_j^* {\bf a}_p^* {\bf a}_p^T(z) \big)}
 {\delta \nu_{p,j}(z') } \right>   \sigma^2_\beta  {\cal R}\Big( \frac{z-z'}{l_\beta}\Big) dz'
$$
The variational derivative can be computed by following \cite{konotop94}.For $z>z'$ it is the solution of 
\begin{eqnarray}
\nonumber
\partial_z  \frac{\delta \big( {\bf a}_p^* {\bf a}_p^T(z) \big)}
 {\delta \nu_{p,j}(z') } = i  {\bf D}_p^*  \frac{\delta \big( {\bf a}_p^* {\bf a}_p^T(z) \big)}
 {\delta \nu_{p,j}(z') } -i  \frac{\delta \big( {\bf a}_p^* {\bf a}_p^T(z) \big)}
 {\delta \nu_{p,j}(z') }  {\bf D}_p^T 
 - i\gamma \frac{\delta\big( {\bf G}_{pp}({\bf a}(z))\big)}{\delta \nu_{p,j}(z') }   
\label{eq:varder1}
\end{eqnarray}
starting from
$$
 \frac{\delta \big( {\bf a}_p^* {\bf a}_p^T(z) \big)}
 {\delta \nu_{p,j}(z') }  \mid_{z=z'}= 
 i \bsigma_j^*  {\bf a}_p^* {\bf a}_p^T(z) - i {\bf a}_p^* {\bf a}_p^T(z) \bsigma_j^T.
$$
We need to know the form of the variational derivative for  $z'<z$ and  $|z-z'| = O(l_\beta)$.
As $\sigma_\beta l_\beta \ll1 $ and $l_\beta \ll L_{nl}$, all terms in the right-hand side of the differential equation (\ref{eq:varder1}) satisfied by the variational derivative are negligible for $|z-z'| = O(l_\beta)$,
so that the leading-order expression of the  variational derivative for $z'<z$ and  $|z-z'| = O(l_\beta)$ is
\begin{eqnarray}
 \frac{\delta \big(  {\bf a}_p^* {\bf a}_p^T(z) \big)}
 {\delta \nu_{p,j}(z') } = i \bsigma_j^*  {\bf a}_p^* {\bf a}_p^T(z') 
-i   {\bf a}_p^* {\bf a}_p^T(z') \bsigma_j^T,
\end{eqnarray}
and therefore
\begin{align*}
 \left<  \nu_{p,j} \bsigma_j^* {\bf a}_p^* {\bf a}_p^T\right> =& 
 i 
 \int_0^z \left< \bsigma_j^*\bsigma_j^*  {\bf a}_p^* {\bf a}_p^T(z') \right>   \sigma^2_\beta  {\cal R}\Big( \frac{z-z'}{l_\beta}\Big) dz' 
 \\
&-i 
 \int_0^z \left< \bsigma_j^* {\bf a}_p^* {\bf a}_p^T(z') \bsigma_j^T \right>   \sigma^2_\beta  {\cal R}\Big( \frac{z-z'}{l_\beta}\Big) dz'.
\end{align*}
For $j=0$ this is zero and for $j=1,2,3$ this can be approximated by (using $l_\beta \ll z$):
$$
 \left<  \nu_{p,j} \bsigma_j^* {\bf a}_p^* {\bf a}_p^T\right> =
\frac{ i \Delta \beta }{2} 
\big(
\left< \bsigma_j^*\bsigma_j^*  {\bf a}_p^* {\bf a}_p^T \right>  
-  \left< \bsigma_j^* {\bf a}_p^* {\bf a}_p^T  \bsigma_j^T \right> \big) 
,
$$
and we find
$$
\partial_z \left< {\bf a}_p^* {\bf a}_p^T\right> = -  
 \Delta \beta 
 \Big( 3  \left< {\bf a}_p^* {\bf a}_p^T \right> 
-\sum_{j=1}^3 \bsigma_j^* \left< {\bf a}_p^* {\bf a}_p^T \right> \bsigma_j^T \Big)     
- i\gamma \big< {\bf G}_{pp}({\bf a})\big>  .
$$
The mean $2\times 2$ matrix $\left< {\bf a}_p^* {\bf a}_p^T\right> $ is Hermitian; therefore, it can be expanded as
$$
\left< {\bf a}_p^* {\bf a}_p^T\right> =  w_p(z) \bsigma_0 + \sum_{j=1}^3 w_{p,j}(z) \bsigma_j .
$$
The real-valued functions $w_p$ and $(w_{p,j})_{j=1}^3$ satisfy
\begin{align}
\partial_z w_p =& - \gamma \bPi_0 \big\{\big< i {\bf P}_p({\bf a})^* {\bf a}_p^T
- i {\bf a}_p^* {\bf P}_p({\bf a})^T \big> \big\},
\label{eq:wp}\\
\partial_z w_{p,j} =& 
- 4\Delta \beta   w_{p,j} 
- \gamma \bPi_j \big\{\big< i {\bf G}_{pp}({\bf a}) \big>\big\}, 
\label{eq:wpj}
\end{align}
where $\bPi_j {\bf W}=$ coefficient of the decomposition of the Hermitian matrix ${\bf W}$ on $\bsigma_j$. In particular, $\bPi_0 {\bf W} = {\rm Tr}({\bf W})/2 $ so that
$$
\bPi_0 \big\{\big< i {\bf G}_{pp}({\bf a})\big> \big\} 
%=  \frac{1}{2} \big< i {\bf P}_p({\bf a})^\dag {\bf a}_p- i{\bf a}_p^\dag {\bf P}_p({\bf a}) \big> 
= - {\rm Im} \big<   {\bf P}_p({\bf a})^\dag {\bf a}_p \big>   .
$$
The coefficients $w_{p,j} $ for $j=1,2,3$ satisfy the damped equations
$$
\partial_z w_{p,j} =
- 4\Delta \beta   w_{p,j} 
- \gamma \bPi_j \big\{\big< i {\bf G}_{pp}({\bf a}) \big>\big\}.
$$
They are of the form
\begin{align*}
w_{p,j}(z) =& w_{p,j}(0) \exp\big(-4 \Delta \beta   z\big) 
\\
&- \gamma \int_0^z\exp\big(-4 \Delta \beta   (z-z')\big) 
 \bPi_j \big\{\big< i {\bf G}_{pp}({\bf a}) \big>\big\}(z') dz'  .
\label{express:wpj1}
\end{align*}
As we have  $z,L_{nl} \gg 1/\Delta \beta$, 
the initial condition $w_{p,j}(0)$ is forgotten, and the second term in the right-hand side can be simplified
and we get for $j=1,2,3$ 
\begin{equation}
w_{p,j}
= - \frac{ \gamma}{4 \Delta \beta}  
\bPi_j \big\{  \big< i {\bf G}_{pp}({\bf a}(z)) \big>\big\} .
\label{eq:w_pj_0}
\end{equation}
Using the assumption $L_d=1/\Delta \beta \ll L_{nl}$,  we have
\begin{equation}
w_{p,j} \simeq 0
\label{eq:mom2pq2b}
\end{equation}
to leading order.

Let us now study correlations among distinct modes ($p \neq q$).
% we derive an equation for the moments of the $2\times 2$ matrix $\left<{\bf a}_p^* {\bf a}_q^T\right> $.
% which depend on fourth-order moments as a consequence of the nonlinearity. 
%The moments that involve disorder (of the form $\left< {\bf W}_p^* {\bf a}_p^* {\bf a}_q^T\right>$) are treated by making use of the Furutsu-Novikov theorem:
%\begin{eqnarray*}
% \left<  \nu_{p,j} \bsigma_j^* {\bf a}_p^* {\bf a}_q^T\right> &=& \int_0^z \left< \frac{\delta \big( \bsigma_j^* {\bf a}_p^* {\bf a}_q^T(z) \big)}
% {\delta \nu_{p,j}(z') } \right>   \sigma^2_\beta  {\cal R}\Big( \frac{z-z'}{l_\beta}\Big) dz'.
%\end{eqnarray*}
The variational derivatives satisfy for $z>z'$
\begin{eqnarray}
\nonumber
\partial_z  \frac{\delta \big( {\bf a}_p^* {\bf a}_q^T(z) \big)}
 {\delta \nu_{p,j}(z') } = i  {\bf D}_p^*  \frac{\delta \big( {\bf a}_p^* {\bf a}_q^T(z) \big)}
 {\delta \nu_{p,j}(z') } -i  \frac{\delta \big( {\bf a}_p^* {\bf a}_q^T(z) \big)}
 {\delta \nu_{p,j}(z') }  {\bf D}_q^T \\
\nonumber
+ i (\beta_p-\beta_q) \frac{\delta \big( {\bf a}_p^* {\bf a}_q^T(z) \big)}
 {\delta \nu_{p,j}(z') }  
 - i\gamma \frac{\delta \big( {\bf G}_{pq}({\bf a}(z))\big)}{\delta \nu_{p,j}(z') }, \quad \quad   
% - i\gamma \frac{\delta \big( {\bf P}_p({\bf a})^* {\bf a}_q^T(z)
%- {\bf a}_p^* {\bf P}_q({\bf a})^T (z)\big)}{\delta \nu_{p,j}(z') }   
\label{eq:varder1b}
\end{eqnarray}
with ${\bf G}_{pq}({\bf a}(z))={\bf P}_p({\bf a})^* {\bf a}_q^T(z)
- {\bf a}_p^* {\bf P}_q({\bf a})^T (z)$ and starting from
$ \frac{\delta ( {\bf a}_p^* {\bf a}_q^T(z) )}
 {\delta \nu_{p,j}(z') }  \mid_{z=z'}=  \bsigma_j^*  {\bf a}_p^* {\bf a}_q^T(z')$.
%We need to know the form of the variational derivative for  $z'<z$ and  $|z-z'| = O(l_\beta)$.
%Assuming that $L_{disor}=(\Delta \beta)^{-1} \ll L_{nl}$, the analysis shows that  the modes are not correlated with each other:
%\begin{equation}
% $\left< {\bf a}_p^* {\bf a}_q^T\right> \simeq 0$.
%\label{eq:mom2pq1b}
%\end{equation}
%This result is also valid for degenerate modes ($\beta_p = \beta_q$).
Assuming $\sigma_\beta l_\beta \ll 1 $ and $l_\beta \ll L_{nl}$, the stochastic and nonlinear terms in the equation for the variational derivatives are negligible for $|z-z'| = O(l_\beta)$,
so that to leading-order:
$ \delta \big( {\bf a}_p^* {\bf a}_q^T(z) \big)/
 \delta \nu_{p,j}(z')    =  \bsigma_j^*  {\bf a}_p^* {\bf a}_q^T(z') 
 \exp\big(i (\beta_p-\beta_q) (z-z') \big)$.
% and therefore
%\begin{eqnarray*}
% \left<  \nu_{p,j} \bsigma_j^* {\bf a}_p^* {\bf a}_q^T\right>  & &= i 
% \int_0^z \left< \bsigma_j^* \bsigma_j^*  {\bf a}_p^* {\bf a}_q^T(z') \right>   \sigma^2_\beta 
%  {\cal R}\Big( \frac{z-z'}{l_\beta}\Big)  \exp\big(i (\beta_p-\beta_q) (z-z') \big) dz'  .
%\end{eqnarray*}
Then using the involution property $\bsigma_j^2 =\bsigma_0$ we obtain
$
 \left< {\bf D}_p^*  {\bf a}_p^* {\bf a}_q^T\right> = 3 i
 \int_0^z \left<  {\bf a}_p^* {\bf a}_q^T(z') \right>   \sigma^2_\beta  {\cal R}\Big( \frac{z-z'}{l_\beta}\Big) 
  \exp\big(i (\beta_p-\beta_q) (z-z') \big)dz',
$
which can be approximated by 
$$
 \left< {\bf D}_p^*  {\bf a}_p^* {\bf a}_q^T\right> 
% = 3 i \left<  {\bf a}_p^* {\bf a}_q^T(z) \right>
%   \sigma^2_\beta \int_0^\infty  {\cal R}\Big( \frac{z'}{l_\beta}\Big) dz'  
   =
   \frac{3 i \Delta \beta}{2}  \left<  {\bf a}_p^* {\bf a}_q^T(z) \right>,
$$
by using $\sigma_\beta l_\beta\ll 1$ and $l_\beta\ll L_{nl}$ (${\bf a}_p^* {\bf a}_q^T(z) \simeq {\bf a}_p^* {\bf a}_q^T(z')
  \exp\big(i (\beta_p-\beta_q) (z-z') \big)$ 
for $|z-z'|=O(l_\beta)$).
Therefore we obtain 
\begin{eqnarray*}
\partial_z \left< {\bf a}_p^* {\bf a}_q^T\right> = 
(- 3 \Delta \beta+ i (\beta_p-\beta_q) ) \left< {\bf a}_p^* {\bf a}_q^T  \right>   - i\gamma \big<  {\bf G}_{pq} \big>  .
\end{eqnarray*}
Writing the solution for $\left< {\bf a}_p^* {\bf a}_q^T\right>(z)$ and using $ z, L_{nl} \gg 1/\Delta \beta$, the initial condition $\left< {\bf a}_p^* {\bf a}_q^T\right>(0)$ is forgotten  and the solution can be written
$$
\left< {\bf a}_p^* {\bf a}_q^T\right>(z) = 
\frac{i \gamma  \big< {\bf G}_{pq} \big>}{3\Delta \beta-i(\beta_p-\beta_q)}.
$$
Since $L_{nl} \gg 1/\Delta \beta$, we obtain $\left< {\bf a}_p^* {\bf a}_q^T\right>(z) \simeq 0$.

\subsection{Closure of the moments equations}
\label{app:weak_disorder_2}

The diagonal modal components $w_p$ in (\ref{eq:wp}) satisfy the undamped equation
\begin{eqnarray*}
\partial_z w_p = 
\gamma {\rm Im} \big<   {\bf P}_p({\bf a})^\dag {\bf a}_p \big>  .
\end{eqnarray*}
With the expression (\ref{def:nlterm0}) of the nonlinear term, we get that the mode occupancies $w_p$ satisfy the coupled equations (\ref{eq:pzwp1}-\ref{def:Xp2}).
%\begin{eqnarray}
%\label{eq:pzwp1}
%\partial_z w_p &=&  \frac{1}{3}  \gamma \left< X_p\right> 
%+ \frac{2}{3}  \gamma \left<  \tilde{X}_p\right> ,\\
%X_p &=&{\rm Im}\Big\{ \sum_{l,m,n} S_{plmn}^* ({\bf a}_l^\dag {\bf a}_m^*) ({\bf a}_n^T {\bf a}_p) \Big\}  ,
%\label{def:Xp}\\
%\tilde{X}_p &=&{\rm Im}\Big\{ \sum_{l,m,n} S_{plmn}^* ({\bf a}_n^T {\bf a}_m^*) ({\bf a}_l^\dag {\bf a}_p) \Big\}  .
%\label{def:tildeXp}
%\end{eqnarray}
We now derive the equations governing the evolutions of the fourth-order moments $\left< X_p^{(1)}\right> $
and $\left<  {X}_p^{(2)}\right>$ given by Eqs.(\ref{def:Xp1}-\ref{def:Xp2}).

%\begin{widetext}
\subsubsection{Computation of the term $\left<  {X}_p^{(1)}\right> $:}

We first write the equation satisfied by the product of four vector fields:
\begin{eqnarray}
\nonumber
\partial_z ({\bf a}_l^\dag {\bf a}_m^*) ( {\bf a}_n^T {\bf a}_p)=&&\hspace*{-0.2in} 
i(\beta_l+\beta_m-\beta_n-\beta_p) ({\bf a}_l^\dag {\bf a}_m^*) ({\bf a}_n^T {\bf a}_p)
+ i ({\bf a}_l^\dag {\bf D}_l^\dag {\bf a}_m^*) ( {\bf a}_n^T {\bf a}_p)\\
\nonumber
&&\hspace*{-0.2in} 
+ i ({\bf a}_l^\dag {\bf D}_m^* {\bf a}_m^*) ( {\bf a}_n^T {\bf a}_p)
- i ({\bf a}_l^\dag {\bf a}_m^*) ( {\bf a}_n^T {\bf D}_n^T {\bf a}_p)
\\
&&\hspace*{-0.2in} 
- i ({\bf a}_l^\dag {\bf a}_m^*) ( {\bf a}_n^T {\bf D}_p {\bf a}_p)+ i \gamma {Y}_{lmnp}^{(1)} ,  
\label{eq:4f_Xp1}
\end{eqnarray}
where
\begin{eqnarray}
\nonumber
{Y}_{lmnp}^{(1)}=&& \hspace*{-0.2in} -\sum_{l',m',n'} S_{l'm'n'l}^* \Big[ \frac{1}{3} ({\bf a}_{l'}^\dag {\bf a}_{m'}^*) ({\bf a}_{n'}^T {\bf a}_m^*)
+\frac{2}{3}({\bf a}_{n'}^T {\bf a}_{m'}^*) ({\bf a}_{l'}^\dag {\bf a}_m^*)\Big]  ({\bf a}_n^T{\bf a}_p) \\
\nonumber
&&\hspace*{-0.5in} -\sum_{l',m',n'} S_{l'm'n'm}^* \Big[ \frac{1}{3} ({\bf a}_{l'}^\dag {\bf a}_{m'}^*) ({\bf a}_{l}^\dag {\bf a}_{n'})
+\frac{2}{3}({\bf a}_{n'}^T {\bf a}_{m'}^*) ( {\bf a}_{l}^\dag {\bf a}_{l'}^*)\Big]  ({\bf a}_n^T {\bf a}_p) \\
\nonumber
&&\hspace*{-0.5in} +\sum_{l',m',n'} S_{l'm'n'n} \Big[ \frac{1}{3}  ({\bf a}_{l'}^T {\bf a}_{m'})
({\bf a}_{n'}^\dag {\bf a}_p)
+\frac{2}{3}(  ( {\bf a}_{n'}^\dag {\bf a}_{m'}) ({\bf a}_{l'}^T {\bf a}_p) \Big] ({\bf a}_{l}^\dag {\bf a}_{m}^*)\\
&&\hspace*{-0.5in} +\sum_{l',m',n'} S_{l'm'n'p} \Big[ \frac{1}{3}  ({\bf a}_{l'}^T {\bf a}_{m'})
({\bf a}_{n}^T {\bf a}_{n'}^*)
+\frac{2}{3} ( {\bf a}_{n'}^\dag {\bf a}_{m'}) ({\bf a}_{n}^T {\bf a}_{l'}) \Big]({\bf a}_{l}^\dag {\bf a}_{m}^*) .
\label{express:tildeY1}
\end{eqnarray}
We take the expectation and we apply the Gaussian summation rule to the sixth-order moments in the expression of $\left< {Y}_{lmnp}^{(1)}\right>$:
%\begin{eqnarray}
%\nonumber
%\left< {Y}_{lmnp}\right> &=& \frac{16}{3} S_{plmn} \big( w_l w_m w_p+w_lw_mw_n-w_nw_pw_m-w_nw_pw_l\big)  \\
%\nonumber
%&&+ \frac{16}{3} \delta_{mp} s_{nl}({\bf w}) w_p (w_l-w_n)+ \frac{16}{3} \delta_{mn} s_{pl}({\bf w}) w_n (w_l-w_p)\\
%&&+ \frac{16}{3} \delta_{lp} s_{nm}({\bf w}) w_p (w_m-w_n)+ \frac{16}{3} \delta_{ln} s_{pm}({\bf w}) w_n (w_m-w_p) , \\
%s_{nl}({\bf w})  &=& \sum_{n'}S_{nln'n'} w_{n'}.
%\end{eqnarray}
\begin{eqnarray}
\nonumber
\left< {Y}_{lmnp}^{(1)}\right> =
&& \hspace*{-0.2in} \frac{16}{3} S_{lmnp} \big( n_l n_m n_p+n_ln_mn_n-n_nn_pn_m-n_nn_pn_l\big)  \\
\nonumber
&& \hspace*{-0.2in}
+ \frac{16}{3} \delta^K_{mp} s_{ln}({\bf w}) n_p (n_l-n_n)
+ \frac{16}{3} \delta^K_{mn} s_{lp}({\bf w}) n_n (n_l-n_p)\\
&& \hspace*{-0.2in}
+ \frac{16}{3} \delta^K_{lp} s_{mn}({\bf w}) n_p (n_m-n_n)+ \frac{16}{3} \delta^K_{ln} s_{mp}({\bf w}) n_n (n_m-n_p) , 
\label{eq:Y_lmnp_1}\\
s_{ln}({\bf w})  =&& \hspace*{-0.2in} \sum_{n'}S_{ln'n'n} n_{n'}.
\label{eq:s_nl_1}
\end{eqnarray}

We now extend the procedure of Sec.~\ref{app:weak_disorder} to the computation of fourth-order modes considered here. 
Making use of the Furutsu-Novikov theorem, and considering the general case $l\neq m\neq n\neq p$ in the regime $\sigma_\beta l_\beta \ll 1 $ and $l_\beta \ll L_{nl}$, we obtain
\begin{eqnarray}
\nonumber
\partial_z \left< ({\bf a}_l^\dag {\bf a}_m^*) ( {\bf a}_n^T {\bf a}_p) \right> =&&\hspace*{-0.2in}
\big( - 8  \Delta \beta + i  (\beta_l+\beta_m-\beta_n-\beta_p)  \big)
\left<({\bf a}_l^\dag {\bf a}_m^*) ({\bf a}_n^T {\bf a}_p) \right>   \\
&&\hspace*{-0.2in} + i \gamma \left< {Y}_{lmnp}^{(1)} \right>,
\label{eq_app:J_lmnp_1}
\end{eqnarray}
whose solution has the form
\begin{eqnarray}
\nonumber
&& \hspace*{-0.6in} \left< ({\bf a}_l^\dag {\bf a}_m^*) ( {\bf a}_n^T {\bf a}_p) \right>(z) \\
\nonumber
&&\hspace*{-0.6in}= 
\left< ({\bf a}_l^\dag {\bf a}_m^*) ( {\bf a}_n^T {\bf a}_p) \right>(0) \exp\big( 
\big( - 8  \Delta \beta + i  (\beta_l+\beta_m-\beta_n-\beta_p)  \big)z\big)\\
&&\hspace*{-0.5in}
+ i \gamma \int_0^z\left< {Y}_{lmnp}^{(1)}\right> (z') \exp\big( 
\big( - 8  \Delta \beta + i  (\beta_l+\beta_m-\beta_n-\beta_p)  \big)(z-z')\big) dz'  .
\end{eqnarray}
These equations correspond to those reported in Eq.(\ref{eq:4th_order_moment}) and Eq.(\ref{eq:mom4pq}) for the fourth-order moment $J_{lmnp}^{(1)}$.

%If $z,L_{nl} \gg 1/ {\Delta \beta}$ then we find that the initial condition is forgotten in (\ref{eq:mom4pq}) and the second term in the right-hand side can be simplified into 
%\begin{eqnarray}
%\left< ({\bf a}_l^\dag {\bf a}_m^*) ( {\bf a}_n^T {\bf a}_p) \right>(z) &=&
%\frac{i\gamma }{8\Delta \beta - i (\beta_l+\beta_m-\beta_n-\beta_p)} \left< {Y}_{lmnp}^{(1)}\right>  (z) .
%\label{eq:mom4pq1}
%\end{eqnarray}
%As, moreover, $\beta_0 \gg \Delta \beta$, this gives
%\begin{eqnarray}
%\left< ({\bf a}_l^\dag {\bf a}_m^*) ( {\bf a}_n^T {\bf a}_p) \right>(z) =
%\frac{i\gamma }{8\Delta \beta} \left< {Y}_{lmnp}^{(1)}\right>(z)  \delta^K(\beta_l+\beta_m-\beta_n-\beta_p).
%\label{eq:mom4pq1b}
%\end{eqnarray}
%The hypothesis $\beta_0 \gg \Delta \beta$ is the one that makes it possible to reduce the equation to a discrete one.

In the other cases when (at least) two indices are equal, i.e., the fourth-order moments involve degenerate modes, the calculation shows that there is still a damping in the moment equation, although the damping factor can be different from $8 \Delta \beta$.
However we will neglect this change because: (i) these terms are negligible in the triple sum and when the number of modes is large ($M \gg 1$), and (ii) this change only affects the multiplicative coefficient $8\Delta \beta$  in  (\ref{eq_app:J_lmnp_1}).
Finally we give the expression of $\left< X_p^{(1)}\right>$ in the discrete wave turbulence regime where $\beta_0 \gg \Delta \beta$. Collecting all terms we obtain
\begin{eqnarray}
\nonumber
\left< X_p^{(1)}\right> =&&\hspace*{-0.2in} \frac{2 \gamma}{3 \Delta \beta} \sum_{l,m,n}
 |S_{lmnp}|^2 \delta^K (\beta_l+\beta_m- \beta_n - \beta_p)\\
 \nonumber &&\times
 \big( n_l n_m n_p+n_ln_mn_n-n_nn_pn_m-n_nn_pn_l\big)  \\
&&\hspace*{-0.2in}
+    \frac{4\gamma}{3 \Delta \beta}  \sum_l  |  s_{lp}({\bf w}) |^2 \delta^K (\beta_l- \beta_p) (n_l-n_p) .
\label{eq:momX_p1}
  \end{eqnarray}

\subsubsection{Computation of the term $\left< {X}_p^{(2)}\right> $:}

The equation satisfied by the product of four vector fields reads
\begin{eqnarray}
\nonumber
\partial_z ({\bf a}_n^T {\bf a}_m^*) ( {\bf a}_l^\dag {\bf a}_p)=&&\hspace*{-0.2in}
i(\beta_l+\beta_m-\beta_n-\beta_p) ({\bf a}_n^T {\bf a}_m^*) ({\bf a}_l^\dag {\bf a}_p)
+ i ({\bf a}_n^T  {\bf a}_m^*) ( {\bf a}_l^\dag {\bf D}_l^\dag {\bf a}_p)\\
\nonumber
&&\hspace*{-0.2in}
+ i ({\bf a}_n^T {\bf D}_m^* {\bf a}_m^*) ( {\bf a}_l^\dag  {\bf a}_p)
- i ({\bf a}_n^T {\bf D}_n^T {\bf a}_m^*) ( {\bf a}_l^\dag  {\bf a}_p)\\
&&\hspace*{-0.2in}
- i ({\bf a}_n^T {\bf a}_m^*) ( {\bf a}_l^\dag {\bf D}_p {\bf a}_p)+ i \gamma {Y}_{lmnp}^{(2)} ,  
\label{eq:4f_Xp2}
\end{eqnarray}
where
\begin{eqnarray}
\nonumber
{Y}_{lmnp}^{(2)}=&&\hspace*{-0.2in} -\sum_{l',m',n'} S_{l'm'n'l}^* 
\Big[ \frac{1}{3} ({\bf a}_{l'}^\dag {\bf a}_{m'}^*) ({\bf a}_{n'}^T {\bf a}_p)
+\frac{2}{3}({\bf a}_{n'}^T {\bf a}_{m'}^*) ({\bf a}_{l'}^\dag {\bf a}_p)\Big] 
 ({\bf a}_m^\dag{\bf a}_n) \\
\nonumber
&&\hspace*{-0.5in}-\sum_{l',m',n'} S_{l'm'n'm}^* \Big[ 
\frac{1}{3} ({\bf a}_{l'}^\dag {\bf a}_{m'}^*)
 ({\bf a}_{n'}^T {\bf a}_{n})
+\frac{2}{3}({\bf a}_{n'}^T {\bf a}_{m'}^*) 
( {\bf a}_{l'}^\dag {\bf a}_n)\Big]  
({\bf a}_l^\dag {\bf a}_p) \\
\nonumber
&&\hspace*{-0.5in}+\sum_{l',m',n'} S_{l'm'n'n} \Big[ \frac{1}{3} ({\bf a}_{l'}^T {\bf a}_{m'})
({\bf a}_{n'}^\dag {\bf a}_m^*)
+\frac{2}{3}  ( {\bf a}_{n'}^\dag {\bf a}_{m'}) ({\bf a}_{l'}^T {\bf a}_m^*) \Big]  ({\bf a}_l^\dag {\bf a}_p)\\
&&\hspace*{-0.5in}+\sum_{l',m',n'} S_{l'm'n'p} \Big[ \frac{1}{3} 
 ({\bf a}_{l'}^T {\bf a}_{m'}) ({\bf a}_{n'}^\dag {\bf a}_{l}^*)
+\frac{2}{3}  ( {\bf a}_{n'}^\dag {\bf a}_{m'}) ({\bf a}_{l'}^T {\bf a}_{l}^*) \Big] ({\bf a}_m^\dag {\bf a}_n).
\label{express:tildeY}
\end{eqnarray}
We take the expectation 
and we apply the Gaussian summation rule to the sixth-order moments in the expression of $\left< {Y}_{lmnp}^{(2)}\right>$:
\begin{eqnarray}
\nonumber
\left< {Y}_{lmnp}^{(2)}\right> =&&\hspace*{-0.2in}
\frac{16}{3} S_{lmnp} \big( n_l n_m n_p+n_ln_mn_n-n_nn_pn_m-n_nn_pn_l\big) \\
\nonumber
&&\hspace*{-0.2in} 
+ \frac{16}{3} \delta^K_{mp} s_{ln}({\bf w}) n_p (n_l-n_n) +\frac{32}{3} \delta^K_{mn} s_{lp}({\bf w}) n_n (n_l-n_p)\\
&&\hspace*{-0.2in} 
+ \frac{32}{3} \delta^K_{lp} s_{mn}({\bf w}) n_p (n_m-n_n)+ \frac{16}{3} \delta^K_{ln} s_{mp}({\bf w}) n_n (n_m-n_p) .
\label{eq:Y_lmnp_2}
\end{eqnarray}
If $l\neq m\neq n\neq p$, then Furutsu-Novikov formula gives
$$
\partial_z \left< ({\bf a}_n^T {\bf a}_m^*) ( {\bf a}_l^\dag {\bf a}_p)\right> =
\big( -8 \Delta \beta  + i(\beta_l+\beta_m-\beta_n-\beta_p)\big)
\left<({\bf a}_n^T {\bf a}_m^*) ( {\bf a}_l^\dag {\bf a}_p)\right>   
 + i \gamma \left< {Y}_{lmnp}^{(2)}\right>  .
$$
%We proceed as for the term $\left< ({\bf a}_l^\dag {\bf a}_m^*) ( {\bf a}_n^T {\bf a}_p) \right>(z)$ and we find that
%\begin{equation}
%\left<({\bf a}_n^T {\bf a}_m^*) ( {\bf a}_l^\dag {\bf a}_p)\right>
%=
% \frac{ i\gamma }{8 {\Delta \beta} }
%\left< {Y}_{lmnp}^{(2)} \right> \delta^K(\beta_l+\beta_m-\beta_n-\beta_p).
%\end{equation}
This equation corresponds to that reported in Eq.(\ref{eq:4th_order_moment}) for the fourth-order moment $J_{lmnp}^{(2)}$.

Following the same procedure, we have also derived the following equation
\begin{eqnarray}
\partial_z \left< ({\bf a}_l^T {\bf a}_l^*) ( {\bf a}_m^\dag {\bf a}_m) \right> =0,
\label{eq:4_l_m}
\end{eqnarray}
for any $l,m$. 
This implies, in particular, that the variance of the intensity fluctuations of each modal component $p$ is preserved $\left< |{\bf a}_p|^4\right> =$const, i.e., the Gaussian statistics is preserved during propagation.\\
Note that Eq.(\ref{eq:4_l_m}) is undamped, but this does not affect our results. 
Indeed, 
$ \left< ({\bf a}_l^T {\bf a}_l^*) ( {\bf a}_m^\dag {\bf a}_m) \right>$ is real-valued
so that  it does not contribute when it is substituted  into Eq.(\ref{def:Xp2}) because $S_{lmml}$ is real-valued as well.

By neglecting the small corrections that appear in  the cases when (at least) two indices are equal (i.e., fourth-order modes involving degenerate modes), we obtain in the discrete wave turbulence regime ($\beta_0 \gg \Delta \beta$):
\begin{eqnarray}
\nonumber
\left< {X}_p^{(2)}\right> =&& \hspace*{-0.2in} \frac{2 \gamma}{3   {\Delta \beta} } \sum_{l,m,n}
\delta^K (\beta_l+\beta_m- \beta_n - \beta_p)  |S_{lmnp}|^2 \\
\nonumber
&& \times
 \big( n_l n_m n_p+n_ln_mn_n-n_nn_pn_m-n_nn_pn_l\big) \\
&&\hspace*{-0.2in}
+   \frac{2\gamma}{  {\Delta \beta} } \sum_l 
\delta^K (\beta_l - \beta_p)  |  s_{lp}({\bf w}) |^2 (n_l-n_p)  .
\label{eq:momX_p2}
\end{eqnarray}
By replacing the expressions of $\left< {X}_p^{(j)}\right>$ ($j=1,2$) given in (\ref{eq:momX_p1}) and (\ref{eq:momX_p2}) into the equation for the evolution of the modal components (\ref{eq:pzwp1}), we obtain the discrete kinetic Eq.(\ref{eq:kin_np_disc}).
%\end{widetext}

%%%%%%%%%%%%%%%%%%%%%%%%%%%%%%%%%%%%%%%%%%%
\subsection{Impact of a correlated noise model of disorder on the kinetic equation}
\label{app:weak_disorder_3}

\subsubsection{Computation of the moment $\left< ({\bf a}_l^\dag {\bf a}_m^*) ( {\bf a}_n^T {\bf a}_p) \right>$}

In the model of correlated disorder, all modes experience the same noise, ${\bf D}_n\equiv {\bf D} = \sum_{j=0}^3 \nu_j \bsigma_j$.
The equation (\ref{eq:4f_Xp1}) for the evolution of the product of four fields now reads
\begin{eqnarray}
\nonumber
\partial_z ({\bf a}_l^\dag {\bf a}_m^*) ( {\bf a}_n^T {\bf a}_p)=&& \hspace*{-0.2in}
i(\beta_l+\beta_m-\beta_n-\beta_p) ({\bf a}_l^\dag {\bf a}_m^*) ({\bf a}_n^T {\bf a}_p)
+ i ({\bf a}_l^\dag {\bf D}^\dag {\bf a}_m^*) ( {\bf a}_n^T {\bf a}_p)\\
\nonumber
&&\hspace*{-0.2in}
+ i ({\bf a}_l^\dag {\bf D}^* {\bf a}_m^*) ( {\bf a}_n^T {\bf a}_p)
- i ({\bf a}_l^\dag {\bf a}_m^*) ( {\bf a}_n^T {\bf D}^T {\bf a}_p)\\
&&\hspace*{-0.2in}
- i ({\bf a}_l^\dag {\bf a}_m^*) ( {\bf a}_n^T {\bf D} {\bf a}_p)+ i \gamma {Y}_{lmnp}^{(1)} .
\end{eqnarray}
We note that ${\bf a}_l^\dag {\bf D}^\dag {\bf a}_m^* + {\bf a}_l^\dag {\bf D}^* {\bf a}_m^* = 2\sum_{j\in \{1,3\}}
\nu_j {\bf a}_l^\dag \bsigma_j {\bf a}_m^*$.
We follow the procedure outlined above in sections \ref{app:weak_disorder_1}-\ref{app:weak_disorder_2}.
Using the Furutsu-Novikov theorem and assuming $\sigma_\beta l_\beta \ll 1 $ and $l_\beta \ll L_{nl}$ we obtain
\begin{align*}
\frac{\delta {\bf a}_l^\dag \bsigma_j {\bf a}_m^* {\bf a}_n^T {\bf a}_p(z)}{\delta \nu_j(z')}
=& 2i
\Big[
{\bf a}_l^\dag {\bf a}_m^* {\bf a}_n^T  {\bf a}_p(z') 
-
{\bf a}_l^\dag \bsigma_j {\bf a}_m^* {\bf a}_n^T\bsigma_j   {\bf a}_p(z') \big] \\
&\times 
\exp\big[ i (\beta_l+\beta_m-\beta_n-\beta_p)(z-z')\big]
\end{align*}
and therefore
\begin{align}
\nonumber
\partial_z \left< ({\bf a}_l^\dag {\bf a}_m^*) ( {\bf a}_n^T {\bf a}_p) \right> =&
\big( - 4  \Delta \beta + i  (\beta_l+\beta_m-\beta_n-\beta_p)  \big)
\left<({\bf a}_l^\dag {\bf a}_m^*) ({\bf a}_n^T {\bf a}_p) \right>   
 \\
& \hspace*{-0.5in} + i \gamma \left< {Y}_{lmnp}^{(1)} \right> +2 \Delta \beta \sum_{j\in \{1,3\}}  \left< ({\bf a}_l^\dag \bsigma_j {\bf a}_m^*) ( {\bf a}_n^T  \bsigma_j{\bf a}_p) \right>.
\label{eq:4m_corr_disor}
\end{align}
The last term can also be written as:
$$
 \sum_{j\in \{1,3\}}  \left< ({\bf a}_l^\dag \bsigma_j {\bf a}_m^*) ( {\bf a}_n^T  \bsigma_j{\bf a}_p) \right>
 =
  \left< ({\bf a}_l^\dag {\bf a}_n) ( {\bf a}_m^\dag {\bf a}_p) \right>
+
  \left< ({\bf a}_l^\dag \bsigma_2 {\bf a}_n) ( {\bf a}_m^\dag \bsigma_2 {\bf a}_p) \right>.
$$
The analysis reveals that the terms involving the dissipation that are proportional to $\Delta \beta$ in (\ref{eq:4m_corr_disor}) essentially vanish.
Indeed, using the factorizability property of statistical Gaussian fields to split the fourth-order moments into products of second-order moments, then we get
$\left<({\bf a}_l^\dag {\bf a}_m^*) ({\bf a}_n^T {\bf a}_p) \right>  = \frac{1}{2} w_l w_m ( \delta^K_{ln}\delta^K_{mp}+\delta^K_{lp}\delta^K_{mn})$
and
$\sum_{j\in \{1,3\}}  \left< ({\bf a}_l^\dag \bsigma_j {\bf a}_m^*) ( {\bf a}_n^T  \bsigma_j{\bf a}_p) \right>
= w_l w_m ( \delta^K_{ln}\delta^K_{mp}+\delta^K_{lp}\delta^K_{mn})$.
This shows that, to leading order,  the terms involving the dissipation indeed
compensate with each other.

\subsubsection{Computation of the moment $\left< ({\bf a}_n^T {\bf a}_m^*) ( {\bf a}_l^\dag {\bf a}_p)\right>$}

The equation (\ref{eq:4f_Xp2}) for the evolution of the product of four fields now reads
\begin{align}
\nonumber
\partial_z ({\bf a}_n^T {\bf a}_m^*) ( {\bf a}_l^\dag {\bf a}_p)=&
i(\beta_l+\beta_m-\beta_n-\beta_p) ({\bf a}_n^T {\bf a}_m^*) ({\bf a}_l^\dag {\bf a}_p)
+ i ({\bf a}_n^T  {\bf a}_m^*) ( {\bf a}_l^\dag {\bf D}^\dag {\bf a}_p)\\
\nonumber
&
+ i ({\bf a}_n^T {\bf D}^* {\bf a}_m^*) ( {\bf a}_l^\dag  {\bf a}_p)
- i ({\bf a}_n^T {\bf D}^T {\bf a}_m^*) ( {\bf a}_l^\dag  {\bf a}_p)
\\
&
- i ({\bf a}_n^T {\bf a}_m^*) ( {\bf a}_l^\dag {\bf D} {\bf a}_p)
+ i \gamma {Y}_{lmnp}^{(2)} .
\end{align}
Since the random matrix is Hermitian (${\bf D}^\dag={\bf D}$), the expression can be reduced
$$
\partial_z ({\bf a}_n^T {\bf a}_m^*) ( {\bf a}_l^\dag {\bf a}_p) = 
i(\beta_l+\beta_m-\beta_n-\beta_p) ({\bf a}_n^T {\bf a}_m^*) ({\bf a}_l^\dag {\bf a}_p)
+ i \gamma {Y}_{lmnp}^{(2)} .  
$$
The fourth-order moment $\left< ({\bf a}_n^T {\bf a}_m^*) ( {\bf a}_l^\dag {\bf a}_p)\right>$ then evolves as in the absence of any disorder ($\Delta \beta=0$).

\subsection{Impact of a partially correlated noise model of disorder on the kinetic equation}
\label{app:weak_disorder_4}

In the partially correlated model of disorder, all degenerate modes experience the same realization of disorder.
Groups of degenerate modes with the same reduced eigenvalue $\beta$ are indexed by $p$, the modes within the $p$th group are indexed by $(p,j)$, and the  linear polarization components of the $(p,j)$-th mode are ${\bf a}_{pj}$.
%The reduced wavenumber of the $p$-th group is denoted by $\beta_p$.
The structural disorder induces a linear random coupling between the modes of different groups, as described by the $2\times 2$ matrix ${\bf D}_p(z)= \sum_{l=0}^3 \nu_{p,l}(z) \bsigma_l$ where $p$ labels the mode group number, i.e., degenerate modes that belong to the same group experience the same noise through the random process $\nu_{p,l}(z)$ (this notation should not be confused with the decorrelated model of disorder where $p$ labels individual modes).
%We define ${\bf a}_{pj}(z)={\bf B}_{pj}(z) \exp(-i\beta_p z)$. 
The evolutions of the modal components are governed by 
$$
i \partial_z {\bf a}_{pj} = \beta_p {\bf a}_{pj}
+ {\bf D}_p (z) {\bf a}_{pj} 
- \gamma  {\bf P}_{pj}({\bf a}).
$$
We look at the second-order moments $\left< {\bf a}_{pj}^* {\bf a}_{ql}^T\right>$.
We follow the procedure outlined above in section~\ref{app:weak_disorder_1} by using the Furutsu-Novikov theorem with $\sigma_\beta l_\beta \ll 1 $ and $l_\beta \ll L_{nl}$.

Considering non-degenerate modes ($p\neq q$), we obtain 
$$
\left< {\bf a}_{pj}^* {\bf a}_{ql}^T\right>(z) =
\frac{i \gamma}{4\Delta \beta-i(\beta_p-\beta_q)}
 \big< {\bf P}_{pj}({\bf a})^* {\bf a}_{ql}^T
- {\bf a}_{pj}^* {\bf P}_{ql}({\bf a})^T \big>  .
$$
In the regime $L_d = 1/\Delta \beta \ll L_{nl}$ the correlation is vanishing small, $\left< {\bf a}_{pj}^* {\bf a}_{ql}^T\right>(z) \simeq 0$, as in the model of decorrelated disorder.\\
Let us now consider correlations among the orthogonal polarization components of a mode ($p = q$ and $j=l$), then we find as before
$$
\left< {\bf a}_{pj}^* {\bf a}_{pj}^T\right> (z)=  w_{p,jj}(z) \bsigma_0 .
$$

We consider correlations among distinct degenerate modes ($p = q$ and $j\neq l$).
We define the two Hermitian matrices 
${\bf W}_{p,jl} = \frac{1}{2} \big( \left< {\bf a}_{pj}^* {\bf a}_{pl}^T\right>
+\left< {\bf a}_{pl}^* {\bf a}_{pj}^T\right>\big)$ and $\tilde{\bf W}_{p,jl} = \frac{i}{2} \big( \left< {\bf a}_{pj}^* {\bf a}_{pl}^T\right>
-\left< {\bf a}_{pl}^* {\bf a}_{pj}^T\right>\big)$
and then by carrying out the same calculations as in the case $p = q$ and $j=l$,
we find 
$$
\left< {\bf a}_{pj}^* {\bf a}_{pl}^T\right>(z) =  w_{p,jl}(z) \bsigma_0 .
$$
The coefficients $w_{p,jl}$ satisfy
\begin{align*}
\partial_z w_{p,jl} &=  \frac{\gamma}{2} {\rm Im} \big<   {\bf P}_{pj}({\bf a})^\dag {\bf a}_{pl} + {\bf P}_{pl}({\bf a})^\dag {\bf a}_{pj} \big>\bm \\
&=
\frac{1}{6}  \gamma \left< X_{p,jl}^{(1)}+ X_{p,lj}^{(1)}\right> 
+ \frac{1}{3}  \gamma \left<  {X}_{p,jl}^{(2)} +{X}_{p,lj}^{(2)} \right> ,\\
X_{p,jl}^{(1)} &={\rm Im}\Big\{ \sum_{p_1j_1,p_2j_2,p_3j_3} S_{pl,p_1j_1,p_2j_2,p_3j_3}^* ({\bf a}_{p_1j_1}^\dag {\bf a}_{p_2j_2}^*) ({\bf a}_{p_3j_3}^T {\bf a}_{pj}) \Big\}  ,\\
{X}_{p,jl}^{(2)} &={\rm Im}\Big\{ \sum_{p_1j_1,p_2j_2,p_3j_3} S_{pl,p_1j_1,p_2j_2,p_3j_3}^* ({\bf a}_{p_3j_3}^T {\bf a}_{p_2j_2}^*) ({\bf a}_{p_1j_1}^\dag {\bf a}_{pj}) \Big\}  .
\end{align*}
Let us look at the fourth-order moments $\left< ({\bf a}_{p_1j_1}^\dag {\bf a}_{p_2j_2}^*) ( {\bf a}_{p_3j_3}^T{\bf a}_{p_4j_4})\right>$\\ or 
$\left< ({\bf a}_{p_1j_1}^T {\bf a}_{p_2j_2}^*) ( {\bf a}_{p_3j_3}^\dag {\bf a}_{p_4j_4})\right>$. 
There are different types of such fourth-order moments that depend on the specific modes that they involve. Almost all of them satisfy an evolution equation with damping proportional to $\Delta \beta$, with different coefficients in front of $\Delta \beta$ that depend on the number of equal indices.
These terms are of the same form as those obtained by considering the decorrelated model of disorder, see section~\ref{app:weak_disorder_2}.
There are special cases when $p_j$ are equal by pairs (e.g. $p_1=p_2$ and $p_3=p_4$) where 
$$
\partial_z \left< ({\bf a}_{p_1j_1}^T {\bf a}_{p_1j_2}^*) ( {\bf a}_{p_3j_3}^\dag {\bf a}_{p_3j_4})\right>=
i \gamma {Y}_{p_3j_3,p_1j_2,p_1j_1,p_3j_4}^{(2)},
$$
which shows that there is no damping. From the expression of ${Y}^{(2)}$ (see Eq.(\ref{express:tildeY})), and the evaluation of its expectation in terms of the $w_{p,jl}$ according to the Gaussian rule for sixth-order moments, such terms induce a reversible exchange of energy between the modes within each group.

%%%%%%%%%%%%%%%%%%%%%%%%%%%%%%%%%%%%%%%%%%%%%%%%%%%%%%%%%%%%%%%%%%%%%%%
%%%%%%%%%%%%%%%%%%%%%%%%%%%%%%%%%%%%%%%%%%%%%%%%%%%%%%%%%%%%%%%%%%%%%%%
%%%%%%%%%%%%%%%%%%%%%%%   Condensation curve
%%%%%%%%%%%%%%%%%%%%%%%%%%%%%%%%%%%%%%%%%%%%%%%%%%%%%%%%%%%%%%%%%%%%%%%
%%%%%%%%%%%%%%%%%%%%%%%%%%%%%%%%%%%%%%%%%%%%%%%%%%%%%%%%%%%%%%%%%%%%%%%

\section{Wave condensation and condensate fraction plotted in Fig.~\ref{fig:sec_rj_cond_2}.}
\label{app:condensate_fraction}

In this Appendix, we sketch the derivation of the condensation curve, i.e., the fraction of the power condensed in the fundamental mode vs the temperature, or the energy, in the classical RJ limit. This in relation with Section~\ref{sec4:cond} where the experimental observation of RJ condensation has been reported.

\subsection{Thermodynamic limit}

In order to render this appendix self-contained, we recall that we consider a 2D parabolic potential $V(\br)=q |\br|^2$ for $|\br| \le R$, that is truncated at $V_0=q R^2$, where $R=26 \mu$m is the fiber radius and $q$ a constant determined by the fiber characteristics, $q=k_0(n_{\rm co}^2-n_{\rm cl}^2)/(2 n_{\rm co} R^2)$, $k_0=2\pi/\lambda_0$ being the laser wave-number, $n_{\rm co}-n_{\rm cl}$ the refractive index difference between the core and the cladding of the MMF. The eigenvalues are well approximated by the ones of the ideal harmonic potential $\beta_p=\beta_0(p_x+p_y+1)$, which are solutions of the Schr\"odinger equation $\beta_p u_p(\br)=-\alpha \nabla^2 u_p(\br) + V(\bm r)u_p(\br)$. 
%with the corresponding Gauss-Hermite modes $w_p$ that will be specified below (we recall that $p$ labels the two integers $(p_x,p_y)$  that specify a mode). Note that the eigenvalues retrieved from the Helmholtz equation $\beta_p^H$ are related to those of the Schr\"odinger equation by $\beta_p^H \simeq k_0 n_{\rm co} - \beta_p$. 
The truncation of the potential $V(\br) \le V_0$ and the corresponding finite number of modes $M$ introduce an effective frequency cut-off in the far-field spectrum $k_c=\sqrt{2V_0/\beta_0}/r_0 \simeq 1.15 \mu$m$^{-1}$, where $r_0$ is the radius of the fundamental mode of the MMF \cite{Baudin2020}.

We start from $N=\sum_p n_p^{eq}$ and $E=\sum_p \beta_p n_p^{eq}$ with the Rayleigh-Jeans (RJ) distribution $n_p^{eq}=T/(\beta_p-\mu)$, where $\beta_p=\beta_{p_x,p_y}=\beta_0(p_x+p_y+1)$ are the eigenvalues of the truncated parabolic potential ($V(\br) \le V_0$), and the index $\{p\}$ labels the two integers $(p_x,p_y)$ that specify a mode.
The sum over the modes reads $\sum_p = \sum_{0\le p_x+p_y < g}$,
%= \sum_{p_y=0}^{g-1} \sum_{p_x=0}^{g-1-p_y}$ 
where $g=V_0/\beta_0$ is the number of groups of non-degenerate modes, with $M=g(g+1)/2$ the number of modes (see sec.~III).
%As discussed in the main text (Fig.~1), 
Condensation arises when $\mu \to \beta_0^-$, 
%The denominator of the RJ distribution $n_p^{eq}=T/(\beta_p-\mu)$ tends to vanish, 
which leads to a macroscopic population of the fundamental mode.
In the following we term `thermodynamic limit' the limit defined by $N \to \infty$, $\beta_0 \to 0$ with $N \beta_0^2=$const and $V_0=$const. 
In this limit the discrete sums over the modes are replaced by continuous integrals, $N=\sum_p n_p^{eq} \to (T/\beta_0^2) \int_0^{V_0} dx \int_0^{V_0-x} dy (x+y+{\bar \mu})^{-1}$:
% which gives
\begin{eqnarray}
N= \frac{T}{\beta_0^2} \Big( V_0 + {\bar \mu} \log\big( {\bar \mu}/(V_0+{\bar \mu} )\big)   \Big),
%N= T \Big( V_0 + {\bar \mu} \log\big( {\bar \mu}/(V_0+{\bar \mu} )\big)   \Big)/\beta_0^2,
\label{eq:NvsTmu}
\end{eqnarray}
with ${\bar \mu}=\beta_0-\mu > 0$. Eq.(\ref{eq:NvsTmu}) shows that $\mu \to \beta_0^{-}$ for a non-vanishing critical temperature 
\begin{eqnarray}
T_{\rm crit}^*=N\beta_0^2/V_0.
\label{eq:Tstar_c}
\end{eqnarray}
%$T_c^*=N\beta_0^2/V_0$. 
This means that wave condensation is reestablished in the thermodynamic limit in 2D owing to the truncated parabolic potential.
The same conclusion is reached through the analysis of the energy. 
In the continuous limit we have
\begin{eqnarray}
\frac{E}{N} = \mu + \frac{V_0^2/2}{ V_0 + {\bar \mu} \log\big({\bar \mu}/(V_0+{\bar \mu})\big) }.
\label{eq:EsN_TL}
\end{eqnarray}
We see that $\mu \to \beta_0^{-}$ for a non-vanishing critical energy
\begin{eqnarray}
E_{\rm crit}^*/N=\beta_0+V_0/2 \simeq V_0/2,
\label{eq:Estar_c}
\end{eqnarray}
%$E_{\rm crit}^*/N=\beta_0+V_0/2 \simeq V_0/2$, 
and $\mu=\beta_0$ for $E \le E_{\rm crit}^*$.
% as for the quantum Bose-Einstein transition.

Note that, in contrast to homogeneous condensation (i.e. $V(\br)=0$) where the density of states is constant, in the presence of the parabolic potential $V(\br)=q|\br|^2$ the density of states is a linear function of $\beta$ ($\rho(\beta) = \beta/\beta_0^2$), which prevents the infrared divergence of the integral of the power $N=\int_0^{V_0} \rho(\beta) n_\beta^{eq} d\beta$.

\subsection{Condensate fraction beyond the thermodynamic limit}
 
The experiments are not performed in the thermodynamic limit, and finite size effects are taken into account by considering discrete sums over the modes (instead of continuous integrals): 
$N=\sum_p n_p^{eq}$, $E=\sum_{p} \beta_p n_p^{eq}$.
As in the usual treatment of condensation, we  split the contribution of the fundamental mode:
\begin{align}
&N = n_0^{eq} + T \sum_{p\neq0} 1/(\beta_p - \beta_0),\\
&E = n_0^{eq} \beta_0 + T \sum_{p\neq0} \beta_p/(\beta_p - \beta_0).
\end{align} 
By taking the ratio we eliminate the temperature $T$, which gives
\begin{eqnarray}
\frac{n_0^{eq}}{N} = 1 - \frac{\varrho (E - N\beta_0)}{N\beta_0 (M - 1)},
%n_0^{eq}/N = 1 - \varrho (E - N\beta_0)/\big(N\beta_0 (M - 1)\big),
\label{eq:n0sN_beta0}
\end{eqnarray}
where $\varrho = \sum_{p\neq0} 1/(p_x+p_y)$.
% and $E_{\rm min}=N\beta_0$.
Eq.(\ref{eq:n0sN_beta0}) shows that condensation arises by decreasing the energy below the critical value
\begin{eqnarray}
E_{\rm crit} = N \beta_0 \big(1 + (M - 1)/\varrho  \big).
\label{eq:E_crit}
\end{eqnarray}
Following the same procedure with the temperature instead of the energy, we obtain  
\begin{eqnarray}
n_0^{eq}/N=1-T/T_{\rm crit}, \quad T_{\rm crit}= N \beta_0/\varrho.
\label{eq:T_crit}
\end{eqnarray}
In the continuous limit $\varrho \to V_0/\beta_0$, so that $E_{\rm crit}$ and $T_{\rm crit}$ recover the expressions in the thermodynamic limit $E_{\rm crit}^{*}$ in Eq.(\ref{eq:Estar_c}) and $T_{\rm crit}^{*}$ in Eq.(\ref{eq:Tstar_c}).

%\subsection{Beyond $\mu = \beta_0$ for $E<E_{\rm crit}$}
%\noindent
%{\bf b) Condensate fraction:} 
Beyond the thermodynamic limit, we remove the assumption $\mu = \beta_0$ for $E \le E_{\rm crit}$.
Starting from $N=\sum_p T/(\beta_p-\mu)$ and remarking that $n_0^{eq}=T/(\beta_0-\mu)$, we get:
%the condensate fraction and the energy read
\begin{eqnarray}
\frac{n_0^{eq}}{N}(\mu)  & =&   \frac{1}{-(\mu-\beta_0) \sum_p (\beta_p-  \mu)^{-1}},
\label{eq:n_0RJ}\\
\frac{E}{E_{\rm crit}}(\mu) &=& \frac{ \sum_p \frac{\beta_p}{\beta_p-\mu}  }{ \big(1+(M-1)/\varrho \big)\sum_p \frac{\beta_0}{\beta_p-\mu}  },
\label{eq:E_mu_1}
\end{eqnarray}
where (\ref{eq:E_mu_1}) is obtained from $E=T\sum_p \beta_p/(\beta_p-\mu)$ with $T=N/\sum_p (\beta_p-\mu)^{-1}$.
The parametric plot of Eqs.(\ref{eq:n_0RJ})-(\ref{eq:E_mu_1}) with respect to $\mu$, provides the condensed fraction $n_0^{eq}/N$ vs $E/E_{\rm crit}$.

%%%%%%%%%%%%%%%%%%%%%%%%%%%%%%%%%%%%%%%%%%%%%%%%%%%%%%%%%%%%%%%%%%%%%%%
%%%%%%%%%%%%%%%%%%%%%%%%%%%%%%%%%%%%%%%%%%%%%%%%%%%%%%%%%%%%%%%%%%%%%%%
%%%%%%%%%%%%%%%%%%%%%%%   Appendix Strong Disorder
%%%%%%%%%%%%%%%%%%%%%%%%%%%%%%%%%%%%%%%%%%%%%%%%%%%%%%%%%%%%%%%%%%%%%%%
%%%%%%%%%%%%%%%%%%%%%%%%%%%%%%%%%%%%%%%%%%%%%%%%%%%%%%%%%%%%%%%%%%%%%%%

\section{Derivation of the kinetic equation with strong disorder [Eq.(\ref{eq:kin})].}
\label{app:strong_disorder}

\subsection{Primary asymptotics}

The starting point is the NLS Eq.(\ref{eq:psi}) written in the mode basis, i.e., Eq.(\ref{eq:a_j}).
We consider the regime where linear propagation dominates over disorder, which in turn dominates over the nonlinearity. 
Accordingly, we introduce a small dimensionless parameter $\eps$ and we consider the regime $\beta_p \to \beta_p, {\bf C} \to \eps {\bf C}, \gamma \to \eps^2 \gamma$.
For propagation distances of order $\eps^{-2}$, the rescaled mode amplitudes 
$a_j^\eps(z) = a_j(z/\eps^2)$ satisfy
\begin{equation*}
\partial_z a_j^\eps = -i \frac{\beta_j}{\eps^2} a_j^\eps +i \gamma \sum_{l,m,n=0}^{M-1} S_{jlmn} a_l^\eps a_m^\eps \overline{a_n^\eps} -
\frac{i}{\eps} \sum_{l=0}^{M-1}
C_{jl}(\frac{z}{\eps^2}) a_l^\eps  ,
\end{equation*}
where the bar stands for complex conjugation.
We set 
$c_j^\eps(z) =  a_j^\eps(z)\exp\big( i \frac{\beta_j}{\eps^2} z\big)$. 
The amplitudes $c_j^\eps(z) $ satisfy:
\begin{align}
\partial_z c_j^\eps =
& i \gamma \sum_{l,m,n=0}^{M-1} S_{jlmn} c_l^\eps c_m^\eps \overline{c_n^\eps} \exp\big( i \frac{\beta_j -\beta_l-\beta_m+\beta_n }{\eps^2} z\big)
\nonumber \\
& -
\frac{i}{\eps} \sum_{l=0}^{M-1}
C_{jl}(\frac{z}{\eps^2}) c_l^\eps \exp\big( i \frac{\beta_j - \beta_l}{\eps^2} z\big)  .
\end{align}
This is the usual diffusion approximation framework \cite{fouque07}.
We get the following Proposition.

\medskip
\noindent
{\bf Proposition:}
The  random process 
$
 ({c}_j^\eps(z) )_{j=0}^{M-1}  
$
converges in distribution in ${\cal C}^0([0,\infty), \mathbb{C}^{M} )$, 
the space of continuous functions from $[0,\infty)$ to $\mathbb{C}^{M}$,
to the Markov process  
$
  (\mathfrak{c}_{j}(z) )_{j=0}^{M-1}   
$
with infinitesimal generator
${\cal L}$:
\begin{align} 
 \label{eq:defL1}
{\cal L}= & {\cal L}_1 + {\cal L}_2+{\cal L}_3 +{\cal L}_4 +{\cal L}_5
,
\end{align}
with
\begin{align*} 
\nonumber {\cal L}_1 &= \frac{1}{2} 
\sum_{j,l=0, j\neq l}^{M-1} 
\Gamma^{\rm OD}_{jl} \big( 
\mathfrak{c}_{j} \overline{\mathfrak{c}_{j}} \partial_{\mathfrak{c}_{l}} \partial_{\overline{\mathfrak{c}_{l}}}
+
\mathfrak{c}_{l} \overline{\mathfrak{c}_{l}} \partial_{\mathfrak{c}_{j}} \partial_{\overline{\mathfrak{c}_{j}}}
 -
\mathfrak{c}_{j}  \mathfrak{c}_{l} \partial_{\mathfrak{c}_{j}} \partial_{ \mathfrak{c}_{l}}
-
\overline{\mathfrak{c}_{j}}  \overline{\mathfrak{c}_{l}} \partial_{\overline{\mathfrak{c}_{j}}} \partial_{ \overline{\mathfrak{c}_{l}}}\big)
 ,
%\label{eq:defL11}
\\
\nonumber 
{\cal L}_2 &= 
 \frac{1}{2} 
\sum_{j,l=0}^{M-1} 
\Gamma^{\rm D}_{j l} \big( 
\mathfrak{c}_{j} \overline{\mathfrak{c}_{l}} \partial_{\mathfrak{c}_{j}} \partial_{\overline{\mathfrak{c}_{l}}}
+
\overline{\mathfrak{c}_{j}} \mathfrak{c}_{l} \partial_{\overline{\mathfrak{c}_{j}}} \partial_{\mathfrak{c}_{l}}
 -
\mathfrak{c}_{j}  \mathfrak{c}_{l} \partial_{\mathfrak{c}_{j}} \partial_{ \mathfrak{c}_{l}}
-
\overline{\mathfrak{c}_{j}}  \overline{\mathfrak{c}_{l}} \partial_{\overline{\mathfrak{c}_{j}}} \partial_{ \overline{\mathfrak{c}_{l}}}\big)  ,
\\
\nonumber 
{\cal L}_3 &= \frac{1}{2} 
\sum_{j=0}^{M-1}  \Gamma^{\rm OD}_{jj} 
\big( \mathfrak{c}_{j} \partial_{\mathfrak{c}_{j}} + \overline{\mathfrak{c}_{j}} \partial_{\overline{\mathfrak{c}_{j}}}
\big)
+i  {\hat \Gamma}^{\rm OD}_{jj}  
\big( \mathfrak{c}_{j} \partial_{\mathfrak{c}_{j}} - \overline{\mathfrak{c}_{j}} \partial_{\overline{\mathfrak{c}_{j}}}
\big) ,  
\\ 
{\cal L}_4 &=  - \frac{1}{2} 
\sum_{j=0}^{M-1}    \Gamma^{\rm D}_{jj} 
\big( \mathfrak{c}_{j} \partial_{\mathfrak{c}_{j}} + \overline{\mathfrak{c}_{j}} \partial_{\overline{\mathfrak{c}_{j}}}
\big) ,\\
{\cal L}_5 &=  i \gamma 
\sum_{l,m,n=0}^{M-1}  \delta^K_{jlmn}  {S_{jlmn}}
\big( \mathfrak{c}_l  \mathfrak{c}_m  \overline{\mathfrak{c}_n} 
 \partial_{\mathfrak{c}_{j}} - 
 \overline{\mathfrak{c}_l}  \overline{\mathfrak{c}_m}   {\mathfrak{c}_n} \partial_{\overline{\mathfrak{c}_{j}}}
\big)  ,
%\label{eq:defL15}
\end{align*}
where 
$
%\begin{equation}
\delta^K_{jlmn} =  {\bf 1}_{\beta_j-\beta_l-\beta_m+\beta_n=0}
$.
%\end{equation}
In this definition we use the  classical complex derivative:
if $ \zeta=\zeta_r+i\zeta_i$, then $\partial_\zeta=(1/2)(\partial_{\zeta_r}-i \partial_{\zeta_i})$ and
$\partial_{\overline{\zeta}} =(1/2)(\partial_{\zeta_r} +i \partial_{\zeta_i})$,
and the coefficients of the operator ${\cal L}_k$ ($k=1,...,5$) are defined for 
$j ,l= 0, \ldots, M-1$, as follows: 

 \noindent - For all $j \neq l$, $\Gamma_{jl}$ and ${\hat \Gamma}^{\rm OD}_{j l}$ are given by 
\begin{align}
\Gamma^{\rm OD}_{jl} &= 2
\int_0^\infty{\cal R}_{jl}(z) \cos \big((\beta_l-\beta_j)z\big) dz , 
%\label{def:Gammalj}
\\
{\hat \Gamma}^{\rm OD}_{jl} &=
2
\int_0^\infty 
{\cal R}_{jl}(z)  \sin \big( (\beta_l-\beta_j)z \big) dz ,
\end{align}
with ${\cal R}_{jl}(z) $ defined by
\begin{align}
\nonumber
{\cal R}_{jl}(z) = \EE [C_{jl}(0)C_{jl}(z)] = \iint u_j(\br) u_j(\br') \EE[\delta V(0,\br) \delta V(z,\br')]  {u_l}(\br)   {u_l}(\br')d\br d\br'.
\label{def:calRjl}
%{\cal R}_{jl}(z) &:= \EE [C_{jl}(0)C_{jl}(z)] = \iint u_j(\bx) u_j(\bx') \EE[V(0,\bx) V(z,\bx')] \overline{u_l}(\bx)  \overline{u_l}(\bx')d\bx d\bx'.
\end{align}
\noindent - For all $j,l=0,\ldots,M-1$:
\begin{align}
\Gamma^{\rm D}_{jl} =&
\int_0^\infty \EE\big[ C_{jj} (0) C_{ll}(z) \big]   dz
+
\int_0^\infty \EE\big[ C_{ll} (0) C_{jj}(z) \big]   dz .
\end{align}
\noindent 
- For all $j=0,\ldots,M-1$:
\begin{align}
\Gamma^{\rm OD}_{jj} =& -\hspace{-0.05in}\sum_{l =0,l\neq j}^{M-1} \Gamma^{\rm OD}_{jl} ,\quad \quad
{\hat \Gamma}^{\rm OD}_{jj} = -\hspace{-0.05in}\sum_{l =0,l\neq j}^{M-1} {\hat \Gamma}^{\rm OD}_{jl}  .
\end{align}

\subsection{Secondary asymptotics}

We observe that $\Gamma^{\rm OD}$ and ${\hat \Gamma}^{\rm OD}$ depend on the power spectral density of the 
random index perturbation evaluated at the difference of distinct frequencies $\beta_p-\beta_l$, while $\Gamma^{\rm D}$ depends on the power spectral density of the index perturbation evaluated at zero-frequency. 
Therefore, when 
%the longitudinal correlation radius $l_\beta$ of the random index perturbation   is larger than 
$L_{lin}=1/\beta_0 \ll l_\beta$, then $\Gamma^{\rm D}$ is larger than $\Gamma^{\rm OD}, {\hat \Gamma}^{\rm OD}$. 
We consider this regime by introducing a small dimensionless parameter $\eta$ with $\Gamma^{\rm D}\to \Gamma^{\rm D}$, $\Gamma^{\rm OD} \to \eta^2 \Gamma^{\rm OD}$, ${\hat \Gamma}^{\rm OD} \to \eta^2 {\hat \Gamma}^{\rm OD}$, $\gamma \to \eta \gamma$.

For propagation distances of order $\eta^{-2}$, we introduce the rescaled mode amplitudes $\mathfrak{c}_j^\eta(z) = \mathfrak{c}_j(z/\eta^2)$.
By the Proposition presented above, 
it is a Markov process with infinitesimal generator
${\cal L}^\eta$:
\begin{equation}
{\cal L}^\eta = 
{\cal L}_1 + \eta^{-2} {\cal L}_2+{\cal L}_3 +\eta^{-2} {\cal L}_4 + \eta^{-1} {\cal L}_5 ,
\label{eq:defL1eta}
\end{equation}
where the operators ${\cal L}_k$ ($k=1,..,5$) are given above.
By (\ref{eq:defL1eta}) the second-order moments satisfy for $j\neq j'$:
\begin{align*}
\partial_z \EE[ \mathfrak{c}_j^\eta \overline{\mathfrak{c}_{j'}^\eta}] = &
-\frac{1}{2\eta^2} (\Gamma^{\rm D}_{jj}+\Gamma^{\rm D}_{j'j'}-2\Gamma^{\rm D}_{jj'}) \EE[ \mathfrak{c}_j^\eta \overline{\mathfrak{c}_{j'}^\eta}]   \\
& \hspace*{-0.3in}
 + \frac{1}{2} \big( \Gamma^{\rm OD}_{jj}+ \Gamma^{\rm OD}_{j'j'}\big) \EE[ \mathfrak{c}_j^\eta \overline{\mathfrak{c}_{j'}^\eta}] 
 +\frac{i}{2}\big( {\hat \Gamma}^{\rm OD}_{jj} - {\hat \Gamma}^{\rm OD}_{j'j'}\big) \EE[ \mathfrak{c}_j^\eta \overline{\mathfrak{c}_{j'}^\eta}] 
 \\
& \hspace*{-0.3in}+ i \frac{\gamma}{\eta} \sum_{l,m,n=0}^{M-1} \delta^K_{jlmn} S_{jlmn} \EE[ \overline{\mathfrak{c}_{j'}^\eta} \mathfrak{c}_l^\eta  \mathfrak{c}_m^\eta  \overline{\mathfrak{c}_n^\eta}  ] 
- i\frac{\gamma}{\eta}  \sum_{l,m,n=0}^{M-1} \delta^K_{j'lmn} 
{S_{j'lmn}} 
\EE[{\mathfrak{c}_{j}^\eta} \overline{\mathfrak{c}_l^\eta}  \overline{\mathfrak{c}_m^\eta}   
{\mathfrak{c}_n^\eta}  ]    ,
\end{align*}
up to negligible terms in $\eta$.
Note that 
$
\Gamma^{\rm D}_{jj}+\Gamma^{\rm D}_{j'j'}-2\Gamma^{\rm D}_{jj'} = 
\int_{-\infty}^\infty \EE\big[ (C_{jj} (0) - C_{j'j'}(0) )( C_{jj}(z)-C_{j'j'}(z) )\big]   dz
$
is positive (it is the power spectral density evaluated at $0$ frequency of the stationary process $C_{jj}(z)-C_{j'j'}(z)$ by Bochner's theorem).
%If it is positive (which is the standard case), then 
Therefore $\EE[ \mathfrak{c}_j^\eta \overline{\mathfrak{c}_{j'}^\eta}] $ is exponentially damped and
\begin{equation}
\label{eq:damped2}
\EE[ \mathfrak{c}_j^\eta \overline{\mathfrak{c}_{j'}^\eta}] = O(\eta).
\end{equation}

If $j=j'$, then the mean square amplitudes
$n_p^\eta(z) = \EE[ |\mathfrak{c}_j^\eta(z)|^2] $
satisfy
\begin{align}
%\nonumber
\partial_z n_p^\eta  =
\sum_{l=0,l\neq j}^{M-1} \Gamma_{jl}^{\rm OD} \big(n_l^\eta-n_p^\eta \big)  
-2 \frac{\gamma}{\eta}   \sum_{l,m,n=0}^{M-1} \delta^K_{jlmn}  S_{jlmn}  {\rm Im}\Big\{
 \EE[ \overline{\mathfrak{c}_{j}^\eta} \mathfrak{c}_l^\eta  \mathfrak{c}_m^\eta  \overline{\mathfrak{c}_n^\eta}   ] \Big\}  .
 \label{eq:odewj}
\end{align}

By (\ref{eq:defL1eta})  
the fourth-order moments satisfy
\begin{align}
 \partial_z \EE[ \overline{\mathfrak{c}_{j}^\eta} \mathfrak{c}_l^\eta  \mathfrak{c}_m^\eta  \overline{\mathfrak{c}_n^\eta}  ]  =
-\frac{1}{2\eta^2} 
G^{\rm D}_{jlmn}
\EE[ \overline{\mathfrak{c}_{j}^\eta} \mathfrak{c}_l^\eta  \mathfrak{c}_m^\eta  \overline{\mathfrak{c}_n^\eta}  ] 
+ i \frac{\gamma}{\eta} Y_{jlmn}^\eta  \nonumber \\
+  \sum_{j',l',m',n'} M_{jlmn,j'l'm'n'}  \EE[ \overline{\mathfrak{c}_{j'}^\eta}  \mathfrak{c}_{l'}^\eta  {\mathfrak{c}_{m'}^\eta}  \overline{\mathfrak{c}_{n'}^\eta}],
\label{eq:odec4}
\end{align}
up to negligible terms in $\eta$.
The coefficients $G^{\rm D}_{jlmn}$ and the sixth-order moment $Y_{jlmn}^\eta$ are given by
\begin{align}
%\nonumber
G^{\rm D}_{jlmn}
&= \Gamma^{\rm D}_{ll}+\Gamma^{\rm D}_{mm}+\Gamma^{\rm D}_{nn}+\Gamma^{\rm D}_{jj}+2\Gamma^{\rm D}_{lm}-2\Gamma^{\rm D}_{ln}
 -2\Gamma^{\rm D}_{lj} 
-2\Gamma^{\rm D}_{mn}-2\Gamma^{\rm D}_{mj}+2\Gamma^{\rm D}_{nj},
\label{def:Gammajlmn} \\
\nonumber
Y_{jlmn}^\eta =&    \sum_{l',m',n'=0}^{M-1} \delta^K_{l l'm'n'} S_{ll'm'n'} \EE[ 
  \mathfrak{c}_{l'}^\eta  \mathfrak{c}_{m'}^\eta  \overline{\mathfrak{c}_{n'}^\eta}
  \mathfrak{c}_m^\eta  \overline{\mathfrak{c}_n^\eta}  \overline{\mathfrak{c}_{j}^\eta}]
+
  \delta^K_{ml'm'n'} S_{ml'm'n'} \EE[ 
  \mathfrak{c}_l^\eta  \mathfrak{c}_{l'}^\eta  \mathfrak{c}_{m'}^\eta  \overline{\mathfrak{c}_{n'}^\eta}
 \overline{\mathfrak{c}_n^\eta}  \overline{\mathfrak{c}_{j}^\eta}] 
  \nonumber 
  \\
&
 \quad   -  \delta^K_{n l'm'n'}  
{S_{n l'm'n'}}
 \EE[  \mathfrak{c}_l^\eta  \mathfrak{c}_m^\eta  
  \overline{\mathfrak{c}_{l'}^\eta}  \overline{\mathfrak{c}_{m'}^\eta}   
{\mathfrak{c}_{n'}^\eta} \overline{\mathfrak{c}_{j}^\eta} ]
-
\delta^K_{j l'm'n'}  
{S_{j l'm'n'}} 
\EE[  \mathfrak{c}_l^\eta  \mathfrak{c}_m^\eta  
  \overline{\mathfrak{c}_{n}^\eta}   \overline{\mathfrak{c}_{l'}^\eta}  \overline{\mathfrak{c}_{m'}^\eta}   
{\mathfrak{c}_{n'}^\eta}   ]  ,
\label{def:Yjlmn}
\end{align}
up to negligible terms in $\eta$.
The tensor $M_{jlmn,j'l'm'n'}$ involves the coefficients $\Gamma^{\rm OD}$ and ${\hat \Gamma}^{\rm OD}$.
Note that we have
$ G^{\rm D}_{jlmn}=
\int_{-\infty}^\infty \EE\big[ (C_{ll} (0) + C_{mm}(0) -C_{nn}(0)-C_{jj}(0) )( C_{ll}(z)+C_{mm}(z)-C_{nn}(z)-C_{jj}(z)  )\big]   dz \ge 0
$. 
Therefore, we find from (\ref{eq:odec4}) that 
$$
\EE[ \overline{\mathfrak{c}_{j}^\eta} \mathfrak{c}_l^\eta  \mathfrak{c}_m^\eta  \overline{\mathfrak{c}_n^\eta}  ]  = \frac{2i \eta \gamma}{ G^{\rm D}_{jlmn}} Y_{jlmn}^\eta 
+O(\eta^2)  .
$$
By substituting into (\ref{eq:odewj}) and by using Isserlis formula for the sixth-order moments that appear in the expression (\ref{def:Yjlmn}) of $Y^\eta_{jlmn}$  
we obtain the kinetic Eq.(\ref{eq:kin}):
\begin{align}
\partial_z n_p^\eta  = &\sum_{l=0, l\neq j}^{M-1} \Gamma_{jl}^{\rm OD} \big(n_l^\eta-n_p^\eta \big) \nonumber\\ 
 &+8 \gamma^2   \sum_{l,m,n=0}^{M-1}  \frac{\delta^K_{jlmn}  S_{jlmn}^2}{G^{\rm D}_{jlmn}} \big(n_l^\eta n_m^\eta n_p^\eta + n_l^\eta n_m^\eta n_n^\eta 
-  n_p^\eta n_n^\eta n_m^\eta - n_p^\eta n_n^\eta n_l^\eta  \big)  .
\label{eq:kineq2}
\end{align}
The second term in (\ref{eq:kineq2}) has a form analogous to the conventional collision term of the wave turbulence kinetic equation \cite{Zakharov92}. Exploiting the invariances properties of the tensors $S_{jlmn}$ and $G^{\rm D}_{jlmn}$, as well as the property $G^{\rm D}_{jlmn} \ge 0$, it can be shown that the collision term conserves the particle number $N=\sum_p n_p$, the energy $E=\sum_p \beta_p n_p$, and exhibits a $H-$theorem of entropy growth $\partial_z {\cal S}^{neq}(z) \ge 0$, where the nonequilibrium entropy reads $S^{neq}(z)=\sum_p \log[n_p(z)]$ (note that, for simplicity we omitted to write  the superscript $\eta$). The entropy growth saturates at thermal equilibrium. The  distribution that maximizes the entropy, under the constraints that $N$ and $E$ are conserved, then corresponds to the RJ equilibrium  $n_p^{\rm RJ}=T/(\beta_p - \mu)$.
%where, as usual, $1/T$ and $-\mu/T$ are the Lagrange multipliers associated to the conservation of $E$ and $N$.
%There is a one to one relation relation between the pair $(N,E)$ and $(T,\mu)$.
%The values of the conserved quantities $(N,E)$ determine uniquely $(T,\mu)$, and thus the RJ equilibrium (\ref{eq:rj}).
%The collision term then describes an irreversible evolution to the RJ equilibrium distribution.}

\subsection{Degenerate modes}
In this section we assume that the modes may be degenerate.
% and we revisit the previous results.
The detailed derivation of the kinetic equation accounting for mode degeneracy is cumbersome and will be reported elsewhere.
Here we report the main results.

There are $G$ distinct wavenumbers:
$$
\{ \beta^{(g)} ,\, g=1,\ldots, G\} ,
$$
and the mode indices can be partitioned into $G$ groups ${\cal G}^{(g)}$, $g=1,\ldots,G$:
$$
{\cal G}^{(g)} = \{ p=1,\ldots,N ,\, \beta_p = \beta^{(g)}\} .
$$

We obtain the kinetic equation
\begin{align*}
\nonumber
\partial_z n^{(g)}= 
8 \gamma^2
 \sum_{g_1,g_2,g_3=1}^G  \delta^{(gg_1g_2g_3)}
 q^{(gg_1g_2g_3)} \big( n^{(g)} n^{(g_3)} n^{(g_2)} \\
 +n^{(g)} n^{(g_3)} n^{(g_1)}
- n^{(g_1)} n^{(g_2)} n^{(g)}  - n^{(g_1)} n^{(g_2)} n^{(g_3)} \big)   ,
 \label{eq:wg2}
\end{align*}
where
\begin{equation*}
q^{(gg_1g_2g_3)}=\frac{1}{|{\cal G}^{(g)}|} 
\sum_{j \in {\cal G}^{(g)} , l\in {\cal G}^{(g_1)}, m \in {\cal G}^{(g_2)}, n\in {\cal G}^{(g_3)} }
\hspace*{-0.3in}
  S_{jlmn} {Q}^{(gg_1g_2g_3)}_{jlmn}  
%\Big]    .
\end{equation*}
where
\begin{align*}
\nonumber
& {\bf Q}^{(gg_1g_2g_3)} =\big( {Q}^{(gg_1g_2g_3)}_{jlmn} \big)_{j\in {\cal G}^{(g)},
l\in {\cal G}^{(g_1)}, m \in {\cal G}^{(g_2)}, n\in {\cal G}^{(g_3)} } \\
&=
({\bf M}^{(gg_1g_2g_3)})^{-1} \big(  ( 
{S_{jlmn}} 
)_{ j\in {\cal G}^{(g)},
l\in {\cal G}^{(g_1)}, m \in {\cal G}^{(g_2)}, n\in {\cal G}^{(g_3)} }  \big)  .
\end{align*}
The tensor ${\bf M}^{(gg_1g_2g_3)}$ 
(seen as a $q \times q$ matrix with $q=|{\cal G}^{(g)}| |{\cal G}^{(g_1)}| |{\cal G}^{(g_2)}| |{\cal G}^{(g_3)}| $) is given by
\begin{align*}
& \sum_{ j' \in {\cal G}^{(g)}, l'\in {\cal G}^{(g_1)}, m' \in {\cal G}^{(g_2)}, n'\in {\cal G}^{(g_3)}   }
M^{(gg_1g_2g_3)}_{jlmn,j'l'm'n'}
 n_{j'l'm'n'}   \\
&  =
   \sum_{l'\in {\cal G}^{(g_1)},m'\in {\cal G}^{(g_2)}} 2\gamma_{ll'mm'} 
   n_{jl'm'n}   
 +  \sum_{n'\in {\cal G}^{(g_3)},j'\in {\cal G}^{(g)} } 2\gamma_{nn'jj'} 
   n_{j'lmn'}   \\
&\quad  - \sum_{l'\in {\cal G}^{(g_1)},n'\in {\cal G}^{(g_3)}} 2\gamma_{ll'nn'} 
  n_{jl'mn'}   
 - \sum_{l'\in {\cal G}^{(g_1)},j'\in {\cal G}^{(g)}} 2\gamma_{ll'jj'} 
   n_{j'l'mn} \\
&\quad  
- \sum_{m'\in {\cal G}^{(g_2)},n'\in {\cal G}^{(g_3)}} 2\gamma_{mm'nn'} 
    n_{jlm'n'}     
     - \sum_{m'\in {\cal G}^{(g_2)},j' \in {\cal G}^{(g)}} 2\gamma_{mm'jj'} 
    n_{j'lm'n} \\
&\quad  
+  \sum_{l',l''\in {\cal G}^{(g_1)}} \gamma_{l''l'll''} 
    n_{jl'mn}    
+  \sum_{m',m''\in {\cal G}^{(g_2)}} \gamma_{m''m'mm''} 
  n_{jlm'n} \\
&\quad +  \sum_{n',n''\in {\cal G}^{(g_3)}} \gamma_{n''n'nn''} 
   n_{jlm n'}
+  \sum_{j',j''\in {\cal G}^{(g)}} \gamma_{j''j'jj''} 
   n_{j' l m n}    .
\end{align*}
where 
$$
\gamma_{pqp'q'}
=2
\int_0^\infty \EE\big[ C_{pq}(z)C_{p'q'}(0)\big] e^{i (\beta_p-\beta_q)z} dz .
$$

\subsection{Model of disorder implemented in the simulations}

We have considered in the numerical simulations an elliptical parabolic potential $V(\bx)=q_x x^2 + q_y y^2$, with 
$$
u_{p_x,p_y}(x,y) = \sqrt{\kappa_x \kappa_y} (\pi p_x! \, p_y! \, 2^{p_x+p_y})^{-1/2} \, H_{p_x}(\kappa_x x) \, H_{p_y}(\kappa_y y) \, \exp[-(\kappa_x^2 x^2+\kappa_y^2 y^2)/2],
$$ 
the normalized Hermite-Gaussian functions with corresponding eigenvalues $\beta_p=\beta_{p_x,p_y}=\beta_{0x}(p_x+1/2)+\beta_{0y}(p_y+1/2)$, with $\kappa_x= (q_x/\alpha)^{1/4}$, $\kappa_y= (q_y/\alpha)^{1/4}$, $\beta_{0x} = 2\sqrt{\alpha q_x}$, $\beta_{0y} = 2\sqrt{\alpha q_y}$, and the radii of the fundamental mode $r_{0x}=1/\kappa_x=\sqrt{2\alpha/\beta_{0x}}$, $r_{0y}=1/\kappa_y=\sqrt{2\alpha/\beta_{0y}}$.
%In the following we may consider $\beta_{0y}=\beta_{0x}/2$. Then, if we denote by $V_{0,m}$ the depth of the potential, and $V_{0,m}$ is slightly larger than $\beta_{0x}(p_{x,m}+1/2)$ for some $p_{x,m}$, then $p_x=0,1... p_{x,m}$ and $p_y=0,1... p_{y,m}$ with $p_{y,m}=2p_{x,m}$. 
%However, as will be discussed below, it is preferable to chose $\beta_{0x}/\beta_{0y}$ irrational.
%The total number of modes is given by the number of distinct pairs $(p_x,p_y)$ such that $\beta_{p_x,p_y} \le V_{0,m}$.\\
%Remark: When applied to the 1st model of disorder ($\nu=2$), the elliptical parabolic fiber still gives $\Gamma_{jlmn}=0$ even in 2D.\\

We have considered the following form of model of disorder:
$\delta V(\bx,z)=\mu(z) \cos(\kappa_x b_x x) \cos(\kappa_y b_y y)$, with $\EE[\mu(0) \mu(z)]= \sigma_\beta^2 f(z)$, $f(z)=\exp(-|z|/l_\beta)$. The advantage of this model is that the matrices ${\bf C}, {\bf \Gamma}^{\rm D}, {\bf \Gamma}^{\rm OD}$ can be computed in analytical form.
We have 
$$
C_{nk}(z)=\mu(z) C_{n_x k_x}^0 C_{n_y k_y}^0=\mu(z) \int u_{n_x}(x) \cos(\kappa_x b_x x) u_{k_x}(x) dx \int u_{n_y}(y) \cos(\kappa_y b_y y) u_{k_y}(y) dy.
$$
%Then $C_{nk}(z)$ factorizes into the $x$ and $y$ components.\\
% which are computed independently of each other by directly applying the 1D results.\\
Then we have for $j_{x},j_y,l_x, l_y \geq 0$: $C_{j,j+2l}=\mu(z) C_{j_x,j_x+2l_x}^0 C_{j_y,j_y+2l_y}^0$ where we denote for $s=x$ or $s=y$:\\
\begin{align*}
C_{j_s,j_s+2l_s}^0= (-1)^{l_s} b_s^{2l_s} \exp(-b_s^2/4) L_{j_s}^{2l_s}(b_s^2/2) \frac{\sqrt{j_s!/(j_s+2l_s)!}}{2^{l_s}}
\end{align*}
and $C_{j_s,j_s+2l_s+1}^0=0$, where $L_{j}^{l}$ is the generalized Laguerre poynomial \cite[formula 7.388.7]{gradstein80}.
%Note that $C_{j_x,j_x+2l_x}^0$ is computed by only considering positive values of $l_x \geq 0$, and the corresponding negative values are obtained by using the fact that $C_{n_x,k_x}^0=C_{k_x,n_x}^0$.\\
In particular 
$C_{j_s j_s}^0= \exp(-b_s^2/4) L_{j_s}(b_s^2/2)$, where $L_{j}$ is the  Laguerre poynomial. 
For $j_{x},j_y,l_x, l_y \geq 0$, 
we have $\Gamma^{\rm D}_{jl}=2\sigma_\beta^2 l_\beta  C_{j_x j_x}^0 C_{j_y j_y}^0 C_{l_x l_x}^0 C_{l_y l_y}^0$.\\
For $n_x,k_x,n_y,k_y \geq 0$ we obtain:
$$
\Gamma^{\rm OD}_{n,k}=\frac{2\sigma_\beta^2 l_\beta {\cal R}_{n_x,k_x}^0 {\cal R}_{n_y,k_y}^0}{1+l_\beta^2 [\beta_{0x}(n_x-k_x)+\beta_{0y}(n_y-k_y)]^2},
$$ 
%The matrices ${\cal R}_{n_x,k_x}^0$ are given  
%for $j_{x},j_y,l_x, l_y \geq 0$: 
%${\cal R}_{j,j+2l}(z)= \sigma_\beta^2 f(z) {\cal R}_{j_x,j_x+2l_x}^0 {\cal R}_{j_y,j_y+2l_y}^0$ 
where ${\cal R}_{j_s,j_s+2l_s+1}^0=0$ and 
$$
{\cal R}_{j_s,j_s+2l_s}^0=b_s^{4l_s} \exp(-b_s^2/2) L_{j_s}^{2l_s}(b_s^2/2)^2 ( j_s!/(j_s+2l_s)!) 2^{-2l_s}.
$$

\noindent
%{\it Parameters used in the simulations:} 
In order to avoid high values of $\Gamma^{\rm OD}_{n,k}$, we have considered an irrational ratio $\beta_{0x}/\beta_{0y}=\sqrt{2}$, so that $\beta_{0x}(n_x-k_x)+\beta_{0y}(n_y-k_y) \neq 0$.
Parameters are ($b_x = 0.4, b_y = 0.5$) in Figs.~\ref{fig:sec3_strong1}-\ref{fig:sec3_strong2}, and ($b_x = 0.4, b_y = 0.3$) in Fig.~\ref{fig:sec3_strong3}. In all cases we considered $M=$46 modes.
Note that the value of $L_{nl}=1/(\gamma N/A_{eff}^0)$ in the simulations is computed by considering that all the power $N$ is in the fundamental mode of effective area $A_{eff}^0=1/\int |u_0|^4(\br) d\br$.
%${\cal L}^{\rm RJ}_{\rm kin}$ given in Eq.(8) (main text) involves in the denominator the tensor $S_{jlmn}$, whose zeros lead to a divergence of ${\cal L}^{\rm RJ}_{\rm kin}$. 
%A threshold value for $S_{jlmn}$ has been adjusted in such a way that ${\cal L}^{\rm RJ}_{\rm kin}$ properly describes the thermalization length scale observed in the simulations of the kinetic Eq.(\ref{eq:kineq2}).

To implement the disorder in the simulations of the NLS Eq.(\ref{eq:a_j}), we considered an {\it exact discretization} of the Ornstein-Uhlenbeck process.
The propagation axis is divided in intervals with deterministic lengths $\Delta z$, with $\Delta z < l_\beta$.
% smaller than $l_\beta$ (say, $\Delta z = l_\beta/10$).
The random function $\mu(z)$ is stepwise constant over each elementary interval
%${\bf D}_p(z) =  \sum_{j=1}^3 \nu_{p,j,k} \bsigma_j$    if 
$z \in [k\Delta z, (k+1)\Delta z)$,   
%\mbox{ for } p=1,\ldots,N_*
%$
%if $z \in [k\Delta z, (k+1)\Delta z) $
where $\mu_{0}\sim {\cal N}(0,\sigma_\beta^2/2)$ denotes the Gaussian distribution, 
%for  $p=1,\ldots,N_*$, $j=1,2,3$, and 
$$
\mu_{k} = \sqrt{1-2\Delta z/l_\beta}\mu_{k-1} +  
\sqrt{2\Delta z/l_\beta} {\cal N}( 0,\sigma_\beta^2/2) , 
$$
with ${\cal N}( 0,\sigma_\beta^2/2)$ all independent and identically distributed.
%We recall that 
%${\cal N}( 0,\frac{\sigma_\beta^2}{2})$ denotes the Gaussian distribution $f(x)=\exp(- x^2 / \sigma_\beta^2) / \sqrt{\pi \sigma_\beta^2}$.
%\end{widetext}

%%%%%%%%%%%%%%%%%%%%%%%%%%%%%%%%%%%%%%%%%%%%%%%%%%%
%\section{Backmatter}
%\begin{backmatter}
\section*{Acknowledgments} 

This work was supported by: 
%EU ERC AdG STEMS (740355); EU HORIZON 2020 MSCA (101064614,101023717); Sapienza University Grants AddSapiExcellence (NOSTERDIS); 
the EU - Next Generation EU under the Italian National Recovery and Resilience Plan (NRRP), Mission 4, CUP B53C22004050001, partnership on “Telecommunications of the Future” (PE00000001 - program “RESTART”), the European Innovation Council Pathfinder Open MULTISCOPE (101185664), the Italian Ministerial grant PRIN2022 "SAFE" (2022ESAC3K), Sapienza University of Rome Seed PNR (SP12218480C7D1E9), the Italian Ministry of Health under the Ricerca Finalizzata 2021 program (Project Code RF-2021-12373094), the Russian Science Foundation (21-72-30024-$\Pi$), the Centre national de la recherche scientifique (CNRS), Conseil régional de Bourgogne Franche-Comté, iXCore Research Fondation, Agence Nationale de la Recherche (ANR-23-CE30-0021, ANR-19-CE46-0007, ANR-15-IDEX-0003, ANR-21-ESRE-0040). Some of the calculations were carried out using  HPC resources from DNUM CCUB (Centre de Calcul, Université de Bourgogne). 
\smallskip

\section*{Disclosures} The authors declare no conflicts of interest.

\section*{Data availability} Data underlying the results presented in this paper are not publicly available at this time but may be obtained from the authors upon reasonable request.

%\end{backmatter}
%%%%%%%%%%%%%%%%%%%%%%%%%%%%%%%%%%%%%%%%%%%%%%%%%%%
%\begin{thebibliography}{00}
%\bibliographystyle{elsarticle-num} cas-model2-names
\bibliographystyle{elsarticle-num}
\bibliography{references}

\begin{thebibliography}{100}
\expandafter\ifx\csname url\endcsname\relax
  \def\url#1{\texttt{#1}}\fi
\expandafter\ifx\csname urlprefix\endcsname\relax\def\urlprefix{URL }\fi
\expandafter\ifx\csname href\endcsname\relax
  \def\href#1#2{#2} \def\path#1{#1}\fi

\bibitem{agrawal2012fiber}
G.~P. "Agrawal, "Fiber-optic communication systems", "John Wiley \& Sons", "2012".

\bibitem{lin1978wideband}
C.~Lin, V.~Nguyen, W.~French, Wideband near-ir continuum (0.7--2.1 $\mu$m) generated in low-loss optical fibres, Electronics Letters 14~(25) (1978) 822--823.

\bibitem{cristiani2022roadmap}
I.~Cristiani, C.~Lacava, G.~Rademacher, B.~J. Puttnam, R.~S. Lu{\`\i}s, C.~Antonelli, A.~Mecozzi, M.~Shtaif, D.~Cozzolino, D.~Bacco, et~al., Roadmap on multimode photonics, Journal of Optics 24~(8) (2022) 083001.

\bibitem{horak_multimode_2012}
P.~Horak, F.~Poletti, \href{https://www.intechopen.com/books/recent-progress-in-optical-fiber-research/multimode-nonlinear-fibre-optics-theory-and-applications}{Multimode {Nonlinear} {Fibre} {Optics}: {Theory} and {Applications}}, Recent Progress in Optical Fiber Research (Jan. 2012).
\newblock \href {https://doi.org/10.5772/27489} {\path{doi:10.5772/27489}}.
\newline\urlprefix\url{https://www.intechopen.com/books/recent-progress-in-optical-fiber-research/multimode-nonlinear-fibre-optics-theory-and-applications}

\bibitem{Wright2015R31}
L.~G. Wright, D.~N. Christodoulides, F.~W. Wise, Controllable spatiotemporal nonlinear effects in multimode fibres, Nat. Photonics (2015) 1--5.

\bibitem{Wright2015R29}
L.~G. Wright, S.~Wabnitz, D.~N. Christodoulides, F.~W. Wise, Ultrabroadband dispersive radiation by spatiotemporal oscillation of multimode waves, Phys. Rev. Lett. 115 (2015) 223902.

\bibitem{Hasegawa}
A.~Hasegawa, Self-confinement of multimode optical pulse in a glass fiber, Opt. Lett. 5 (1980) 416--417.

\bibitem{Longhi2003R27}
S.~Longhi, Modulation instability and space-time dynamics in nonlinear parabolic-index optical fibers, Opt. Lett. 28 (2003) 2363--2365.

\bibitem{Renninger2012R31}
W.~H. Renninger, F.~W. Wise, Optical solitons in graded-index multimode fibres, Nat. Commun. 4 (2012) 1719.

\bibitem{KrupaPRLGPI}
K.~Krupa, A.~Tonello, A.~Barth\'el\'emy, V.~Couderc, B.~M. Shalaby, A.~Bendahmane, G.~Millot, S.~Wabnitz, \href{https://link.aps.org/doi/10.1103/PhysRevLett.116.183901}{Observation of geometric parametric instability induced by the periodic spatial self-imaging of multimode waves}, Phys. Rev. Lett. 116 (2016) 183901.
\newblock \href {https://doi.org/10.1103/PhysRevLett.116.183901} {\path{doi:10.1103/PhysRevLett.116.183901}}.
\newline\urlprefix\url{https://link.aps.org/doi/10.1103/PhysRevLett.116.183901}

\bibitem{aschieri2011condensation}
P.~Aschieri, J.~Garnier, C.~Michel, V.~Doya, A.~Picozzi, Condensation and thermalization of classsical optical waves in a waveguide, Physical Review A 83~(3) (2011) 033838.

\bibitem{Baudin2020}
K.~Baudin, A.~Fusaro, K.~Krupa, J.~Garnier, S.~Rica, G.~Millot, A.~Picozzi, \href{https://link.aps.org/doi/10.1103/PhysRevLett.125.244101}{Classical rayleigh-jeans condensation of light waves: Observation and thermodynamic characterization}, Phys. Rev. Lett. 125 (2020) 244101.
\newblock \href {https://doi.org/10.1103/PhysRevLett.125.244101} {\path{doi:10.1103/PhysRevLett.125.244101}}.
\newline\urlprefix\url{https://link.aps.org/doi/10.1103/PhysRevLett.125.244101}

\bibitem{Wu2019}
F.~O. Wu, A.~U. Hassan, D.~N. Christodoulides, \href{http://dx.doi.org/10.1038/s41566-019-0501-8}{Thermodynamic theory of highly multimoded nonlinear optical systems}, Nature Photonics 13 (2019) 776--782.
\newblock \href {https://doi.org/10.1038/s41566-019-0501-8} {\path{doi:10.1038/s41566-019-0501-8}}.
\newline\urlprefix\url{http://dx.doi.org/10.1038/s41566-019-0501-8}

\bibitem{parto19thermodynamic}
M.~Parto, F.~O. Wu, P.~S. Jung, K.~Makris, D.~N. Christodoulides, \href{https://opg.optica.org/ol/abstract.cfm?URI=ol-44-16-3936}{Thermodynamic conditions governing the optical temperature and chemical potential in nonlinear highly multimoded photonic systems}, Opt. Lett. 44~(16) (2019) 3936--3939.
\newblock \href {https://doi.org/10.1364/OL.44.003936} {\path{doi:10.1364/OL.44.003936}}.
\newline\urlprefix\url{https://opg.optica.org/ol/abstract.cfm?URI=ol-44-16-3936}

\bibitem{makris2020statistical}
K.~G. Makris, F.~O. Wu, P.~S. Jung, D.~N. Christodoulides, Statistical mechanics of weakly nonlinear optical multimode gases, Optics letters 45~(7) (2020) 1651--1654.

\bibitem{wright2022physics}
L.~G. Wright, F.~O. Wu, D.~N. Christodoulides, F.~W. Wise, Physics of highly multimode nonlinear optical systems, Nature Physics 18~(9) (2022) 1018--1030.

\bibitem{efremidis21fundamental}
N.~K. Efremidis, D.~N. Christodoulides, \href{https://link.aps.org/doi/10.1103/PhysRevA.103.043517}{Fundamental entropic processes in the theory of optical thermodynamics}, Phys. Rev. A 103 (2021) 043517.
\newblock \href {https://doi.org/10.1103/PhysRevA.103.043517} {\path{doi:10.1103/PhysRevA.103.043517}}.
\newline\urlprefix\url{https://link.aps.org/doi/10.1103/PhysRevA.103.043517}

\bibitem{Selim:23}
M.~A. Selim, G.~G. Pyrialakos, F.~O. Wu, Z.~Musslimani, K.~G. Makris, M.~Khajavikhan, D.~Christodoulides, \href{https://opg.optica.org/ol/abstract.cfm?URI=ol-48-8-2206}{Thermalization of the ablowitz-ladik lattice in the presence of non-integrable perturbations}, Opt. Lett. 48~(8) (2023) 2206--2209.
\newblock \href {https://doi.org/10.1364/OL.489165} {\path{doi:10.1364/OL.489165}}.
\newline\urlprefix\url{https://opg.optica.org/ol/abstract.cfm?URI=ol-48-8-2206}

\bibitem{Pyrialakos22Thermalization}
G.~G. Pyrialakos, H.~Ren, P.~S. Jung, M.~Khajavikhan, D.~N. Christodoulides, \href{https://link.aps.org/doi/10.1103/PhysRevLett.128.213901}{Thermalization dynamics of nonlinear non-hermitian optical lattices}, Phys. Rev. Lett. 128 (2022) 213901.
\newblock \href {https://doi.org/10.1103/PhysRevLett.128.213901} {\path{doi:10.1103/PhysRevLett.128.213901}}.
\newline\urlprefix\url{https://link.aps.org/doi/10.1103/PhysRevLett.128.213901}

\bibitem{wu20Thermodynamic}
F.~O. Wu, P.~S. Jung, M.~Parto, M.~Khajavikhan, D.~N. Christodoulides, Thermodynamic optical pressures in tight-binding nonlinear multimode photonic systems, Communications Physics 3 (2020) 216.

\bibitem{efremidis22simple}
N.~Efremidis, D.~Christodoulides, Thermodynamic optical pressures in tight-binding nonlinear multimode photonic systems, Communications Physics 5 (2022) 286.

\bibitem{ren23nature}
H.~Ren, G.~G. Pyrialakos, F.~O. Wu, P.~S. Jung, N.~K. Efremidis, M.~Khajavikhan, D.~N. Christodoulides, \href{https://link.aps.org/doi/10.1103/PhysRevLett.131.193802}{Nature of optical thermodynamic pressure exerted in highly multimoded nonlinear systems}, Phys. Rev. Lett. 131 (2023) 193802.
\newblock \href {https://doi.org/10.1103/PhysRevLett.131.193802} {\path{doi:10.1103/PhysRevLett.131.193802}}.
\newline\urlprefix\url{https://link.aps.org/doi/10.1103/PhysRevLett.131.193802}

\bibitem{Efremidis:24Statistical}
N.~K. Efremidis, D.~N. Christodoulides, \href{https://opg.optica.org/ol/abstract.cfm?URI=ol-49-10-2777}{Statistical mechanics and pressure of composite multimoded weakly nonlinear optical systems}, Opt. Lett. 49~(10) (2024) 2777--2780.
\newblock \href {https://doi.org/10.1364/OL.511787} {\path{doi:10.1364/OL.511787}}.
\newline\urlprefix\url{https://opg.optica.org/ol/abstract.cfm?URI=ol-49-10-2777}

\bibitem{Ren:24dalton}
H.~Ren, G.~G. Pyrialakos, F.~O. Wu, N.~K. Efremidis, M.~Khajavikhan, D.~N. Christodoulides, \href{https://opg.optica.org/ol/abstract.cfm?URI=ol-49-7-1802}{Dalton's law of partial optical thermodynamic pressures in highly multimoded nonlinear photonic systems}, Opt. Lett. 49~(7) (2024) 1802--1805.
\newblock \href {https://doi.org/10.1364/OL.517715} {\path{doi:10.1364/OL.517715}}.
\newline\urlprefix\url{https://opg.optica.org/ol/abstract.cfm?URI=ol-49-7-1802}

\bibitem{garnier19}
J.~Garnier, A.~Fusaro, K.~Baudin, C.~Michel, K.~Krupa, G.~Millot, A.~Picozzi, \href{https://link.aps.org/doi/10.1103/PhysRevA.100.053835}{Wave condensation with weak disorder versus beam self-cleaning in multimode fibers}, Phys. Rev. A 100 (2019) 053835.
\newblock \href {https://doi.org/10.1103/PhysRevA.100.053835} {\path{doi:10.1103/PhysRevA.100.053835}}.
\newline\urlprefix\url{https://link.aps.org/doi/10.1103/PhysRevA.100.053835}

\bibitem{podivilov2022thermalization}
E.~Podivilov, F.~Mangini, O.~Sidelnikov, M.~Ferraro, M.~Gervaziev, D.~Kharenko, M.~Zitelli, M.~Fedoruk, S.~Babin, S.~Wabnitz, Thermalization of orbital angular momentum beams in multimode optical fibers, Physical Review Letters 128~(24) (2022) 243901.

\bibitem{wu2022thermalization}
F.~O. Wu, Q.~Zhong, H.~Ren, P.~S. Jung, K.~G. Makris, D.~N. Christodoulides, Thermalization of light’s orbital angular momentum in nonlinear multimode waveguide systems, Physical Review Letters 128~(12) (2022) 123901.

\bibitem{Rasmussen00}
K.~Rasmussen, T.~Cretegny, P.~G. Kevrekidis, N.~Gronbech-Jensen, Statistical mechanics of a discrete nonlinear system, Phys. Rev. Lett. 84 (2000) 3740.

\bibitem{Jordan00self}
R.~Jordan, C.~Josserand, Self-organization in nonlinear wave turbulence, Phys. Rev. E 61 (2000) 1527.

\bibitem{Jordan00mean}
R.~Jordan, B.~Turkington, C.~L. Zirbel, A mean-field statistical theory for the nonlinear {S}chr\"odinger equation, Physica D 137 (2000) 353.

\bibitem{Rumpf01}
R.~Rumpf, A.~C. Newell, Coherent structures and entropy in constrained, modulationally unstable, nonintegrable systems, Phys. Rev. Lett. 87 (2001) 054102.

\bibitem{Rumpf03}
R.~Rumpf, A.~C. Newell, Localization and coherence in nonintegrable systems, Physica D 184 (2003) 162.

\bibitem{Rumpf04}
B.~Rumpf, Simple statistical explanation for the localization of energy in nonlinear lattices with two conserved quantities, Phys. Rev. E 69 (2004) 016618.

\bibitem{leuzzi09}
L.~Leuzzi, C.~Conti, V.~Folli, L.~Angelani, , G.~Ruocco, Phase diagram and complexity of mode-locked lasers: From order to disorder, Phys. Rev. Lett. 102 (2009) 083901.

\bibitem{conti11complexity}
C.~Conti, L.~Leuzzi, Complexity of waves in nonlinear disordered media, Phys. Rev. B 83 (2011) 134204.

\bibitem{Ghofraniha15}
N.~Ghofraniha, I.~Viola, F.~Di~Maria, G.~Barbarella, G.~Gigli, L.~L., C.~Conti, Experimental evidence of replica symmetry breaking in random lasers, Nature Commun. 6 (2015) 6058.

\bibitem{Pierangeli17}
D.~Pierangeli, A.~Tavani, F.~Di~Mei, A.~Agranat, C.~Conti, E.~DelRe, Observation of replica symmetry breaking in disordered nonlinear wave propagation, Nature Commun. 8 (2017) 1501.

\bibitem{baudin23}
K.~Baudin, J.~Garnier, A.~Fusaro, N.~Berti, C.~Michel, K.~Krupa, G.~Millot, A.~Picozzi, \href{https://link.aps.org/doi/10.1103/PhysRevLett.130.063801}{Observation of light thermalization to negative-temperature rayleigh-jeans equilibrium states in multimode optical fibers}, Phys. Rev. Lett. 130 (2023) 063801.
\newblock \href {https://doi.org/10.1103/PhysRevLett.130.063801} {\path{doi:10.1103/PhysRevLett.130.063801}}.
\newline\urlprefix\url{https://link.aps.org/doi/10.1103/PhysRevLett.130.063801}

\bibitem{marquesmuniz23}
A.~L.~M. Muniz, F.~O. Wu, P.~S. Jung, M.~Khajavikhan, D.~N. Christodoulides, U.~Peschel, \href{https://www.science.org/doi/abs/10.1126/science.ade6523}{Observation of photon-photon thermodynamic processes under negative optical temperature conditions}, Science 379~(6636) (2023) 1019--1023.
\newblock \href {http://arxiv.org/abs/https://www.science.org/doi/pdf/10.1126/science.ade6523} {\path{arXiv:https://www.science.org/doi/pdf/10.1126/science.ade6523}}, \href {https://doi.org/10.1126/science.ade6523} {\path{doi:10.1126/science.ade6523}}.
\newline\urlprefix\url{https://www.science.org/doi/abs/10.1126/science.ade6523}

\bibitem{Zakharov92}
V.~E. Zakharov, V.~S. L'vov, G.~Falkovich, Kolmogorov Spectra of Turbulence I, Springer; Berlin, 1992.

\bibitem{Nazarenko11}
S.~Nazarenko, Wave Trubulence, Lecture Notes in Physics 825, Springer, 2011.

\bibitem{Newell11}
A.~C. Newell, R.~Rumpf, Wave turbulence, Annual Review of Fluid Mechanics 43 (2011) 59--78.

\bibitem{galtier22}
S.~Galtier, Physics of Wave Trubulence, Cambridge University Press, 2022.

\bibitem{eloy21}
A.~Eloy, O.~Boughdad, M.~Albert, P.-E. Larr\'e, F.~Mortessagne, M.~Bellec, C.~Michel, Experimental observation of turbulent coherent structures in a superfluid of light, Europhysics Letters 134 (2021) 26001.

\bibitem{glorieux23turbulent}
M.~Baker-Rasooli, W.~Liu, T.~Aladjidi, A.~Bramati, Q.~Glorieux, \href{https://link.aps.org/doi/10.1103/PhysRevA.108.063512}{Turbulent dynamics in a two-dimensional paraxial fluid of light}, Phys. Rev. A 108 (2023) 063512.
\newblock \href {https://doi.org/10.1103/PhysRevA.108.063512} {\path{doi:10.1103/PhysRevA.108.063512}}.
\newline\urlprefix\url{https://link.aps.org/doi/10.1103/PhysRevA.108.063512}

\bibitem{hammaniPLA10}
K.~Hammani, B.~Kibler, C.~Finot, A.~Picozzi, Emergence of rogue waves from optical turbulence, Phys. Lett. A 374 (2010) 3585.

\bibitem{kiblerPLA11}
B.~Kibler, K.~Hammani, C.~Finot, A.~Picozzi, Rogue waves, rational solitons and wave turbulence theory, Phys. Lett. A 375 (2011) 3149.

\bibitem{onorato2013rogue}
M.~Onorato, S.~Residori, U.~Bortolozzo, A.~Montina, F.~Arecchi, Rogue waves and their generating mechanisms in different physical contexts, Physics Reports (2013).

\bibitem{dudley19rogue}
J.~Dudley, G.~Genty, A.~Mussot, A.~Chabchoub, F.~Dias, Rogue waves and analogies in optics and oceanography, Nature Reviews Physics 1 (2019) 675.

\bibitem{barviauOE09}
B.~Barviau, B.~Kibler, A.~Kudlinski, A.~Mussot, G.~Millot, A.~Picozzi, Experimental signature of optical wave thermalization through supercontinuum generation in photonic crystal fiber, Opt. Express 17 (2009) 7392.

\bibitem{barviauPRA09}
B.~Barviau, B.~Kibler, A.~Picozzi, Wave turbulence description of supercontinuum generation: influence of self-steepening and higher-order dispersion, Opt. Express 79 (2009) 063840.

\bibitem{kiblerPRE11}
B.~Kibler, C.~Michel, A.~Kudlinski, B.~Barviau, G.~Millot, A.~Picozzi, Emergence of spectral incoherent solitons through supercontinuum generation in a photonic crystal fiber, Phys. Rev. E 84 (2011) 066605.

\bibitem{barviauPRA13}
B.~Barviau, J.~Garnier, G.~Xu, B.~Kibler, G.~Millot, A.~Picozzi, Truncated thermalization of incoherent optical waves through supercontinuum generation in photonic crystal fibers, Physical Review A 87~(3) (2013) 035803.

\bibitem{meng21intracavity}
F.~Meng, C.~Lapre, C.~Billet, T.~Sylvestre, J.-M. Merolla, C.~Finot, S.~K. Turitsyn, G.~Genty, J.~Dudley, Intracavity incoherent supercontinuum dynamics and rogue waves in a broadband dissipative soliton laser, Nat Commun 12 (2021) 5567.

\bibitem{congy24statistics}
T.~Congy, G.~A. El, G.~Roberti, A.~Tovbis, S.~Randoux, P.~Suret, \href{https://link.aps.org/doi/10.1103/PhysRevLett.132.207201}{Statistics of extreme events in integrable turbulence}, Phys. Rev. Lett. 132 (2024) 207201.
\newblock \href {https://doi.org/10.1103/PhysRevLett.132.207201} {\path{doi:10.1103/PhysRevLett.132.207201}}.
\newline\urlprefix\url{https://link.aps.org/doi/10.1103/PhysRevLett.132.207201}

\bibitem{suret24soliton}
P.~Suret, S.~Randoux, A.~Gelash, D.~Agafontsev, B.~Doyon, G.~El, \href{https://link.aps.org/doi/10.1103/PhysRevE.109.061001}{Soliton gas: Theory, numerics, and experiments}, Phys. Rev. E 109 (2024) 061001.
\newblock \href {https://doi.org/10.1103/PhysRevE.109.061001} {\path{doi:10.1103/PhysRevE.109.061001}}.
\newline\urlprefix\url{https://link.aps.org/doi/10.1103/PhysRevE.109.061001}

\bibitem{connaughton05}
C.~Connaughton, C.~Josserand, A.~Picozzi, Y.~Pomeau, S.~Rica, Condensation of classical nonlinear waves, Phys. Rev. Lett. 95 (2005) 263901.

\bibitem{sun2012observation}
C.~Sun, S.~Jia, C.~Barsi, S.~Rica, A.~Picozzi, J.~Fleischer, Observation of the kinetic condensation of classical waves, Nature Physics 8 (2012) 470.

\bibitem{Bortolozzo09}
U.~Bortolozzo, J.~Laurie, S.~Nazarenko, S.~Residori, Optical wave turbulence and the condensation of light, J. Opt. Soc. Am. B 26 (2009) 2280.

\bibitem{Laurie12}
J.~Laurie, U.~Bortolozzo, S.~Nazarenko, S.~Residori, One-dimensional optical wave turbulence: Experiment and theory, Phys. Rep. 514 (2012) 121.

\bibitem{chiocchetta16}
A.~Chiocchetta, P.~Larr\'e, I.~Carusotto, Thermalization and bose-einstein condensation of quantum light in bulk nonlinear media, Reviews of Modern Physics 115 (2016) 24002.

\bibitem{pitoisPRL06b}
S.~Pitois, S.~Lagrange, H.~R. Jauslin, A.~Picozzi, Velocity locking of incoherent nonlinear wave packets, Phys. Rev. Lett. 97 (2006) 033902.

\bibitem{suretPRL10}
P.~Suret, S.~Randoux, H.~Jauslin, A.~Picozzi, Anomalous thermalization of nonlinear wave systems, Phys. Rev. Lett. 104 (2010) 054101.

\bibitem{santic18nonequilibrium}
N.~\ifmmode \check{S}\else \v{S}\fi{}anti\ifmmode~\acute{c}\else \'{c}\fi{}, A.~Fusaro, S.~Salem, J.~Garnier, A.~Picozzi, R.~Kaiser, \href{https://link.aps.org/doi/10.1103/PhysRevLett.120.055301}{Nonequilibrium precondensation of classical waves in two dimensions propagating through atomic vapors}, Phys. Rev. Lett. 120 (2018) 055301.
\newblock \href {https://doi.org/10.1103/PhysRevLett.120.055301} {\path{doi:10.1103/PhysRevLett.120.055301}}.
\newline\urlprefix\url{https://link.aps.org/doi/10.1103/PhysRevLett.120.055301}

\bibitem{kottos24}
A.~Kurnosov, L.~J. Fern\'andez-Alc\'azar, A.~Ramos, B.~Shapiro, T.~Kottos, \href{https://link.aps.org/doi/10.1103/PhysRevLett.132.193802}{Optical kinetic theory of nonlinear multimode photonic networks}, Phys. Rev. Lett. 132 (2024) 193802.
\newblock \href {https://doi.org/10.1103/PhysRevLett.132.193802} {\path{doi:10.1103/PhysRevLett.132.193802}}.
\newline\urlprefix\url{https://link.aps.org/doi/10.1103/PhysRevLett.132.193802}

\bibitem{lian24coupled}
M.~Lian, Y.~Geng, Y.-J. Chen, Y.~Chen, J.-T. L\"u, \href{https://link.aps.org/doi/10.1103/PhysRevLett.133.116303}{Coupled thermal and power transport of optical waveguide arrays: Photonic wiedemann-franz law and rectification effect}, Phys. Rev. Lett. 133 (2024) 116303.
\newblock \href {https://doi.org/10.1103/PhysRevLett.133.116303} {\path{doi:10.1103/PhysRevLett.133.116303}}.
\newline\urlprefix\url{https://link.aps.org/doi/10.1103/PhysRevLett.133.116303}

\bibitem{michel201111thermalization}
C.~Michel, M.~Haelterman, P.~Suret, S.~Randoux, R.~Kaiser, A.~Picozzi, Thermalization and condensation in an incoherently pumped passive optical cavity, Phys. Rev. A 84 (2011) 033848.

\bibitem{turitsyna2013laminar}
E.~Turitsyna, S.~Smirnov, S.~Sugavanam, N.~Tarasov, X.~Shu, S.~Babin, E.~Podivilov, D.~Churkin, G.~Falkovich, S.~Turitsyn, The laminar-turbulent transition in a fibre laser, Nature Photonics 7~(10) (2013) 783--786.

\bibitem{Turitsyna09}
E.~Turitsyna, G.~Falkovich, V.~Mezentsev, S.~Turitsyn, Optical turbulence and spectral condensate in long-fiber lasers, Phys. Rev. A 80 (2009) 031804.

\bibitem{Turitsyna12}
E.~Turitsyna, G.~Falkovich, A.~El-Taher, X.~Shu, P.~Harper, S.~Turitsyn, Optical turbulence and spectral condensate in long fibre lasers, Proc. R. Soc. A 468 (2012) 2145.

\bibitem{Churkin15}
D.~Churkin, I.~Kolokolov, E.~Podivilov, I.~D. Vatnik, M.~A. Nikulin, S.~S. Vergeles, I.~S. Terekhov, V.~V. Lebedev, G.~Falkovich, S.~A. Babin, S.~K. Turitsyn, Wave kinetics of random fibre lasers, Nature Communications 6 (2015) 6214.

\bibitem{picozzi14}
A.~Picozzi, J.~Garnier, T.~Hansson, P.~Suret, S.~Randoux, G.~Millot, D.~Christodoulides, \href{https://www.sciencedirect.com/science/article/pii/S0370157314001203}{Optical wave turbulence: Towards a unified nonequilibrium thermodynamic formulation of statistical nonlinear optics}, Physics Reports 542~(1) (2014) 1--132, optical wave turbulence: Towards a unified nonequilibrium thermodynamic formulation of statistical nonlinear optics.
\newblock \href {https://doi.org/https://doi.org/10.1016/j.physrep.2014.03.002} {\path{doi:https://doi.org/10.1016/j.physrep.2014.03.002}}.
\newline\urlprefix\url{https://www.sciencedirect.com/science/article/pii/S0370157314001203}

\bibitem{TuritsynOWTbookNazarenko}
S.~K. Turitsyn, S.~A. Babin, E.~G. Turitsyna, G.~E. Falkovich, E.~Podivilov, D.~Churkin, Optical wave turbulence, Eds. V. Shira and S. Nazarenko, Wave Turbulence, World Scientific Series on Nonlinear Science Series A: Volume 83, 2013.

\bibitem{Tsytovich}
V.~N. Tsytovich, Nonlinear effects in plasma, Plenum; New-York, 1970.

\bibitem{Hasselmann1}
K.~Hasselmann, On the non-linear energy transfer in a gravity-wave spectrum. part 1. general theory, J. Fluid Mech. 12 (1962) 481--500.

\bibitem{Hasselmann2}
K.~Hasselmann, On the non-linear energy transfer in a gravity-wave spectrum. part 2. conservation theorems; wave-particle analogy; irreversibility, J. Fluid Mech. 15 (1963) 273--281.

\bibitem{Dyachenko92}
S.~Dyachenko, A.~C. Newell, A.~Pushkarev, V.~E. Zakharov, Optical turbulence: weak turbulence, condensates and collapsing filaments in the nonlinear {S}chr\"{o}dinger equation, Physica D 57 (1992) 96.

\bibitem{Dias04}
V.~Zakharov, F.~Dias, A.~Pushkarev, One-dimensional wave turbulence, Phys. Rep. 398 (2004) 1.

\bibitem{Benney}
D.~J. Benney, P.~G. Saffman, Nonlinear interactions of random waves in dispersive medium, Proc. R. Soc. London Ser. A 289 (1966) 301.

\bibitem{NewellWT}
A.~Newell, The closure problem in a system of random gravity waves, Rev. of Geophys. 6 (1968) 1.

\bibitem{BenNew}
D.~J. Benney, A.~Newell, Random wave closure, Stud. Appl. Math. 48 (1969) 29.

\bibitem{Newell01}
A.~C. Newell, S.~Nazarenko, L.~Biven, Wave turbulence and intermittency, Physica D 152-153 (2001) 520--550.

\bibitem{Huang}
K.~Huang, Statistical Mechanics, Wiley, 1963.

\bibitem{Krupa:16}
K.~Krupa, C.~Louot, V.~Couderc, M.~Fabert, R.~Guenard, B.~M. Shalaby, A.~Tonello, D.~Pagnoux, P.~Leproux, A.~Bendahmane, R.~Dupiol, G.~Millot, S.~Wabnitz, Spatiotemporal characterization of supercontinuum extending from the visible to the mid-infrared in a multimode graded-index optical fiber, Opt. Lett. 41~(24) (2016) 5785--5788.
\newblock \href {https://doi.org/10.1364/OL.41.005785} {\path{doi:10.1364/OL.41.005785}}.

\bibitem{Krupanatphotonics}
K.~Krupa, A.~Tonello, B.~M. Shalaby, M.~Fabert, A.~Barth{\'e}l{\'e}my, G.~Millot, S.~Wabnitz, V.~Couderc, Spatial beam self-cleaning in multimode fibres, Nat. Photonics 11 (2017) 234--241.

\bibitem{LiuKerr}
Z.~Liu, L.~G. Wright, D.~N. Christodoulides, F.~W. Wise, Kerr self-cleaning of femtosecond-pulsed beams in graded-index multimode fiber, Opt. Lett. 41~(16) (2016) 3675--3678.
\newblock \href {https://doi.org/10.1364/OL.41.003675} {\path{doi:10.1364/OL.41.003675}}.

\bibitem{WrightNP2016}
L.~G. Wright, Z.~Liu, D.~A. Nolan, M.-J. Li, D.~N. Christodoulides, F.~W. Wise, Self-organized instability in graded-index multimode fibres, Nat. Photonics 10 (2016) 771--776.

\bibitem{Ferraro2023apx}
M.~Ferraro, F.~Mangini, M.~Zitelli, S.~Wabnitz, {On spatial beam self-cleaning from the perspective of optical wave thermalization in multimode graded-index fibers}, Advances in Physics: X 8~(1) (2023).
\newblock \href {https://doi.org/10.1080/23746149.2023.2228018} {\path{doi:10.1080/23746149.2023.2228018}}.

\bibitem{teugin2020single}
U.~Te{\u{g}}in, B.~Rahmani, E.~Kakkava, D.~Psaltis, C.~Moser, Single-mode output by controlling the spatiotemporal nonlinearities in mode-locked femtosecond multimode fiber lasers, Advanced Photonics 2~(5) (2020) 056005--056005.

\bibitem{moussa2021spatiotemporal}
N.~O. Moussa, T.~Mansuryan, C.-H. Hage, M.~Fabert, K.~Krupa, A.~Tonello, M.~Ferraro, L.~Leggio, M.~Zitelli, F.~Mangini, et~al., Spatiotemporal beam self-cleaning for high-resolution nonlinear fluorescence imaging with multimode fiber, Scientific Reports 11~(1) (2021) 18240.

\bibitem{wehbi2022continuous}
S.~Wehbi, T.~Mansuryan, K.~Krupa, M.~Fabert, A.~Tonello, M.~Zitelli, M.~Ferraro, F.~Mangini, Y.~Sun, S.~Vergnole, et~al., Continuous spatial self-cleaning in grin multimode fiber for self-referenced multiplex cars imaging, Optics Express 30~(10) (2022) 16104--16114.

\bibitem{Hansson2020}
T.~Hansson, A.~Tonello, T.~Mansuryan, F.~Mangini, M.~Zitelli, M.~Ferraro, A.~Niang, R.~Crescenzi, S.~Wabnitz, V.~Couderc, {Nonlinear beam self-imaging and self-focusing dynamics in a GRIN multimode optical fiber: theory and experiments}, Optics Express 28~(16) (2020) 24005.
\newblock \href {http://arxiv.org/abs/2005.07280} {\path{arXiv:2005.07280}}, \href {https://doi.org/10.1364/oe.398531} {\path{doi:10.1364/oe.398531}}.

\bibitem{leventoux20203d}
Y.~Leventoux, G.~Granger, K.~Krupa, A.~Tonello, G.~Millot, M.~Ferraro, F.~Mangini, M.~Zitelli, S.~Wabnitz, S.~F{\'e}vrier, V.~Couderc, 3d time-domain beam mapping for studying nonlinear dynamics in multimode optical fibers, Optics Letters 46~(1) (2020) 66--69.

\bibitem{sidelnikov19}
O.~S. Sidelnikov, E.~V. Podivilov, M.~P. Fedoruk, S.~Wabnitz, \href{https://www.sciencedirect.com/science/article/pii/S1068520019301580}{Random mode coupling assists kerr beam self-cleaning in a graded-index multimode optical fiber}, Optical Fiber Technology 53 (2019) 101994.
\newblock \href {https://doi.org/https://doi.org/10.1016/j.yofte.2019.101994} {\path{doi:https://doi.org/10.1016/j.yofte.2019.101994}}.
\newline\urlprefix\url{https://www.sciencedirect.com/science/article/pii/S1068520019301580}

\bibitem{degueldreNP16}
H.~Degueldre, J.~Metzger, T.~Geisel, F.~R., Random focusing of tsunami waves, Nature Physics 259 (2016) 12.

\bibitem{carmazzaNC19}
P.~Caramazza, O.~Moran, R.~Murray-Smith, D.~Faccio, Transmission of natural scene images through a multimode fibre, Nature Communications 2029 (2019) 10.

\bibitem{cherroret15}
N.~Cherroret, T.~Karpiuk, B.~Gr\'emaud, C.~Miniatura, \href{https://link.aps.org/doi/10.1103/PhysRevA.92.063614}{Thermalization of matter waves in speckle potentials}, Phys. Rev. A 92 (2015) 063614.
\newblock \href {https://doi.org/10.1103/PhysRevA.92.063614} {\path{doi:10.1103/PhysRevA.92.063614}}.
\newline\urlprefix\url{https://link.aps.org/doi/10.1103/PhysRevA.92.063614}

\bibitem{cherroret21}
N.~Cherroret, T.~Scoquart, D.~Delande, \href{https://www.sciencedirect.com/science/article/pii/S0003491621001494}{Coherent multiple scattering of out-of-equilibrium interacting bose gases}, Annals of Physics 435 (2021) 168543, special Issue on Localisation 2020.
\newblock \href {https://doi.org/https://doi.org/10.1016/j.aop.2021.168543} {\path{doi:https://doi.org/10.1016/j.aop.2021.168543}}.
\newline\urlprefix\url{https://www.sciencedirect.com/science/article/pii/S0003491621001494}

\bibitem{wang20}
Z.~Wang, W.~Fu, Y.~Zhang, H.~Zhao, \href{https://link.aps.org/doi/10.1103/PhysRevLett.124.186401}{Wave-turbulence origin of the instability of anderson localization against many-body interactions}, Phys. Rev. Lett. 124 (2020) 186401.
\newblock \href {https://doi.org/10.1103/PhysRevLett.124.186401} {\path{doi:10.1103/PhysRevLett.124.186401}}.
\newline\urlprefix\url{https://link.aps.org/doi/10.1103/PhysRevLett.124.186401}

\bibitem{nazarenko19}
S.~Nazarenko, A.~Soffer, M.-B. Tran, \href{https://www.mdpi.com/1099-4300/21/9/823}{On the wave turbulence theory for the nonlinear schrödinger equation with random potentials}, Entropy 21~(9) (2019).
\newblock \href {https://doi.org/10.3390/e21090823} {\path{doi:10.3390/e21090823}}.
\newline\urlprefix\url{https://www.mdpi.com/1099-4300/21/9/823}

\bibitem{kottos23}
A.~Ramos, C.~Shi, L.~Fern\'andez-Alc\'azar, D.~Christodoulides, T.~Kottos, Theory of localization-hindered thermalization in nonlinear multimode photonics, Communications Physics 6 (2023) 189.

\bibitem{Pyrialakos24}
G.~G. Pyrialakos, F.~O. Wu, P.~S. Jung, H.~Ren, K.~G. Makris, Z.~H. Musslimani, M.~Khajavikhan, T.~Kottos, D.~Christodoulides, \href{https://link.aps.org/doi/10.1103/PhysRevResearch.6.013072}{Slowdown of thermalization and the emergence of prethermal dynamics in disordered optical lattices}, Phys. Rev. Res. 6 (2024) 013072.
\newblock \href {https://doi.org/10.1103/PhysRevResearch.6.013072} {\path{doi:10.1103/PhysRevResearch.6.013072}}.
\newline\urlprefix\url{https://link.aps.org/doi/10.1103/PhysRevResearch.6.013072}

\bibitem{Fusaro:PRL:2019}
A.~Fusaro, J.~Garnier, K.~Krupa, G.~Millot, A.~Picozzi, \href{https://link.aps.org/doi/10.1103/PhysRevLett.122.123902}{Dramatic acceleration of wave condensation mediated by disorder in multimode fibers}, Phys. Rev. Lett. 122 (2019) 123902.
\newblock \href {https://doi.org/10.1103/PhysRevLett.122.123902} {\path{doi:10.1103/PhysRevLett.122.123902}}.
\newline\urlprefix\url{https://link.aps.org/doi/10.1103/PhysRevLett.122.123902}

\bibitem{berti22}
N.~Berti, K.~Baudin, A.~Fusaro, G.~Millot, A.~Picozzi, J.~Garnier, \href{https://link.aps.org/doi/10.1103/PhysRevLett.129.063901}{Interplay of thermalization and strong disorder: Wave turbulence theory, numerical simulations, and experiments in multimode optical fibers}, Phys. Rev. Lett. 129 (2022) 063901.
\newblock \href {https://doi.org/10.1103/PhysRevLett.129.063901} {\path{doi:10.1103/PhysRevLett.129.063901}}.
\newline\urlprefix\url{https://link.aps.org/doi/10.1103/PhysRevLett.129.063901}

\bibitem{Pariente2016np}
G.~Pariente, V.~Gallet, A.~Borot, O.~Gobert, F.~Qu{\'{e}}r{\'{e}}, {Space-time characterization of ultra-intense femtosecond laser beams}, Nat. Photonics 10~(8) (2016) 547--553.
\newblock \href {https://doi.org/10.1038/nphoton.2016.140} {\path{doi:10.1038/nphoton.2016.140}}.

\bibitem{Gervaziev2021lpl}
M.~D. Gervaziev, I.~Zhdanov, D.~S. Kharenko, V.~A. Gonta, V.~M. Volosi, E.~V. Podivilov, S.~A. Babin, S.~Wabnitz, \href{https://doi.org/10.1088/1612-202X/abcf27 https://iopscience.iop.org/article/10.1088/1612-202X/abcf27}{{Mode decomposition of multimode optical fiber beams by phase-only spatial light modulator}}, Laser Phys. Lett. 18~(1) (2021) 015101.
\newblock \href {https://doi.org/10.1088/1612-202X/abcf27} {\path{doi:10.1088/1612-202X/abcf27}}.
\newline\urlprefix\url{https://doi.org/10.1088/1612-202X/abcf27 https://iopscience.iop.org/article/10.1088/1612-202X/abcf27}

\bibitem{pourbeyram2022direct}
H.~Pourbeyram, P.~Sidorenko, F.~O. Wu, N.~Bender, L.~Wright, D.~N. Christodoulides, F.~Wise, Direct observations of thermalization to a rayleigh--jeans distribution in multimode optical fibres, Nature Physics 18~(6) (2022) 685--690.

\bibitem{Mangini2022Statistical}
F.~Mangini, M.~Gervaziev, M.~Ferraro, D.~S. Kharenko, M.~Zitelli, Y.~Sun, V.~Couderc, E.~V. Podivilov, S.~A. Babin, S.~Wabnitz, \href{https://opg.optica.org/oe/abstract.cfm?URI=oe-30-7-10850}{Statistical mechanics of beam self-cleaning in grin multimode optical fibers}, Opt. Express 30~(7) (2022) 10850--10865.
\newblock \href {https://doi.org/10.1364/OE.449187} {\path{doi:10.1364/OE.449187}}.
\newline\urlprefix\url{https://opg.optica.org/oe/abstract.cfm?URI=oe-30-7-10850}

\bibitem{mangini2024maximization}
F.~Mangini, M.~Ferraro, W.~A. Gemechu, Y.~Sun, M.~Gervaziev, D.~Kharenko, S.~Babin, V.~Couderc, S.~Wabnitz, On the maximization of entropy in the process of thermalization of highly multimode nonlinear beams, Optics Letters 49~(12) (2024) 3340--3343.

\bibitem{ferraro2024calorimetry}
M.~Ferraro, F.~Mangini, F.~Wu, M.~Zitelli, D.~Christodoulides, S.~Wabnitz, Calorimetry of photon gases in nonlinear multimode optical fibers, Physical Review X 14~(2) (2024) 021020.

\bibitem{mangini2023modal}
F.~Mangini, M.~Ferraro, Y.~Sun, M.~Gervaziev, P.~Parra-Rivas, D.~S. Kharenko, V.~Couderc, S.~Wabnitz, Modal phase-locking in multimode nonlinear optical fibers, Optics Letters 48~(14) (2023) 3677--3680.

\bibitem{Podivilov2022prl}
E.~V. Podivilov, F.~Mangini, O.~S. Sidelnikov, M.~Ferraro, M.~Gervaziev, D.~S. Kharenko, M.~Zitelli, M.~P. Fedoruk, S.~A. Babin, S.~Wabnitz, \href{https://link.aps.org/doi/10.1103/PhysRevLett.128.243901}{{Thermalization of Orbital Angular Momentum Beams in Multimode Optical Fibers}}, Phys. Rev. Lett. 128~(24) (2022) 243901.
\newblock \href {https://doi.org/10.1103/PhysRevLett.128.243901} {\path{doi:10.1103/PhysRevLett.128.243901}}.
\newline\urlprefix\url{https://link.aps.org/doi/10.1103/PhysRevLett.128.243901}

\bibitem{doi:10.1098/rspl.1866.0039}
J.~C. Maxwell, Ii. on the dynamical theory of gases, Proceedings of the Royal Society of London 15 (1867) 167--171.
\newblock \href {https://doi.org/10.1098/rspl.1866.0039} {\path{doi:10.1098/rspl.1866.0039}}.

\bibitem{Mecozzi:12}
A.~Mecozzi, C.~Antonelli, M.~Shtaif, \href{http://www.opticsexpress.org/abstract.cfm?URI=oe-20-21-23436}{Coupled manakov equations in multimode fibers with strongly coupled groups of modes}, Opt. Express 20~(21) (2012) 23436--23441.
\newblock \href {https://doi.org/10.1364/OE.20.023436} {\path{doi:10.1364/OE.20.023436}}.
\newline\urlprefix\url{http://www.opticsexpress.org/abstract.cfm?URI=oe-20-21-23436}

\bibitem{MecozziOpEx1}
A.~Mecozzi, C.~Antonelli, M.~Shtaif, \href{https://opg.optica.org/oe/abstract.cfm?URI=oe-20-11-11673}{Nonlinear propagation in multi-mode fibers in the strong coupling regime}, Opt. Express 20~(11) (2012) 11673--11678.
\newblock \href {https://doi.org/10.1364/OE.20.011673} {\path{doi:10.1364/OE.20.011673}}.
\newline\urlprefix\url{https://opg.optica.org/oe/abstract.cfm?URI=oe-20-11-11673}

\bibitem{MumtazJLT12}
S.~Mumtaz, R.-J. Essiambre, G.~P. Agrawal, {Nonlinear Propagation in Multimode and Multicore Fibers: Generalization of the Manakov Equations}, Journal of Lightwave Technology 31~(3) (2013) 398--406.
\newblock \href {https://doi.org/10.1109/JLT.2012.2231401} {\path{doi:10.1109/JLT.2012.2231401}}.

\bibitem{Xiao:14}
Y.~Xiao, R.-J. Essiambre, M.~Desgroseilliers, A.~M. Tulino, R.~Ryf, S.~Mumtaz, G.~P. Agrawal, \href{http://www.opticsexpress.org/abstract.cfm?URI=oe-22-26-32039}{Theory of intermodal four-wave mixing with random linear mode coupling in few-mode fibers}, Opt. Express 22~(26) (2014) 32039--32059.
\newblock \href {https://doi.org/10.1364/OE.22.032039} {\path{doi:10.1364/OE.22.032039}}.
\newline\urlprefix\url{http://www.opticsexpress.org/abstract.cfm?URI=oe-22-26-32039}

\bibitem{ho14}
K.~Ho, J.~Kahn, Linear propagation effects in mode-division multiplexing systems, J. Lightwave Tech. 32 (2014) 4.

\bibitem{kaminow13}
I.~Kaminow, T.~Li, A.~Willner, Optical Fiber Telecommunications, Systems and Networks, Sixth Ed., Elsevier, 2013.

\bibitem{agrawal_physics_2023}
G.~P. Agrawal, \href{https://www.cambridge.org/core/books/physics-and-engineering-of-gradedindex-media/328D1D61DEB81F78CC7AAD69658A4904}{Physics and {Engineering} of {Graded}-{Index} {Media}}, Cambridge University Press, Cambridge, 2023.
\newblock \href {https://doi.org/10.1017/9781009282086} {\path{doi:10.1017/9781009282086}}.
\newline\urlprefix\url{https://www.cambridge.org/core/books/physics-and-engineering-of-gradedindex-media/328D1D61DEB81F78CC7AAD69658A4904}

\bibitem{Kuznetsov2021sr}
S.~A. Babin, A.~G. Kuznetsov, O.~S. Sidelnikov, A.~A. Wolf, I.~N. Nemov, S.~I. Kablukov, E.~V. Podivilov, M.~P. Fedoruk, S.~Wabnitz, \href{https://doi.org/10.1038/s41598-021-01491-0 https://www.nature.com/articles/s41598-021-01491-0}{{Spatio-spectral beam control in multimode diode-pumped Raman fibre lasers via intracavity filtering and Kerr cleaning}}, Sci. Rep. 11~(1) (2021) 21994.
\newblock \href {https://doi.org/10.1038/s41598-021-01491-0} {\path{doi:10.1038/s41598-021-01491-0}}.
\newline\urlprefix\url{https://doi.org/10.1038/s41598-021-01491-0 https://www.nature.com/articles/s41598-021-01491-0}

\bibitem{Kharenko2022ol}
D.~S. Kharenko, M.~D. Gervaziev, A.~G. Kuznetsov, E.~V. Podivilov, S.~Wabnitz, S.~A. Babin, \href{https://www.osapublishing.org/ol/abstract.cfm?doi=10.1364/OL.449119 https://opg.optica.org/abstract.cfm?URI=ol-47-5-1222}{{Mode-resolved analysis of pump and Stokes beams in LD-pumped GRIN fiber Raman lasers}}, Opt. Lett. 47~(5) (2022) 1222--1225.
\newblock \href {https://doi.org/10.1364/OL.449119} {\path{doi:10.1364/OL.449119}}.
\newline\urlprefix\url{https://www.osapublishing.org/ol/abstract.cfm?doi=10.1364/OL.449119 https://opg.optica.org/abstract.cfm?URI=ol-47-5-1222}

\bibitem{biasi21}
A.~Biasi, O.~Evnin, B.~A. Malomed, \href{https://link.aps.org/doi/10.1103/PhysRevE.104.034210}{Fermi-pasta-ulam phenomena and persistent breathers in the harmonic trap}, Phys. Rev. E 104 (2021) 034210.
\newblock \href {https://doi.org/10.1103/PhysRevE.104.034210} {\path{doi:10.1103/PhysRevE.104.034210}}.
\newline\urlprefix\url{https://link.aps.org/doi/10.1103/PhysRevE.104.034210}

\bibitem{biasi23}
A.~Biasi, O.~Evnin, B.~A. Malomed, \href{https://link.aps.org/doi/10.1103/PhysRevE.108.034204}{Obstruction to ergodicity in nonlinear schr\"odinger equations with resonant potentials}, Phys. Rev. E 108 (2023) 034204.
\newblock \href {https://doi.org/10.1103/PhysRevE.108.034204} {\path{doi:10.1103/PhysRevE.108.034204}}.
\newline\urlprefix\url{https://link.aps.org/doi/10.1103/PhysRevE.108.034204}

\bibitem{Zakharov05}
V.~E. Zakharov, A.~O. Korotkevich, A.~N. Pushkarev, A.~I. Dyachenko, Mesoscopic wave turbulence, JETP Letters 82~(8) (2005) 487--491.
\newblock \href {https://doi.org/10.1134/1.2150867} {\path{doi:10.1134/1.2150867}}.

\bibitem{kartashova08}
E.~Kartashova, S.~Nazarenko, O.~Rudenko, \href{https://link.aps.org/doi/10.1103/PhysRevE.78.016304}{Resonant interactions of nonlinear water waves in a finite basin}, Phys. Rev. E 78 (2008) 016304.
\newblock \href {https://doi.org/10.1103/PhysRevE.78.016304} {\path{doi:10.1103/PhysRevE.78.016304}}.
\newline\urlprefix\url{https://link.aps.org/doi/10.1103/PhysRevE.78.016304}

\bibitem{kartashova09}
E.~Kartashova, \href{https://dx.doi.org/10.1209/0295-5075/87/44001}{Discrete wave turbulence}, Europhysics Letters 87~(4) (2009) 44001.
\newblock \href {https://doi.org/10.1209/0295-5075/87/44001} {\path{doi:10.1209/0295-5075/87/44001}}.
\newline\urlprefix\url{https://dx.doi.org/10.1209/0295-5075/87/44001}

\bibitem{Lvov10}
V.~S. L'vov, S.~Nazarenko, \href{https://link.aps.org/doi/10.1103/PhysRevE.82.056322}{Discrete and mesoscopic regimes of finite-size wave turbulence}, Phys. Rev. E 82 (2010) 056322.
\newblock \href {https://doi.org/10.1103/PhysRevE.82.056322} {\path{doi:10.1103/PhysRevE.82.056322}}.
\newline\urlprefix\url{https://link.aps.org/doi/10.1103/PhysRevE.82.056322}

\bibitem{mordant18}
R.~Hassaini, N.~Mordant, \href{https://link.aps.org/doi/10.1103/PhysRevFluids.3.094805}{Confinement effects on gravity-capillary wave turbulence}, Phys. Rev. Fluids 3 (2018) 094805.
\newblock \href {https://doi.org/10.1103/PhysRevFluids.3.094805} {\path{doi:10.1103/PhysRevFluids.3.094805}}.
\newline\urlprefix\url{https://link.aps.org/doi/10.1103/PhysRevFluids.3.094805}

\bibitem{Chibbaro17}
S.~Chibbaro, G.~Dematteis, C.~Josserand, L.~Rondoni, \href{https://link.aps.org/doi/10.1103/PhysRevE.96.021101}{Wave-turbulence theory of four-wave nonlinear interactions}, Phys. Rev. E 96 (2017) 021101.
\newblock \href {https://doi.org/10.1103/PhysRevE.96.021101} {\path{doi:10.1103/PhysRevE.96.021101}}.
\newline\urlprefix\url{https://link.aps.org/doi/10.1103/PhysRevE.96.021101}

\bibitem{Chibbaro18}
S.~Chibbaro, G.~Dematteis, L.~Rondoni, \href{https://www.sciencedirect.com/science/article/pii/S0167278917301847}{4-wave dynamics in kinetic wave turbulence}, Physica D: Nonlinear Phenomena 362 (2018) 24--59.
\newblock \href {https://doi.org/https://doi.org/10.1016/j.physd.2017.09.001} {\path{doi:https://doi.org/10.1016/j.physd.2017.09.001}}.
\newline\urlprefix\url{https://www.sciencedirect.com/science/article/pii/S0167278917301847}

\bibitem{Kuksin}
S.~Kuksin, A.~Maiocchi, The effective equation method, Eds. E. Tobisch, New Approaches to Nonlinear Waves, Springer, New York, 2015.

\bibitem{during09}
G.~During, A.~Picozzi, S.~Rica, Breakdown of weak-turbulence and nonlinear wave condensation, Physica D 238 (2009) 1524.

\bibitem{lvov03wave}
Y.~Lvov, S.~Nazarenko, R.~West, \href{https://www.sciencedirect.com/science/article/pii/S0167278903002392}{Wave turbulence in bose–einstein condensates}, Physica D: Nonlinear Phenomena 184~(1) (2003) 333--351, complexity and Nonlinearity in Physical Systems -- A Special Issue to Honor Alan Newell.
\newblock \href {https://doi.org/https://doi.org/10.1016/S0167-2789(03)00239-2} {\path{doi:https://doi.org/10.1016/S0167-2789(03)00239-2}}.
\newline\urlprefix\url{https://www.sciencedirect.com/science/article/pii/S0167278903002392}

\bibitem{baudin21}
K.~Baudin, A.~Fusaro, J.~Garnier, N.~Berti, K.~Krupa, I.~Carusotto, S.~Rica, G.~Millot, A.~Picozzi, \href{https://dx.doi.org/10.1209/0295-5075/134/14001}{Energy and wave-action flows underlying rayleigh-jeans thermalization of optical waves propagating in a multimode fiber(a)}, Europhysics Letters 134~(1) (2021) 14001.
\newblock \href {https://doi.org/10.1209/0295-5075/134/14001} {\path{doi:10.1209/0295-5075/134/14001}}.
\newline\urlprefix\url{https://dx.doi.org/10.1209/0295-5075/134/14001}

\bibitem{Podivilov2018prl}
E.~V. Podivilov, D.~S. Kharenko, V.~A. Gonta, K.~Krupa, O.~S. Sidelnikov, S.~Turitsyn, M.~P. Fedoruk, S.~A. Babin, S.~Wabnitz, \href{http://arxiv.org/abs/1811.10239 https://link.aps.org/doi/10.1103/PhysRevLett.122.103902}{{Hydrodynamic 2D Turbulence and Spatial Beam Condensation in Multimode Optical Fibers}}, Phys. Rev. Lett. 122~(10) (2019) 103902.
\newblock \href {http://arxiv.org/abs/1811.10239} {\path{arXiv:1811.10239}}, \href {https://doi.org/10.1103/PhysRevLett.122.103902} {\path{doi:10.1103/PhysRevLett.122.103902}}.
\newline\urlprefix\url{http://arxiv.org/abs/1811.10239 https://link.aps.org/doi/10.1103/PhysRevLett.122.103902}

\bibitem{zakharov65weak}
V.~Zakharov, Weak turbulence in media with a decay spectrum, Journal of Applied Mechanics and Technical Physics 6 (1965) 22.

\bibitem{zakharov67on}
V.~Zakharov, On the weak turbulence spectrum in plasma without magnetic field, Sov. Phys. JETP 24 (1967) 455.

\bibitem{kolmogorov41the}
A.~Kolmogorov, The local structure of turbulence in incompressible viscous fluid for very large reynolds numbers, Doklady Akademiia Nauk SSSR 30 (1941) 301.

\bibitem{zhu2023self}
Y.~Zhu, B.~Semisalov, G.~Krstulovic, S.~Nazarenko, \href{https://link.aps.org/doi/10.1103/PhysRevE.108.064207}{Self-similar evolution of wave turbulence in gross-pitaevskii system}, Phys. Rev. E 108 (2023) 064207.
\newblock \href {https://doi.org/10.1103/PhysRevE.108.064207} {\path{doi:10.1103/PhysRevE.108.064207}}.
\newline\urlprefix\url{https://link.aps.org/doi/10.1103/PhysRevE.108.064207}

\bibitem{baudin2023rayleigh}
K.~Baudin, J.~Garnier, A.~Fusaro, C.~Michel, K.~Krupa, G.~Millot, A.~Picozzi, Rayleigh--jeans thermalization vs beam cleaning in multimode optical fibers, Optics Communications 545 (2023) 129716.

\bibitem{cao18}
W.~Xiong, C.~Hsu, Y.~e.~a. Bromberg, Complete polarization control in multimode fibers with polarization and mode coupling, Light Sci Appl 7 (2018) 54.

\bibitem{shemirani09principal}
M.~B. Shemirani, W.~Mao, R.~A. Panicker, J.~M. Kahn, Principal modes in graded-index multimode fiber in presence of spatial- and polarization-mode coupling, Journal of Lightwave Technology 27~(10) (2009) 1248--1261.
\newblock \href {https://doi.org/10.1109/JLT.2008.2005066} {\path{doi:10.1109/JLT.2008.2005066}}.

\bibitem{shen05compensation}
X.~Shen, J.~M. Kahn, M.~A. Horowitz, \href{https://opg.optica.org/ol/abstract.cfm?URI=ol-30-22-2985}{Compensation for multimode fiber dispersion by adaptive optics}, Opt. Lett. 30~(22) (2005) 2985--2987.
\newblock \href {https://doi.org/10.1364/OL.30.002985} {\path{doi:10.1364/OL.30.002985}}.
\newline\urlprefix\url{https://opg.optica.org/ol/abstract.cfm?URI=ol-30-22-2985}

\bibitem{fouque07}
J.-P. Fouque, J.~Garnier, G.~Papanicolaou, K.~S\o~lna, Wave Propagation and Time Reversal in Randomly Layered Media, Springer, 2007.

\bibitem{haldar24rayleigh}
P.~Haldar, S.~Mu, B.~Georgeot, J.~Gong, C.~Miniatura, G.~Lemari\'e, Rayleigh-jeans prethermalization and wave condensation in a nonlinear disordered floquet system, EPL 144 (2024) 63001.

\bibitem{Krupa:19}
K.~Krupa, G.~G. Casta{\~{n}}eda, A.~Tonello, A.~Niang, D.~S. Kharenko, M.~Fabert, V.~Couderc, G.~Millot, U.~Minoni, D.~Modotto, S.~Wabnitz, Nonlinear polarization dynamics of {K}err beam self-cleaning in a graded-index multimode optical fiber, Opt. Lett. 44~(1) (2019) 171--174.
\newblock \href {https://doi.org/10.1364/OL.44.000171} {\path{doi:10.1364/OL.44.000171}}.

\bibitem{cao16}
W.~Xiong, P.~Ambichl, Y.~Bromberg, B.~Redding, S.~Rotter, H.~Cao, \href{https://link.aps.org/doi/10.1103/PhysRevLett.117.053901}{Spatiotemporal control of light transmission through a multimode fiber with strong mode coupling}, Phys. Rev. Lett. 117 (2016) 053901.
\newblock \href {https://doi.org/10.1103/PhysRevLett.117.053901} {\path{doi:10.1103/PhysRevLett.117.053901}}.
\newline\urlprefix\url{https://link.aps.org/doi/10.1103/PhysRevLett.117.053901}

\bibitem{Zitelli:23}
M.~Zitelli, V.~Couderc, M.~Ferraro, F.~Mangini, P.~Parra-Rivas, Y.~Sun, S.~Wabnitz, \href{https://opg.optica.org/prj/abstract.cfm?URI=prj-11-5-750}{Spatiotemporal mode decomposition of ultrashort pulses in linear and nonlinear graded-index multimode fibers}, Photon. Res. 11~(5) (2023) 750--756.
\newblock \href {https://doi.org/10.1364/PRJ.484271} {\path{doi:10.1364/PRJ.484271}}.
\newline\urlprefix\url{https://opg.optica.org/prj/abstract.cfm?URI=prj-11-5-750}

\bibitem{olshansky75}
R.~Olshansky, Mode coupling effects in graded-index optical fibers, Applied Optics 14 (1975) 935.

\bibitem{marcuse74}
D.~Marcuse, Theory of dielectric optical waveguides, Academic Press, New York, 1974.

\bibitem{garnier20}
J.~Garnier, Intensity fluctuations in random waveguides, Commun. Math. Sci. 18 (2020) 947.

\bibitem{Agrawal2013}
G.~P. Agrawal, \href{https://linkinghub.elsevier.com/retrieve/pii/B9780123970237000012}{Nonlinear Fiber Optics}, sixth Edition, Elsevier, 2019.
\newblock \href {https://doi.org/10.1016/B978-0-12-397023-7.00001-2} {\path{doi:10.1016/B978-0-12-397023-7.00001-2}}.
\newline\urlprefix\url{https://linkinghub.elsevier.com/retrieve/pii/B9780123970237000012}

\bibitem{ferraro2021femtosecond}
M.~Ferraro, F.~Mangini, M.~Zitelli, A.~Tonello, A.~De~Luca, V.~Couderc, S.~Wabnitz, Femtosecond nonlinear losses in multimode optical fibers, Photonics Research 9~(12) (2021) 2443--2453.

\bibitem{gervaziev2020mode}
M.~Gervaziev, I.~Zhdanov, D.~Kharenko, V.~Gonta, V.~Volosi, E.~Podivilov, S.~Babin, S.~Wabnitz, Mode decomposition of multimode optical fiber beams by phase-only spatial light modulator, Laser Physics Letters 18~(1) (2020) 015101.

\bibitem{mangini2023high}
F.~Mangini, M.~Ferraro, A.~Tonello, V.~Couderc, S.~Wabnitz, High-temperature wave thermalization spoils beam self-cleaning in nonlinear multimode grin fibers, Optics Letters 48~(18) (2023) 4741--4744.

\bibitem{ferraro2023spatial}
M.~Ferraro, F.~Mangini, M.~Zitelli, R.~Jauberteau, Y.~Sun, P.~Parra-Rivas, K.~Krupa, A.~Tonello, V.~Couderc, S.~Wabnitz, Spatial beam cleaning in multimode grin fibers: Polarization effects, IEEE Photonics Journal (2023).

\bibitem{ferraro2023multimode}
M.~Ferraro, F.~Mangini, Y.~Leventoux, A.~Tonello, M.~Zitelli, T.~Mansuryan, Y.~Sun, S.~Fevrier, K.~Krupa, D.~Kharenko, et~al., Multimode optical fiber beam-by-beam cleanup, Journal of Lightwave Technology 41~(10) (2023) 3164--3174.

\bibitem{steinmeyer2023generalized}
G.~Steinmeyer, Generalized thermodynamics of optical multimode systems and the breach of the second law of thermodynamics, in: 2023 Conference on Lasers and Electro-Optics Europe \& European Quantum Electronics Conference (CLEO/Europe-EQEC), IEEE, 2023, pp. 1--1.

\bibitem{zanaglia2024bridging}
L.~Zanaglia, J.~Garnier, S.~Rica, R.~Kaiser, S.~Wabnitz, C.~Michel, V.~Doya, A.~Picozzi, \href{https://link.aps.org/doi/10.1103/PhysRevA.110.063530}{Bridging rayleigh-jeans and bose-einstein condensation of a guided fluid of light with positive and negative temperatures}, Phys. Rev. A 110 (2024) 063530.
\newblock \href {https://doi.org/10.1103/PhysRevA.110.063530} {\path{doi:10.1103/PhysRevA.110.063530}}.
\newline\urlprefix\url{https://link.aps.org/doi/10.1103/PhysRevA.110.063530}

\bibitem{davis2001simulations}
M.~J. Davis, S.~A. Morgan, K.~Burnett, \href{https://link.aps.org/doi/10.1103/PhysRevLett.87.160402}{Simulations of bose fields at finite temperature}, Phys. Rev. Lett. 87 (2001) 160402.
\newblock \href {https://doi.org/10.1103/PhysRevLett.87.160402} {\path{doi:10.1103/PhysRevLett.87.160402}}.
\newline\urlprefix\url{https://link.aps.org/doi/10.1103/PhysRevLett.87.160402}

\bibitem{blakie2005projected}
P.~B. Blakie, M.~J. Davis, \href{https://link.aps.org/doi/10.1103/PhysRevA.72.063608}{Projected gross-pitaevskii equation for harmonically confined bose gases at finite temperature}, Phys. Rev. A 72 (2005) 063608.
\newblock \href {https://doi.org/10.1103/PhysRevA.72.063608} {\path{doi:10.1103/PhysRevA.72.063608}}.
\newline\urlprefix\url{https://link.aps.org/doi/10.1103/PhysRevA.72.063608}

\bibitem{onsager49}
L.~Onsager, Statistical hydrodynamics, Il Nuovo Cimento 6 (1949) 279.

\bibitem{ramsey56}
N.~Ramsey, Thermodynamics and statistical mechanics at negative absolute temperatures, Physical Review 103 (1956) 20.

\bibitem{baldovin21}
M.~Baldovin, S.~Iubini, R.~Livi, A.~Vulpiani, \href{https://www.sciencedirect.com/science/article/pii/S0370157321001204}{Statistical mechanics of systems with negative temperature}, Physics Reports 923 (2021) 1--50, statistical mechanics of systems with negative temperature.
\newblock \href {https://doi.org/https://doi.org/10.1016/j.physrep.2021.03.007} {\path{doi:https://doi.org/10.1016/j.physrep.2021.03.007}}.
\newline\urlprefix\url{https://www.sciencedirect.com/science/article/pii/S0370157321001204}

\bibitem{onorato23}
M.~Onorato, Y.~Lvov, G.~Dematteis, S.~Chibbaro, \href{https://www.sciencedirect.com/science/article/pii/S0370157323003046}{Wave turbulence and thermalization in one-dimensional chains}, Physics Reports 1040 (2023) 1--36, wave Turbulence and thermalization in one-dimensional chains.
\newblock \href {https://doi.org/https://doi.org/10.1016/j.physrep.2023.09.006} {\path{doi:https://doi.org/10.1016/j.physrep.2023.09.006}}.
\newline\urlprefix\url{https://www.sciencedirect.com/science/article/pii/S0370157323003046}

\bibitem{braun13}
S.~Braun, J.~P. Ronzheimer, M.~Schreiber, S.~S. Hodgman, T.~Rom, I.~Bloch, U.~Schneider, \href{https://www.science.org/doi/abs/10.1126/science.1227831}{Negative absolute temperature for motional degrees of freedom}, Science 339~(6115) (2013) 52--55.
\newblock \href {http://arxiv.org/abs/https://www.science.org/doi/pdf/10.1126/science.1227831} {\path{arXiv:https://www.science.org/doi/pdf/10.1126/science.1227831}}, \href {https://doi.org/10.1126/science.1227831} {\path{doi:10.1126/science.1227831}}.
\newline\urlprefix\url{https://www.science.org/doi/abs/10.1126/science.1227831}

\bibitem{gauthier19}
G.~Gauthier, M.~T. Reeves, X.~Yu, A.~S. Bradley, M.~A. Baker, T.~A. Bell, H.~Rubinsztein-Dunlop, M.~J. Davis, T.~W. Neely, \href{https://www.science.org/doi/abs/10.1126/science.aat5718}{Giant vortex clusters in a two-dimensional quantum fluid}, Science 364~(6447) (2019) 1264--1267.
\newblock \href {http://arxiv.org/abs/https://www.science.org/doi/pdf/10.1126/science.aat5718} {\path{arXiv:https://www.science.org/doi/pdf/10.1126/science.aat5718}}, \href {https://doi.org/10.1126/science.aat5718} {\path{doi:10.1126/science.aat5718}}.
\newline\urlprefix\url{https://www.science.org/doi/abs/10.1126/science.aat5718}

\bibitem{johnstone19}
S.~P. Johnstone, A.~J. Groszek, P.~T. Starkey, C.~J. Billington, T.~P. Simula, K.~Helmerson, \href{https://www.science.org/doi/abs/10.1126/science.aat5793}{Evolution of large-scale flow from turbulence in a two-dimensional superfluid}, Science 364~(6447) (2019) 1267--1271.
\newblock \href {http://arxiv.org/abs/https://www.science.org/doi/pdf/10.1126/science.aat5793} {\path{arXiv:https://www.science.org/doi/pdf/10.1126/science.aat5793}}, \href {https://doi.org/10.1126/science.aat5793} {\path{doi:10.1126/science.aat5793}}.
\newline\urlprefix\url{https://www.science.org/doi/abs/10.1126/science.aat5793}

\bibitem{fienup82phase}
J.~Fienup, Phase retrieval algorithms: a comparison, Applied Optics 21 (1982) 2758.

\bibitem{shechtman15phase}
Y.~Shechtman, Y.~C. Eldar, O.~Cohen, H.~N. Chapman, J.~Miao, M.~Segev, Phase retrieval with application to optical imaging: A contemporary overview, IEEE Signal Processing Magazine 32~(3) (2015) 87--109.
\newblock \href {https://doi.org/10.1109/MSP.2014.2352673} {\path{doi:10.1109/MSP.2014.2352673}}.

\bibitem{fabert2020coherent}
M.~Fabert, M.~S{\u{a}}p{\^a}nțan, K.~Krupa, A.~Tonello, Y.~Leventoux, S.~F{\'e}vrier, T.~Mansuryan, A.~Niang, B.~Wetzel, G.~Millot, et~al., Coherent combining of self-cleaned multimode beams, Scientific reports 10~(1) (2020) 20481.

\bibitem{guasoni17incoherent}
M.~Guasoni, J.~Garnier, B.~Rumpf, D.~Sugny, J.~Fatome, F.~Amrani, G.~Millot, A.~Picozzi, \href{https://link.aps.org/doi/10.1103/PhysRevX.7.011025}{Incoherent fermi-pasta-ulam recurrences and unconstrained thermalization mediated by strong phase correlations}, Phys. Rev. X 7 (2017) 011025.
\newblock \href {https://doi.org/10.1103/PhysRevX.7.011025} {\path{doi:10.1103/PhysRevX.7.011025}}.
\newline\urlprefix\url{https://link.aps.org/doi/10.1103/PhysRevX.7.011025}

\bibitem{pyrialakos2024optical}
G.~G. Pyrialakos, M.~S. Kirsch, R.~M. Altenkirch, J.~Beck, H.~Ren, M.~A. Selim, P.~S. Jung, M.~Khajavikhan, A.~Szameit, M.~Heinrich, et~al., Optical condensation of light via joule-thomson expansion, in: CLEO: Fundamental Science, Optica Publishing Group, 2024, pp. FTu3R--3.

\bibitem{bogoliubov1947on}
N.~Bogoliubov, On the theory of superfluidity, J. Phys. (USSR) 11 (1947) 23.

\bibitem{fontaine2018observation}
Q.~Fontaine, T.~Bienaim\'e, S.~Pigeon, E.~Giacobino, A.~Bramati, Q.~Glorieux, \href{https://link.aps.org/doi/10.1103/PhysRevLett.121.183604}{Observation of the bogoliubov dispersion in a fluid of light}, Phys. Rev. Lett. 121 (2018) 183604.
\newblock \href {https://doi.org/10.1103/PhysRevLett.121.183604} {\path{doi:10.1103/PhysRevLett.121.183604}}.
\newline\urlprefix\url{https://link.aps.org/doi/10.1103/PhysRevLett.121.183604}

\bibitem{frisch1992transition}
T.~Frisch, Y.~Pomeau, S.~Rica, \href{https://link.aps.org/doi/10.1103/PhysRevLett.69.1644}{Transition to dissipation in a model of superflow}, Phys. Rev. Lett. 69 (1992) 1644--1647.
\newblock \href {https://doi.org/10.1103/PhysRevLett.69.1644} {\path{doi:10.1103/PhysRevLett.69.1644}}.
\newline\urlprefix\url{https://link.aps.org/doi/10.1103/PhysRevLett.69.1644}

\bibitem{carusotto2014superfluid}
I.~Carusotto, Superfluid light in bulk nonlinear media, Proceedings of the Royal Society A 470 (2014) 2169.

\bibitem{michel2018}
C.~Michel, O.~Boughdad, M.~Albert, P.-E. Larr\'e, M.~Bellec, Superfluid motion and drag-force cancellation in a fluid of light, Nature Communications 9 (2018) 2108.

\bibitem{braidotti22}
M.~C. Braidotti, R.~Prizia, C.~Maitland, F.~Marino, A.~Prain, I.~Starshynov, N.~Westerberg, E.~M. Wright, D.~Faccio, \href{https://link.aps.org/doi/10.1103/PhysRevLett.128.013901}{Measurement of penrose superradiance in a photon superfluid}, Phys. Rev. Lett. 128 (2022) 013901.
\newblock \href {https://doi.org/10.1103/PhysRevLett.128.013901} {\path{doi:10.1103/PhysRevLett.128.013901}}.
\newline\urlprefix\url{https://link.aps.org/doi/10.1103/PhysRevLett.128.013901}

\bibitem{situ2020dynamics}
G.~Situ, J.~Fleischer, Dynamics of the berezinskii-kosterlitz-thouless transition in a photon fluid, Nature Photonics 14 (2020) 517.

\bibitem{Rumpf09}
B.~Rumpf, A.~C. Newell, V.~E. Zakharov, Turbulent transfer of energy by radiating pulses, Phys. Rev. Lett. 103 (2009) 074502.

\bibitem{PhysRevA.97.043836}
K.~Krupa, A.~Tonello, V.~Couderc, A.~Barth\'el\'emy, G.~Millot, D.~Modotto, S.~Wabnitz, Spatiotemporal light-beam compression from nonlinear mode coupling, Phys. Rev. A 97 (2018) 043836.
\newblock \href {https://doi.org/10.1103/PhysRevA.97.043836} {\path{doi:10.1103/PhysRevA.97.043836}}.

\bibitem{labaz2024spatial}
M.~Labaz, P.~Sidorenko, Spatial--spectral complexity in kerr beam self-cleaning, Optics Letters 49~(11) (2024) 2902--2905.

\bibitem{genty2025real}
G.~Genty, J.~Li, J.~Ruelle, J.~M. Dudley, Real-time full-field spatial measurement of multimode fibers in the nonlinear regime, in: Real-time Measurements, Rogue Phenomena, and Single-Shot Applications X, SPIE, 2025, p. PC1334803.

\bibitem{PhysRevLett.115.223902}
L.~G. Wright, S.~Wabnitz, D.~N. Christodoulides, F.~W. Wise, Ultrabroadband dispersive radiation by spatiotemporal oscillation of multimode waves, Phys. Rev. Lett. 115 (2015) 223902.
\newblock \href {https://doi.org/10.1103/PhysRevLett.115.223902} {\path{doi:10.1103/PhysRevLett.115.223902}}.

\bibitem{Shtyrina2018}
O.~V. Shtyrina, M.~P. Fedoruk, Y.~S. Kivshar, S.~K. Turitsyn, Coexistence of collapse and stable spatiotemporal solitons in multimode fibers, Physical Review A 97 (1 2018).
\newblock \href {https://doi.org/10.1103/PhysRevA.97.013841} {\path{doi:10.1103/PhysRevA.97.013841}}.

\bibitem{PhysRevA.97.023803}
C.~Mas~Arab\'{\i}, A.~Kudlinski, A.~Mussot, M.~Conforti, \href{https://link.aps.org/doi/10.1103/PhysRevA.97.023803}{Geometric parametric instability in periodically modulated graded-index multimode fibers}, Phys. Rev. A 97 (2018) 023803.
\newblock \href {https://doi.org/10.1103/PhysRevA.97.023803} {\path{doi:10.1103/PhysRevA.97.023803}}.
\newline\urlprefix\url{https://link.aps.org/doi/10.1103/PhysRevA.97.023803}

\bibitem{Eftekhar:17}
M.~A. Eftekhar, L.~G. Wright, M.~S. Mills, M.~Kolesik, R.~A. Correa, F.~W. Wise, D.~N. Christodoulides, Versatile supercontinuum generation in parabolic multimode optical fibers, Opt. Express 25~(8) (2017) 9078--9087.
\newblock \href {https://doi.org/10.1364/OE.25.009078} {\path{doi:10.1364/OE.25.009078}}.

\bibitem{KrupaLuot:16}
K.~Krupa, C.~Louot, V.~Couderc, M.~Fabert, R.~Guenard, B.~M. Shalaby, A.~Tonello, D.~Pagnoux, P.~Leproux, A.~Bendahmane, R.~Dupiol, G.~Millot, S.~Wabnitz, \href{http://ol.osa.org/abstract.cfm?URI=ol-41-24-5785}{Spatiotemporal characterization of supercontinuum extending from the visible to the mid-infrared in a multimode graded-index optical fiber}, Opt. Lett. 41~(24) (2016) 5785--5788.
\newblock \href {https://doi.org/10.1364/OL.41.005785} {\path{doi:10.1364/OL.41.005785}}.
\newline\urlprefix\url{http://ol.osa.org/abstract.cfm?URI=ol-41-24-5785}

\bibitem{Galmiche:16}
G.~L. Galmiche, Z.~S. Eznaveh, M.~A. Eftekhar, J.~A. Lopez, L.~G. Wright, F.~Wise, D.~Christodoulides, R.~A. Correa, Visible supercontinuum generation in a graded index multimode fiber pumped at 1064 nm, Opt. Lett. 41~(11) (2016) 2553--2556.
\newblock \href {https://doi.org/10.1364/OL.41.002553} {\path{doi:10.1364/OL.41.002553}}.

\bibitem{xu2017origin}
G.~Xu, J.~Garnier, B.~Rumpf, A.~Fusaro, P.~Suret, S.~Randoux, A.~Kudlinski, G.~Millot, A.~Picozzi, \href{https://link.aps.org/doi/10.1103/PhysRevA.96.023817}{Origins of spectral broadening of incoherent waves: Catastrophic process of coherence degradation}, Phys. Rev. A 96 (2017) 023817.
\newblock \href {https://doi.org/10.1103/PhysRevA.96.023817} {\path{doi:10.1103/PhysRevA.96.023817}}.
\newline\urlprefix\url{https://link.aps.org/doi/10.1103/PhysRevA.96.023817}

\bibitem{konotop94}
K.~V.V., L.~Vazquez, Nonlinear Random Waves, Singapore, 1994.

\bibitem{gradstein80}
I.~Gradstein, I.~Ryzhik, Tables of Integrals, Sums, Series, and Products, Academic Press, New York, 1980.

\end{thebibliography}

%%%%%%%%%%%%%%%%%%%%%%%%%%%%%%%%%%%%%%%%%%%%%%%%%%%
\end{document}